\documentclass[10pt,twocolumn,romanappendices,final]{IEEEtran}
\pdfoutput=0

\usepackage{styles/rnd_fields_cs}

\acrodef{AAPM}{American Association of Physicists in Medicine}		%001
\acrodef{ACB}{asyn\-chro\-nous compressed beamformer}			%002
\acrodef{ACA}{adaptive cross approximation}				%003
\acrodef{ACR}{American College of Radiology}				%004
\acrodef{ADC}{analog-to-digital conversion}				%005
\acrodef{AIC}{analog-to-information conversion}				%006
\acrodef{BIBO}{bounded-input, bounded-output}				%007
\acrodef{BIM}{\name{Born} iterative method}				%008
\acrodef{BP}{basis pursuit}						%009
\acrodef{BPDN}{basis pursuit denoise}					%010

\acrodef{BVP}{boundary value problem}					%011
\acrodef{CAR}{Canadian Association of Radiologists}			%012
\acrodef{CDF}{cumulative distribution function}				%013
\acrodef{CGP}{conjugate gradient pursuit}				%014
\acrodef{CNR}{contrast-to-noise ratio}					%015
\acrodef{CoSaMP}{compressive sampling matching pursuit}			%016
\acrodef{CPU}{central processing unit}					%017
\acrodef{CPV}{\name{Cauchy} principal value}				%018
\acrodef{CS}{compressed sensing}					%019
\acrodef{CSI}{contrast source inversion}				%020

\acrodef{CT}{x-ray computed tomography}					%021
\acrodef{CTR}{cyst-to-tissue ratio}					%022
\acrodef{CTTE}[CT-TE]{continuous-time ternary encoding}			%023
\acrodef{DAC}{digital-to-analog conversion}				%024
\acrodef{DAIC}{diagnostically acceptable irreversible compression}	%025
\acrodef{DAS}{delay-and-sum}						%026
\acrodef{DBIM}{distorted \name{Born} iterative method}			%027
\acrodef{DCT}{discrete cosine transform}				%028
\acrodef{DICOM}{Digital Imaging and Communications in Medicine}		%029
\acrodef{DFT}{discrete \name{Fourier} transform}			%030

\acrodef{DNA}{deoxyribonucleic acid}					%031
\acrodef{DRG}{German \name{Röntgen} Society}				%032
\acrodef{DRIM}{distorted \name{Rytov} iterative method}			%033
\acrodef{DRS}{digital random sampler}					%034
\acrodef{ERM}{exploding reflector model}				%035
\acrodef{ESR}{European Society of Radiology}				%036
\acrodef{EU}{European Union}						%037
\acrodef{FBP}{filtered backpropagation}					%038
\acrodef{FDA}{US Food and Drug Administration}				%039 %U.S. Food and Drug Administration
\acrodef{FDBF}{frequency domain beamforming}				%040

\acrodef{FDT}{\name{Fourier} diffraction theorem}			%041
\acrodef{FEHM}{full extent at half maximum}				%042.s
\acrodefplural{FEHM}{full extents at half maximum}			%042.p
\acrodef{FFT}{fast \name{Fourier} transform}				%043
\acrodef{FIR}{finite impulse response}					%044
\acrodef{FLOP}{floating point operation}				%045
\acrodef{FMM}{fast multipole method}					%046
\acrodef{FOCUS}{fast object-oriented C++ ultrasound simulator}		%047
\acrodef{FOV}{field of view}						%048.s
\acrodefplural{FOV}{fields of view}					%048.p
\acrodef{FPDE}{fractional partial differential equation}		%049
\acrodef{FRI}{finite rate of innovation}				%050

\acrodef{FWHM}{full width at half maximum}				%051.s
\acrodefplural{FWHM}{full widths at half maximum}			%051.p
\acrodef{FWT}{fast wavelet transform}					%052
\acrodef{GPU}{graphics processing unit}					%053
\acrodef{GWN}{Gaussian white noise}					%054
\acrodef{HTP}{hard thresholding pursuit}				%055
\acrodef{IHT}{iterative hard thresholding}				%056
\acrodef{IID}[i.i.d.]{independent and identically distributed}		%057
\acrodef{IQ}{in-phase and quadrature}					%058
\acrodef{IRLS}{iteratively reweighted least squares}			%059
\acrodef{ISP}{inverse scattering problem}				%060

\acrodef{JPEG}{Joint Photographic Experts Group}			%061
\acrodef{JSM}{joint sparsity model}					%062
\acrodef{LASSO}{least absolute shrinkage and selection operator}	%063
\acrodef{LS}{\name{Lippmann}-\name{Schwinger}}				%064
\acrodef{LSB}{least significant bit}					%065
\acrodef{LTI}{linear time-shift invariant}				%066
\acrodef{MAP}{maximum \emph{a posteriori}}				%067
\acrodef{MCMC}{\name{Markov} chain Monte Carlo}				%068
\acrodef{MIRF}{material impulse response function}			%069
\acrodef{MLFMA}{multilevel fast multipole algorithm}			%070

\acrodef{MMV}{multiple measurement vectors}				%071
\acrodef{MOM}[MoM]{method of moments}					%072
\acrodef{MP}{matching pursuit}						%073
\acrodef{MP3}{MPEG Audio Layer III}					%074
\acrodef{MRI}{magnetic resonance imaging}				%075
\acrodef{MTF}{material transfer function}				%076
\acrodef{MV}{minimum variance}						%077
\acrodef{MVDR}{minimum-variance distortionless response}		%078
\acrodef{MWC}{modulated wideband converter}				%079
\acrodef{NESTA}{\name{Nesterov}'s algorithm}				%080

\acrodef{NSP}{null space property}					%081
\acrodef{NNZC}{number of nonzero components}				%082
\acrodef{OMP}{orthogonal matching pursuit}				%083
\acrodef{PAM}{pulse-amplitude modulation}				%084
\acrodef{PAT}{photoacoustic tomography}					%085
\acrodef{PC}{personal computer}						%086
\acrodef{PDE}{partial differential equation}				%087
\acrodef{PDF}{probability density function}				%088
\acrodef{PET}{positron emission tomography}				%089
\acrodef{PICMUS}{plane-wave imaging challenge in medical ultrasound}	%090

\acrodef{PSF}{point spread function}					%091
\acrodef{PSNR}{peak signal-to-noise ratio}				%092
\acrodef{PW}{plane wave}						%093
\acrodef{QCW}{quasi-cylindrical wave}					%094
\acrodef{QPW}{quasi-plane wave}						%095
\acrodef{QQ}{quantile-quantile}						%096
\acrodef{RADAR}[RADAR]{radio detection and ranging}			%097
\acrodef{RAM}{random-access memory}					%098
\acrodef{RCR}{Royal College of Radiologists}				%099
\acrodef{RF}{radio frequency}						%100

\acrodef{RIC}{restricted isometry constant}				%101
\acrodef{RIP}{restricted isometry property}				%102
\acrodef{RMSE}{root mean-squared error}					%103
\acrodef{ROI}{region of interest}					%104
\acrodef{ROMP}{regularized \ac{OMP}}					%105
\acrodef{RS}{\name{Rayleigh}-\name{Sommerfeld}}				%106
\acrodef{SaS}[S$\alpha$S]{symmetric $\alpha$-stable}			%107
\acrodef{SA}{synthetic aperture}					%108
\acrodef{SBL}{sparse Bayesian learning}					%109
\acrodef{SDK}{software development kit}					%110

\acrodef{SDR}{software-defined radio}					%111
\acrodef{SIIM}{Society for Imaging Informatics in Medicine}		%112
\acrodef{SISO}{single-input, single-output}				%113
\acrodef{SINR}{signal-to-interference-plus-noise ratio}			%114
\acrodef{SMI}{sample matrix inversion}					%115
\acrodef{SNR}{signal-to-noise ratio}					%116
\acrodef{SOMP}{simultaneous orthogonal matching pursuit}		%117
\acrodef{SoS}{sum of sincs}						%118
\acrodef{SP}{subspace pursuit}						%119
\acrodef{spark}{sparse rank}						%120

\acrodef{SPECT}{single photon emission computed tomography}		%121
\acrodef{SPG}{spectral projected gradient}				%122
\acrodef{SPGL1}[SPG$\ell_{1}$]{spectral projected gradient for $\ell_{1}$-minimization}	%123
\acrodef{SPLS}{split-projection least squares}				%124
\acrodef{SR}{sparse recovery}						%125
\acrodef{SRC}{\name{Sommerfeld} radiation condition}			%126
\acrodef{SSI}{supersonic shear imaging}					%127
\acrodef{SSIM}{structural similarity}					%128
\acrodef{STFT}{short-time \name{Fourier} transform}			%129
\acrodef{StOMP}{stagewise \ac{OMP}}					%130

\acrodef{StWGP}{stagewise weak \ac{CGP}}				%131
\acrodef{SVD}{singular value decomposition}				%132
\acrodef{TDMA}{time-division multiple access}				%133
\acrodef{TMSBL}[T-MSBL]{temporally correlated \ac{MMV}-based \ac{SBL}}  %134
\acrodef{TOF}{time-of-flight}						%135.s
\acrodefplural{TOF}{times-of-flight}					%135.p
\acrodef{TGC}{time gain compensation}					%136
\acrodef{THI}{tissue harmonic imaging}					%137
\acrodef{TPSF}{transform point spread function}				%138
\acrodef{TR}{\name{Tikhonov} regularization}				%139
\acrodef{TV}{total variation}						%140

\acrodef{UCA}{ultrasound contrast agents}				%141
\acrodef{UDT}{ultrasound diffraction tomography}			%142
\acrodef{UI}{ultrasound imaging}					%143
\acrodef{UoS}{union of subspaces}					%144
\acrodef{USA}{United States of America}					%145
\acrodef{UWB}{ultra-wideband}						%146
\acrodef{Xampling}{CS-Sampling}						%147

\acresetall

\title{Random Incident Waves for Fast Compressed Pulse-Echo Ultrasound Imaging}
\author{%
Martin~F.~Schiffner%
\thanks{M.~F.~Schiffner is with the Lehrstuhl für Medizintechnik, Ruhr-Universität Bochum, D-44801 Bochum, Germany (e-mail: martin.schiffner@rub.de).}%
}

\begin{document}

%%%%%%%%%%%%%%%%%%%%%%%%%%%%%%%%%%%%%%%%%%%%%%%%%%%%%%%%%%%%%%%%%%%%%%%%%%%%%%%%%%%%%%%%%%%%%%%%%%%%%%%%%%%%%%%%
% title
%%%%%%%%%%%%%%%%%%%%%%%%%%%%%%%%%%%%%%%%%%%%%%%%%%%%%%%%%%%%%%%%%%%%%%%%%%%%%%%%%%%%%%%%%%%%%%%%%%%%%%%%%%%%%%%%
\maketitle

%%%%%%%%%%%%%%%%%%%%%%%%%%%%%%%%%%%%%%%%%%%%%%%%%%%%%%%%%%%%%%%%%%%%%%%%%%%%%%%%%%%%%%%%%%%%%%%%%%%%%%%%%%%%%%%%
% abstract: less than 200 words and index terms
%%%%%%%%%%%%%%%%%%%%%%%%%%%%%%%%%%%%%%%%%%%%%%%%%%%%%%%%%%%%%%%%%%%%%%%%%%%%%%%%%%%%%%%%%%%%%%%%%%%%%%%%%%%%%%%%
\begin{abstract}
  \boldmath
  %---------------------------------------------------------------------------------------------------------------
% 1.) state of the art in fast pulse-echo ultrasound imaging
%---------------------------------------------------------------------------------------------------------------
% a) established image recovery methods in fast UI trade the image quality for the high frame rate
Established image recovery methods in
fast \acl{UI}, e.g.
% 1.) delay-and-sum
\acl{DAS}, trade
% 2.) image quality
the image quality for
% 3.) high frame rate
the high frame rate.
\TODO{model inaccuracies}
% b) cutting-edge inverse scattering methods based on CS disrupt this tradeoff via a priori information
Cutting-edge inverse scattering methods based on
% 1.) compressed sensing
\ac{CS} disrupt
% 2.) tradeoff
this tradeoff via
% 3.) a priori information
\emph{a priori} information.
% c) cutting-edge inverse scattering methods based on CS iteratively recover a high-quality image from only a few sequential pulse-echo measurements or less echo signals
They iteratively recover
% 1.) high-quality image
a high-quality image from
% 2.) only a few sequential pulse-echo measurements
only a few sequential pulse-echo measurements or
% 3.) less echo signals
less echo signals, if
% 4.) known dictionary of structural building blocks represents the image almost sparsely
(i) a known dictionary of
structural building blocks represents
the image almost sparsely, and
% 5.) individual pulse echoes are sufficiently uncorrelated
(ii) their individual pulse echoes, which are predicted by
a linear model, are
sufficiently uncorrelated.
% d) exclusive modeling of the incident waves as steered PWs or cylindrical waves has so far limited the convergence speed and the image quality by violating condition (ii)
The exclusive modeling of
% 1.) incident waves
the incident waves as
% 2.) steered PWs
steered \aclp{PW} or
% 3.) outgoing 1-spherical waves (cylindrical waves)
cylindrical waves, however, has so far limited
% 3.) convergence speed
the convergence speed,
% 4.) image quality
the image quality, and
%by violating
% 5.) potential
the potential to meet
% 6.) condition (ii) [ individual pulse echoes are sufficiently uncorrelated ]
condition (ii).

%---------------------------------------------------------------------------------------------------------------
% 2.) contributions of the paper
%---------------------------------------------------------------------------------------------------------------
% a) novel method for the fast compressed acquisition and the subsequent recovery of images is proposed to overcome these limitations
Motivated by
% 1.) benefits
the benefits of
% 2.) randomness
randomness in
% 3.) compressed sensing (CS)
\ac{CS},
% 4.) novel method
a novel method for
% 5.) fast compressed acquisition
the fast compressed acquisition and
% 6.) subsequent recovery
the subsequent recovery of
images is proposed to overcome
% 6.) limitations [convergence speed and image quality]
these limitations.
% b) novel method recovers the spatial compressibility fluctuations in weakly-scattering soft tissue structures by a sparsity-promoting lq-minimization method
It recovers
% 1.) spatial compressibility fluctuations
the spatial compressibility fluctuations in
% 2.) weakly-scattering soft tissue structures
weakly-scattering soft tissue structures, where
% 3.) orthonormal basis represents the image almost sparsely
an orthonormal basis meets
% 4.) condition (i) [known dictionary of structural building blocks represents the image almost sparsely]
condition (i), by
% 5.) sparsity-promoting lq-minimization method
a sparsity-promoting $\ell_{q}$-minimization method,
% 5.) parameter q \in [ 0; 1 ]
$q \in [ 0; 1 ]$.
% c) realistic d-dimensional model predicts the pulse echoes of the individual basis functions
A realistic $d$-dimensional model, $d \in \{ 2, 3 \}$, accounting for
% 1.) diffraction
diffraction,
% 2.) single monopole scattering within the first Born approximation
single monopole scattering,
% 3.) combination of power-law absorption and dispersion
the combination of
power-law absorption and
dispersion, and
% 4.) specifications of a planar transducer array
the specifications of
a planar transducer array, predicts
% 5.) pulse echoes
the pulse echoes of
% 6.) individual basis functions
the individual basis functions.
% d) three innovative types of incident waves aid in meeting condition (ii)
Three innovative types of
% 1.) incident waves
incident waves, whose
% 2.) syntheses
syntheses leverage
% 3.) random apodization weights
random apodization weights,
% 4.) random time delays
time delays, or
% 5.) combinations thereof
combinations thereof, aid in meeting
% 9.) condition (ii) [ individual pulse echoes are sufficiently uncorrelated ]
condition (ii).

%---------------------------------------------------------------------------------------------------------------
% 3.) results of the numerical simulations
%---------------------------------------------------------------------------------------------------------------
% a) single realizations of the random waves outperform the prevalent QPW for both the canonical and the Fourier bases
In
% 1.) two-dimensional numerical simulations
two-dimensional numerical simulations,
% 2.) single realizations
single realizations of
% 3.) random waves
these waves outperform
% 4.) quasi-plane wave (QPW)
the prevalent \acl{QPW} for both
% 5.) canonical basis
the canonical and
% 6.) Fourier basis
the \name{Fourier} bases.
% b) single realizations of the random waves significantly decorrelate the pulse echoes
% article:Schiffner2018, Sect. VIII. Results / Sect. VIII-B. Tissue-Mimicking Phantom / Sect. VIII-B.2) Transform Point Spread Functions (subsubsec:results_phantom_tissue_tpsf)
% - In fact, the MAXIMUM NORMALIZED DIFFERENCES ranged from
%   \SI{41.2}{\percent} at the fixed spatial frequencies $s \in \{ 1, 9 \}$ to
%   \SI{62.5}{\percent} at the fixed spatial frequencies $s \in \{ 4, 6 \}$.
% article:Schiffner2018, Sect. VIII. Results / Sect. VIII-A. Wire Phantom / Sect. VIII-A.2) Point Spread Functions (subsubsec:results_phantom_wire_psf)
% - The MAXIMUM NORMALIZED DIFFERENCES ranged from
%   \SI{23.5}{\percent} for the superposition of both randomly-apodized and randomly-delayed \acp{QCW} at the ninth fixed position, i.e. $s = 9$, to
%   \SI{73.7}{\percent} for the superposition of randomly-delayed \acp{QCW} at the first fixed position, i.e. $s = 1$.
They significantly decorrelate
% 1.) pulse echoes
the pulse echoes, e.g.
% 2.) single realizations
they reduce
% 3.) full extents at half maximum (FEHMs)
the \aclp{FEHM} of
% 4.) point spread functions (PSFs)
the \aclp{PSF} by
% 5.) up to 73.7 %
up to \SI{73.7}{\percent}.
% c) single realizations of the random waves improve the convergence speed and the image quality in terms of the mean SSIM indices and the relative RMSEs
For
% 1.) tissue-mimicking phantom
a tissue-mimicking phantom and
% 2.) parameter q \in \{ 1, 0.5 \}
$q \in \{ 1, 0.5 \}$,
% 3.) single realizations of the random waves
they improve
the convergence speed and
% 5.) image quality
% article:Schiffner2018, Sect. VIII. Results / Sect. VIII-B. Tissue-Mimicking Phantom / Sect. VIII-B.4) Recovery by lq-Minimization (subsubsec:results_phantom_tissue_lq_minimization)
% - Using the NONCONVEX $\ell_{0.5}$-minimization method \eqreflqmin{eqn:recovery_reg_norm_lq_minimization}{ 0.5 },
%   the RANDOM WAVES consistently achieved
%   MEAN \ac{SSIM} INDICES CLOSE TO UNITY and
%   RELATIVE \acp{RMSE} BELOW \SI{6.9}{\percent} for
%   all reference \acp{SNR}.
the image quality in terms of
% 6.) mean SSIM indices
% MATLAB:
% SSIM_index_spgl1_l1_qpw(1,:)*1e2	   = 36.6000   41.2488   46.4364   90.4708   99.9325   99.9368
% SSIM_index_spgl1_lq_qpw(1,:)*1e2	   = 43.6973   49.3404   56.1582   93.0138   99.8934   99.8323
%
% SSIM_index_spgl1_l1_rnd_apo(1,:)*1e2	   = 80.6549   81.5283   83.8071   86.1310   86.2028   86.1152
% SSIM_index_spgl1_lq_rnd_apo(1,:)*1e2	   = 99.1838   99.7243   99.8458   99.9954   99.9958   99.9934
%
% SSIM_index_spgl1_l1_rnd_del(1,:)*1e2	   = 80.2673   82.3483   84.2717   86.3884   90.6166   78.3656
% SSIM_index_spgl1_lq_rnd_del(1,:)*1e2	   = 98.7991   99.7009   99.8962   99.9935   99.9957   99.9928
%
% SSIM_index_spgl1_l1_rnd_apo_del(1,:)*1e2 = 80.2278   81.4986   85.3900   85.2345   86.2158   80.0985
% SSIM_index_spgl1_lq_rnd_apo_del(1,:)*1e2 = 98.6458   99.6973   99.8945   99.9930   99.9962   99.9940
%
% differences to QPW (q = 1):
% rnd. apo.:		44.0549   40.2794   37.3707   -4.3397  -13.7297  -13.8217
% rnd. del.:		43.6673   41.0994   37.8353   -4.0823   -9.3159  -21.5713
% rnd. apo. del.:	43.6279   40.2498   38.9537   -5.2362  -13.7167  -19.8383
% minimum: -13.7297 (rnd. apo. @ 30 dB)
% maximum:  44.0549 (rnd. apo. @ 3 dB)
%
% differences to QPW (q = 0.5):
% rnd. apo.:		55.4865   50.3839   43.6877    6.9816    0.1024    0.1611
% rnd. del.:		55.1018   50.3605   43.7381    6.9797    0.1023    0.1605
% rnd. apo. del.:	54.9485   50.3569   43.7363    6.9792    0.1029    0.1618
% minimum:  0.1023 (rnd. del. @ 30 dB)
% maximum: 55.4865 (rnd. apo. @ 3 dB)
the mean \acl{SSIM} indices and
% 7.) relative RMSEs
% MATLAB:
% rel_RMSE_spgl1_l1_qpw(1,:)*1e2	 = 74.9216   69.2613   63.4289   20.9912    1.3762    2.0671
% rel_RMSE_spgl1_lq_qpw(1,:)*1e2         = 65.8391   60.6768   54.3489   16.3019    1.4052    2.8359
%
% rel_RMSE_spgl1_l1_rnd_apo(1,:)*1e2	 = 32.8928   31.8203   29.1923   26.2924   26.1477   26.3357
% rel_RMSE_spgl1_lq_rnd_apo(1,:)*1e2	 =  5.3555    3.0675    2.0639    0.4661    0.3106    0.6444
%
% rel_RMSE_spgl1_l1_rnd_del(1,:)*1e2	 = 33.2994   31.0622   28.7238   26.0569   20.5849   36.2633
% rel_RMSE_spgl1_lq_rnd_del(1,:)*1e2	 =  6.5540    3.2644    1.7489    0.5210    0.3171    0.6548
%
% rel_RMSE_spgl1_l1_rnd_apo_del(1,:)*1e2 = 33.2734   31.8067   26.8410   27.4907   26.2436   32.9772
% rel_RMSE_spgl1_lq_rnd_apo_del(1,:)*1e2 =  6.8596    3.2637    1.7899    0.5512    0.2990    0.6206
%
% differences to QPW (q = 1):
% rnd. apo.:		42.0289   37.4409   34.2366   -5.3012  -24.7715  -24.2686
% rnd. del.:		41.6222   38.1990   34.7052   -5.0657  -19.2087  -34.1962
% rnd. apo. del.:	41.6482   37.4546   36.5880   -6.4995  -24.8674  -30.9101
% minimum: -24.8674 (rnd. apo. del. @ 30 dB)
% maximum:  42.0289 (rnd. apo. @ 3 dB)
%
% differences to QPW (q = 0.5):
% rnd. apo.:		60.4836   57.6093   52.2851   15.8357    1.0946    2.1916
% rnd. del.:		59.2850   57.4124   52.6000   15.7809    1.0882    2.1811
% rnd. apo. del.:	58.9795   57.4131   52.5590   15.7507    1.1062    2.2153
% minimum:  1.0882 (rnd. del. @ 30 dB)
% maximum: 60.4836 (rnd. apo. @ 3 dB)
the relative \aclp{RMSE} by
% 8.) up to 2.7 and 22.9 %
up to
$\{ \SI{2.7}{\percent}, \SI{22.9}{\percent} \}$,
% 9.) up to 44.1 and 55.5 %
$\{ \SI{44.1}{\percent}, \SI{55.5}{\percent} \}$, and
% 10.) up to 42 and 60.5 %
$\{ \SI{42}{\percent}, \SI{60.5}{\percent} \}$,
respectively.
%---------------------------------------------------------------------------------------------------------------
% 4.) results of the experiments
%---------------------------------------------------------------------------------------------------------------
% a) experiment with a wire phantom validates the feasibility of the proposed method
%An experiment with
%a wire phantom validates
%the feasibility of
%the proposed method.

\end{abstract}
\acresetall

%%%%%%%%%%%%%%%%%%%%%%%%%%%%%%%%%%%%%%%%%%%%%%%%%%%%%%%%%%%%%%%%%%%%%%%%%%%%%%%%%%%%%%%%%%%%%%%%%%%%%%%%%%%%%%%%
% keywords
%%%%%%%%%%%%%%%%%%%%%%%%%%%%%%%%%%%%%%%%%%%%%%%%%%%%%%%%%%%%%%%%%%%%%%%%%%%%%%%%%%%%%%%%%%%%%%%%%%%%%%%%%%%%%%%%

%%%%%%%%%%%%%%%%%%%%%%%%%%%%%%%%%%%%%%%%%%%%%%%%%%%%%%%%%%%%%%%%%%%%%%%%%%%%%%%%%%%%%%%%%%%%%%%%%%%%%%%%%%%%%%%%
% 1.) introduction
%%%%%%%%%%%%%%%%%%%%%%%%%%%%%%%%%%%%%%%%%%%%%%%%%%%%%%%%%%%%%%%%%%%%%%%%%%%%%%%%%%%%%%%%%%%%%%%%%%%%%%%%%%%%%%%%
\section{Introduction}
\label{sec:introduction}
\IEEEPARstart{M}{edical} \ac{UI} is
% 1.) safe
% book:Szabo2013, Chapter 1: Introduction / Sect. 1.8: Ultrasound and Other Diagnostic Imaging Modalities / Sect. 1.8.2: Ultrasound
% - Ultrasound is regarded as [1.)] SAFE and does not have any cumulative biological side effects (more on this topic can be found in Chapter 15). (p. 27)
a safe,
% 2.) cost-effective
cost-effective,
% 3.) portable
% book:Szabo2013, Chapter 1: Introduction / Sect. 1.8: Ultrasound and Other Diagnostic Imaging Modalities / Sect. 1.8.2: Ultrasound
% - With the widespread availability of
%   [3.)] MINIATURE PORTABLE and POCKET ultrasound systems for SCREENING and IMAGING,
%   these two factors will continue to improve. (p. 27)
portable, and
% 4.) frequently-used
frequently-used modality that provides
% 5.) sub-millimeter spatial resolutions
sub-millimeter spatial resolutions in
% 6.) real-time
% book:Szabo2013, Chapter 1: Introduction / Sect. 1.11: Conclusion
% - KEY STRENGTHS of ULTRASOUND are its abilities to reveal
%   ANATOMY, the DYNAMIC MOVEMENT OF ORGANS, and DETAILS OF BLOOD FLOW IN REAL TIME. (p. 34)
% book:Szabo2013, Chapter 1: Introduction / Sect. 1.8: Ultrasound and Other Diagnostic Imaging Modalities / Sect. 1.8.2: Ultrasound
% - The DYNAMIC MOTION OF ORGANS such as the heart can be revealed by ultrasound operating up to HUNDREDS of FRAMES PER SECOND. (p. 26)
% article:JensenProgBMB2007: Medical ultrasound imaging
% 3. Anatomic ultrasound imaging
% - The imaging is performed in REAL TIME with 20–100 images/s. (p. 155)
% - The available frame rate is closely linked to the image depth. (p. 155)
% - Dynamic imaging is, thus, possible, and the anatomy and its dynamics can be easily visualized in REAL TIME. (p. 155)
% coll:Jensen2002: Ultrasound Imaging and Its Modeling
% ABSTRACT
% - The imaging is performed in [6.)] REAL TIME with 20 to 100 images per second. (p. 135)
real-time
\cite[Fig. 1.15, Table 1.2]{book:Szabo2013},
\cite{article:BierigJDMS2009}.
The typical progressive scanning of
% 1.) specified FOV
a specified \ac{FOV} by
% 2.) focused beams
focused beams, however, requires
% 3.) hundreds of sequential pulse-echo measurements per image
% book:Bushberg2011, Chapter 14: Ultrasound / Sect. 14.6: Two-Dimensional Image Display and Storage / Electronic Scanning and Real-Time Display
% - The ULTRASOUND BEAM SWEEPS ACROSS THE VOLUME OF INTEREST IN A SEQUENTIAL FASHION, with
%   the NUMBER OF A-LINES APPROXIMATELY EQUAL TO THE NUMBER OF TRANSDUCER ELEMENTS. (p. 537)
% article:JensenProgBMB2007: Medical ultrasound imaging
% 3. Anatomic ultrasound imaging
% - N_{l} [number of lines in the image] is typically 200 lines and D = 10cm results in f_{r} = 38.5 Hz. (p. 155)
hundreds of
sequential pulse-echo measurements per
image and, owing to
% 4.) finite sound speed
the finite sound speed, limits
% 5.) frame rate
% article:JensenUlt2006: Synthetic aperture ultrasound imaging
% Abstract
% - This [linewise acquisition] puts a strict limit on
%   [1.)] the FRAME RATE and
%   [2.)] the possibility of acquiring a sufficient amount of data for HIGH PRECISION FLOW ESTIMATION. (p. e5)
the frame rate
\cite[536--539]{book:Bushberg2011},
\cite{article:JensenProgBMB2007,article:WellsPMB2006}.
Advances in
% 1.) electronic miniaturization
electronic miniaturization and
% 2.) processing power
processing power have recently led to
% 3.) freely programmable UI systems
freely programmable \ac{UI} systems and
% 4.) software-based ultrafast imaging modes
software-based \term{ultrafast} imaging modes, e.g.
% 5.) coherent plane-wave compounding
% article:ChernyakovaITUFFC2018: Fourier-Domain Beamforming and Structure-Based Reconstruction for Plane-Wave Imaging
% Abstract
% - Ultrafast imaging based on COHERENT PLANE-WAVE COMPOUNDING IS
%   ONE OF THE MOST IMPORTANT RECENT DEVELOPMENTS in medical ultrasound. (p. 1)
% - It significantly improves image quality and allows for MUCH FASTER IMAGE ACQUISITION. (p. 1)
% article:MontaldoITUFFC2009: Coherent plane-wave compounding for very high frame rate ultrasonography and transient elastography
coherent plane-wave compounding
\cite{article:MontaldoITUFFC2009},
% 6.) synthetic aperture imaging
% article:MoghimiradITUFFC2016: Synthetic Aperture Ultrasound Fourier Beamformation Using Virtual Sources
% article:JensenUlt2006: Synthetic aperture ultrasound imaging
% Abstract
% - SA imaging is a radical break with today’s commercial systems [...]. (p. e5)
% - These constrictions [limited frame rate, insufficient data for high precision flow estimation] can be lifted by employing SA imaging. (p. e5)
% - It is also possible to improve penetration depth by employing codes during ultrasound transmission. (p. e5)
% 3. Introduction to synthetic aperture imaging
% - SA imaging makes it possible to decouple frame rate and pulse repetition time, as only
%   a sparse set of emissions can be used for creating a full image. (p. e7)
% - Very fast imaging can, therefore, be made albeit with a lower resolution and higher side-lobes. (p. e7)
\ac{SA} imaging
\cite{article:MoghimiradITUFFC2016,article:JensenUlt2006}, or
% 7.) limited-diffraction beam imaging
% article:ChengITUFFC2006: Extended high-frame rate imaging method with limited-diffraction beams
% Abstract
% - Previously, a HIGH-FRAME RATE (HFR) IMAGING THEORY was developed in which
%   a pulsed plane wave was used in transmission, and
%   LIMITED-DIFFRACTION ARRAY BEAM WEIGHTINGS WERE APPLIED TO RECEIVED ECHO SIGNALS to produce
%   a spatial Fourier transform of object function for 3-D image reconstruction. (p. 880)
% - In this paper, THE THEORY IS EXTENDED TO INCLUDE EXPLICITLY VARIOUS TRANSMISSION SCHEMES [...]. (p. 880)
% article:LuITUFFC1997: 2D and 3D High Frame Rate Imaging with Limited Diffraction Beams
limited-diffraction beam imaging
\cite{article:ChengITUFFC2006,article:LuITUFFC1997}, that overcome
% 8.) limited frame rate
this limitation and capture
% 9.) large FOVs
% article:BerthonPMB2018: Spatiotemporal matrix image formation for programmable ultrasound scanners
% 1. Introduction
% - This enables frame rates approximately 100 times faster than standard ultrasound sequences, even in three dimensions (Provost et al 2014). (p. 1)
% - Images can then be formed in real-time, i.e. at frame rates higher than 30 images per second, or in post-processing using software-based approaches
%   (Lu 1997, Montaldo et al 2009, Garcia et al 2013). (p. 1)
% article:TanterITUFFC2014: Ultrafast imaging in biomedical ultrasound
% III. The concept of Plane-Wave Imaging and Plane-Wave compounding
% - It is important to keep in mind that
%   the ULTRAFAST TERMINOLOGY describes any acquisition sequence enabling
%   2-D OR 3-D IMAGING OVER A LARGE FIELD OF VIEW AT VERY HIGH FRAME RATES, typically in the KILOHERTZ RANGE
%   (i.e., the COMBINATION OF WIDE-FIELD TRANSMISSIONS AND PARALLEL RECEIVE BEAMFORMING). (p. 106)
large \acp{FOV} at
% 10.) rates in the kilohertz range
% article:GarciaITUFFC2013: Stolt's f-k Migration for Plane Wave Ultrasound Imaging
% I. Introduction
% - Theoretically, frame rates up to 15000 Hz can thus be attained for a 5-cm imaging depth [2]. (p. 1853)
rates in
the kilohertz range
\cite{article:TanterITUFFC2014},
\cite[Sect. 1.11]{book:Szabo2013}.
Using
% 1.) fully-sampled transducer array
a fully-sampled transducer array,
% 2.) sequential insonification of the entire FOV by only a few unfocused waves
they sequentially insonify
% 3.) entire FOV
the entire \ac{FOV} by
% 4.) a few unfocused waves
only a few unfocused waves and process
% 5.) RF voltage signals
the echo signals generated by
%the \ac{RF} voltage signals generated by
% 6.) all elements
all elements.
Besides rendering \ac{UI}
% 1.) safety
% safety issues: as flat transmitted beams impose much less constraints than classical focused beams
% article:BercoffITUFFC2004: Supersonic Shear Imaging: A New Technique for Soft Tissue Elasticity Mapping
% less \emph{spatial peak-temporal average acoustic intensity} (cf. \cite[Sect. 14.11]{book:Bushberg2011}) than focused sound beams.
%[15] Food and Drug Administration, “Information for Manufacturers Seeking Marketing Clearance of Diagnostic Ultrasound Systems and Transducer,” U. S. Dept. Health and Human Services, Food and Drug Administration, Center for Devices and Radiological Health., 1997.
%\cite{guidelines:FDAUltrasound2008}. % standard:IEC60601-2-37-2007,
safer
\cite{article:BercoffITUFFC2004} and
% 2.) sensitivity
% article:MaceITUFFC2013: Functional Ultrasound Imaging of the Brain: Theory and Basic Principles
% - Here, we present a µDoppler ultrasound method able to detect and map the cerebral blood volume (CBV) over the entire brain with
%   an IMPORTANT INCREASE IN SENSITIVITY. (p. 492)
% article:BercoffITUFFC2011: Ultrafast Compound Doppler Imaging: Providing Full Blood Flow Characterization
more sensitive
\cite{article:BercoffITUFFC2011},
% 4.) ultrafast imaging modes
they have enabled
% 5.) observation
the observation of
% 6.) time-variant objects
moving objects and
% 7.) transient phenomena
% 1.) cardiac electrophysiology
%     article:PapadacciITUFFC2014: High-Contrast Ultrafast Imaging of the Heart
% 2.) functional brain imaging
%     article:MaceITUFFC2013: Functional Ultrasound Imaging of the Brain: Theory and Basic Principles
%     article:MaceNatMeth2011: Functional ultrasound imaging of the brain
%     article:BercoffITUFFC2011: Ultrafast Compound Doppler Imaging: Providing Full Blood Flow Characterization
% 3.) soft tissue elasticity mapping
%     article:BercoffITUFFC2004: Supersonic Shear Imaging: A New Technique for Soft Tissue Elasticity Mapping
%     article:BercoffApplPhysLett2004: Sonic boom in soft materials: The elastic Cerenkov effect
transient phenomena
\cite{article:PapadacciITUFFC2014,article:MaceITUFFC2013,article:BercoffITUFFC2004}, even in
% 8.) three-dimensional ultrafast UI
% article:ProvostPMB2014: 3D ultrafast ultrasound imaging in vivo
% article:JensenProgBMB2007: Medical ultrasound imaging
% 3. Anatomic ultrasound imaging / 3.2. Three-dimensional imaging
% - Often the 3D volume consists of 64x64 image lines and the time for one image to a depth of 15 cm is then
%   T_{i} = N_{l} 2 D / c = 0.53 s and
%   the frame rate is then less than 2 Hz. (p. 159)
% - This is insufficient for CARDIAC IMAGING, and parallel beam formation is then employed. (p. 159)
% article:ChengITUFFC2006: Extended high-frame rate imaging method with limited-diffraction beams
% I. Introduction
% - High-frame rate imaging is important for IMAGING OF FAST MOVING OBJECTS SUCH AS THE HEART, especially,
%   in 3-D imaging in which many 2-D image frames are needed to form a 3-D volume that
%   may reduce image frame rate dramatically with conventional imaging methods. (p. 880)
three dimensions
\cite{article:ProvostPMB2014}.

The established image recovery methods, which are
% 1.) explicit (closed-form expressions)
% article:BerthonPMB2018: Spatiotemporal matrix image formation for programmable ultrasound scanners
% 1. Introduction
% - Both these STANDARD APPROACHES [DAS, Fourier-based] are based on
%   the IMPLEMENTATION OF CLOSED-FORM SOLUTIONS TO THE INVERSE PROBLEM, i.e.
%   finding the object to image from a collection of measurements. (p. 1)
% - While such an approach is IDEAL TO GENERATE FAST ALGORITHMS [...]. (p. 1)
explicit and
% 2.) computationally efficient
computationally efficient, gradually trade
% 3.) image quality
the image quality for
% 4.) frame rate
the frame rate
\cite{article:GarciaITUFFC2013,article:MontaldoITUFFC2009,article:JensenUlt2006,article:ChengITUFFC2006}.
% b) physical models neglect various effects and basic abilities of programmable UI systems
% article:IlovitshNatComBio2018: Acoustical structured illumination for super-resolution ultrasound imaging
% - Despite these advances [user-programmable systems, schemes for image reconstruction],
%   ULTRASOUND STILL SUFFERS FROM LIMITATIONS in
%   RESOLUTION, CONTRAST and SIGNAL TO NOISE RATIO (SNR), and from ARTIFACTS [4].
%   [4] 4. Kremkau, F. W. & Taylor, K. J. Artifacts in ultrasound imaging. J Ultrasound Med. 5, 227–237 (1986).
Their physical models neglect
% 1.) various effects
various effects, e.g.
% 2.) finite number of array elements
% article:GarciaITUFFC2013: Stolt's f-k Migration for Plane Wave Ultrasound Imaging
% IV. Discussion / C. PWI Using f-k Migration: Limitations and Perspectives
% - Ultrasound PWI needs the WAVEFRONTS TO BE PLANAR AND TILTED with the desired incident angle. (p. 1863)
% - To get high-quality images by PWI, one must ensure that
%   a PLANAR WAVEFIELD IS SYNTHESIZED PROPERLY. (p. 1863)
% - An unlimited number of coplanar elementary sources can produce a perfect slant plane wave. (p. 1863)
% - In practice though,
%   the AMOUNT OF ELEMENTS IN A LINEAR-ARRAY TRANSDUCER IS LIMITED to 64, 128, or 192. (p. 1863)
% - This technical limitation may cause ADVERSE EFFECTS that may negatively affect the resulting images. (p. 1863)
% - More importantly, a more disturbing effect may rise from
%   the GRATING LOBES WHICH ARE INDUCED BY THE REGULAR SPACING of the individual transducer elements. (p. 1863)
% - It is known that the grating lobes are of larger magnitude as the steering angle increases [47]. (p. 1863)
the finite number of
array elements and
% 3.) anisotropic directivities
their anisotropic directivities, and
% 4.) basic abilities
basic abilities of
% 5.) freely programmable UI systems
programmable \ac{UI} systems, e.g.
% 6.) syntheses of complex incident waves
the syntheses of
complex incident waves.
The popular \ac{DAS} method, for example, focuses
% 1.) recorded RF voltage signals
the echo signals on
% 2.) specified points
specified points in
% 3.) field-of-view
the \ac{FOV} to quantify
% 4.) echogeneity
their echogeneity.
Emitting
steered \acp{PW}
\cite{article:ProvostPMB2014,article:MaceITUFFC2013,article:BercoffITUFFC2011,article:MontaldoITUFFC2009,article:BercoffITUFFC2004}, whose
% 2.) spatial extent and energy content are unlimited
% article:ChengITUFFC2006: Extended high-frame rate imaging method with limited-diffraction beams
% I. Introduction
% - The advantage of LIMITED DIFFRACTION BEAMS is that,
%   even if they are produced with FINITE APERTURE AND ENERGY, they have
%   A VERY LARGE DEPTH OF FIELD. (p. 880)
% II. Theory / A. Extension of High-Frame Rate Imaging Theory
% - As shown in Fig. 1, a 2-D array transducer located at z = 0 plane is excited to generate
%   a broadband, limited-diffraction array beam or a steered pulsed plane wave. (p. 882)
% - The same transducer also is used to receive echoes scattered from objects. (p. 882)
% - The APERTURE OF THE TRANSDUCER IS ASSUMED TO BE INFINITELY LARGE, and
%   the SIZE OF EACH TRANSDUCER ELEMENT IS INFINITELY SMALL. (p. 882)
spatial extent and
energy content are
unlimited, or
% 3.) outgoing (d-1)-spherical waves
% single array element:
%   article:JensenUlt2006: Synthetic aperture ultrasound imaging
%   4. Penetration problem
%   - A major problem in SA imaging is the limited penetration depth, since
%     AN UN-FOCUSED WAVE IS USED IN TRANSMIT and
%     ONLY A SINGLE ELEMENT EMITS ENERGY. (p. e8)
% virtual point sources:
%   article:PapadacciITUFFC2014: High-Contrast Ultrafast Imaging of the Heart [Feb. 2014]
%   Abstract
%   - In this paper, we propose ultrafast imaging of the heart with adapted sector size by coherently compounding
%     diverging waves emitted from a standard transthoracic cardiac phased-array probe. (p. 288)
%   - As in ultrafast imaging with plane wave coherent compounding,
%     diverging waves can be summed coherently to obtain high-quality images of the entire heart at high frame rate in a full field of view. (p. 288)
%   article:JensenUlt2006: Synthetic aperture ultrasound imaging
%   4. Penetration problem
%   - The problem [limited penetration depth] can be solved by
%     COMBINING SEVERAL ELEMENTS FOR TRANSMISSION and
%     USING LONGER WAVEFORMS EMITTING MORE ENERGY. (p. e8)
%   - Karman et al. [10] suggested COMBINING SEVERAL ELEMENTS N_{t} IN TRANSMIT, with
%     a DELAY CURVE TO DE-FOCUS THE EMISSION TO EMULATE A SPHERICAL WAVE.  (p. e8)
%   - This can increase the emitted amplitude be a factor of \sqrt{ N_{t} }. (p. e8)
outgoing $\{ 1, 2 \}$-spherical waves
\cite{article:ProvostPMB2014,article:PapadacciITUFFC2014,article:JensenUlt2006}, whose
% 4.) isotropic sources are points
% article:MoghimiradITUFFC2016: Synthetic Aperture Ultrasound Fourier Beamformation Using Virtual Sources
% II. MULTISTATIC WAVENUMBER ALGORITHM
% - This essentially assumes that
%   the ARRAY CONSISTS OF POINT SOURCES AND RECEIVERS. (p. 2020)
% - It has been proved that the fields can be assumed spherical for a distance from the source [44],
%   so the model is acceptable for medical ultrasound field. (p. 2020)
isotropic sources are
points,
% 5.) popular DAS method
it adds
% 6.) signal samples
the signal samples at
% 7.) round-trip times-of-flight
the round-trip \acp{TOF}
\cite[(2), (6)]{article:MontaldoITUFFC2009},
\cite[(4), (5)]{article:JensenUlt2006}.
The competing \name{Fourier} methods, in contrast, invert
% 1.) wave equation
the wave equation and potentially improve
the image quality at
reduced computational costs
\cite{article:ZhangITUFFC2016,article:MoghimiradITUFFC2016,article:GarciaITUFFC2013,article:KruizingaITUFFC2012,proc:SchiffnerAI2012,article:ChengITUFFC2006,article:LuITUFFC1997}.
Detecting
% 1.) scattered waves
the scattered waves on
% 2.) infinite planes
infinite planes,
% 3.) competing Fourier methods
they derive
% 4.) d-dimensional spatial Fourier transform
% article:GarciaITUFFC2013: Stolt's f-k Migration for Plane Wave Ultrasound Imaging
% IV. Discussion
% - Although the theory has been derived in TWO DIMENSIONS,
%   the f-k migration for PWI can be generalized in three dimensions. (p. 1861)
the $\{ 2, 3 \}$-dimensional spatial \name{Fourier} transform of
% 5.) image
the image using either
% 6.) exploding reflector model for steered PWs
% article:GarciaITUFFC2013: Stolt's f-k Migration for Plane Wave Ultrasound Imaging
% Abstract
% - As a PLANE WAVE reaches a given scatterer, the latter becomes a secondary source emitting upward spherical waves and
%   CREATING A DIFFRACTION HYPERBOLA IN THE RECEIVED RF SIGNALS. (p. 1853)
% - To produce an image of the scatterers,
%   ALL THE HYPERBOLAS MUST BE MIGRATED BACK TO THEIR APEXES. (p. 1853)
% - The f-k migration for PWI has been ADAPTED FROM THE STOLT MIGRATION FOR SEISMIC IMAGING. (p. 1853)
% - This migration technique is BASED ON THE EXPLODING REFLECTOR MODEL (ERM), which consists in
%   ASSUMING THAT ALL THE SCATTERERS EXPLODE IN CONCERT AND BECOME ACOUSTIC SOURCES. (p. 1853)
% I. Introduction
% - The f-k MIGRATION FOR PWI is inspired by the
%   ORIGINAL FOURIER MIGRATION INTRODUCED BY STOLT FOR SEISMIC IMAGING [10], [11]. (p. 1854)
% - In this manuscript,
%   WE DERIVE A NEW FOURIER f-k MIGRATION TECHNIQUE FOR PLANE WAVE ULTRASOUND IMAGING by
%   MODIFYING the so-called EXPLODING REFLECTOR MODEL. (p. 1854)
% IV. Discussion
% - In the present paper,
%   A WELL-ESTABLISHED SEISMIC MIGRATION METHOD - the Stolt’s f-k migration - HAS BEEN MODIFIED FOR PWI. (p. 1861)
% - Because we are NOT IN THE SPECIFIC ZERO-OFFSET CONDITION of seismic imaging,
%   WE NEEDED TO ADAPT THE EXPLODING REFLECTOR MODEL (ERM). (p. 1861)
% - We demonstrated that
%   THE ERM CAN BE MADE SUITABLE TO PWI BY FINE-TUNING THE DIFFRACTION HYPERBOLAS PRESENT IN THE RF DATA. (p. 1861)
an \acl{ERM} for
steered \acp{PW}
\cite{article:GarciaITUFFC2013} or
variants of
the \acl{FDT}%
\footnote{
  The \acs{FDT}
  (cf. e.g.
  \cite[Thm. 8.4]{book:Devaney2012},
  \cite[Thm. 3.1]{book:Natterer2001},
  \cite[Sect. 6.3]{book:Kak2001}%
  ), which is also referred to as
  % 1.) generalized projection-slice theorem
  the \term{generalized projection-slice theorem}, adapts
  % 2.) Fourier slice theorem
  the \name{Fourier} slice theorem underlying
  % 3.) image recovery
  the image recovery in
  % 4.) x-ray computed tomography (CT)
  \acl{CT} from
  % 5.) rays
  rays to
  % 6.) diffracting waves
  diffracting waves.
  % b) FDT expands both the incident and scattered waves into steered PWs
  Using
  % 1.) Weyl expansions
  % book:Devaney2012, Chapter 4: Angular-spectrum and multipole expansions / Sect. 4.1: The Weyl expansion
  % - Although there are several ways of deriving the ANGULAR-SPECTRUM EXPANSION of the field U_{+}, the most direct procedure is to expand
  %   the OUTGOING-WAVE GREEN FUNCTION G_{+} in an angular-spectrum expansion and then substitute
  %   this expansion into the Green-function solution for U_{+} given in Eq. (2.23). (pp. 118, 119)
  % - The ANGULAR-SPECTRUM EXPANSION OF THE OUTGOING-WAVE GREEN FUNCTION, originally due to Weyl (Weyl, 1919) and called the WEYL EXPANSION, is derived directly from
  %   the Fourier-integral representation of the outgoing-wave Green function given in Eq. (2.15). (p. 119)
  % - Our goal here, however, is not to obtain a closed-form expression for G_{+} (which we already have) but, rather, to express
  %   G_{+} as a superposition of plane waves all of which satisfy the homogeneous Helmholtz equation. (p. 119)
  the \name{Weyl} expansions of
  % 2.) outgoing free-space Green's functions (two- and three-dimensional Euclidean spaces)
  the outgoing free-space \name{Green}'s functions
  \cite[Sect. 4.1]{book:Devaney2012}, it decomposes
  % 3.) all waves
  all waves into
  % 4.) steered PWs
  steered \acp{PW} and, thus, facilitates
  % 5.) treatment
  their treatment.
} (\acs{FDT})\acused{FDT} for
% 8.) non-diffracting beams (steered PWs, X waves)
% article:ChengITUFFC2006: Extended high-frame rate imaging method with limited-diffraction beams
% Abstract
% - In this paper, THE THEORY [article:LuITUFFC1997] IS EXTENDED TO INCLUDE EXPLICITLY VARIOUS TRANSMISSION SCHEMES such as
%   [1.)] MULTIPLE LIMITED-DIFFRACTION ARRAY BEAMS and
%   [2.)] STEERED PLANE WAVES. (p. 880)
% I. Introduction
% - Forty-six years later, Durnin [2] and Durnin et al. [3] studied the Bessel beam again and termed the beam
%   “NONDIFFRACTING BEAM” or “DIFFRACTION-FREE BEAM”. (p. 880)
% - Because Durnin’s terminologies are controversial in the scientific community and
%   practical beams will eventually diffract,
%   we termed the propagation-invariant beams or waves “LIMITED DIFFRACTION BEAMS” [4]. (p. 880)
% - One class of limited diffraction beams is called X WAVE [44], [45] and has been investigated by many physicists [49]–[56]. (p. 880)
% - In this paper, the theory of high-frame rate imaging [27], [28] is extended to include explicitly
%   various transmission schemes such as
%   [1.)] MULTIPLE LIMITED-DIFFRACTION ARRAY BEAMS and
%   [2.)] STEERED PLANE WAVES [61], [62] (the first report was given in [61]). (p. 880)
non-diffracting beams, e.g.
% 8.a) steered PWs
steered \acp{PW}
\cite{article:KruizingaITUFFC2012,proc:SchiffnerAI2012,article:ChengITUFFC2006,article:LuITUFFC1997} or
% 8.b) X waves
% article:ChengITUFFC2006: Extended high-frame rate imaging method with limited-diffraction beams
% III. Relationships Between Fourier Domains of Echoes and Object Function / A. Image Reconstruction with Limited-Diffraction Array Beams
% - For LIMITED-DIFFRACTION ARRAY BEAM TRANSMISSIONS, BOTH SINE AND COSINE WEIGHTINGS ARE APPLIED; thus
%   the ECHOES NEED TO BE COMBINED using a 2-D version of (30)–(33) to get
%   TWO NEW SETS OF ECHOES before the mapping process above. (p. 887)
\TODO{exact name of wave}
X waves
\cite{article:ChengITUFFC2006}, and
outgoing $1$-spherical (cylindrical) waves
\cite{article:MoghimiradITUFFC2016}.
\TODO{\cite{article:ZhangITUFFC2016}}

Novel inverse scattering methods, which increase both
% 1.) complexity
the complexity and
% 2.) computational costs
% article:BessonITUFFC2018: Ultrafast Ultrasound Imaging as an Inverse Problem: Matrix-Free Sparse Image Reconstruction [Mar. 2018]
% I. INTRODUCTION
% - The main problem of these models resides in their COMPUTATIONAL COMPLEXITY, usually translated in
%   STORAGE REQUIREMENTS OF THE CORRESPONDING MATRIX REPRESENTATION. (p. 340)
% - The models proposed by David et al. [10] and Schiffner and Schmitz [15] require
%   the storage of several hundreds of gigabytes for matrix coefficients in 2-D. (p. 340)
% - Zhang et al. [13] have divided the image in stripes in order to make the problem tractable. (p. 340)
% - This issue severely limits the appeal of ITERATIVE METHODS AGAINST CLASSICAL APPROACHES. (p. 340)
% article:BerthonPMB2018: Spatiotemporal matrix image formation for programmable ultrasound scanners [Feb. 2018]
% 1. Introduction
% - Yet, THIS TYPE OF IMAGE FORMATION STRATEGY [inverse scattering] WAS NOT CONSIDERED PRACTICAL FOR STANDARD IMAGE FORMATION, perhaps because of
%   the LARGE COMPUTATIONAL OVERHEADS THAT LIMITED THEIR REAL-TIME APPLICATION. (p. 2)
% - While constructing the dictionary can be more or less time consuming depending on the level of sophistication of the simulation software or experiments,
%   the inverse problem can be solved in a few milliseconds as it consists of simple matrix-vector products and thus it enables real-time imaging. (p. 2)
the computational costs, improve
% 3.) image quality
the image quality
\cite{article:OzkanITUFFC2018,article:BessonITUFFC2018,article:BerthonPMB2018,article:BessonITUFFC2016,article:DavidJASA2015,article:ZhangUlt2013,proc:SchiffnerIUS2013a,proc:SchiffnerIUS2013b,proc:SchiffnerIUS2012,article:SchiffnerBMT2012,proc:SchiffnerIUS2011}.
% b) novel inverse scattering methods regularize the ill-posed recovery problem given both a set of echo signals and a linear model for its prediction
% article:GarciaITUFFC2013: Stolt's f-k Migration for Plane Wave Ultrasound Imaging
% IV. Discussion / A. Differences Among the Three Migration Methods
% - One must be aware that migration is an ILL-POSED INVERSE PROBLEM. (p. 1862)
% - Available information (acoustic pressure at z = 0, only) is INSUFFICIENT to recover the insonified medium. (p. 1862)
% - Additional assumptions are thus required to close the problem. (p. 1862)
They regularize
% 1.) ill-posed image recovery problem [linear inverse problem]
% book:Hansen2010, Chapter 1: Introduction and Motivation
% - INVERSE PROBLEMS, in turn, belong to the CLASS OF ILL-POSED PROBLEMS. (p. 2)
% - Hadamard’s definition says that a LINEAR PROBLEM IS WELL-POSED if it satisfies the following three requirements:
%   [1.)] EXISTENCE: The problem must have a solution.
%   [2.)] UNIQUENESS: There must be only one solution to the problem.
%   [3.)] STABILITY: The solution must depend continuously on the data. (p. 2)
% - If the problem violates one or more of these requirements, it is said to be ILL-POSED. (p. 2)
% - The stability condition is much harder to “deal with” because a violation implies that
%   arbitrarily small perturbations of data can produce arbitrarily large perturbations of the solution. (p. 3)
% - Again the key is to REFORMULATE THE PROBLEM such that
%   THE SOLUTION TO THE NEW PROBLEM IS LESS SENSITIVE TO THE PERTURBATIONS. (p. 3)
% - We say that
%   WE STABILIZE OR REGULARIZE THE PROBLEM, such that
%   the solution becomes more stable and regular. (p. 3)
the ill-posed recovery problem given both
% 2.) set of echo signals
a set of
echo signals and
% 3.) linear model for its [set of echo signals] prediction
a linear model for
its prediction.
% c) universality optimally complements the programmable UI systems
% article:BerthonPMB2018: Spatiotemporal matrix image formation for programmable ultrasound scanners
% 1. Introduction
% - The NOVELTY OF THIS APPROACH RESIDES IN ITS GENERALITY:
%   SPATIOTEMPORAL MATRIX IMAGE FORMATION (SMIF) can act as
%   [1.)] a REAL-TIME, UNIVERSAL IMAGE FORMATION FRAMEWORK based on an INVERSE PROBLEM FORMULATION that can handle
%   [2.)] ANY TYPE OF EMISSIONS AND PROBE CHARACTERISTICS, regardless of their complexity. (p. 2)
% Merriam-Webster: universal = adapted or adjustable to meet varied requirements (as of use, shape, or size)
Their universality, which stems from
% 1.) arbitrary sophistication
the arbitrary sophistication of
% 2.) linear model for the prediction of the set of echo signals
this model and
% 3.) optional calibration
its optional calibration via
% 4.) experimental measurements
experimental measurements, optimally complements
% 5.) freely programmable UI systems
the programmable \ac{UI} systems.
%The universality of
% 1.) linear model for the prediction of the set of echo signals
%this model, which includes
% 2.) optional calibration
%its optional calibration via
% 3.) experimental measurements
%experimental measurements, optimally complements
% 4.) freely programmable UI systems
%the programmable \ac{UI} systems.
% d) features of the novel inverse scattering methods
% article:BerthonPMB2018: Spatiotemporal matrix image formation for programmable ultrasound scanners
% 1. Introduction
% - While such an approach [implementation of closed-form solution to the inverse problem] is IDEAL TO GENERATE FAST ALGORITHMS, it is impractical to
%   [1.)] GENERALIZE THEM TO NEW TYPES OF ULTRASOUND WAVE EMISSIONS or to
%   [2.)] ENHANCE THEM BY INCLUDING ADDITIONAL A PRIORI KNOWLEDGE. (p. 1)
% - Indeed, in practice, it means that
%   A DIFFERENT IMAGE FORMATION ALGORITHM must be designed, written, tested, and optimized to include, e.g.
%   different types of waves (plane, diverging, focused, or a combination thereof), coherent compounding, chirps,
%   attenuation, different probe geometries, element angular sensitivity, or the lens characteristics. (p. 2)
For instance,
% 1.) novel inverse scattering methods
they theoretically support
% 2.) incident waves
incident waves and
% 3.) array geometries
array geometries of
% 4.) any complexity
any complexity,
% 5.) separate recovery
the separate recovery of
% 6.) multiple acoustic material parameters
multiple acoustic material parameters,
efficient spatiotemporal sampling concepts for
data rate reduction,
% 8.) denoising
denoising, and
% 9.) inclusion of a priori information about the image
the inclusion of
\emph{a priori} information about
the image.
% e) cutting-edge variants have recently adopted CS to disrupt the tradeoff between the image quality and the frame rate
% article:BessonITUFFC2018: Ultrafast Ultrasound Imaging as an Inverse Problem: Matrix-Free Sparse Image Reconstruction
% Abstract
% - We present two different techniques, which
%   [1.)] take advantage of fast and matrix-free formulations derived for the measurement model and its adjoint, and
%   [2.)] rely on sparsity of US images in well-chosen models. (p. 339)
% - Compressed beamforming exploits the compressed sensing framework to restore
%   HIGH-QUALITY IMAGES FROM FEWER RAW DATA than state-of-the-art approaches. (p. 339)
% - Sparse regularization is used for ENHANCED IMAGE RECONSTRUCTION. (p. 339)
% I. INTRODUCTION
% - An ALTERNATIVE TO BACKPROJECTION METHODS consists of
%   SPARSE REGULARIZATION (SR) TECHNIQUES [5]. (p. 339)
% - Medical imaging is well suited to SR methods. (p. 339)
Cutting-edge variants
\cite{article:OzkanITUFFC2018,article:BessonITUFFC2018,article:BessonITUFFC2016,article:DavidJASA2015,proc:SchiffnerIUS2013a,article:ZhangUlt2013,proc:SchiffnerIUS2012,article:SchiffnerBMT2012,proc:SchiffnerIUS2011}
have recently adopted
% 1.) compressed sensing
% book:Foucart2013, Chapter 1: An Invitation to Compressive Sensing / Sect. 1.1: What is Compressive Sensing? (Aug. 2013)
% - Thus, it came as a surprise that under certain assumptions it is actually possible to reconstruct signals when
%   the number m of available measurements is smaller than the signal length N. (p. 2)
% - Even more surprisingly, efficient algorithms do exist for the reconstruction. (p. 2)
% - The UNDERLYING ASSUMPTION which makes all this possible is SPARSITY. (p. 2)
% - THE RESEARCH AREA ASSOCIATED TO THIS PHENOMENON HAS BECOME KNOWN AS
%   COMPRESSIVE SENSING, COMPRESSED SENSING, COMPRESSIVE SAMPLING, OR SPARSE RECOVERY. (p. 2)
% - This whole book is devoted to the mathematics underlying this field. (p. 2)
% book:Eldar2012:
% article:DonohoITIT2006: Compressed Sensing
\term{\acl{CS}}%
\footnote{
  The initial distinction between
  the terms
  \term{\acl{SR}} and
  \term{\acl{CS}} (\acs{CS}), which was based on
  the usage of either
  % 1.) deterministic sensing matrices
  deterministic or
  % 2.) random sensing matrices
  random sensing matrices and their
  theoretical guarantees, has been abandoned
  \cite[2]{book:Foucart2013},
  \cite{article:KutyniokGAMM2013}.
  % b) additional names include
  Additional names include
  % 1.) compressive sensing
  % article:BaraniukSPM2007: Compressive Sensing [Lecture Notes]
  \term{compressive sensing}
  \cite{article:BaraniukSPM2007} and
  % 2.) compressive sampling
  % article:CandesSPM2008: An Introduction To Compressive Sampling
  \term{compressive sampling}
  \cite{article:CandesSPM2008}.
} (\acs{CS})\acused{CS},
% 2.) data acquisition method
a data acquisition and
% 3.) recovery method
recovery method providing
% 4.) essential benefits
essential benefits in
% 5.) other imaging modalities [magnetic resonance imaging, x-ray computed tomography, photoacoustic tomography]
% article:BessonITUFFC2018: Ultrafast Ultrasound Imaging as an Inverse Problem: Matrix-Free Sparse Image Reconstruction
% I. INTRODUCTION
% - Indeed, in many medical imaging modalities,
%   the IMAGE RECONSTRUCTION PROCESS AMOUNTS TO SOLVING A LINEAR INVERSE PROBLEM. (p. 339)
% 	- In magnetic resonance imaging (MRI), the image is reconstructed from k-space samples and the measurement model is an inverse Fourier transform [6]. (p. 339)
%	- In X-ray-computed tomography (CT), the sinogram is related to the measurements by the Beer–Lambert law, which can be expressed as a linear inverse problem in the discrete domain [7]. (p. 339)
% - Moreover, sparsity priors have been expressed for medical images. (p. 339)
%	- Lustig et al. [6] have exploited sparsity of MRI images in the wavelet domain and under the total-variation (TV) transform. (p. 339)
% 	- In X-ray CT, sparsity priors under the TV-norm have been extensively used [8], [9]. (p. 339)
% article:ProvostITMI2009: The Application of Compressed Sensing for Photo-Acoustic Tomography
% article:ChenMedPhys2008: Prior image constrained compressed sensing (PICCS): A method to accurately reconstruct dynamic CT images from highly undersampled projection data sets
% [article:SongMedPhys2007,article:SidkyJXRST2006]
% article:LustigMRM2007: Sparse MRI: The application of compressed sensing for rapid MR imaging
other modalities
\cite{article:ProvostITMI2009,article:ChenMedPhys2008,article:LustigMRM2007}, to disrupt
% 6.) tradeoff
the tradeoff between
% 7.) image quality
the image quality and
% 8.) frame rate
the frame rate.
% f) cutting-edge variants of the numerical inverse scattering methods iteratively recovered a high-quality image from only a few sequential pulse-echo measurements
% article:BessonITUFFC2018: Ultrafast Ultrasound Imaging as an Inverse Problem: Matrix-Free Sparse Image Reconstruction
% Abstract
% - Using SIMULATED DATA and IN VIVO EXPERIMENTAL ACQUISITIONS, we show that
%   the proposed approach is THREE ORDERS OF MAGNITUDE FASTER than
%   non-DAS state-of-the-art methods, with comparable or better image quality. (p. 339)
They iteratively recover
% 1.) high-quality image
a high-quality image from
% 2.) single pulse-echo measurement
only a single pulse-echo measurement or
% 3.) less echo signals
less echo signals, if
% 4.) known dictionary of structural building blocks represents the image almost sparsely
(i) a known dictionary of
structural building blocks represents
the image almost sparsely, and
% 5.) individual pulse echoes are sufficiently uncorrelated
(ii) their individual pulse echoes, which are predicted by
% 6.) linear model for the prediction of the set of echo signals
the linear model, are
sufficiently uncorrelated.

The linear models, which are formulated in
% 1.) time domain
% article:OzkanITUFFC2018: Inverse Problem of Ultrasound Beamforming With Sparsity Constraints and Regularization
% article:BessonITUFFC2018: Ultrafast Ultrasound Imaging as an Inverse Problem: Matrix-Free Sparse Image Reconstruction
% article:DavidJASA2015: Time domain compressive beam forming of ultrasound signals
% article:ZhangUlt2013: A measurement-domain adaptive beamforming approach for ultrasound instrument based on distributed compressed sensing: Initial development
the time domain
\cite{article:OzkanITUFFC2018,article:BessonITUFFC2018,article:DavidJASA2015,article:ZhangUlt2013} or
% 2.) temporal Fourier domain
% proc:SchiffnerIUS2013a: Compensating the Combined Effects of Absorption and Dispersion in Plane Wave Pulse-Echo Ultrasound Imaging Using Sparse Recovery
% proc:SchiffnerIUS2013b: The Separate Recovery of Spatial Fluctuations in Compressibility and Mass Density in Plane Wave Pulse-Echo Ultrasound Imaging
% proc:SchiffnerIUS2012: Fast Image Acquisition in Pulse-Echo Ultrasound Imaging Using Compressed Sensing
% article:SchiffnerBMT2012: Compressed Sensing for Fast Image Acquisition in Pulse-Echo Ultrasound
% proc:SchiffnerIUS2011: Fast Pulse-Echo Ultrasound Imaging Employing Compressive Sensing
the temporal
\cite{proc:SchiffnerIUS2013a,proc:SchiffnerIUS2013b,proc:SchiffnerIUS2012,article:SchiffnerBMT2012,proc:SchiffnerIUS2011} and
% 3.) spatiotemporal Fourier domain
% article:BessonITUFFC2016: A Sparse Reconstruction Framework for Fourier-Based Plane-Wave Imaging
spatiotemporal
\cite{article:BessonITUFFC2016} \name{Fourier} domains, however, currently limit
% 4.) convergence speed
the convergence speed,
% 5.) image quality
the image quality, and
% 6.) potential
the potential to meet
% 7.) condition (ii) [ individual pulse echoes are sufficiently uncorrelated ]
condition (ii).
% b) linear models partly neglect diffraction, the combination of frequency-dependent absorption and dispersion, and the specifications of the instrumentation
% article:Schiffner2018, Sect. I: Introduction (sec:introduction)
% - Their [novel inverse scattering methods] UNIVERSALITY, which stems from
%   the arbitrary sophistication of this model including its optional calibration via experimental measurements, OPTIMALLY COMPLEMENTS
%   THE PROGRAMMABLE \ac{UI} SYSTEMS.
% article:Schiffner2018, Sect. I: Introduction (sec:introduction)
% - The ESTABLISHED IMAGE RECOVERY METHODS, which are explicit and computationally efficient, gradually trade
%   the image quality for the high frame rate \cite{article:GarciaITUFFC2013,article:MontaldoITUFFC2009,article:JensenUlt2006,article:ChengITUFFC2006}.
% - Their PHYSICAL MODELS NEGLECT VARIOUS EFFECTS, e.g.
%   the finite number of array elements and their anisotropic directivities, and
%   BASIC ABILITIES OF PROGRAMMABLE \ac{UI} SYSTEMS, e.g. the syntheses of complex incident waves.
Like
% 1.) linear model for the prediction of the set of echo signals
%those underlying
% 2.) established image recovery methods
the established methods,
% 3.) linear model for the prediction of the set of echo signals
they partly neglect
% 4.) diffraction
diffraction,
% 5.) combination of frequency-dependent absorption and dispersion
the combination of
frequency-dependent absorption and
dispersion, and
% 6.) specifications of the instrumentation
the specifications of
the instrumentation, including
% 7.) array geometry
the array geometry,
% 8.) acoustic lens
the acoustic lens, and
% 9.) electromechanical transfer behavior
% article:BessonITUFFC2018: Ultrafast Ultrasound Imaging as an Inverse Problem: Matrix-Free Sparse Image Reconstruction
% II. PARAMETRIC MATRIX-FREE FORMULATIONS OF THE MEASUREMENT MODEL AND ITS ADJOINT / C. Inverse Problem of Interest
% - While the model described in Section II can be used for inverse scattering problems,
%   WE FOCUS, in this paper, ON REGULARIZED BEAMFORMING, which means that
%   WE NEGLECT THE EFFECT OF THE PULSE-ECHO WAVEFORM v_{pe} in the model. (p. 342)
the electromechanical transfer behavior.
Moreover,
% 1.) limit the number of spatial dimensions to only two
they partly limit
the number of
spatial dimensions to
only two, intermix
% 2.) intermix results for the two- and three-dimensional Euclidean spaces
results for
the two- and
three-dimensional Euclidean spaces, and replace
the acoustic material parameters by
an abstract reflectivity function.
% d) exclusive usage of steered PWs or outgoing cylindrical waves ignores the syntheses of the incident waves by the UI system
% d) exclusive usage of steered PWs or outgoing cylindrical waves ignores basic abilities of programmable \ac{UI} systems
% article:BessonITUFFC2018: Ultrafast Ultrasound Imaging as an Inverse Problem: Matrix-Free Sparse Image Reconstruction
% I. INTRODUCTION
% - Firstly, parametric, fast and matrix-free formulations of the measurement model and its adjoint are described for both
%   PLANE-WAVE (PW) and DIVERGING-WAVE (DW) compounding. (p. 2)
% II. PARAMETRIC MATRIX-FREE FORMULATIONS OF THE MEASUREMENT MODEL AND ITS ADJOINT / A. Formulation of the Measurement Model
% - Ultrafast US imaging involves transmission of either steered PW (SPW) or DWs. (p. 340)
The exclusive usage of
steered \acp{PW}
\cite{article:OzkanITUFFC2018,article:BessonITUFFC2018,article:BessonITUFFC2016,article:DavidJASA2015,proc:SchiffnerIUS2013a,proc:SchiffnerIUS2013b,article:ZhangUlt2013,proc:SchiffnerIUS2012,article:SchiffnerBMT2012,proc:SchiffnerIUS2011} or
% 2.) outgoing 1-spherical waves (cylindrical waves)
% article:BessonITUFFC2018: Ultrafast Ultrasound Imaging as an Inverse Problem: Matrix-Free Sparse Image Reconstruction
outgoing cylindrical waves
\cite{article:BessonITUFFC2018} ignores
% 3.) basic abilities
basic abilities of
% 4.) freely programmable UI systems
programmable \ac{UI} systems.
% e) UI system typically specifies pulse shapes, apodization weights, and time delays for the waves emitted by the individual array elements and provides hundreds of degrees of freedom to synthesize alternative types of incident waves
% article:IlovitshNatComBio2018: Acoustical structured illumination for super-resolution ultrasound imaging
% - Advances in ultrasound technologies have now led to user-programmable systems, capable of
%   [1.)] a NEARLY INFINITE VARIETY OF TRANSMITTED PULSE TRAINS, and
%   [2.)] schemes for image reconstruction. (p. 2)
The latter typically specifies
% 1.) pulse shapes
pulse shapes,
% 2.) apodization weights
apodization weights, and
% 3.) time delays
time delays for
% 4.) waves emitted by the individual array elements
the waves emitted by
the individual array elements and provides
% 5.) hundreds
hundreds of
% 6.) degrees of freedom
degrees of
freedom to synthesize
% 7.) alternative types of incident waves
alternative types of
incident waves.
Although
% 1.) nonconvex recovery methods
nonconvex recovery methods require
% 2.) less data
less data
\cite{article:FoucartACHA2009,proc:ChartrandICASSP2008,article:ChartrandISPL2007},
convex variants are
prevalent
\cite{article:BessonITUFFC2018,article:BessonITUFFC2016,article:DavidJASA2015,proc:SchiffnerIUS2013a,proc:SchiffnerIUS2012,article:SchiffnerBMT2012,proc:SchiffnerIUS2011}.
% Besides \acl{MAP} estimates \cite{article:ZhangUlt2013},
% article:ZhangUlt2013: A measurement-domain adaptive beamforming approach for ultrasound instrument based on distributed compressed sensing: Initial development
% 3. The measurement-domain adaptive beamforming (MABF) based on DCS / 3.2. The measurement-domain adaptive beamformer (MABF)
% - The maximum likelihood estimate of f is given by [21] (14). (p. 259)
% - The (14) becomes the MAP optimization problem (15). (p. 259)
% - Therefore, to find the MAP estimate of target vector f, we must maximize the priori PDF p(f) subject to the constraint that y = V_{DCS} f. (p. 259)

%---------------------------------------------------------------------------------------------------------------
% 5.) randomization of the apodization weights and the time delays / random incident waves
%---------------------------------------------------------------------------------------------------------------
% a) randomization of the apodization weights and the time delays in the syntheses of the incident waves potentially improves the conformity with condition (ii)
% article:Ghanbarzadeh-DagheyanSensors2018: Holey-Cavity-Based Compressive Sensing for Ultrasound Imaging
% 1. Introduction
% - When the sensing matrix is built through RANDOM PROJECTIONS in the undersampled signal space,
%   the RESULTING COHERENCE BETWEEN EACH OF TWO OF ITS COLUMNS IS SMALL, with a high probability. (p. 1)
% - This is the reason why
%   MEASUREMENT RANDOMIZATION HAS THE POTENTIAL TO ENHANCE THE ACCURACY OF THE RECOVERED UNKNOWN SIGNAL [2]. (p. 1)
The randomization of
% 1.) apodization weights
the apodization weights and
% 2.) time delays
the time delays in
% 3.) syntheses of the incident waves
the syntheses of
the incident waves, which is motivated by
% 4.) essential theorems underlying CS
the essential theorems underlying
\ac{CS}, potentially improves
% 5.) conformity
the conformity with
% 6.) condition (ii) [ individual pulse echoes are sufficiently uncorrelated ]
condition (ii).
% b) LTI measurement process physically superimposes the weighted and delayed echo signals induced by the individual array elements
In fact,
% 1.) LTI measurement process
the \ac{LTI} measurement process physically superimposes
% 2.) weighted and delayed echo signals
the weighted and
delayed echo signals induced by
% 3.) individual array elements
the individual array elements.
% c) each incident wave randomly projects a complete SA acquisition sequence on a single pulse-echo measurement and strongly compresses both the acquisition time and the data volume
% article:KruizingaSciAdv2017: Compressive 3D ultrasound imaging using a single sensor
% INTRODUCTION
% - Our work follows the concept central to compressive imaging: that of
%   PROJECTING THE OBJECT/IMAGE INFORMATION THROUGH A SET OF INCOHERENT FUNCTIONS ONTO A SINGLE MEASUREMENT. (p. 2)
% DISCUSSION
% - Our work differs in the fact that rather than focusing our ultrasound waves to a single point,
%   WE ENCODE THE WHOLE 3D VOLUME ONTO ONE SPATIOTEMPORAL VARYING ULTRASOUND FIELD, from which
%   we then reconstruct an image based on solving a linear signal model. (p. 8)
% - This has the advantage of requiring only a few measurements to image an entire 3D volume, in contrast to time-reversal work. (p. 8)
% - This difference in acquisition time is crucial for medical imaging where the object changes over time, for example, in a beating heart. (p. 8)
Each incident wave randomly projects
% 1.) complete SA acquisition sequence
a complete \ac{SA} acquisition sequence on
% 2.) single pulse-echo measurement
a single pulse-echo measurement and, thus, strongly compresses both
% 3.) acquisition time
the acquisition time and
% 4.) data volume
the data volume.
% d) few studies of such waves in compressed UI and their origins will now be reviewed
The few studies of
% 1.) random waves
such waves in
% 2.) compressed UI
compressed \ac{UI} and
% 3.) origins
their origins will now be reviewed.

%%%%%%%%%%%%%%%%%%%%%%%%%%%%%%%%%%%%%%%%%%%%%%%%%%%%%%%%%%%%%%%%%%%%%%%%%%%%%%%%%%%%%%%%%%%%%%%%%%%%%%%%%%%%%%%%
% 1.) related work
%%%%%%%%%%%%%%%%%%%%%%%%%%%%%%%%%%%%%%%%%%%%%%%%%%%%%%%%%%%%%%%%%%%%%%%%%%%%%%%%%%%%%%%%%%%%%%%%%%%%%%%%%%%%%%%%
\subsection{Related Work}
\label{subsec:intro_related_work}
%---------------------------------------------------------------------------------------------------------------
% 1.) random waves in computational microwave imaging
%---------------------------------------------------------------------------------------------------------------
% a) syntheses of random waves in UI traces back to computational microwave imaging
% article:Ghanbarzadeh-DagheyanSensors2018: Holey-Cavity-Based Compressive Sensing for Ultrasound Imaging
% 1. Introduction
% - SEVERAL SENSING AND IMAGING APPLICATIONS [4] have been able to take advantage of CS by using
%   [1.)] PSEUDO-RANDOM ILLUMINATION IN THE OUTGOING WAVES from the transmitters and collecting
%   [2.)] PSEUDO-RANDOM MEASUREMENTS FROM THE INCOMING WAVES to the receivers. (pp. 1, 2)
% - Carin et al. used random positions for the elements of a sensing array to make use of CS. (p. 2)
% - They also showed that PLACING SPHERICAL SCATTERING OBJECTS IN FRONT OF THE WAVES increases
%   the RANDOMNESS AND INCOHERENCE of the measurements [5]. (p. 2)
%   [5] Carin, L.; Liu, D.; Guo, B. Coherence, compressive sensing, and random sensor arrays. IEEE Antennas Propag. Mag. 2011, 53, 28–39.
The syntheses of
% 1.) random waves
random waves in
% 2.) ultrasound imaging (UI)
\ac{UI} partly trace back to
% 3.) time-reversal methods
% article:MontaldoITUFFC2005: Building three-dimensional images using a time-reversal chaotic cavity
time-reversal
\cite{article:MontaldoITUFFC2005} and
% 4.) computational microwave imaging
% article:FromentezeOptExp2017: Computational polarimetric microwave imaging (leaky reverberant cavity)
% article:GollubSciRep2017: Large Metasurface Aperture for Millimeter Wave Computational Imaging at the Human-Scale (complex metamaterial)
% article:FromentezeApplPhysLett2015: Computational imaging using a mode-mixing cavity at microwave frequencies (leaky reverberant cavity)
% article:LipworthJOSAA2013: Metamaterial apertures for coherent computational imaging on the physical layer (complex metamaterial)
% article:HuntScience2013: Metamaterial Apertures for Computational Imaging (complex metamaterial)
computational microwave imaging
\cite{article:FromentezeOptExp2017,article:GollubSciRep2017,article:FromentezeApplPhysLett2015,article:LipworthJOSAA2013,article:HuntScience2013}.
% b) computational microwave imaging trades the hardware complexity and costs for the computational complexity of the image recovery
% article:FromentezeOptExp2017: Computational polarimetric microwave imaging (microwave imaging)
% 1. Introduction
% - The FREQUENCY-DIVERSE APERTURE LIMITS
%   THE COMPLEXITY OF THE HARDWARE ARCHITECTURE required for real-time high-resolution imaging, obviating
%   the ACTIVELY CONTROLLED COMPONENTS or
%   the NEED FOR MECHANICAL MOTION that is typically required in conventional systems. (p. 3)
% article:ChewITAP1997: Fast solution methods in electromagnetics
% I. Introduction
% - However,
%   [1.)] the RECENT PHENOMENAL GROWTH IN COMPUTER TECHNOLOGY, coupled with
%   [2.)] the DEVELOPMENT OF FAST ALGORITHMS with reduced computational complexity and memory requirements,
%   have made a RIGOROUS NUMERICAL SOLUTION OF THE PROBLEM OF SCATTERING FROM ELECTRICALLY LARGE OBJECTS FEASIBLE. (p. 533)
The latter trades
% 1.) hardware complexity
the hardware complexity and
% 2.) hardware costs
costs, which are raised by
% 3.) fully-sampled transceiver arrays
transceiver arrays or
% 4.) mechanical scanning
mechanical scans, for
% 5.) computational costs
the computational costs of
% 6.) image recovery
the image recovery.
% c) highly dispersive customized apertures convert excitations at different frequencies into spatially diverse and distinct emitted fields
% article:Ghanbarzadeh-DagheyanSensors2018: Holey-Cavity-Based Compressive Sensing for Ultrasound Imaging
% 1. Introduction
% - Incoherence between each of two measurements can also be achieved by using
%   PHYSICAL STRUCTURES that exhibit DIFFERENT WAVE-MATTER RESPONSES AT DIFFERENT INSTANTANEOUS FREQUENCIES, without
%   the need to CHANGE THE ARRANGEMENT OF THE SENSING ARRAY. (p. 2)
Highly dispersive customized apertures, e.g.
complex metamaterials
\cite{article:GollubSciRep2017,article:LipworthJOSAA2013,article:HuntScience2013} or
leaky reverberant cavities
\cite{article:FromentezeOptExp2017,article:FromentezeApplPhysLett2015}, form
% 3.) virtual transceiver arrays
% article:KruizingaSciAdv2017: Compressive 3D ultrasound imaging using a single sensor
% RESULTS / Ultrasound wave field diversity using a coded aperture
% - By MODELING THE APERTURE MASK AS A COLLECTION OF POINT SENSORS, each having different transmit/receive delays,
%   the MASK CAN STILL BE REGARDED AS A SENSOR ARRAY. (p. 3)
virtual transceiver arrays.
% c) highly dispersive customized apertures form virtual transceiver arrays that expand excitations at sufficiently different frequencies into distinct spatial codes and mix the scattered waves in reception
% article:Ghanbarzadeh-DagheyanSensors2018: Holey-Cavity-Based Compressive Sensing for Ultrasound Imaging
% 1. Introduction
% - More specifically, the natural reverberation of the wave propagation inside the medium emulates random illuminations of a scene at different frequencies.
% - These systems exploit the conversion between spatial and spectral degrees of freedom, which are
%   also at the core of the time reversal technique in complex media for spatio-temporal focusing with a single broadband antenna. [29,30]
% article:KruizingaSciAdv2017: Compressive 3D ultrasound imaging using a single sensor
% RESULTS / Ultrasound wave field diversity using a coded aperture
% - However, all these point sensor signals are SUBSEQUENTLY SUMMED BY THE PIEZO SENSOR just after they have passed though the mask, resulting in
%   a SINGLE COMPRESSED MEASUREMENT, as depicted in the right panel of Fig. 1
%   (for a further comparison between a sensor array and a single sensor with a coding mask, see text S2 and figs. S1 and S2). (p. 3)
% article:FromentezeOptExp2017: Computational polarimetric microwave imaging (microwave imaging)
% 1. Introduction
% - These systems RADIATE PSEUDO-ORTHOGONAL FIELD DISTRIBUTIONS IN TRANSMISSION and - by exploitation of the reciprocity principle - IN RECEPTION,
%   to MULTIPLEX INFORMATION AND RECONSTRUCT AN IMAGE. (p. 3)
These expand
% 1.) excitations
excitations at
% 2.) sufficiently different frequencies
sufficiently different frequencies into
% 3.) distinct spatial codes
distinct spatial codes, i.e.
% 4.) spatially erratic incident fields
% TODO: complex erratic spatiotemporal interference patterns
spatially erratic incident fields, and, reciprocally, mix
% 5.) scattered fields
the scattered fields in
% 6.) reception
reception, providing
% 7.) frequency-diverse projections
frequency-diverse projections of
% 8.) field-of-view (FOV)
the \ac{FOV}.
% c)
The knowledge of
% 1.) two-way transfer function
the two-way transfer function enables
% 2.) image recovery
the image recovery.
Unlike
the random sampling
\cite{article:BessonITUFFC2018,proc:BessonICIP2016,article:DavidJASA2015} or
% 3.) mixing
mixing
\cite{article:BessonITUFFC2018,article:LiutkusSciRep2014} of
% 4.) scattered waves
the scattered waves, which reduce
the number of
transceivers and measurements,
% 5.)
the random waves improve
the conformity with
condition (ii).

\name{Kruizinga} \emph{et al.} \cite{article:KruizingaSciAdv2017} equipped
a single transducer with
a plastic delay mask to enable
compressive three-dimensional \ac{UI} with
cheap and simple hardware.
The varying thickness of
the mask introduced random time delays into both
the emitted and
the received waves to decorrelate
the pulse echoes received from
distinct voxels.
Adopting
a simple calibration procedure, which measured
the random sound field in
water and, thus, limited
its range of
validity,
% article:KruizingaSciAdv2017: Compressive 3D ultrasound imaging using a single sensor
% RESULTS / Signal model and image reconstruction
% - For the full 3D reconstruction shown in Fig. 6F, we made use of the sparsity of
%   these two letters in water by applying the sparsity-promoting basis pursuit denoising (BPDN) algorithm (27). (p. 6)
% - As can be seen, this prior knowledge about the image could be effectively exploited to improve image quality,
%   significantly improving the dynamic range from 15 to 40 dB. (p. 6)
% - Fig. 6. Compressive 3D ultrasound imaging using a single sensor. [...]
%   The images shown in (B) and (D) were obtained using 72 EVENLY SPACED MASK ROTATIONS, and
%   the full 3D image in (F) was obtained using only 50 EVENLY SPACED ROTATIONS to reduce the total matrix size. (p. 7)
% article:KruizingaSciAdv2017: Compressive 3D ultrasound imaging using a single sensor
% RESULTS / Signal model and image reconstruction
% - Thus, instead of applying a standard geometric operation to multiple sensor observations,
%   we attempt to explain the received signal as a linear combination of point scatterer echo signals [23, 24]. (p. 5)
the recovery of
sparse objects required
$50$ sequential pulse-echo measurements at
evenly spaced angles of
rotation.
\name{Ghanbarzadeh-Dagheyan} \emph{et al.} \cite{article:Ghanbarzadeh-DagheyanSensors2018} optimized
a holey cavity with
respect to
its opening sizes and
the materials to enable
compressive two-dimensional \ac{UI} with
only a few transceivers.
Encasing only two point-like transceivers that perform
a complete \ac{SA} acquisition sequence,
its presence significantly improved
the lateral resolution of
two point-like targets in
a lossless homogeneous fluid.
Unlike
% 1.) time-reversal technique
the time-reversal technique
\cite{article:MontaldoITUFFC2005}, which uses
% 2.) leaky reverberant cavity
a leaky cavity to generate
% 3.) focused beams
% TODO: Sinai Billiard
focused beams,
% 4.) field-of-view (FOV)
the \ac{FOV} is not progressively scanned.

\name{Van Sloun} \emph{et al.}
\cite{letter:VanSlounITBME2015} proposed
randomly-apodized monofrequent emissions from
a circular array in
two-dimensional tomography.
% b) randomly-apodized emissions outperformed sparse SA acquisition sequences
These outperformed
sparse \ac{SA} acquisition sequences using
only a few emissions from
random elements.
\name{Liu} \emph{et al.} \cite{article:LiuITUFFC2018,article:LiuITMI2017} realized
uniformly-distributed apodization weights for
linear and convex arrays.
Unlike
% 1.) inverse scattering methods
the inverse scattering methods,
% 2.) Liu et al.
they recovered
% 3.) recorded RF voltage signals
the echo signals induced by
% 4.) complete SA acquisition sequence
a complete \ac{SA} acquisition sequence, which were represented almost sparsely by
% 5.) sym8 wavelet basis
a sym8 wavelet basis, and subsequently applied
% 6.) popular DAS method
the popular \ac{DAS} method for
% 7.) image formation
image formation.
% c) large number of unknown temporal samples required on the order of 30 sequential pulse-echo measurements per image
% article:LiuITUFFC2018: Compressed Sensing Based Synthetic Transmit Aperture Imaging: Validation in a Convex Array Configuration
% Abstract
% - The experimental results showed that STA and CS-STA performed better than ME-STA and the focused method at small depths. (p. 300)
% - At the depth of 110 mm, CS-STA, ME-STA, and the focused methods improved the contrast and contrast-to-noise ratio of STA. (p. 300)
% - The improvements in CS-STA are higher than those in ME-STA but lower than those in the focused mode. (p. 300)
% - These results can also be observed qualitatively in the in vivo experiments on the liver of a healthy male volunteer. (p. 300)
% - The CS-STA method is thus proved to increase the frame rate and achieve high image quality at full depth in the convex array configuration. (p. 300)
% article:LiuITMI2017: A Compressed Sensing Strategy for Synthetic Transmit Aperture Ultrasound Imaging
% Abstract
% - In addition, the CONTRAST OF THE STA IMAGE CAN BE IMPROVED at the same time owing to the higher energy of plane wave firing in CS-STA. (p. 878)
% - The results demonstrate that, implemented with the same frame rate, CS-STA achieves HIGHER OR COMPARABLE RESOLUTION AND CONTRAST. (p. 878)
Despite
% 1.) improvements
the improvements in
% 2.) contrast
contrast and
% 3.) spatial resolution
spatial resolution,
% 4.) large number of unknown temporal samples
the large number of
unknown temporal samples required
% 5.) tens of sequential pulse-echo measurements per image
tens of
sequential pulse-echo measurements per
image.

%%%%%%%%%%%%%%%%%%%%%%%%%%%%%%%%%%%%%%%%%%%%%%%%%%%%%%%%%%%%%%%%%%%%%%%%%%%%%%%%%%%%%%%%%%%%%%%%%%%%%%%%%%%%%%%%
% 2.) specific contributions
%%%%%%%%%%%%%%%%%%%%%%%%%%%%%%%%%%%%%%%%%%%%%%%%%%%%%%%%%%%%%%%%%%%%%%%%%%%%%%%%%%%%%%%%%%%%%%%%%%%%%%%%%%%%%%%%
\subsection{Specific Contributions}
A method for
% 1.) fast compressed acquisition
the fast compressed acquisition and
% 2.) subsequent recovery
the subsequent recovery of
images is proposed that features
% 3.) three major innovations
three major innovations.
% b) proposed method significantly enhances current inverse scattering methods by realistic d-dimensional physical models
First,
% 1.) realistic d-dimensional physical models
realistic $d$-dimensional physical models for
% 2.) linear physical model for the pulse-echo measurement process
the pulse-echo measurement process and
% 3.) syntheses of the incident waves
the syntheses of
the incident waves minimize
% 4.) minimize model inaccuracies
inaccuracies and leverage
% 5.) abilities
the abilities of
% 6.) programmable UI systems
programmable \ac{UI} systems.
% c) proposed method recovers the compressibility fluctuations by a sparsity-promoting lq-minimization method
%   What are the SCIENTIFIC MERITS of this particular model system?
They linearly relate
% 1.) spatial compressibility fluctuations
the spatial compressibility fluctuations in
% 2.) weakly-scattering soft tissue structures
weakly-scattering soft tissue structures to
% 3.) recorded RF voltage signals
the \ac{RF} voltage signals provided by
% 4.) instrumentation
the instrumentation.
% d) proposed physical models are universal and permit additional applications beyond ultrafast UI
They readily support
% 1.) calibration procedures
calibration procedures,
% 5.) usage
the usage of
% 6.) measured incident fields
measured incident fields, and
% 7.) applications
applications beyond
% 8.) ultrafast UI
ultrafast \ac{UI}, e.g.
% 5.) conventional UI based on progressive scanning
progressive scanning,
% 6.) structured insonification
structured insonification, or
% 7.) simulation studies
simulation studies.
% a) three innovative types of random incident waves aid in meeting condition (ii)
Second,
% 1.) three innovative types of energy equivalent random waves
three innovative types of
energy equivalent random waves are synthesized using
% 2.) pseudo-random apodization weights
random apodization weights,
% 3.) random time delays
time delays, or
% 4.) combinations thereof
combinations thereof.
% b) associated spatial codes decorrelate the pulse echoes of the structural building blocks defined by an orthonormal basis and facilitate their discrimination in the image recovery
The associated
% 1.) spatial codes
spatial codes decorrelate
% 2.) pulse echoes
the pulse echoes of
% 3.) structural building blocks
the structural building blocks defined by
% 4.) orthonormal basis
an orthonormal basis meeting
% 5.) condition (i) [known dictionary of structural building blocks represents the image almost sparsely]
condition (i) and, thus, improve
% 6.) conformity
the conformity with
% 7.) condition (ii): [individual pulse echoes are sufficiently uncorrelated]
condition (ii).
% c) convex and nonconvex variants of a sparsity-promoting lq-minimization method enable the quantitative recovery of the compressibility fluctuations
Third, both
% 1.) convex
convex and
% 2.) nonconvex
nonconvex variants of
% 3.) sparsity-promoting lq-minimization method
a sparsity-promoting $\ell_{q}$-minimization method, $q \in [ 0; 1 ]$, enable
% 4.) quantitative recovery
the quantitative recovery of
% 5.) compressibility fluctuations
the compressibility fluctuations.

%---------------------------------------------------------------------------------------------------------------
% 2.) results of the simulation study
%---------------------------------------------------------------------------------------------------------------
% - Why did you choose this kind of experiment or experimental design?
% a) numerical simulation of a pulse-echo in the two-dimensional Euclidean space validates the emissions of single random incident waves
Two-dimensional numerical simulations validate
the proposed method using
single realizations of
the random waves for
% 1.) wire phantom
a wire phantom and
% 2.) tissue-mimicking phantom
a tissue-mimicking phantom.
% -> basis transform: Fourier
% d) Liu et al. require on the order of 30 sequential pulse-echo measurements per image, whereas the method proposed in this paper only requires the minimum number of a single wave emission
%In contrast to Liu and Kruizinga,
%the proposed method only requires
%the minimum number of
%a single pulse-echo measurement per
%image and uses additional types of
%random waves.
% b) former phantom permits a sparse representation of its spatial compressibility fluctuations in the canonical basis, whereas the latter phantom requires the Fourier basis
%The former phantom permits
%a sparse representation of
%its spatial compressibility fluctuations in
% 1.) canonical basis
%the canonical basis, whereas
%the latter phantom requires
% 2.) Fourier basis
%the \name{Fourier} basis.
% TODO: insight into the transfer behavior of the UI system
% TODO: amount of information collected by the array through different excited spatial frequencies is increased compared to conventional waves
% c) random incident waves decreased the robustness to
%   What ADVANTAGES DOES IT CONFER in answering the particular question(s) you are posing?
Although
the random waves decrease
% 1.) robustness against additive errors
the robustness against
additive errors for
the wire phantom,
they significantly increase both
% 2.) image quality
the image quality and
% 3.) speed of convergence
the convergence speed for
the tissue-mimicking phantom.
% a) author published two abstracts outlining the fundamental ideas of this paper in connection with oral presentations at two conferences
The study significantly expands
the initial results published in
two abstracts
\cite{proc:SchiffnerIUS2017,article:SchiffnerJASA2017}.
%% c) recovery experiments confirm better performance
%Numerical recovery experiments additionally demonstrate
%% 1.) increased SSIM indices
%increased \ac{SSIM} indices,
%% 2.) reduced relative RMSEs
%reduced relative \acp{RMSE}, and
%% 3.) faster convergence
%faster convergence.

%---------------------------------------------------------------------------------------------------------------
% 3.) results of the experimental validation
%---------------------------------------------------------------------------------------------------------------
% For the experimental validation of
% the random incident waves,
% we acquired pulse-echo measurement data from
% a real phantom consisting of nine wires.

%%%%%%%%%%%%%%%%%%%%%%%%%%%%%%%%%%%%%%%%%%%%%%%%%%%%%%%%%%%%%%%%%%%%%%%%%%%%%%%%%%%%%%%%%%%%%%%%%%%%%%%%%%%%%%%%
% table: summary of the mathematical symbols used throughout the paper
%%%%%%%%%%%%%%%%%%%%%%%%%%%%%%%%%%%%%%%%%%%%%%%%%%%%%%%%%%%%%%%%%%%%%%%%%%%%%%%%%%%%%%%%%%%%%%%%%%%%%%%%%%%%%%%%
\begin{table*}[tb]
 \centering
 \caption{%
  Summary of
  the mathematical symbols used
  throughout the paper.
 }
 \label{tab:list_symbols_math}
 \small
 \begin{tabular}{%
  @{}%
  >{$}l<{$}%		01.) symbol
  p{0.9\textwidth}%	02.) meaning
  @{}%
 }
 \toprule
  \multicolumn{1}{@{}H}{Symbol} &
  \multicolumn{1}{H@{}}{Meaning}\\
  \cmidrule(r){1-1}\cmidrule(l){2-2}
 \addlinespace
 %--------------------------------------------------------------------------------------------------------------
 % a) number sets
 %--------------------------------------------------------------------------------------------------------------
  % 1.) set of consecutive positive integers
  \setcons{ N } &
  Set of consecutive positive integers,
  $\setcons{ N } = \{ 1, 2, \dotsc, N \}$ for $N \in \N$\\
  % 2.) set of consecutive nonnegative integers
  \setconsnonneg{ N } &
  Set of consecutive nonnegative integers,
  $\setconsnonneg{ N } = \{ 0, 1, \dotsc, N \}$ for $N \in \Nnonneg$\\
 %--------------------------------------------------------------------------------------------------------------
 % b) inner product
 %--------------------------------------------------------------------------------------------------------------
  % 1.) inner product
  \inprod{ \vect{a} }{ \vect{b} } &
  Inner product of
  the vectors
  $\vect{a} = \trans{ ( a_{1}, \dotsc, a_{N} ) } \in \C^{ N }$ and
  $\vect{b} = \trans{ ( b_{1}, \dotsc, b_{N} ) } \in \C^{ N }$,
  $\inprod{ \vect{a} }{ \vect{b} } = \sum_{ n = 1 }^{ N } a_{n} \conj{ b }_{n}$\\
 %--------------------------------------------------------------------------------------------------------------
 % c) norms, quasinorms, and NNZC
 %--------------------------------------------------------------------------------------------------------------
  % 1.) lq-norm or lq-quasinorm
  \tnorm{ \vect{a} }{q} &
  $\ell_{q}$-norm, $q \in [ 1; \infty )$, or
  $\ell_{q}$-quasinorm, $q \in ( 0; 1 )$, of
  the vector
  $\vect{a} \in \C^{ N }$,
  $\tnorm{ \vect{a} }{q}^{q} = \sum_{ n = 1 }^{ N } \tabs{ a_{n} }^{q}$\\
  % 2.) number of nonzero components
  % book:Foucart2013: A Mathematical Introduction to Compressive Sensing / Chapter 2: Sparse Solutions of Underdetermined Systems / Sect. 2.1: Sparsity and Compressibility
  % - The customary notation \tnorm{ \vect{x} }{0} - the notation \tnorm{ \vect{x} }{0}^{0} would in fact be more appropriate - comes from the observation that
  %   \norm{ \vect{x} }{p}^{p} = \sum_{ j = 1 }^{ N } \abs{ x_{j} }^{p} \rightarrow \sum_{ j = 1 }^{ N } \indicator{ x_{j} \neq 0 }. (p. 41)
  % - In other words the quantity \tnorm{ \vect{x} }{0} is
  %   the limit as p decreases to zero of the pth power of the \ell_{p}-quasinorm of x. (p. 42)
  \tnorm{ \vect{a} }{0} &
  Number of nonzero components,
  $\tnorm{ \vect{a} }{0} := \tnorm{ \vect{a} }{0}^{0} = \lim_{q \rightarrow 0} \tnorm{ \vect{a} }{q}^{q} = \tabs{ \{ n \in \setcons{ N }: a_{n} \neq 0 \} }$\\
 %--------------------------------------------------------------------------------------------------------------
 % c) spatial coordinates
 %--------------------------------------------------------------------------------------------------------------
  % 1.) spatial position
  \vect{r} &
  Spatial position in
  the $d$-dimensional Euclidean space,
  $\vect{r} = \trans{ ( r_{1}, \dotsc, r_{d} ) } \in \R^{d}$\\
  % 2.) lateral coordinates and axial coordinate
  \vect{r}_{\rho}, r_{d} &
  % 2.1) lateral coordinates
  Lateral coordinates
  $\vect{r}_{\rho} = \trans{ ( r_{1}, \dotsc, r_{d-1} ) } \in \R^{d-1}$ and
  % 2.2) axial coordinate
  axial coordinate
  $r_{d} \in \R$ of
  the spatial position
  $\vect{r} = \trans{ ( \trans{ \vect{r}_{\rho} }, r_{d} ) }$\\
  % 3.) unit (d-1)-sphere
  \usphere{d-1} &
  Unit $(d-1)$-sphere,
  $\usphere{d-1} = \{ \vect{r} \in \R^{d}: \norm{ \vect{r} }{2} = 1 \}$\\
  % 4.) unit (d-1)-hemisphere
  \uhemisphere{d-1} &
  Unit $(d-1)$-hemisphere,
  $\uhemisphere{d-1} = \{ \vect{r} \in \R^{d}: \norm{ \vect{r} }{2} = 1, r_{d} \in \Rplus \}$\\
  % 5.) unit vector indicates the direction of the r_{\delta}-axis in a d-dimensional Cartesian coordinate system
  \uvect{\delta} &
  Unit vector indicating
  the direction of
  the $r_{\delta}$-axis, $\delta \in \setcons{ d }$, in
  a $d$-dimensional Cartesian coordinate system,
  $\uvect{\delta} \in \usphere{d-1}$\\
 %--------------------------------------------------------------------------------------------------------------
 % d) matrices
 %--------------------------------------------------------------------------------------------------------------
  % 1.) superscript H indicates an adjoint matrix
  ^{\hermsymbol} &
  Superscript indicating
  an adjoint (conjugate transpose) matrix\\
  % 2.) I denotes the identity matrix
  \mat{I} &
  Identity matrix\\
 %--------------------------------------------------------------------------------------------------------------
 % e) signal
 %--------------------------------------------------------------------------------------------------------------
  % 1.) tilde symbol identifies time-domain signals
  % TODO: recorded \ac{RF} voltage signals in the time domain.
  \tilde{ \cdot } &
  Tilde accent indicating
  a time-domain signal\\%, e.g. voltage signals
 \addlinespace
 \bottomrule
 \end{tabular}
\end{table*}

%---------------------------------------------------------------------------------------------------------------
% 3.) structure of the presentation and list of mathematical symbols
%---------------------------------------------------------------------------------------------------------------
% a) contributions are organized as follows
These contributions are organized as
follows.
% 2.) compressed sensing
\Cref{sec:compressed_sensing} briefly reviews
the \ac{CS} framework.
% 3.) linear physical model for the pulse-echo measurement process
% 4.) syntheses of the incident waves
\Cref{sec:linear_model,sec:syn_p_in} present
the physical models for
the pulse-echo measurement process and
the syntheses of
the incident waves.
% 5.) image recovery based on compressed sensing
\Cref{sec:recovery} details
the image recovery based on
\ac{CS}, and
% 6.) implementation
\cref{sec:implementation} adds
an efficient matrix-free implementation.
% 7.) simulation study
% 8.) experimental validation
%\cref{sec:experimental_validation}
% 9.) results
\Cref{sec:simulation_study} summarizes
the parameters of
the numerical simulations, and
\cref{sec:results} presents
the results.
% 10.) discussion
\Cref{sec:discussion} discusses
these results and
the proposed method.
% 11.) conclusion and outlook
Eventually,
\cref{sec:conclusion_outlook} concludes
the paper.
% b) table summarizes the mathematical symbols
\Cref{tab:list_symbols_math} summarizes
the mathematical symbols.

\section{Compressed Sensing in a Nutshell}
\label{sec:compressed_sensing}
%---------------------------------------------------------------------------------------------------------------
% 1.) mathematical description of compressed sensing
%---------------------------------------------------------------------------------------------------------------
% a) CS deals with the stable recovery of a high-dimensional vector from a low-dimensional vector of potentially corrupted observations
% article:KutyniokGAMM2013: Theory and Applications of Compressed Sensing
% 1 Introduction
% - Taking a different viewpoint [key idea of CS],
%   it concerns the EXACT RECOVERY OF A HIGH-DIMENSIONAL SPARSE VECTOR AFTER A DIMENSION REDUCTION STEP. (p. 79)
\ac{CS} deals with
the stable%
\footnote{
  % a) adjective "stable" indicates that neither inaccurate observations nor a sparsity defect result in huge recovery errors
  % book:Foucart2013, Chapter 1: An Invitation to Compressive Sensing / Sect. 1.1: What is Compressive Sensing?
  % Stability.
  % - COMPRESSIVE SENSING FEATURES ANOTHER CRUCIAL ASPECT, namely,
  %   ITS RECONSTRUCTION ALGORITHMS ARE STABLE. (p. 7)
  % - This means that THE RECONSTRUCTION ERROR STAYS UNDER CONTROL when
  %   THE VECTORS ARE NOT EXACTLY SPARSE and when
  %   THE MEASUREMENTS y ARE SLIGHTLY INACCURATE. (pp. 7, 8)
  % - Without the stability requirement,
  %   the compressive sensing problem would be swiftly resolved and would not present much interest since
  %   MOST PRACTICAL APPLICATIONS INVOLVE NOISE AND COMPRESSIBILITY RATHER THAN SPARSITY. (p. 8)
  The adjective \adjective{stable} indicates that
  neither inaccurate observations nor
  a sparsity defect result in
  huge recovery errors
  \cite[7, 8]{book:Foucart2013}.
}
recovery of
% 1.) high-dimensional vector to be recovered
a high-dimensional vector
$\vect{x} \in \C^{ N }$ from
% 2.) low-dimensional vector of potentially corrupted observations
a low-dimensional vector of
potentially corrupted observations
$\vect{y}^{(\eta)} \in \C^{ M }$, $M \ll N$.
% b) known nonadaptive linear map \Phi from the high-dimensional space of vectors to be recovered to the low-dimensional space of observations provides the pristine observations
% article:KutyniokGAMM2013: Theory and Applications of Compressed Sensing
% 1 Introduction
% - The key idea of compressed sensing is to recover a sparse signal from VERY FEW NONADAPTIVE, LINEAR MEASUREMENTS by convex optimization. (p. 79)
Both vectors satisfy
the underdetermined linear algebraic system
\begin{equation}
 %--------------------------------------------------------------------------------------------------------------
 % underdetermined linear algebraic system / potentially corrupted observations
 %--------------------------------------------------------------------------------------------------------------
  \vect{y}^{(\eta)}
  =
  \mat{\Phi}
  \vect{x}
  +
  \vectsym{\eta},
 \label{eqn:cs_math_prob_general_obs_error}
\end{equation}
where
% 1.) known matrix representing the observation process
the known matrix
$\mat{\Phi} \in \C^{ M \times N }$ represents
the nonadaptive observation process, and
% 2.) unknown additive errors of bounded l2-norm
\TODO{upper bound is known!}
the unknown vector
$\vectsym{\eta} \in \C^{ M }$ denotes
additive errors of
bounded $\ell_{2}$-norm
$\tnorm{ \vectsym{\eta} }{2} \leq \eta$.

%---------------------------------------------------------------------------------------------------------------
% 2.) CS as a regularization method based on sparsity
%---------------------------------------------------------------------------------------------------------------
% a) basic linear algebra either negates the existence of any solution to the underdetermined linear algebraic system or predicts infinitely many solutions
Since
basic linear algebra either negates
the existence of
any solution to
% 1.) underdetermined linear algebraic system
the underdetermined linear algebraic system
\eqref{eqn:cs_math_prob_general_obs_error} or predicts
infinitely many solutions,
% b) CS replaces the identity by an inequality using the known upper bound on the l2-error and postulates that a known dictionary of structural building blocks represents the high-dimensional vector almost sparsely
% article:TroppPIEEE2010: Computational Methods for Sparse Solution of Linear Inverse Problems
% I. INTRODUCTION
% - In most applications, these problems [linear inverse problems] are
%   ILL-CONDITIONED OR UNDERDETERMINED, so
%   one MUST APPLY ADDITIONAL REGULARIZING CONSTRAINTS in order to obtain interesting or useful solutions. (p. 948)
% - Over the last two decades,
%   SPARSITY CONSTRAINTS HAVE EMERGED AS A FUNDAMENTAL TYPE OF REGULARIZER. (p. 948)
% - THIS APPROACH SEEKS
%   AN APPROXIMATE SOLUTION TO A LINEAR SYSTEM WHILE REQUIRING THAT
%   THE UNKNOWN HAS FEW NONZERO ENTRIES RELATIVE TO ITS DIMENSION
%   [Find sparse \vect{x} such that \mat{\Phi} \vect{x} \approx \vect{u}]
%   where \vect{u} is a target signal and \mat{\Phi} is a known matrix. (p. 948)
% - Generically, THIS FORMULATION IS REFERRED TO AS SPARSE APPROXIMATION [1]. (p. 948)
% - COMPRESSIVE SAMPLING REFERS TO A SPECIFIC TYPE OF SPARSE APPROXIMATION PROBLEM FIRST STUDIED IN [2] and [3]. (p. 948)
%   [2] article:CandesITIT2006_1, [3] article:DonohoITIT2006
\ac{CS} replaces
% 1.) CS replaces the identity by an inequality using the known upper bound on the l2-norm of the additive errors
the identity by
an inequality using
the known upper bound on
the $\ell_{2}$-norm of
the additive errors and postulates that
% 2.) CS postulates that a known dictionary of structural building blocks represents the high-dimensional vector almost sparsely
a known dictionary of
structural building blocks, e.g.
an orthonormal basis or
a frame, represents
the high-dimensional vector almost sparsely
\cite{article:TroppPIEEE2010}.
% c) latter constraint effectively reduces the total number of unknown components to a relatively small number of unknown coefficients associated with the relevant structural building blocks
The latter constraint effectively reduces
the total number of
unknown components to
a relatively small number of
unknown coefficients associated with
the relevant structural building blocks.
% d) identification of these building blocks and the subsequent estimation of their coefficients then enable the approximate recovery of the high-dimensional vector
The identification of
these building blocks and
the subsequent estimation of
their coefficients then enable
the approximate recovery of
the high-dimensional vector.
% e) both measures render CS a unique type of regularization method
% book:Hansen2010, Chapter 1: Introduction and Motivation
% book:Hansen1998, Sect. 1.3: Prelude to Regularization
Both measures render
\ac{CS} a unique type of
regularization method
(cf. e.g.
\cite[Chapt. 1]{book:Hansen2010},
\cite[Sect. 1.3]{book:Hansen1998}%
).

%---------------------------------------------------------------------------------------------------------------
% 3.) nearly-sparse representation
%---------------------------------------------------------------------------------------------------------------
% a) column vectors \vectsym{\psi}_{n} of the unitary matrix \mat{\Psi} define the admissible structural building blocks
Let
the column vectors
$\vectsym{\psi}_{n} \in \C^{ N }$,
$n \in \setcons{ N }$, of
the unitary matrix
$\mat{\Psi} \in \C^{ N \times N }$, which represents
a suitable orthonormal basis of
$\C^{ N }$, e.g.
the \name{Fourier},
a wavelet, or
the canonical basis, define
the admissible structural building blocks.
%, i.e. $\mat{\Psi} \herm{ \mat{\Psi} } = \herm{ \mat{\Psi} } \mat{\Psi} = \mat{I}$
% b) vector of transform coefficients constitutes a nearly-sparse representation of the high-dimensional vector, if the sorted absolute values of its components decay rapidly
The vector of
transform coefficients
\begin{equation}
 %--------------------------------------------------------------------------------------------------------------
 % nearly-sparse representation / vector of transform coefficients
 %--------------------------------------------------------------------------------------------------------------
  \vectsym{\theta}
  =
  \herm{ \mat{\Psi} }
  \vect{x}
 \label{eqn:def_transform_coefficients}
\end{equation}
constitutes
a nearly-sparse representation of
%\footnote{
  % paper avoids the adjective "compressible" to prevent confusion with the acoustic material parameter of "compressibility"
  % article:TroppPIEEE2010: Computational Methods for Sparse Solution of Linear Inverse Problems
  % I. INTRODUCTION / A. Formulations
  % - IN PRACTICE, SIGNALS TEND TO BE COMPRESSIBLE, RATHER THAN SPARSE. (p. 949)
  % - Mathematically,
  %   A COMPRESSIBLE SIGNAL HAS A REPRESENTATION WHOSE ENTRIES DECAY RAPIDLY WHEN SORTED IN ORDER OF DECREASING MAGNITUDE. (p. 949)
%  This paper avoids
%  the adjective \adjective{compressible} to
%  prevent confusion with
%  the acoustic material parameter of
%  \term{compressibility}.
%} of
the high-dimensional vector, if
the sorted absolute values of
its components decay rapidly.
% c) exact indices of these significant components are typically unknown a priori
The exact indices of
the significant components, which exceed
a specified absolute value, however, are
typically unknown \emph{a priori}.
% e) exactly sparse representations constitute excellent approximations of nearly-sparse representations
% article:TroppPIEEE2010: Computational Methods for Sparse Solution of Linear Inverse Problems
% I. INTRODUCTION / A. Formulations
% - COMPRESSIBLE SIGNALS ARE WELL APPROXIMATED BY SPARSE SIGNALS, so
%   the sparse approximation framework applies to this class. (p. 949)
% Exactly-sparse representations constitute
% excellent approximations of
% nearly-sparse representations.

%---------------------------------------------------------------------------------------------------------------
% 4.) CS problem and sparsity-promoting methods for its solution
%---------------------------------------------------------------------------------------------------------------
% a) insertions of the nearly-sparse representation and the sensing matrix into the underdetermined linear algebraic system
The insertions of
% 1.) nearly-sparse representation
the nearly-sparse representation
\eqref{eqn:def_transform_coefficients} and
% 2.) sensing matrix
the sensing matrix
\begin{equation}
 %--------------------------------------------------------------------------------------------------------------
 % sensing matrix
 %--------------------------------------------------------------------------------------------------------------
  \mat{A}
  =
  \mat{\Phi}
  \mat{\Psi},
 \label{eqn:cs_math_prob_general_sensing_matrix}
\end{equation}
which
% 3.) sensing matrix does not include any zero columns
% article:KutyniokGAMM2013: Theory and applications of compressed sensing
% 1 Introduction / 1.1 The Compressed Sensing Problem
% - Throughout we will always assume that
%   m < n and that
%   A DOES NOT POSSESS ANY ZERO COLUMNS, even if not explicitly mentioned. (p. 80)
is assumed not to contain
any zero columns, i.e.
$\vect{a}_{n} \in \C^{ M } \setminus \{ \vect{0} \}$ for
all $n \in \setcons{ N }$, into
% 4.) underdetermined linear algebraic system
the underdetermined linear algebraic system
\eqref{eqn:cs_math_prob_general_obs_error} yield
\begin{equation}
 %--------------------------------------------------------------------------------------------------------------
 % corrupted observations from the nearly-sparse representation
 %--------------------------------------------------------------------------------------------------------------
  \vect{y}^{(\eta)}
  =
  \underbrace{
    \mat{\Phi}
    \mat{\Psi}
  }_{ = \mat{A} }
  \vectsym{\theta}
  +
  \vectsym{\eta}
  =
  \mat{A}
  \vectsym{\theta}
  +
  \vectsym{\eta}.
 \label{eqn:cs_math_prob_general_obs_trans_coef_error}
\end{equation}
% b) CS problem associated with the corrupted observations
The associated \ac{CS} problem reads
\begin{equation}
\begin{alignedat}{2}
 %--------------------------------------------------------------------------------------------------------------
 % CS problem associated with the corrupted observations
 %--------------------------------------------------------------------------------------------------------------
  &
  \text{Recover}
  &
  \text{nearly-sparse }
  \vectsym{\theta}
  \in
  \C^{ N }\\
  &
  \text{subject to}
  \quad
  &
  \dnorm{ \vect{y}^{(\eta)} - \mat{A} \vectsym{\theta} }{2}{1}
  &\leq
  \eta
\end{alignedat}
\label{eqn:cs_math_prob_general}
\end{equation}
and
the methods for
its stable solution coalesce into
the sparsity-promoting $\ell_{q}$-minimization method
\cite[(1.11) and (1.12)]{book:Eldar2012},
\cite[(2) and (11)]{article:TroppPIEEE2010},
\cite[($\text{P}_{q, \theta}$)]{article:FoucartACHA2009}%
\footnote{
  % article:FoucartACHA2010: A note on guaranteed sparse recovery via l1-minimization
  % - Let us note that THE RESULTS OF [2, 5, 1], EVEN THOUGH STATED FOR \R RATHER THAN \C, ARE VALID IN BOTH SETTINGS. (p. 97)
  %   [5] article:FoucartACHA2009
  The first author confirms
  the validity of
  the results, which are exclusively stated for
  $\R^{N}$, for
  $\C^{N}$ in
  \cite{article:FoucartACHA2010}.
}
\begin{equation}
\begin{alignedat}{2}
 %--------------------------------------------------------------------------------------------------------------
 % sparsity-promoting lq-minimization method
 %--------------------------------------------------------------------------------------------------------------
  \hat{\vectsym{\theta}}^{(q, \eta)}
  &\in
  \underset{ \tilde{\vectsym{\theta}} \in \C^{ N } }{ \arg\min }
  \dnorm{ \tilde{\vectsym{\theta}} }{q}{1}\\
  &
  \mspace{24.5mu}
  \text{subject to}
  &
  \dnorm{ \vect{y}^{(\eta)} - \mat{A} \tilde{\vectsym{\theta}} }{2}{1}
  &\leq
  \eta,
\end{alignedat}
\tag{$\text{P}_{q, \eta}$}
\label{eqn:cs_lq_minimization}
\end{equation}
where
% 1.) parameter q of quasinorm
the parameter
$q \in [ 0; 1 ]$ determines
the type of
optimization method.
% d) parameter q = 1 induces the convex l1-minimization method
% book:Eldar2012, Chapter 1: Introduction to compressed sensing / Sect. 1.5 Signal recovery via l1 minimization
% - One avenue for translating this problem into something more tractable is to REPLACE
%   \norm{ . }{0} WITH ITS CONVEX APPROXIMATION \norm{ . }{1}. (p. 27)
% - Specifically, we consider
%   [ \hat{x} = argmin \norm{ z }{1} subject to z \in B(y) ] (1.12). (p. 27)
% - Provided that B(y) is convex, (1.12) IS COMPUTATIONALLY FEASIBLE. (p. 27)
% article:TroppPIEEE2010: Computational Methods for Sparse Solution of Linear Inverse Problems
% III. OPTIMIZATION
% - Another fundamental approach to sparse approximation replaces
%   the combinatorial l0 function in the mathematical programs from Section I-A with
%   the l1-norm, yielding CONVEX OPTIMIZATION PROBLEMS THAT ADMIT TRACTABLE ALGORITHMS. (p. 953)
% - In a concrete sense [48], the l1-NORM IS THE CLOSEST CONVEX FUNCTION TO THE l0 FUNCTION,
%   so this ``relaxation'' is quite natural. (p. 953)
%   [48] R. Gribonval and M. Nielsen, "Highly sparse representations from dictionaries are unique and independent of the sparseness measure," Aalborg Univ., Aalborg, Denmark, Tech. Rep., Oct. 2003.
% article:ChartrandISPL2007: Exact Reconstruction of Sparse Signals via Nonconvex Minimization
% I. INTRODUCTION
% - Importantly, this result [exact recovery w/ very high probability] continues to hold if
%   the l0 norm is replaced by the l1 norm, RESULTING IN A CONVEX PROBLEM as follows:
%   [min_{ \vect{u} } \norm{ \vect{u} }{1}, subject to \mat{Phi} \vect{u} = \vect{y}] (2) (p. 707)
The parameter $q = 1$ induces
% 1.) convex l1-minimization method
the convex $\ell_{1}$-minimization method, whose
implementation permits
computationally efficient algorithms, whereas
the half-open parameter interval
$q \in [ 0; 1 )$ induces
% 1.) nonconvex lq-minimization method
the nonconvex $\ell_{q}$-minimization method, whose
global intractability necessitates
local approximations.

%---------------------------------------------------------------------------------------------------------------
% 5.) sufficient conditions on the sensing matrix for the stable recovery and their verifiability
%---------------------------------------------------------------------------------------------------------------
% a) multiple sufficient conditions on the sensing matrix ensure the stable recovery of the nearly-sparse representation in the CS problem by the sparsity-promoting lq-minimization method
Multiple sufficient conditions on
% 1.) sensing matrix
the sensing matrix
\eqref{eqn:cs_math_prob_general_sensing_matrix} ensure
the stable recovery of
% 2.) nearly-sparse representation
the nearly-sparse representation
\eqref{eqn:def_transform_coefficients} in
% 3.) CS problem
the \ac{CS} problem
\eqref{eqn:cs_math_prob_general} by
% 4.) sparsity-promoting lq-minimization method
the sparsity-promoting $\ell_{q}$-minimization method
\eqref{eqn:cs_lq_minimization}.
% b) sufficient conditions specify upper bounds for various characteristic measures quantifying the suitability of the sensing matrix
They specify
% 1.) upper bounds
upper bounds for
% 2.) various characteristic measures
various characteristic measures quantifying
% 3.) suitability
the suitability of
% 4.) sensing matrix
the sensing matrix
\eqref{eqn:cs_math_prob_general_sensing_matrix}, e.g.
the null space constants
\cite[Def. 4.21]{book:Foucart2013},
\cite[Def. 1.2]{book:Eldar2012},
% 6.) restricted isometry ratio
% article:FoucartACHA2009: Sparsest solutions of underdetermined linear systems via lq-minimization for 0 < q \leq 1
% 2. Exact recovery via lq-minimization
% - Our results are to be stated in terms of a quantity invariant under the change A ← cA, namely
%   [ \gamma_{2s} := {alpha_{2s}}^{-2} {\beta_{2s}}^{2} \geq 1 ]. (p. 396)
the restricted isometry ratio
\cite{article:FoucartACHA2009}, and
% 7.) restricted isometry constant (RIC)
% article:FoucartACHA2010: A note on guaranteed sparse recovery via l1-minimization
% - A much favored tool in the analysis of (P_{1}) has been
%   the RESTRICTED ISOMETRY CONSTANTS δ_{k} of the m × N measurement matrix \mat{A}, defined as
%   the SMALLEST POSITIVE CONSTANTS δ such that
%   [(1 - δ) \norm{ \vect{z} }{2}^{2} \leq \norm{ \mat{A} \vect{z} }{2}^{2} \leq (1 + δ) \norm{ \vect{z} }{2}^{2}] for
%   all k-sparse vector \vect{z} \in \C^{ N }. (1) (p. 97)
% - This notion was introduced by Candès and Tao in [3], where
%   it was shown that all s-sparse vectors are recovered as unique solutions of (P_{1}) as soon as δ_{3s} + 3 δ_{4s} < 2. (p. 97)
%   [3] article:CandesITIT2005
% - Candès showed in [2] that s-sparse recovery is guaranteed as soon as δ_{2s} < √2 − 1 ≈ 0.4142. (p. 97)
%   [2] article:CandesCRAS2008
% - The purpose of this note is to show that the threshold on δ_{2s} can be pushed further —
%   we point out that Davies and Gribonval proved that it cannot be pushed further than 1 / √2 ≈ 0.7071 in [4].
%   [4] article:DaviesITIT2009
% article:CandesCRAS2008: The Restricted Isometry Property and Its Implications For Compressed Sensing [May]
% article:CandesSPM2008: An Introduction To Compressive Sampling [Mar.]
the \ac{RIC}
\cite{article:FoucartACHA2010,article:CandesCRAS2008,article:CandesSPM2008}.
The evaluation of
% 1.) various characteristic measures
these measures for
% 2.) sensing matrix
a fixed sensing matrix
\eqref{eqn:cs_math_prob_general_sensing_matrix}, however, is
% 3.) intractable combinatorial problem
a computationally-intractable combinatorial problem
\cite{article:TillmannITIT2014}.
% d) complexity [evaluation of measures] impedes both the deterministic construction of high-dimensional sensing matrices and the verification of the sufficient conditions
% article:KutyniokGAMM2013: Theory and applications of compressed sensing
% - It is still an OPEN QUESTION (cf. Section 4 for more details) whether
%   DETERMINISTIC MATRICES CAN BE CAREFULLY CONSTRUCTED TO HAVE SIMILAR PROPERTIES with respect to compressed sensing problems.
Its complexity impedes both
% 1.) deterministic construction of high-dimensional sensing matrices
the deterministic construction of
high-dimensional sensing matrices
\eqref{eqn:cs_math_prob_general_sensing_matrix}, whose
characteristic measures meet
the upper bounds, and
% 2.) verification of the sufficient conditions
the verification of
the sufficient conditions.
% e) evaluation of a characteristic measure named worst-case coherence is relatively simple
% book:Foucart2013, Chapter 5: Coherence / Sect. 5.1 Definitions and Basic Properties
% - We start with the definition of the COHERENCE OF A MATRIX. (p. 111)
% - Definition 5.1.
%	- The COHERENCE µ = µ(A) of the matrix A is defined as (5.1). (p. 111)
% book:Foucart2013, Chapter 5: Coherence
% - In compressive sensing, the analysis of recovery algorithms usually involves
%   a QUANTITY THAT MEASURES THE SUITABILITY OF THE MEASUREMENT MATRIX. (p. 111)
% - THE COHERENCE IS A VERY SIMPLE SUCH MEASURE OF QUALITY. (p. 111)
% - In general, THE SMALLER THE COHERENCE, THE BETTER THE RECOVERY ALGORITHMS PERFORM. (p. 111)
% - In Sects. 5.3, 5.4,and 5.5, we give
%   some SUFFICIENT CONDITIONS EXPRESSED IN TERMS OF THE COHERENCE that guarantee
%   the success of orthogonal matching pursuit, basis pursuit, and thresholding algorithms. (p. 111)
% article:KutyniokGAMM2013: Theory and applications of compressed sensing
% Sect. 3: Conditions for Sparse Recovery / Sect. 3.2: Sufficient Conditions / Sect. 3.2.1: Mutual Coherence
% - The MUTUAL COHERENCE OF A MATRIX, initially introduced in [21], measures the smallest angle between each pair of its columns. (p. 88)
%   [21] D.L. Donoho and X. Huo. Uncertainty principles and ideal atomic decomposition. IEEE Trans. Inform. Theory, 47:2845–2862, 2001.
% - The maximal mutual coherence of a matrix certainly equals 1 in the case when two columns are linearly dependent. (p. 88)
In contrast,
% 1.) evaluation
the evaluation of
% 2.) characteristic measure
a characteristic measure named
% 3.) worst-case coherence
worst-case coherence
\cite[Def. 5.1]{book:Foucart2013} is
% 4.) relatively simple
relatively simple.
% f) worst-case coherence loosely bounds from above the RIC and ensures the RIP for s nonzero components, if the number of observations meets M \in \bigomega{ s^{2} }
% article:TillmannITIT2014: The Computational Complexity of the Restricted Isometry Property, the Nullspace Property, and Related Concepts in Compressed Sensing
% - However, the sparsity levels for which the mutual coherence can guarantee recoverability are
%   quite often too small to be of practical use.
% - This emphasizes the importance of other concepts.
% book:Foucart2013, Chapter 6: Restricted Isometry Property / Sect. 6.1 Definitions and Basic Properties
% - As with the coherence, small restricted isometry constants are desired. (p. 133)
% - Proposition 6.2.:
%	- If the matrix A has l2-normalized columns a_{1}, ..., a_{N}, i.e., \norm{ a_{j} }{2} = 1 for all j ∈ [N], then
%	  \delta_{1} = 0, \delta_{2} = \mu, \delta_{s} \leq \mu_{1} (s - 1) \leq \mu (s - 1), s \geq 2. (p. 134)
% book:Foucart2013, Chapter 6: Restricted Isometry Property
% - THE COHERENCE IS A SIMPLE AND USEFUL MEASURE OF THE QUALITY OF A MEASUREMENT MATRIX.
%   However, THE LOWER BOUND ON THE COHERENCE in Theorem 5.7 limits
%   the performance analysis of recovery algorithms to RATHER SMALL SPARSITY LEVELS. (p. 133)
% - A finer measure of the quality of a measurement matrix is needed to overcome this limitation. (p. 133)
% - This is provided by the concept of restricted isometry property, also known as uniform uncertainty principle. (p. 133)
It loosely bounds from above
% 1.) restricted isometry constant (RIC)
the \ac{RIC}
\cite[Prop. 6.2]{book:Foucart2013} and, by
% 2.) Welch lower bound
% article:KutyniokGAMM2013: Theory and applications of compressed sensing
% Sect. 3: Conditions for Sparse Recovery / Sect. 3.2: Sufficient Conditions / Sect. 3.2.1: Mutual Coherence
% - The LOWER BOUND presented in the next result, also KNOWN AS THE WELCH BOUND, is more interesting. (p. 88)
% - Lemma 3.7:
%	- Let A be an m×n matrix. Then we have [\mu(\mat{A}) \in [ \sqrt{ \frac{ n - m }{ m (n - 1) } } ; 1 ]]. (p. 88)
its \name{Welch} lower bound
\cite[Thm. 5.7]{book:Foucart2013},
\cite[Lem. 3.7]{article:KutyniokGAMM2013}, ensures
% 3.) restricted isometry property (RIP)
the \ac{RIP} for
% 4.) s-sparse representations
$s$ nonzero components, if
% 5.) number of observations
the number of
observations meets
% 6.) M \in \bigomega{ s^{2} }
$M \in \bigomega{ s^{2} }$.
% g) certain types of random sensing matrices also obey the RIP with very high probability, if the number of observations is sufficiently large
% book:Foucart2013, Chapter 1: An Invitation to Compressive Sensing / Sect. 1.1: What is Compressive Sensing?
% - A BREAKTHROUGH IS ACHIEVED BY RESORTING TO RANDOM MATRICES - THIS DISCOVERY CAN BE VIEWED AS THE BIRTH OF COMPRESSIVE SENSING. (p. 6)
% article:TroppPIEEE2010: Computational Methods for Sparse Solution of Linear Inverse Problems
% I. INTRODUCTION / C. Verifying Correctness
% - CERTAIN RANDOM MATRICES, HOWEVER, SATISFY MUCH STRONGER RIP BOUNDS WITH HIGH PROBABILITY. (p. 950)
% - This fact explains the BENEFIT OF RANDOMNESS IN COMPRESSIVE SAMPLING. (p. 950)
% - Establishing the RIP for a random matrix requires techniques more sophisticated than
%   the simple coherence arguments; see [14] for discussion. (p. 950)
Fortunately,
certain types of
% 1.) random sensing matrices
random sensing matrices
\eqref{eqn:cs_math_prob_general_sensing_matrix} also obey
the \ac{RIP} with
very high probability, if
the number of
observations is
sufficiently large
\cite[6]{book:Foucart2013},
\cite{article:TroppPIEEE2010}.
Realizations of
\ac{IID} random variables governed by
certain distributions, e.g.
Gaussian or
\name{Bernoulli}, as
entries
\cite[Thm. 5.2]{article:BaraniukCA2008}%
\footnote{
  The result holds universally under any
  unitary transform, i.e.
  the right multiplication of
  % 1.) random observation process
  a random observation process by
  % 2.) N_{\text{lat}}-dimensional DFT matrix
  any complex-valued unitary
  $N_{\text{lat}} \times N_{\text{lat}}$ matrix preserves
  the \ac{RIP}
  \cite[222]{coll:Fornasier2015},
  %\cite[280, 281]{book:Foucart2013},
  \cite[Sect. 6]{article:BaraniukCA2008}.
} and
% 2.) randomly and uniformly chosen scaled rows of a Fourier basis
% article:TroppPIEEE2010: Computational Methods for Sparse Solution of Linear Inverse Problems
% I. INTRODUCTION / C. Verifying Correctness
% - For GAUSSIAN AND BERNOULLI MATRICES, RIP holds when K \approx m / log( N / m ). (p. 950)
% - For MORE STRUCTURED MATRICES, such as a RANDOM SECTION OF A DISCRETE FOURIER TRANSFORM,
%   RIP often holds when K \approx m / log^{p}( N ) for a small integer p. (p. 950)
% article:RudelsonCPAM2008: On Sparse Reconstruction from Fourier and Gaussian Measurements
%
\TODO{scaled?}
randomly and uniformly chosen scaled rows of
a \name{Fourier} basis
\cite[Thm. 3.3]{article:RudelsonCPAM2008}, for example, require
$M \in \bigomega{ s \ln( N / s ) }$ and
$M \in \bigomega{ s \ln^{4}( N ) }$ observations,
respectively.
% i) orders of growth are significantly better than that guaranteed by the worst-case coherence
These orders of
growth are
almost linear in $s$ and, thus, significantly better than
that guaranteed by
the worst-case coherence.

%---------------------------------------------------------------------------------------------------------------
% 6.) consequences of the underlying physical models for the sensing matrix / random structure
%---------------------------------------------------------------------------------------------------------------
% a) individual entries of the observation process and those of the sensing matrix depend on the underlying physical model
% book:Foucart2013, Chapter 12: Random Sampling in Bounded Orthonormal Systems
% - While this [subgaussian random matrices are optimal] is a very important insight for the theory,
%   THE USE OF SUCH “COMPLETELY RANDOM” MATRICES, WHERE ALL ENTRIES ARE INDEPENDENT,
%   IS LIMITED FOR PRACTICAL PURPOSES. (p. 367)
% - However, structure is important for several reasons:
%	1.) APPLICATIONS MAY IMPOSE CERTAIN STRUCTURE ON THE MEASUREMENT MATRIX DUE TO
%	    PHYSICAL OR OTHER CONSTRAINTS. (p. 367)
% article:KutyniokGAMM2013: Theory and applications of compressed sensing
% - Moreover,
%   MOST APPLICATIONS DO NOT ALLOW FOR A FREE CHOICE OF THE SENSING MATRIX AND
%   ENFORCE A PARTICULARLY STRUCTURED MATRIX.
% - Exemplary situations are
%   the application of data separation, in which
%   the sensing matrix has to consist of two or more orthonormal bases or frames [32, Chapter 11], or
%   HIGH RESOLUTION RADAR, for which
%   THE SENSING MATRIX HAS TO BEAR A PARTICULAR TIME-FREQUENCY STRUCTURE [38].
In
medical imaging,
% 1.) individual entries of the observation process
the individual entries of
the observation process and, in conjunction with
the orthonormal basis,
% 2.) individual entries of the sensing matrix
those of
the sensing matrix
\eqref{eqn:cs_math_prob_general_sensing_matrix} depend on
the underlying physical model.
% b) various imaging parameters controlling the instrumentation enable the systematic manipulation of groups of these entries within technologically and physiologically tolerable limits
% book:Foucart2013, Chapter 12: Random Sampling in Bounded Orthonormal Systems
% - By a STRUCTURED RANDOM MATRIX, we mean a STRUCTURED MATRIX THAT IS GENERATED BY A RANDOM CHOICE OF PARAMETERS. (p. 367)
Various imaging parameters controlling
the instrumentation, however, enable
the systematic manipulation of
groups of
these entries within
technologically and
physiologically tolerable limits.
% c) randomizations of these degrees of freedom [imaging parameters within tolerable limits] generate sensing matrices with random structures
The randomizations of
these degrees of
freedom generate
% 1.) sensing matrices with random structures
sensing matrices
\eqref{eqn:cs_math_prob_general_sensing_matrix} with
random structures that
% 2.) random structures potentially improve the aforementioned characteristic measures
potentially improve
the aforementioned characteristic measures and, consequently, aid in
% 3.) random structures aid in meeting the associated sufficient conditions
meeting the associated sufficient conditions.
% d) degrees of freedom in the physical models underlying MRI and compressed beamforming in UI specify subsets of scaled Fourier coefficients to be processed
In fact,
the degrees of
freedom in
the physical models underlying
\ac{MRI}
\cite[11, 12]{book:Foucart2013},
\cite{article:LustigMRM2007} and
compressed beamforming in
\ac{UI}
\cite{article:ChernyakovaITUFFC2018,proc:SchiffnerIUS2016a,article:BurshteinITUFFC2016,article:ChernyakovaITUFFC2014,article:WagnerITSP2012}, for example, specify
subsets of
scaled \name{Fourier} coefficients to
be processed.
% e) random and uniform selection [Fourier coefficients] generates the aforementioned random sensing matrix meeting the RIP with relatively few observations
Their random and uniform selection generates
the aforementioned random sensing matrix meeting
the \ac{RIP} with
relatively few observations.
% f) introduction of diagonal weighting matrices into the underdetermined linear algebraic system always enables the l2-normalization of the sensing matrix's column vectors
In addition,
the introduction of
% 1.) diagonal weighting matrices
diagonal weighting matrices, whose
entries equal
% 2.) l2-norms of the sensing matrix's column vectors
the $\ell_{2}$-norms of
the sensing matrix's column vectors
$\tnorm{ \vect{a}_{n} }{2}$ or
% 3.) reciprocal l2-norms of the sensing matrix's column vectors
their reciprocals, into
% 4.) underdetermined linear algebraic system
the underdetermined linear algebraic system
\eqref{eqn:cs_math_prob_general_obs_trans_coef_error} always enables
% 5.) l2-normalization of the sensing matrix's column vectors
the $\ell_{2}$-normalization of
the sensing matrix's column vectors without violating
% 6.) mathematical equivalence
the mathematical equivalence.
% g) resulting normalized sensing matrix minimizes both the RIR and the RIC and meets the associated sufficient conditions
% book:Foucart2013, Sect. 6.1 Definitions and Basic Properties
% - As with the coherence, small restricted isometry constants are desired. (p. 133)
% - Proposition 6.2.:
%	- If the matrix A has L2-NORMALIZED COLUMNS a_{1}, ..., a_{N}, i.e., \norm{ a_{j} }{2} = 1 for all j ∈ [N], then
%	  \delta_{1} = 0, \delta_{2} = \mu, \delta_{s} \leq \mu_{1} (s - 1) \leq \mu (s - 1), s \geq 2. (p. 134)
% article:KutyniokGAMM2013: Theory and applications of compressed sensing
% - Ideally, we aim for a matrix which has high spark, low mutual coherence, and a small RIP constant.
The resulting normalized sensing matrix minimizes both
% 1.) restricted isometry ratio
the restricted isometry ratio and
% 2.) restricted isometry constant (RIC)
the \ac{RIC} for
% 3.) nearly-sparse representations
$1$-sparse representations
\eqref{eqn:def_transform_coefficients} and better conforms with
% 4.) associated sufficient conditions
%\TODO{really? minimum $2s = 2$?}
the associated sufficient conditions
\cite[Prop. 6.2]{book:Foucart2013}.

The \ac{TPSF} frequently quantifies
% 1.) coherence
the coherence of
% 2.) sensing matrices
the sensing matrices
\eqref{eqn:cs_math_prob_general_sensing_matrix} in
% 3.) medical imaging
medical imaging
(cf. e.g.
\cite{article:ProvostITMI2009,article:LustigMRM2007}%
).
% b) TPSF equals the mutual correlation coefficient of the column vectors
% article:ProvostITMI2009: The Application of Compressed Sensing for Photo-Acoustic Tomography
% VI. COMPRESSED SENSING / B. PA Forward Operator as a CS-Matrix
% - The TPSF is defined as (23). (p. 589)
% article:LustigMRM2007: Sparse MRI: The application of compressed sensing for rapid MR imaging
% THEORY / Point Spread Function and Transform Point Spread Function Analysis
% - Let \Psi be an orthogonal sparsifying transform (nonorthogonal TPSF analysis is beyond our scope and is not discussed here). (p. 1185)
% - The TPSF(i; j) IS GIVEN BY THE FOLLOWING EQUATION,
%   [ TPSF(i; j) = e_{j}^{∗} \Psi F_{u}^{∗} F_{u} \Psi^{*} e_{i} ]. [2] (p. 1185)
It equals
% 1.) mutual correlation coefficient
the mutual correlation coefficient of
% 2.) column vectors
the column vectors given by
\cite[(23)]{article:ProvostITMI2009},
\cite[(2)]{article:LustigMRM2007}
\begin{equation}
 %--------------------------------------------------------------------------------------------------------------
 % transform point spread function (TPSF)
 %--------------------------------------------------------------------------------------------------------------
  \tpsf{ \mat{A} }{ n_{1} }{ n_{2} }
  =
  \frac{
    \inprod{ \vect{a}_{ n_{1} } }{ \vect{a}_{ n_{2} } }
  }{
    \norm{ \vect{a}_{ n_{1} } }{2}
    \norm{ \vect{a}_{ n_{2} } }{2}
  }
 \label{eqn:cs_math_tpsf}
\end{equation}
for
% 3.) all pairs of indices
all $( n_{1}, n_{2} ) \in \setcons{ N }^{2}$.
% c) TPSF reduces to the PSF and exclusively quantifies the coherence of the observation process
% article:LustigMRM2007: Sparse MRI: The application of compressed sensing for rapid MR imaging
% THEORY / Point Spread Function and Transform Point Spread Function Analysis
% - Let F_{u} be the UNDERSAMPLED FOURIER OPERATOR and let e_{i} be the ith vector of the NATURAL BASIS
%   (i.e, having “1” at the ith location and zeroes elsewhere). (p. 1185)
% - Then PSF(i; j) = e_{j}^{∗} F_{u}^{∗} F_{u} e_{i} measures
%   the CONTRIBUTION OF A UNIT-INTENSITY PIXEL AT THE iTH POSITION TO A PIXEL AT THE jTH POSITION. (p. 1185)
% - Under Nyquist sampling there is no interference between pixels and PSF(i; j) | i \neq j = 0. (p. 1185)
% - UNDERSAMPLING CAUSES PIXELS TO INTERFERE and PSF(i; j) | i \neq j TO ASSUME NONZERO VALUES. (p. 1185)
% - The PSF of pure 2D random sampling, where samples are chosen at random from a Cartesian grid, offers a standard for comparison. (p. 1185)
%	-> Empirically, the real and the imaginary parts separately behave much like zero-mean random white Gaussian noise. (p. 1185)
%	-> The standard deviation of the observed SPR depends on the number, N, of samples taken and the number, D, of grid points defining the underlying image. (p. 1185)
If
% 1.) orthonormal basis
the orthonormal basis is
% 2.) canonical
canonical, i.e.
$\mat{\Psi} = \mat{I}$,
% 3.) transform point spread function (TPSF)
the \ac{TPSF}
\eqref{eqn:cs_math_tpsf} reduces to
% 4.) point spread function (PSF)
the \ac{PSF} and exclusively quantifies
% 5.) coherence
the coherence of
% 6.) observation process
the observation process
\cite{article:LustigMRM2007}.
% d) TPSF trivially attains its maximum absolute value of unity
For
% 1.) n_{1} = n_{2}
$n_{1} = n_{2}$,
% 2.) both column vectors match
both column vectors match, and
% 3.) transform point spread function (TPSF)
the \ac{TPSF}
\eqref{eqn:cs_math_tpsf} trivially attains
% 4.) maximum absolute value
its maximum absolute value of
% 5.) unity
unity.
% e) absolute value of the TPSF ideally approaches zero with noise-like statistics
% article:ProvostITMI2009: The Application of Compressed Sensing for Photo-Acoustic Tomography
% VI. COMPRESSED SENSING / B. PA Forward Operator as a CS-Matrix
% - We would like this TPSF to have the FOLLOWING PROPERTIES for m \neq n.
%	1.) First, TPSF(m, n) SHOULD BE MUCH SMALLER THAN 1 [absolute value].
%	    -> This property characterizes the fact that two different basis vectors output different measurements. (p. 589)
%	2.) Second, TPSF(m, n) SHOULD BE noise-like to characterize the fact that
%	    one measurement of a basis vector is not correlated to the measurement of another basis vector. (p. 589)
% article:LustigMRM2007: Sparse MRI: The application of compressed sensing for rapid MR imaging
% THEORY / Point Spread Function and Transform Point Spread Function Analysis
% - WE WOULD LIKE TPSF(i; j) | i \neq j TO BE AS SMALL AS POSSIBLE, AND HAVE RANDOM NOISE-LIKE STATISTICS. (p. 1185)
For
% 1.) n_{1} \neq n_{2}
$n_{1} \neq n_{2}$, however,
% 2.) both column vectors typically differ
both column vectors typically differ, and
% 3.) absolute value
the absolute value of
% 4.) transform point spread function (TPSF)
the \ac{TPSF}
\eqref{eqn:cs_math_tpsf} ideally approaches
% 5.) zero
zero with
% 6.) noise-like statistics
noise-like statistics
\cite{article:ProvostITMI2009,article:LustigMRM2007}.
% f) properties indicate the reliable discrimination of the admissible structural building blocks by the observation process and guide the sparsity-promoting lq-minimization method
These properties, which are referred to as
% 1.) incoherent aliasing
incoherent aliasing, indicate
% 2.) reliable discrimination
the reliable discrimination of
% 3.) admissible structural building blocks
the admissible structural building blocks by
% 4.) observation process
the observation process and guide
% 5.) sparsity-promoting lq-minimization method
the sparsity-promoting $\ell_{q}$-minimization method
\eqref{eqn:cs_lq_minimization}.
% g) practical evaluations of the TPSF typically fix the second index according to the expected support of the nearly-sparse representation
% article:ProvostITMI2009: The Application of Compressed Sensing for Photo-Acoustic Tomography
% VI. COMPRESSED SENSING / B. PA Forward Operator as a CS-Matrix
% - Fig. 1 shows TPSFs evaluated for ONE BASIS COEFFICIENT m and all n's: TPSF(m, n). (p. 589)
% article:LustigMRM2007: Sparse MRI: The application of compressed sensing for rapid MR imaging
% THEORY / Point Spread Function and Transform Point Spread Function Analysis
% - In this case PSF(i; j) | i \neq j looks random as illustrated in Fig. 4a.
% - An example [TPSF] using an orthogonal wavelet transform is illustrated by Fig. 4b. (p. 1185)
% article:LustigMRM2007: Sparse MRI: The application of compressed sensing for rapid MR imaging
% RESULTS / Multislice Fast Spin-Echo Brain Imaging
% - The reason is that some of the coarse-scale wavelet components in these reconstructions were not recovered correctly because of
%   the large peak interference of coarse-scale components that was documented in the TPSF theoretical analysis (see Fig. 5a). (p. 1192)
% - This is because the theoretical TPSF peak interference in such sampling scheme is significantly smaller (see Fig. 5b), which enables
%   better recovery of these components. (p. 1192)
Owing to
% 1.) high dimensionality
the high dimensionality of
% 2.) sensing matrices
the sensing matrices
\eqref{eqn:cs_math_prob_general_sensing_matrix},
% 3.) practical evaluations
practical evaluations of
% 4.) transform point spread function (TPSF)
the \ac{TPSF}
\eqref{eqn:cs_math_tpsf} usually select
% 5.) one index
one index from
% 6.) expected support
the expected support of
% 7.) nearly-sparse representation
the nearly-sparse representation
\eqref{eqn:def_transform_coefficients}, i.e.
$n_{2} \in \supp( \vectsym{\theta} )$
(cf. e.g.
\cite[Fig. 1]{article:ProvostITMI2009},
\cite[Figs. 4 and 5]{article:LustigMRM2007}%
).

\section{Linear Physical Model for the Pulse-Echo Measurement Process}
\label{sec:linear_model}
The proposed $d$-dimensional, $d \in \{ 2, 3 \}$, physical model uses
% 1.) interactions of arbitrary incident waves with the human body
% article:JensenJASA1991: A model for the propagation and scattering of ultrasound in tissue
% Abstract
% - Analytic expressions are found in the literature for a number of transducers, and
%   ANY TRANSDUCER EXCITATION CAN BE INCORPORATED INTO THE MODEL. (p. 182)
% INTRODUCTION
% - NO RESTRICTIONS are enforced on the transducer geometry or its excitation, and
%   ANALYTIC EXPRESSIONS FOR A NUMBER OF GEOMETRIES CAN BE INCORPORATED INTO THE MODEL. (p. 182)
the interactions of
arbitrary incident waves with
the human body to predict
% 2.) RF voltage signals provided by the instrumentation
the \ac{RF} voltage signals provided by
the instrumentation.
% c) instrumentation consists of a planar transducer array, connecting cables, and driving and receiving electric circuits
%The instrumentation consists of
% 1.) planar transducer array
%a planar transducer array,
% 2.) connecting cables
%connecting cables, and
% 3.) driving and receiving electric circuits
%driving and
%receiving electric circuits.
% d) interactions include diffraction, single monopole scattering, and the combined effects of power-law absorption and dispersion
The interactions include
% 1.) diffraction
diffraction,
% 2.) single monopole scattering within the Born approximation
single monopole scattering, and
% 3.) combined effects of power-law absorption and dispersion
the combined effects of
power-law absorption and
dispersion.
A rigid baffle on
the hyperplane $r_{d} = 0$, which embeds
% 1.) vibrating faces of the transducer elements
the vibrating faces of
the transducer elements and bounds
% 2.) half-space r_{d} > 0
the half-space $r_{d} > 0$, approximates
the complex boundary conditions.
The temporal \name{Fourier} domain simplifies
% 1.) mathematical formulation of the model
the mathematical formulation of
the model and enables
% 2.) inclusions of absorption and the bandpass transfer behavior
%the inclusions of
% 3.) combined effects of power-law absorption and dispersion
%absorption and
% 4.) bandpass transfer behavior
%the bandpass transfer behavior.
% b) temporal Fourier domain additionally enables the parallel processing of distinct discrete frequencies in numerical evaluations
%It additionally enables
% 1.) parallel processing of distinct discrete frequencies in numerical evaluations
the parallel processing of
distinct discrete frequencies in
numerical evaluations.
\subsection{Pulse-Echo Measurement Process}
\label{subsec:lin_mod_measurement_process}
%%%%%%%%%%%%%%%%%%%%%%%%%%%%%%%%%%%%%%%%%%%%%%%%%%%%%%%%%%%%%%%%%%%%%%%%%%%%%%%%%%%%%%%%%%%%%%%%%%%%%%%%%%%%%%%%
% graphic: pulse-echo measurement process (two-dimensional Euclidean space)
%%%%%%%%%%%%%%%%%%%%%%%%%%%%%%%%%%%%%%%%%%%%%%%%%%%%%%%%%%%%%%%%%%%%%%%%%%%%%%%%%%%%%%%%%%%%%%%%%%%%%%%%%%%%%%%%
%
\begin{figure}[t!]
 \centering%
  \input{linear_model/figures/latex/lin_mod_scan_configuration_fov.tex}
 \caption{}
 \label{fig:V}
\end{figure}
C
{% a) illustration of the pulse-echo measurement process in the two-dimensional Euclidean space
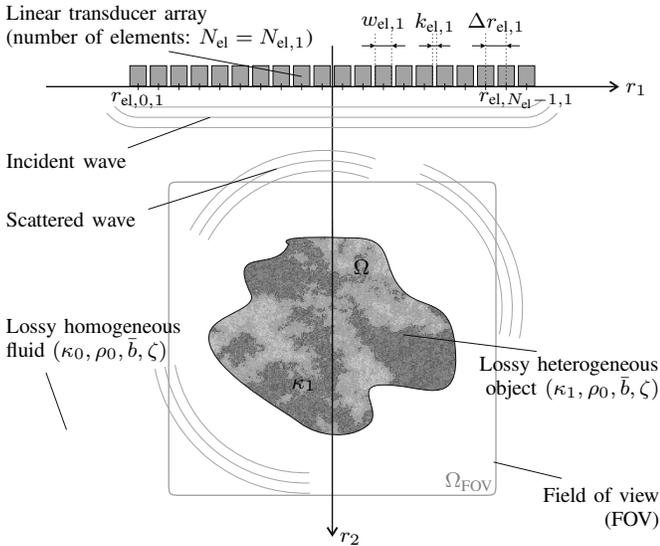
 Pulse-echo measurement process in
 the two-dimensional Euclidean space, i.e. $d = 2$.
 % b) linear transducer array emits a broadband incident wave into a lossy homogeneous fluid
 A linear transducer array emits
 % 1.) broadband incident wave
 a broadband incident wave into
 % 2.) lossy homogeneous fluid with the unperturbed values of the compressibility \kappa_{0} and the mass density \rho_{0}
 a lossy homogeneous fluid with
 the unperturbed values of
 % 3.) unperturbed compressibility
 the compressibility
 $\kappa_{0} \in \Rplus$ and
 % 4.) mass density
 the mass density
 $\rho_{0} \in \Rplus$.
 % c) broadband incident wave penetrates an embedded lossy heterogeneous object, and its interactions with the unperturbed compressibility induce a scattered wave
 This wave penetrates
 % 1.) embedded lossy heterogeneous object
 an embedded lossy heterogeneous object represented by
 the bounded set
 $\Omega \subset \{ \vect{r} \in \R^{d}: r_{d} > 0 \}$, and
 its interactions with
 % 2.) unperturbed compressibility
 the unperturbed compressibility
 $\kappa_{1}: \Omega \mapsto \Rplus$ induce
 % 3.) scattered wave
 a scattered wave.
 % d) a portion of the scattered wave mechanically excites the faces of the array elements
 % TODO: transduce
 A portion of
 the latter mechanically excites
 % 1.) faces of the array elements
 the faces of
 the array elements that generate
 % 2.) RF voltage signals
 \ac{RF} voltage signals.
 % e) RF voltage signals enable the imaging of the specified FOV represented by the bounded set \Omega_{\text{FOV}}
 These enable
 the imaging of
 the specified \ac{FOV} represented by
 the bounded set
 $\Omega_{\text{FOV}} \subset \{ \vect{r} \in \R^{d}: r_{d} > 0 \}$.
}%
{lin_mod_scan_configuration}

%%%%%%%%%%%%%%%%%%%%%%%%%%%%%%%%%%%%%%%%%%%%%%%%%%%%%%%%%%%%%%%%%%%%%%%%%%%%%%%%%%%%%%%%%%%%%%%%%%%%%%%%%%%%%%%%
% table: geometric and electromechanical parameters of the instrumentation
%%%%%%%%%%%%%%%%%%%%%%%%%%%%%%%%%%%%%%%%%%%%%%%%%%%%%%%%%%%%%%%%%%%%%%%%%%%%%%%%%%%%%%%%%%%%%%%%%%%%%%%%%%%%%%%%
\begin{table*}[t!]
 \centering
 \caption{%
  Geometric and
  electromechanical parameters of
  the instrumentation for
  all $\delta \in \setcons{ d - 1 }$,
  $m \in \setconsnonneg{ N_{\text{el}} - 1 }$,
  $l \in \setsymbol{L}_{ \text{BP} }^{(n)}$.
 }
 \label{tab:lin_mod_scan_config_instrum_params}
 \small
 \begin{tabular}{%
  @{}%
  >{$}l<{$}%		01.) symbol
  p{0.925\textwidth}%	02.) meaning
  @{}%
 }
 \toprule
  \multicolumn{1}{@{}H}{Symbol} &
  \multicolumn{1}{H@{}}{Meaning}\\
  \cmidrule(r){1-1}\cmidrule(l){2-2}
 \addlinespace
 %--------------------------------------------------------------------------------------------------------------
 % a) geometric parameters of the planar transducer array
 %--------------------------------------------------------------------------------------------------------------
  % 1.) number of elements along the r_{\delta}-axis
  N_{\text{el}, \delta} &
  Number of
  elements along
  the $r_{\delta}$-axis,
  $N_{\text{el}, \delta} \in \N$\\
  % 2.) width of the vibrating faces along the r_{\delta}-axis
  w_{\text{el}, \delta} &
  Width of
  the vibrating faces along
  the $r_{\delta}$-axis,
  $w_{\text{el}, \delta} \in \Rplus$\\
  % 3.) width of the kerfs separating the elements along the r_{\delta}-axis
  k_{\text{el}, \delta} &
  Width of
  the kerfs separating
  the elements along
  the $r_{\delta}$-axis,
  $k_{\text{el}, \delta} \in \Rnonneg$\\
  % 4.) constant spacing between the centers of the adjacent vibrating faces along the r_{\delta}-axis (element pitch)
  \Delta r_{\text{el}, \delta} &
  Element pitch, i.e.
  constant spacing between
  the centers of
  the adjacent vibrating faces along
  the $r_{\delta}$-axis,
  $\Delta r_{\text{el}, \delta} = w_{\text{el}, \delta} + k_{\text{el}, \delta}$\\
  % 5.) center coordinates of the vibrating faces
  \vect{r}_{\text{el}, m} &
  Center coordinates of
  the vibrating faces,\par
  $\mathcal{M} = \{ \vect{r}_{\text{el}, m} \in \R^{d}: \vect{r}_{\text{el}, m} = \sum_{ \delta = 1 }^{ d - 1 } ( m_{\delta} - M_{\text{el}, \delta} ) \Delta r_{\text{el}, \delta} \uvect{\delta}, m_{\delta} \in \setconsnonneg{ N_{\text{el}, \delta} - 1 }, m = \mathcal{I}( \vect{m}, \vect{N}_{\text{el}} ) \}$, where\par
  % 5.a) shift of index along the r_{\delta}-axis
  $M_{\text{el}, \delta} = ( N_{\text{el}, \delta} - 1 ) / 2$ and
  % 5.b) forward index transform
  $\mathcal{I}( \vect{m}, \vect{N}_{\text{el}} ) = \sum_{ \delta = 1 }^{ d - 2 } m_{\delta} \prod_{ k = \delta + 1 }^{ d - 1 } N_{\text{el}, k} + m_{d-1}$\\
  % 6.) total number of elements
  N_{\text{el}} &
  Total number of
  elements,
  $N_{\text{el}} = \tabs{ \mathcal{M} } = \prod_{ \delta = 1 }^{ d - 1 } N_{\text{el}, \delta}$\\
  % 7.) coplanar compact sets specifying the (d-1)-dimensional vibrating faces on the hyperplane r_{d} = 0
  L_{m} &
  Coplanar compact sets specifying
  the $(d-1)$-dimensional vibrating faces on
  the hyperplane $r_{d} = 0$,\par
  $L_{m} = \prod_{ \delta = 1 }^{ d - 1 } [ r_{\text{el}, m, \delta} - 0.5 w_{\text{el}, \delta}; r_{\text{el}, m, \delta} + 0.5 w_{\text{el}, \delta} ] \subset \R^{d-1}$\\
  % 8.) transmitter apodization functions
  \chi_{m, l}^{(\text{tx})} &
  Transmitter apodization functions accounting for
  the heterogeneous normal velocities and
  the acoustic lens,
  $\chi_{m, l}^{(\text{tx})}: L_{m} \mapsto \C$\\
  % 9.) receiver apodization functions
  \chi_{m, l}^{(\text{rx})} &
  Receiver apodization functions accounting for
  the heterogeneous sensitivities and
  the acoustic lens,
  $\chi_{m, l}^{(\text{rx})}: L_{m} \mapsto \C$\\
 %--------------------------------------------------------------------------------------------------------------
 % b) electromechanical parameters of the instrumentation
 %--------------------------------------------------------------------------------------------------------------
  % 1.) transmitter electromechanical transfer functions
  h_{m, l}^{(\text{tx})} &
  Transmitter electromechanical transfer functions accounting for
  the driving circuits,
  the cables, and
  the radiating elements,
  $h_{m, l}^{(\text{tx})} \in \C$\\
  % 2.) receiver electromechanical transfer functions
  h_{m, l}^{(\text{rx})} &
  Receiver electromechanical transfer functions accounting for
  the receiving elements,
  the cables, and
  the amplifiers,
  $h_{m, l}^{(\text{rx})} \in \C$\\
 \addlinespace
 \bottomrule
 \end{tabular}
\end{table*}

The \ac{UI} system sequentially performs
% 1.) N_{\text{in}} independent pulse-echo measurements
$N_{\text{in}} \in \N$ independent pulse-echo measurements using
% 2.) planar transducer array
a planar transducer array
(cf. \cref{fig:lin_mod_scan_configuration,tab:lin_mod_scan_config_instrum_params}).
% b) each measurement begins at the time instant t = 0 and triggers the concurrent recording of the RF voltage signals in the specified time interval
Each measurement begins at
% 1.) time instant t = 0
the time instant
$t = 0$ and triggers
% 2.) concurrent recording
the concurrent recording of
% 3.) RF voltage signals
the \ac{RF} voltage signals
$\tilde{u}_{m}^{(\text{rx}, n)}: \setsymbol{T}_{ \text{rec} }^{(n)} \mapsto \R$ generated by
% 4.) all array elements
all array elements
$m \in \setconsnonneg{ N_{\text{el}} - 1 }$ in
% 5.) specified time interval
% book:Briggs1995, Chapter 2: The Discrete Fourier Transform / Sect. 2.4.: DFT Approximations to Fourier Series Coefficients [NONPERIODIC FUNCTIONS]
% - There seems to be NO AGREEMENT IN THE LITERATURE about whether the INTERVAL FOR DEFINING FOURIER SERIES should be
%   [1.)] the CLOSED INTERVAL [-A/2, A/2],
%   [2.)] a HALF-OPEN INTERVAL (-A/2, A/2], or
%   [3.)] the OPEN INTERVAL (-A/2, A/2). (p. 38)
% - Arguments can be made for or against any of these choices. (p. 38)
% - We will use the CLOSED INTERVAL [-A/2, A/2] throughout the book to emphasize the point (the subject of sermons to come!) that
%   IN DEFINING THE INPUT TO THE DFT, VALUES OF THE SAMPLED FUNCTION AT BOTH ENDPOINTS CONTRIBUTE TO THE INPUT. (p. 38)
the specified time interval
\begin{equation}
 %--------------------------------------------------------------------------------------------------------------
 % specified recording time intervals for the RF voltage signals generated by all array elements
 %--------------------------------------------------------------------------------------------------------------
  \setsymbol{T}_{ \text{rec} }^{(n)}
  =
  \bigl[ t_{\text{lb}}^{(n)}; t_{\text{ub}}^{(n)} \bigr],
 \label{eqn:lin_mod_scan_config_volt_rx_obs_interval}
\end{equation}
where
% 6.) lower bounds in the specified recording time intervals
$t_{\text{lb}}^{(n)} \in \Rnonneg$ and
% 7.) upper bounds in the specified recording time intervals
$t_{\text{ub}}^{(n)} > t_{\text{lb}}^{(n)}$ denote
its lower and
upper bounds,
respectively.
The finite recording times
$T_{ \text{rec} }^{(n)} = \tabs{ \setsymbol{T}_{ \text{rec} }^{(n)} } = t_{\text{ub}}^{(n)} - t_{\text{lb}}^{(n)}$ enable
the representation of
% 1.) RF voltage signals
these signals by
% 2.) Fourier series
the \name{Fourier} series
%\footnote{
  % a) adjective "stable" indicates that neither inaccurate observations nor a sparsity defect result in huge recovery errors
%  validity
%}
(cf. e.g.
%\cite[(3.39/40)]{book:Mallat2009},
\cite[(2.2.1/2)]{book:Manolakis2005},
\cite[(2.12/13)]{book:Briggs1995}%
)
\begin{subequations}
\label{eqn:recovery_disc_freq_v_rx_Fourier_series}
\begin{equation}
 %--------------------------------------------------------------------------------------------------------------
 % Fourier series representation of the recorded RF voltage signals (time domain)
 %--------------------------------------------------------------------------------------------------------------
  \tilde{u}_{m}^{(\text{rx}, n)}( t )
  =
  u_{m, 0}^{(\text{rx}, n)}
  +
  2
  \dreal{
    \sum_{ l = 1 }^{ \infty }
      u_{m, l}^{(\text{rx}, n)}
      e^{ j \omega_{l} t }
  }{2}
 \label{eqn:recovery_disc_freq_v_rx_Fourier_series_sum}
\end{equation}
for
% 3.) all sequential pulse-echo measurements and all array elements
all $( n, m ) \in \setconsnonneg{ N_{\text{in}} - 1 } \times \setconsnonneg{ N_{\text{el}} - 1 }$, where
% 4.) discrete angular frequencies
$\omega_{l} = 2 \pi f_{l} = 2 \pi l / T_{ \text{rec} }^{(n)}$ denote
the discrete angular frequencies, and
% 5.) Fourier coefficients of the recorded RF voltage signals
\begin{equation}
 %--------------------------------------------------------------------------------------------------------------
 % Fourier coefficients of the recorded RF voltage signals
 %--------------------------------------------------------------------------------------------------------------
  u_{m, l}^{(\text{rx}, n)}
  =
  \frac{ 1 }{ T_{ \text{rec} }^{(n)} }
  \int_{ \setsymbol{T}_{ \text{rec} }^{(n)} }
    \tilde{u}_{m}^{(\text{rx}, n)}( t )
    e^{ -j \omega_{l} t }
  \text{d} t
 \label{eqn:recovery_disc_freq_v_rx_Fourier_series_coef}
\end{equation}
\end{subequations}
are
the complex-valued coefficients, whose
% 6.) conjugate even symmetry
conjugate even symmetry renders
% 7.) negative frequency indices
the negative frequency indices redundant.

%---------------------------------------------------------------------------------------------------------------
% 2.) bandpass characters of the recorded RF voltage signals / truncation of the Fourier series
%---------------------------------------------------------------------------------------------------------------
% a) bandpass characters of the recorded RF voltage signals define the sets of relevant discrete frequencies
The bandpass characters of
% 1.) recorded RF voltage signals
the recorded \ac{RF} voltage signals, which are described by
the lower and
upper frequency bounds
% 2.) lower frequency bounds
$f_{\text{lb}}^{(n)} \in \Rplus$ and
% 3.) upper frequency bounds
$f_{\text{ub}}^{(n)} \geq f_{\text{lb}}^{(n)} + 1 / T_{ \text{rec} }^{(n)}$,
respectively, define
% 4.) sets of relevant discrete frequencies
the sets of
relevant discrete frequencies
\begin{subequations}
\label{eqn:recon_disc_axis_f_discrete_BP}
\begin{equation}
 %--------------------------------------------------------------------------------------------------------------
 % sets of relevant discrete frequencies
 %--------------------------------------------------------------------------------------------------------------
  \setsymbol{F}_{ \text{BP} }^{(n)}
  =
  \Bigl\{
    f_{l} \in \Rplus:
    f_{l} = \frac{ l }{ T_{ \text{rec} }^{(n)} },
    l \in \setsymbol{L}_{ \text{BP} }^{(n)}
  \Bigr\}
 \label{eqn:recon_disc_axis_f_discrete_BP_set}
\end{equation}
for
% 5.) all sequential pulse-echo measurements
all $n \in \setconsnonneg{ N_{\text{in}} - 1 }$, where
% 6.) sets of admissible frequency indices
the admissible index sets are
\begin{equation}
 %--------------------------------------------------------------------------------------------------------------
 % sets of admissible frequency indices
 %--------------------------------------------------------------------------------------------------------------
  \setsymbol{L}_{ \text{BP} }^{(n)}
  =
  \left\{
    l \in \N:
    l_{\text{lb}}^{(n)} \leq l \leq l_{\text{ub}}^{(n)}
  \right\}
 \label{eqn:recon_disc_axis_f_discrete_BP_indices}
\end{equation}
with
the lower and
upper bounds
% a) lower bounds on the admissible frequency indices
% 1.) t_{\text{lb}}^{(n)} \in \Rnonneg and t_{\text{ub}}^{(n)} > t_{\text{lb}}^{(n)}
% => T_{ \text{rec} }^{(n)} = t_{\text{ub}}^{(n)} - t_{\text{lb}}^{(n)} > 0
% 2.) f_{\text{lb}}^{(n)} \in \Rplus and f_{\text{ub}}^{(n)} \geq f_{\text{lb}}^{(n)} + 1 / T_{ \text{rec} }^{(n)} > f_{\text{lb}}^{(n)} > 0
% => T_{ \text{rec} }^{(n)} f_{\text{lb}}^{(n)} > 0
% => l_{\text{lb}}^{(n)} = \dceil{ T_{ \text{rec} }^{(n)} f_{\text{lb}}^{(n)} }{1} \in \N
% b) upper bounds on the admissible frequency indices
% => T_{ \text{rec} }^{(n)} f_{\text{ub}}^{(n)} \geq T_{ \text{rec} }^{(n)} f_{\text{lb}}^{(n)} + 1
% => l_{\text{ub}}^{(n)} = \dfloor{ T_{ \text{rec} }^{(n)} f_{\text{ub}}^{(n)} }{1} \geq \dfloor{ T_{ \text{rec} }^{(n)} f_{\text{lb}}^{(n)} }{1} + 1 \geq \dceil{ T_{ \text{rec} }^{(n)} f_{\text{lb}}^{(n)} }{1} = l_{\text{lb}}^{(n)}
\begin{align}
 %--------------------------------------------------------------------------------------------------------------
 % a) lower bounds on the admissible frequency indices
 %--------------------------------------------------------------------------------------------------------------
  l_{\text{lb}}^{(n)}
  &=
  \dceil{ T_{ \text{rec} }^{(n)} f_{\text{lb}}^{(n)} }{1}
  & \text{and} & &
 %--------------------------------------------------------------------------------------------------------------
 % b) upper bounds on the admissible frequency indices
 %--------------------------------------------------------------------------------------------------------------
  l_{\text{ub}}^{(n)}
  &=
  \dfloor{ T_{ \text{rec} }^{(n)} f_{\text{ub}}^{(n)} }{1},
 \label{eqn:recon_disc_axis_f_discrete_BP_indices_lb_ub}
\end{align}
\end{subequations}
respectively.
% b) sets of relevant discrete frequencies enable the truncation of each Fourier series and the representation of each pulse-echo measurement by N_{\text{el}} N_{f, \text{BP}}^{(n)} complex-valued coefficients
These enable
the truncation of
% 1.) Fourier series representation of the recorded RF voltage signals (time domain)
each \name{Fourier} series
\eqref{eqn:recovery_disc_freq_v_rx_Fourier_series_sum}, where, defining
% 2.) effective bandwidths
the effective bandwidths
$B_{ u }^{(n)} = f_{\text{ub}}^{(n)} - f_{\text{lb}}^{(n)}$,
% 3.) number of relevant discrete frequencies
the number of
relevant discrete frequencies approximates
% 4.) effective time-bandwidth products
the effective time-bandwidth products
\begin{equation}
 %--------------------------------------------------------------------------------------------------------------
 % numbers of relevant discrete frequencies (effective time-bandwidth products)
 %--------------------------------------------------------------------------------------------------------------
  N_{f, \text{BP}}^{(n)}
  =
  \dabs{ \setsymbol{L}_{ \text{BP} }^{(n)} }{1}
  =
  l_{\text{ub}}^{(n)} - l_{\text{lb}}^{(n)} + 1
  \approx
  T_{ \text{rec} }^{(n)} B_{ u }^{(n)}
 \label{eqn:recon_disc_axis_f_discrete_BP_TB_product}
\end{equation}
for
% 5.) all sequential pulse-echo measurements
all $n \in \setconsnonneg{ N_{\text{in}} - 1 }$, and
the representation of
each pulse-echo measurement by
% 6.) Fourier coefficients of the recorded RF voltage signals
$N_{\text{el}} N_{f, \text{BP}}^{(n)}$ coefficients
\eqref{eqn:recovery_disc_freq_v_rx_Fourier_series_coef}.
% c) let the subscript l indicate an admissible frequency index in the sets of relevant discrete frequencies in the following
%In the following,
%the subscript
%$l \in \setsymbol{L}_{ \text{BP} }^{(n)}$ indicates
%an admissible frequency index.
% d) dependence on the superscript n is implicitly understood
%For the sake of
%notational lucidity,
%its dependence on
%the superscript $n$, which identifies
%the sequential pulse-echo measurement, is
%implicitly understood.

%%%%%%%%%%%%%%%%%%%%%%%%%%%%%%%%%%%%%%%%%%%%%%%%%%%%%%%%%%%%%%%%%%%%%%%%%%%%%%%%%%%%%%%%%%%%%%%%%%%%%%%%%%%%%%%%
% 2.) acoustic model for human soft tissues
%%%%%%%%%%%%%%%%%%%%%%%%%%%%%%%%%%%%%%%%%%%%%%%%%%%%%%%%%%%%%%%%%%%%%%%%%%%%%%%%%%%%%%%%%%%%%%%%%%%%%%%%%%%%%%%%
\subsection{Acoustic Model for Human Soft Tissues}
Medical \ac{UI} usually models
soft tissue structures%
\footnote{
  % a) strict definition of the term "soft tissue" (anatomy)
  The anatomic term \term{soft tissue} refers to
  tendons, ligaments, skin, nerves,
  % website:NCIDictionary2017: NCI Dictionary of Cancer Terms
  % - 'soft tissue': Refers to MUSCLE, FAT, FIBROUS TISSUE, blood vessels, or other SUPPORTING TISSUE of the BODY.
  % fibrous tissue =  the common connective tissue of the body, composed of yellow or white parallel elastic and collagen fibers.
  muscle, fat, fibrous tissue, blood vessels, or
  other supporting tissue of
  the body
  \cite[\term{soft tissue}]{website:NCIDictionary2017}.
  % b) loose definition of the term "soft tissue" (UI)
  In the context of \ac{UI}, however,
  the term additionally includes
  organs like
  liver, kidney, thyroid, brain, and
  the heart.
} as
% 1.) quiescent
% article:MastJASA1997:
% - The TISSUE WAS ASSUMED TO BE MOTIONLESS except for small acoustic perturbations.
quiescent,
% 2.) lossless
% article:JensenJASA1991: A model for the propagation and scattering of ultrasound in tissue
% INTRODUCTION
% - The model includes attenuation due to propagation and scattering, but not
%   THE DISPERSIVE ATTENUATION OBSERVED FOR PROPAGATION IN TISSUE. (p. 182)
% - This can, however, be incorporated into the model as indicated in Sec. VI. (p. 182)
% I. DERIVATION OF THE WAVE EQUATION
% - Our second assumption is that
%   NO HEAT CONDUCTION OR CONVERSION OF ULTRASOUND TO THERMAL ENERGY TAKE PLACE. (p. 182)
% article:GorePMB1977a: Ultrasonic backscattering from human tissue: A realistic model
% 1. Introduction
% - For simplicity, however, such ABSORPTION EFFECTS ARE NOT CONSIDERED HERE,
%   but the introduction of simple exponential absorption leads to only minor changes in the theory. (p. 318)
lossless, and
% 3.) heterogeneous
% article:NgITUFFC2006: Modeling ultrasound imaging as a linear, shift-variant system
% III. The Wave Equation
% - Our analysis necessarily begins by considering
%   the PARTIAL DIFFERENTIAL EQUATION (PDE) that describes
%   the PROPAGATION OF ACOUSTIC WAVES IN A NONUNIFORM MEDIUM. (p. 550)
% article:JensenJASA1991: A model for the propagation and scattering of ultrasound in tissue
% Abstract
% - An INHOMOGENEOUS WAVE EQUATION is derived describing PROPAGATION AND SCATTERING OF ULTRASOUND IN AN INHOMOGENEOUS MEDIUM. (p. 182)
heterogeneous fluids that linearly propagate
% 4.) small-amplitude approximation
% article:NgITUFFC2006: Modeling ultrasound imaging as a linear, shift-variant system
% I. Introduction
% - We restrict ourselves to consider LINEAR WAVE PROPAGATION ONLY. (p. 549)
% article:JensenJASA1991: A model for the propagation and scattering of ultrasound in tissue
% I. DERIVATION OF THE WAVE EQUATION
% - To obtain a solvable wave equation, some ASSUMPTIONS AND APPROXIMATIONS MUST BE MADE. (p. 182)
% - The first one states that the instantaneous acoustic pressure and density can be written as
%   [ P_{\text{ins}}( \vect{r}, t ) = P + p_{1}( \vect{r}, t ) ], (1)
%   [ \rho_{\text{ins}}( \vect{r}, t ) = \rho( \vect{r} ) + \rho_{1}( \vect{r}, t ) ], (2)
%   in which P is the mean pressure of the medium and \rho is the density of the undisturbed medium. (p. 182)
% - The PRESSURE VARIATION p_{1} IS CAUSED BY THE ULTRASOUND WAVE AND IS CONSIDERED SMALL compared to P. (p. 182)
% - The density change caused by the wave is \rho_{1}. (p. 182)
% - Both p_{1} and \rho_{1} are SMALL QUANTITIES OF FIRST ORDER. (p. 182)
small-amplitude disturbances of
the stationary state as
% 5.) longitudinal waves
% article:GorePMB1977a: Ultrasonic backscattering from human tissue: A realistic model
% 2. The wave equation for ultrasound propagation through tissue
% - ULTRASOUND PROPAGATION BY MODES OTHER THAN PURELY LONGITUDINAL IS NEGLECTED,
%   not only for reasons of simplicity, but also because
%   their significance is not well documented or understood for scattering from tissue. (p. 320)
longitudinal waves
\cite{article:NgITUFFC2006,coll:Jensen2002,article:JensenJASA1991,article:GorePMB1977a}.
The relevant acoustic material parameters are
the unperturbed values of both
% 1.) unperturbed compressibility
the compressibility and
% 2.) unperturbed mass density
the mass density, which are typically normalized by
% 3.) spatial averages
% book:Kak2001, Sect. 6.1.2: Inhomogeneous Wave Equation
% - \kappa_{0} and \rho_{0} are either
%   the compressibility and the density of the medium in which the object is immersed, or
%   THE AVERAGE COMPRESSIBILITY AND THE DENSITY OF THE OBJECT, depending upon how the process of imaging is modeled. (p. 210)
% article:JensenJASA1991: A model for the propagation and scattering of ultrasound in tissue
% I. DERIVATION OF THE WAVE EQUATION
% - We now assume that
%   [1.)] the PROPAGATION VELOCITY AND
%   [2.)] THE DENSITY ONLY VARY SLIGHTLY FROM
%   THEIR MEAN VALUES, so that
%   [ \rho( \vect{r} ) = \rho_{0} + \Delta \rho( \vect{r} ) ],
%   [ c( \vect{r} ) = c_{0}( \vect{r} ) + \Delta c( \vect{r} ) ], (12)
%   where \rho_{0} \gg \Delta \rho and c_{0} \gg \Delta c. (p. 183)
% article:GorePMB1977a: Ultrasonic backscattering from human tissue: A realistic model
% 2. The wave equation for ultrasound propagation through tissue
% - It is convenient to consider formally
%   the scattering region V to be embedded in SOME NON-DISPERSIVE MEDIUM WITH
%   CONSTANT DENSITY, \rho_{0}, and
%   COMPRESSIBILITY \kappa_{0} = ( \rho_{0} {c_{0}}^{2} )^{-1}, with
%   c_{0} the acoustic velocity in the embedding medium. (p. 319)
% - THE VALUES OF THESE PARAMETERS ARE CHOSEN TO BE THE MEAN VALUES THEY ASSUME INSIDE V. (p. 319)
their spatial averages
% 4.) reference value for the unperturbed compressibility
$\kappa_{0} \in \Rplus$ and
% 5.) reference value for the unperturbed mass density
$\rho_{0} \in \Rplus$,
respectively
\cite{article:NgITUFFC2006,article:JensenJASA1991,article:GorePMB1977a}.
% c) proposed model considers a homogeneous unperturbed mass density
For
the sake of
simplicity,
the proposed model considers
a homogeneous unperturbed mass density.
% d) global unperturbed compressibility / associated relative spatial fluctuations
The relative spatial fluctuations in
the unperturbed compressibility read
(cf. \cref{fig:lin_mod_scan_configuration})
\begin{equation}
 %--------------------------------------------------------------------------------------------------------------
 % relative spatial fluctuations in the unperturbed compressibility
 %--------------------------------------------------------------------------------------------------------------
  \gamma^{(\kappa)}( \vect{r} )
  =
  \begin{cases}
    1 - \kappa_{1}( \vect{r} ) / \kappa_{0} & \text{for } \vect{r} \in \Omega,\\
    0 & \text{for } \vect{r} \notin \Omega.\\
  \end{cases}
 \label{eqn:lin_mod_mech_model_tis_simple_rel_fluctuations}
\end{equation}

The spatial amplitude absorption coefficient
$\alpha_{l} \in \Rnonneg$, which is
% 1.) commonly neglected
% article:GorePMB1977a: Ultrasonic backscattering from human tissue: A realistic model
% 1. Introduction
% - Ideally, the scattering from a particular region should be specified in a way which allows
%   the EFFECT OF THE OVERLYING TISSUE TO BE EASILY QUANTIFIED; in particular,
%   the EFFECT OF FREQUENCY DEPENDENT ATTENUATION SHOULD BE INCLUDED. (p. 318)
% 4. Discussion and implications of results
% - The TISSUE MODEL DISCUSSED HERE MAY BE EXTENDED TO INCLUDE EFFECTS SUCH AS ABSORPTION, and
%   further study of this is under way. (p. 325)
commonly neglected, obeys
% 2.) power law
the power law
(cf. e.g.
\cite[Sect. 4.3.8]{book:Duck1990},
\cite[(4)]{article:WellsUMB1975}%
)
\begin{equation}
 %--------------------------------------------------------------------------------------------------------------
 % spatial amplitude absorption coefficient
 %--------------------------------------------------------------------------------------------------------------
  \alpha_{l}
  =
  \bar{b} \abs{ \omega_{l} }^{ \zeta }
 \label{eqn:lin_mod_mech_model_tis_abs_power_law}
\end{equation}
in
% 3.) entire Euclidean space
the entire Euclidean space for
% 4.) all relevant discrete frequencies
all relevant discrete frequencies
$l \in \setsymbol{L}_{ \text{BP} }^{(n)}$, where
% 5.) pair of absorption parameters
the parameter pair
$( \bar{b}, \zeta ) \in \Rnonneg \times \Rnonneg$ depends on both
% 6.) specific type of tissue
the specific type of
tissue and
% 7.) ambient conditions
the ambient conditions.
The exponent $\zeta$ usually ranges between
$1.0$ and $1.5$
\cite{article:KellyJASA2008b},
\cite[112]{book:Duck1990},
\cite{article:WellsUMB1975}.
\TODO{check exponent}
Given
reference values of
% 1.) reference angular frequency
the angular frequency
$\omega_{\text{ref}} \in \Rplus$ and
% 2.) reference phase velocity
the associated phase velocity
$c_{\text{ref}} \in \Rplus$,
% 3.) complex-valued wavenumber with respect to k_{\text{ref}}
the complex-valued wavenumber
% TODO: Kelly?
\cite{article:WatersITUFFC2005,article:SzaboJASA1995}
\begin{subequations}
\label{eqn:lin_mod_mech_model_tis_abs_time_causal_wavenumber_complex_kref}
\begin{equation}
 %--------------------------------------------------------------------------------------------------------------
 % complex-valued wavenumber with respect to k_{\text{ref}}
 %--------------------------------------------------------------------------------------------------------------
  \munderbar{k}_{l}
  =
  \underbrace{
    \frac{ \omega_{l} }{ c_{\text{ref}} }
    +
    \beta_{\text{E,ref}, l}
  }_{ = \beta_{l} = \omega_{l} / c_{l} }
  - j
  \underbrace{
    \bar{b} \abs{ \omega_{l} }^{ \zeta }
  }_{ = \alpha_{l} },
 \label{eqn:lin_mod_mech_model_tis_abs_time_causal_wavenumber_complex_kref_sum}
\end{equation}
where
% 4.) phase term
the phase term
$\beta_{l} \in \R$ sums
% 5.) real-valued wavenumber with respect to c_{\text{ref}}
the real-valued wavenumber
$k_{\text{ref}, l} = \omega_{l} / c_{\text{ref}}$ and
% 6.) excess dispersion term with respect to k_{\text{ref}}
the excess dispersion term
\begin{equation}
 %--------------------------------------------------------------------------------------------------------------
 % excess dispersion term with respect to k_{\text{ref}}
 %--------------------------------------------------------------------------------------------------------------
  \beta_{\text{E,ref}, l}
  =
  \begin{cases}
   %------------------------------------------------------------------------------------------------------------
   % a) exponent of unity
   %------------------------------------------------------------------------------------------------------------
    -
    2 \bar{b} \omega_{l}
    \ln\bigl( \abs{ \omega_{l} / \omega_{\text{ref}} } \bigr) / \pi
    &
    \text{for } \zeta = 1,\\
   %------------------------------------------------------------------------------------------------------------
   % b) even integer or noninteger
   %------------------------------------------------------------------------------------------------------------
    \bar{b}
    \tan\bigl( \zeta \pi / 2 \bigr)
    \omega_{l}
    \bigl( \abs{ \omega_{l} }^{ \zeta - 1 } - \abs{ \omega_{\text{ref}} }^{ \zeta - 1 } \bigr)
    &
    \text{else},
  \end{cases}
 \label{eqn:lin_mod_mech_model_tis_abs_time_causal_wavenumber_complex_kref_excess_dispersion}
\end{equation}
\end{subequations}
combines
% 1.) power-law absorption
power-law absorption with
% 2.) anomalous dispersion [ phase velocity increases w/ frequency ]
% book:Cobbold2006, Sect. 3.10: Effects of Attenuation / Sect. 3.10.1: Kramers-Kronig Relationships
% - It will be noted that FOR n > 2 the SPEED DECREASES WITH INCREASING FREQUENCY and NORMAL DISPERSION is said to be present. (p. 207)
% - FOR n < 2 the OPPOSITE IS TRUE, and the DISPERSION is said to be ANOMALOUS. (p. 207)
% article:SzaboJASA1995: Causal theories and data for acoustic attenuation obeying a frequency power law
% I. TIME DOMAIN CAUSAL RELATIONSHIPS / B. Anomalous dispersion
% - Before proceeding, it is necessary to be MORE PRECISE ABOUT THE DEFINITION OF c_{0} and whether
%   VELOCITY DISPERSION INCREASES OR DECREASES WITH FREQUENCY. (p. 15)
% - According to conventions established in electromagnetic theory,
%   "NORMAL" DISPERSION is that in which PHASE VELOCITY DECREASES WITH AN INCREASE IN FREQUENCY. (p. 15)
% - "ANOMALOUS" DISPERSION is defined as the condition in which PHASE VELOCITY INCREASES WITH FREQUENCY
%   (Stratton, 1941; Jackson, 1975; Gurumurthy and Arthur, 1982). (pp. 15, 16)
% - For THESE CASES [power-law absorption], when phase velocity dispersion has been observed,
%   it was found to INCREASE WITH FREQUENCY in accordance with the ANOMALOUS CATEGORY. (p. 16)
% - The fact that DISPERSION is MOST OFTEN ANOMALOUS DETERMINES THE VALUE OF c_{0} used in the propagation factor \beta_{0} = \omega / c_{0}. (p. 16)
anomalous dispersion.

%%%%%%%%%%%%%%%%%%%%%%%%%%%%%%%%%%%%%%%%%%%%%%%%%%%%%%%%%%%%%%%%%%%%%%%%%%%%%%%%%%%%%%%%%%%%%%%%%%%%%%%%%%%%%%%%
% 3.) recorded radio frequency voltage signals
%%%%%%%%%%%%%%%%%%%%%%%%%%%%%%%%%%%%%%%%%%%%%%%%%%%%%%%%%%%%%%%%%%%%%%%%%%%%%%%%%%%%%%%%%%%%%%%%%%%%%%%%%%%%%%%%
\subsection{Recorded Radio Frequency Voltage Signals}
%\label{subsec:lin_mod_exc_sup_qsw_p_sc_born}
%%%%%%%%%%%%%%%%%%%%%%%%%%%%%%%%%%%%%%%%%%%%%%%%%%%%%%%%%%%%%%%%%%%%%%%%%%%%%%%%%%%%%%%%%%%%%%%%%%%%%%%%%%%%%%%%
% graphic: block diagram of the pulse-echo measurement process
%%%%%%%%%%%%%%%%%%%%%%%%%%%%%%%%%%%%%%%%%%%%%%%%%%%%%%%%%%%%%%%%%%%%%%%%%%%%%%%%%%%%%%%%%%%%%%%%%%%%%%%%%%%%%%%%
%
\begin{figure*}[t!]
 \centering%
  \input{linear_model/figures/latex/lin_mod_v_rx_sgn_proc_chain.tex}
 \caption{}
 \label{fig:V}
\end{figure*}
C
{% a) block diagram of the pulse-echo measurement process
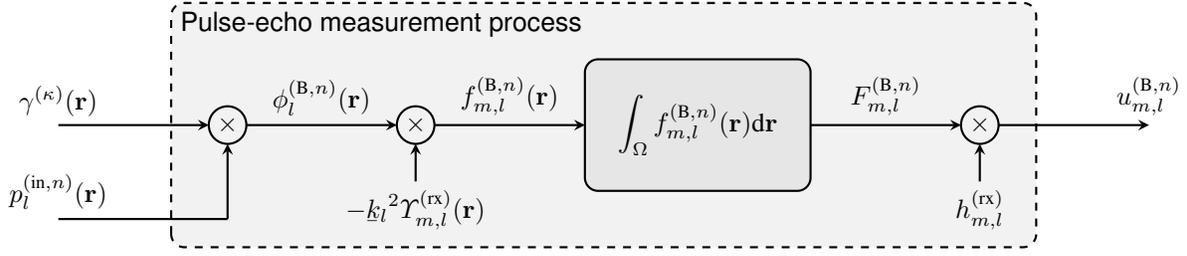
 Block diagram of
 % 1.) pulse-echo measurement process (single pulse-echo measurement, monofrequent, single array element)
 the pulse-echo measurement process
 \eqref{eqn:lin_mod_v_rx_born_obs_proc}.
 % b) observation operators map the relative spatial fluctuations in compressibility to the Born approximation of of the recorded RF voltage signals
 Given
 % 1.) incident acoustic pressure fields
 the incident acoustic pressure fields
 $p_{l}^{(\text{in}, n)}: \R^{d} \mapsto \C$, which satisfy
 % 2.) Helmholtz equations for the incident acoustic pressure fields
 the \name{Helmholtz} equations
 \eqref{eqn:lin_mod_sol_wave_eq_pde_p_in},
 % 3.) observation operators
 the \name{Born} approximation linearly maps
 % 4.) relative spatial fluctuations in the unperturbed compressibility
 the compressibility fluctuations
 \eqref{eqn:lin_mod_mech_model_tis_simple_rel_fluctuations} to
 % 5.) Born approximation of the recorded RF voltage signals
 the recorded \ac{RF} voltage signals
 \eqref{eqn:lin_mod_v_rx_born}.
 %--------------------------------------------------------------------------------------------------------------
 % decomposition of the observation operators
 %--------------------------------------------------------------------------------------------------------------
 % a) first two multiplications yield the Born approximations of the contrast sources and the force densities
 %The first two multiplications yield
 %the \name{Born} approximations of
 % 1.) Born approximation of the contrast sources
 %the contrast sources
 %$\phi_{l}^{(\text{B}, n)}: \R^{d} \mapsto \C$ and
 % 2.) Born approximation of the force densities
 %the force densities
 %$f_{m, l}^{(\text{B}, n)}: \R^{d} \mapsto \C$.
 % b) receiver electromechanical transfer function
 %The multiplications of
 %their spatial integrals, i.e.
 % 1.) Born approximation of the compressive blocked forces
 %the \name{Born} approximation of
 %the compressive blocked forces
 %\eqref{eqn:lin_mod_exc_sup_qsw_blocked_force_born}, by
 % 2.) receiver electromechanical transfer functions
 %the receiver electromechanical transfer functions
 %$h_{m}^{(\text{rx})}$ according to
 %the transfer relations
 %\eqref{eqn:lin_mod_scan_config_trans_array_transfer_v_rx} yield
 % 3.) Born approximation of the recorded RF voltage signals
 %the desired signals.
}%
{lin_mod_v_rx_sgn_proc_chain}
The array elements transduce
the compressive blocked forces exerted by
% 2.) free-space scattered acoustic pressure fields
the free-space scattered acoustic pressure fields
$p_{l}^{(\text{sc}, n)}: \R^{d} \mapsto \C$ on
% 3.) planar faces L_{m}
their planar faces
$L_{m} \subset \R^{d-1}$ into
the \ac{RF} voltage signals
(cf. e.g.
\cite[Sect. 9.2]{book:Schmerr2015},
\cite{article:LabyedITUFFC2014,article:NgITUFFC2006,article:JensenJASA1991}%
)
\begin{equation}
 %--------------------------------------------------------------------------------------------------------------
 % recorded RF voltage signals provided by the receiving amplification networks
 %--------------------------------------------------------------------------------------------------------------
  u_{m, l}^{(\text{rx}, n)}
  =
  2 h_{m, l}^{(\text{rx})}
  \int_{ L_{m} }
    \chi_{m, l}^{(\text{rx})}( \vect{r}_{\rho} )
    p_{l}^{(\text{sc}, n)}( \vect{r}_{\rho}, 0 )
  \text{d} \vect{r}_{\rho}
 \label{eqn:lin_mod_scan_config_trans_array_transfer_v_rx}
\end{equation}
for
% 4.) all sequential pulse-echo measurements, all relevant discrete frequencies, and all array elements
all $( n, l, m ) \in \setconsnonneg{ N_{\text{in}} - 1 } \times \setsymbol{L}_{ \text{BP} }^{(n)} \times \setconsnonneg{ N_{\text{el}} - 1 }$, where
% 5.) receiver electromechanical transfer functions
$h_{m, l}^{(\text{rx})} \in \C$ denote
the electromechanical transfer functions and
% 6.) receiver apodization functions
$\chi_{m, l}^{(\text{rx})}: L_{m} \mapsto \C$ are
the apodization functions
(cf. \cref{tab:lin_mod_scan_config_instrum_params}).

The \name{Born} approximation, which drives
% 1.) established image recovery methods in ultrafast UI
% article:ChernyakovaITUFFC2018: Fourier-Domain Beamforming and Structure-Based Reconstruction for Plane-Wave Imaging
% article:MoghimiradITUFFC2016: Synthetic Aperture Ultrasound Fourier Beamformation Using Virtual Sources
% proc:SchiffnerIUS2016a: A low-rate parallel Fourier domain beamforming method for ultrafast pulse-echo imaging
% article:LabyedITUFFC2014: TR-MUSIC inversion of the density and compressibility contrasts of point scatterers
% article:MontaldoITUFFC2009: Coherent plane-wave compounding for very high frame rate ultrasonography and transient elastography
% article:JensenUlt2006: Synthetic aperture ultrasound imaging [Dec.]
% article:ChengITUFFC2006: Extended high-frame rate imaging method with limited-diffraction beams [May]
% article:WalkerITUFFC2001: C- and D-weighted ultrasonic imaging using the translating apertures algorithm
% - The LACK OF MULTIPLE SCATTERING IN SOFT TISSUES IS COMMONLY ASSUMED WITH GOOD RESULTS [13].
% - [13] M. F. Insana and D. G. Brown, “Acoustic scattering theory applied to soft biological tissues,”
%   in Ultrasonic Scattering in Biological Tissues. K. K. Shung, Ed. Ann Arbor, MI: CRC Press, 1993, pp. 75–124.
% article:LuITUFFC1997: 2D and 3D High Frame Rate Imaging with Limited Diffraction Beams
the established image recovery methods in
ultrafast \ac{UI}
\cite{article:MoghimiradITUFFC2016,article:LabyedITUFFC2014,article:MontaldoITUFFC2009,article:JensenUlt2006,article:ChengITUFFC2006,article:LuITUFFC1997}, uses
% 2.) incident acoustic pressure fields
% article:JensenJASA1991: A model for the propagation and scattering of ultrasound in tissue
% III. CALCULATION OF THE INCIDENT FIELD
% - The INCIDENT FIELD is generated by the ultrasound transducer, assuming NO OTHER SOURCES EXIST IN THE TISSUE. (p. 184)
% - By this method [spatial impulse response]
%   the INCIDENT FIELD IS FOUND BY SOLVING THE WAVE EQUATION FOR THE HOMOGENEOUS CASE:
%   [ \nabla^{2} p_{1} - \frac{ 1 }{ c_{0}^{2} } \frac{ \partial^{2} p_{1} }{ \partial^{2} t } = 0 ]. (25) (p. 184)
the incident acoustic pressure fields
$p_{l}^{(\text{in}, n)}: \R^{d} \mapsto \C$ induced by
% 3.) transducer array
the transducer array in
the homogeneous fluid and governed by
the \name{Helmholtz} equations
(cf. e.g.
\cite[(6.4)]{book:Devaney2012},    %
%\cite{article:NgITUFFC2006},		%
\cite[47]{book:Natterer2001},           %
\cite[(30)]{book:Kak2001}%
)
\begin{equation}
 %--------------------------------------------------------------------------------------------------------------
 % Helmholtz equations for the incident acoustic pressure fields
 %--------------------------------------------------------------------------------------------------------------
  \left( \Delta + {\munderbar{k}_{l}}^{2} \right)
  p_{l}^{(\text{in}, n)}( \vect{r} )
  = 0
 \label{eqn:lin_mod_sol_wave_eq_pde_p_in}
\end{equation}
to estimate
% 5.) free-space scattered acoustic pressure fields
% article:NgITUFFC2006: Modeling ultrasound imaging as a linear, shift-variant system
% III. The Wave Equation / A. The Total Pressure Field
% - We also know that the SCATTERED PRESSURE FIELD must obey (2), and so we can assign our particular solution P_{s}( \vect{x}, ω ) to
%   be the SCATTERED PRESSURE FIELD. (p. 550)
% book:Kak2001, Chapter 6: Tomographic Imaging with Diffracting Sources / Sect. 6.1: Diffracted Projections / Sect. 6.1.2: Inhomogeneous Wave Equation
% - The component u_{s}(r), known as the SCATTERED FIELD, will be that part of the total field that can be ATTRIBUTED SOLELY TO THE INHOMOGENEITIES. (p. 210)
the free-space scattered acoustic pressure fields as
(cf. e.g.
\cite[(6.53)]{book:Devaney2012},		% term: "Born approximation" (checked!)
%\cite[268, 287]{book:Cobbold2006},		% term: "Born approximation" (checked!)
\cite[(3.16)]{book:Natterer2001},		% term: "Born approximation", plane-wave insonification
\cite[(40)]{book:Kak2001},			% term: "first Born approximation"
\cite[(57)]{book:Born1999}%			% terms: first Born approximation, first-order Born approximation, Born approximation (checked!)
)
\begin{equation}
 %--------------------------------------------------------------------------------------------------------------
 % Born approximations of the free-space scattered acoustic pressure fields
 %--------------------------------------------------------------------------------------------------------------
  p_{l}^{(\text{sc}, n)}( \vect{r} )
  \approx
  {\munderbar{k}_{l}}^{2}
  \int_{ \Omega }
    \gamma^{(\kappa)}( \vect{r}' )
    p_{l}^{(\text{in}, n)}( \vect{r}' )
    g_{l}( \vect{r} - \vect{r}' )
  \text{d} \vect{r}'
 \label{eqn:lin_mod_v_rx_p_sc_born}
\end{equation}
for
% 6.) all sequential pulse-echo measurements and all relevant discrete frequencies
all $( n, l ) \in \setconsnonneg{ N_{\text{in}} - 1 } \times \setsymbol{L}_{ \text{BP} }^{(n)}$, where
% 7.) outgoing free-space Green's functions (two- and three-dimensional Euclidean spaces)
the outgoing free-space \name{Green}'s functions
\eqref{eqn:app_helmholtz_green_free_space_2_3_dim} account for
% 8.) diffraction
diffraction and
% 9.) monopole scattering
monopole scattering
(cf. Appendix \ref{app:helmholtz_green}).
The resulting single scattering is valid for
% 1.) weakly-scattering lossy heterogeneous objects
weakly-scattering heterogeneous objects, i.e.
% 2.) validity condition
% article:DevaneyJASA1985: Variable density acoustic tomography
% - The validity of the FIRST BORN APPROXIMATION clearly requires that
%   the STRENGTH OF THE SCATTERED FIELD COMPONENT OF THE PRESSURE FIELD (6) REMAINS SMALL THROUGHOUT THE VOLUME OF THE OBJECT (scatterer).
$\tabs{ p_{l}^{(\text{sc}, n)}( \vect{r} ) } \ll \tabs{ p_{l}^{(\text{in}, n)}( \vect{r} ) }$ for
% 3.) all object points
all $\vect{r} \in \Omega$.
These feature both
% 1.) small absolute value of the relative spatial fluctuations in compressibility
small absolute values of
% 1.) relative spatial fluctuations in the unperturbed compressibility
the compressibility fluctuations
\eqref{eqn:lin_mod_mech_model_tis_simple_rel_fluctuations} and
% 2.) small spatial extent of the lossy heterogeneous object
small acoustic sizes
\cite{article:LiPIER2010},
\cite[708]{book:Born1999}.

%---------------------------------------------------------------------------------------------------------------
% 3.) Born approximation of the recorded RF voltage signals
%---------------------------------------------------------------------------------------------------------------
% a) insertion into the receiver electromechanical transfer relations yield the recorded RF voltage signals
The \name{Born} approximation of
the scattered acoustic pressure fields
\eqref{eqn:lin_mod_v_rx_p_sc_born} estimates
% 2.) recorded RF voltage signals
the recorded \ac{RF} voltage signals
\eqref{eqn:lin_mod_scan_config_trans_array_transfer_v_rx} as
the \name{Fredholm} integral equations of
the first kind
%(cf. e.g.
%\cite[(1.1)]{article:AltuerkJIASF2017},
%\cite[(2.2)]{book:Hansen2010}%
%)
\begin{subequations}
\label{eqn:lin_mod_v_rx_born}
\begin{equation}
 %--------------------------------------------------------------------------------------------------------------
 % a) Born approximation of the recorded RF voltage signals
 %--------------------------------------------------------------------------------------------------------------
  u_{m, l}^{(\text{rx}, n)}
  \approx
  u_{m, l}^{(\text{B}, n)}
  =
  \dopobservation{ m, l }{ p_{l}^{(\text{in}, n)} }{ \gamma^{(\kappa)} }{1}
 \label{eqn:lin_mod_v_rx_born_expression}
\end{equation}
for
% 5.) all sequential pulse-echo measurements, all relevant discrete frequencies, and all array elements
all $( n, l, m ) \in \setconsnonneg{ N_{\text{in}} - 1 } \times \setsymbol{L}_{ \text{BP} }^{(n)} \times \setconsnonneg{ N_{\text{el}} - 1 }$, where
% 6.)
the properties of
% 1.) pulse-echo measurement process (single pulse-echo measurement, monofrequent, single transducer element)
the pulse-echo measurement process
\begin{equation}
 %--------------------------------------------------------------------------------------------------------------
 % b) pulse-echo measurement process (single pulse-echo measurement, monofrequent, single transducer element)
 %--------------------------------------------------------------------------------------------------------------
  \dopobservation{ m, l }{ p_{l}^{(\text{in}, n)} }{ \gamma^{(\kappa)} }{1}
  =
  - {\munderbar{k}_{l}}^{2} h_{m, l}^{(\text{rx})}
  \int_{ \Omega }
    \gamma^{(\kappa)}( \vect{r} )
    p_{l}^{(\text{in}, n)}( \vect{r} )
    \varUpsilon_{m, l}^{(\text{rx})}( \vect{r} )
  \text{d} \vect{r}
 \label{eqn:lin_mod_v_rx_born_obs_proc}
\end{equation}
significantly depend on
% 2.) incident waves
the incident waves
(cf. \cref{fig:lin_mod_v_rx_sgn_proc_chain}), and
% 4.) apodized spatial receive functions
the apodized spatial receive functions
\begin{equation}
 %--------------------------------------------------------------------------------------------------------------
 % c) apodized spatial receive functions
 %--------------------------------------------------------------------------------------------------------------
  \varUpsilon_{m, l}^{(\text{rx})}( \vect{r} )
  =
  - 2
  \int_{ L_{m} }
    \chi_{m, l}^{(\text{rx})}( \vect{r}_{\rho}' )
    g_{l}( \vect{r}_{\rho}' - \vect{r}_{\rho}, - r_{d} )
  \text{d} \vect{r}_{\rho}',
 \label{eqn:lin_mod_exc_sup_qsw_volt_rx_spat_trans}
\end{equation}
\end{subequations}
which correspond to
the spatial impulse responses in
the time domain
\cite{coll:Jensen2002,article:JensenJASA1991}, characterize
% 8.) anisotropic directivities of the planar faces
the anisotropic directivities of
the planar faces.

% book:Devaney2012, Chapter 6: Scattering theory / Sect. 6.1: Potential scattering theory
% - We should note that the INCIDENT FIELD WILL, in fact, BE PRODUCED BY SOME SOURCE RADIATING IN THE BACKGROUND MEDIUM and,
%   hence, will actually SATISFY THE INHOMOGENEOUS HELMHOLTZ EQUATION. (p. 232)
% - However, WE ASSUME THAT THIS SOURCE IS WELL SEPARATED FROM THE SCATTERER so that
%   THE FIELD U^{(in)} WILL SATISFY THE HOMOGENEOUS HELMHOLTZ EQUATION AT LEAST WITHIN THE SCATTERER VOLUME τ0. (p. 232)
% - Moreover, as we discussed in our treatment of the angular-spectrum expansion in Section 4.2 of Chapter 4,
%   the FIELD RADIATED BY A COMPACTLY SUPPORTED SOURCE CAN BE ACCURATELY APPROXIMATED AS
%   A FREE FIELD at distances that are MORE THAN A FEW WAVELENGTHS FROM THE SOURCE SUPPORT VOLUME. (p. 232)
% - Thus, insofar as the POTENTIAL SCATTERING PROBLEM is concerned
%   the INCIDENT FIELD CAN BE MODELED AS A FREE FIELD that
%   SATISFIES THE HOMOGENEOUS HELMHOLTZ EQUATION OVER ALL OF SPACE. (p. 232)
%the lossy heterogeneous object exceeds
%a few wavelengths
%\cite[232]{book:Devaney2012}.

%%%%%%%%%%%%%%%%%%%%%%%%%%%%%%%%%%%%%%%%%%%%%%%%%%%%%%%%%%%%%%%%%%%%%%%%%%%%%%%%%%%%%%%%%%%%%%%%%%%%%%%%%%%%%%%%
% 4.) syntheses of the incident waves
%%%%%%%%%%%%%%%%%%%%%%%%%%%%%%%%%%%%%%%%%%%%%%%%%%%%%%%%%%%%%%%%%%%%%%%%%%%%%%%%%%%%%%%%%%%%%%%%%%%%%%%%%%%%%%%%
\section{Syntheses of the Incident Waves}
\label{sec:syn_p_in}
%---------------------------------------------------------------------------------------------------------------
% 1.) synthesis process and superposition of pulse echoes
%---------------------------------------------------------------------------------------------------------------
% a) modern programmable UI systems synthesize various types of incident waves by superimposing the weighted and delayed waves emitted by the individual elements of the planar transducer array
% coll:Jensen2002, Sect. 5: Spatial Impulse Responses / Sect. 5.3: Geometric Considerations
% - The calculation of the spatial impulse response assumes LINEARITY, and
%   any complex-shaped transducer can therefore be divided into smaller apertures, and
%   THE RESPONSE CAN BE FOUND BY ADDING THE RESPONSES FROM THE SUBAPERTURES. (p. 153)
% - The integral is, as mentioned before, a statement of Huygens’ principle of summing contributions from all areas of the aperture. (p. 153)
% coll:Jensen2002, Sect. 5: Spatial Impulse Responses / Sect. 6: Fields from Array Transducers
% - MOST MODERN SCANNERS USE ARRAYS TO GENERATE AND RECEIVE THE ULTRASOUND FIELDS. (p. 159)
% - These FIELDS ARE QUITE SIMPLE TO CALCULATE, when the SPATIAL IMPULSE RESPONSE FOR A SINGLE ELEMENT IS KNOWN. (p. 159)
% - Since ULTRASOUND PROPAGATION IS ASSUMED TO BE LINEAR,
%   the INDIVIDUAL SPATIAL IMPULSE RESPONSES CAN SIMPLY BE ADDED. (p. 159)
Modern programmable \ac{UI} systems synthesize
% 1.) various types of incident waves
various types of
incident waves by superimposing
% 2.) weighted and delayed waves emitted by the individual elements of the planar transducer array
the weighted and
delayed waves emitted by
the individual elements of
the planar transducer array.
% b) quasi-(d-1)-spherical waves reflect the anisotropic directivities of the vibrating faces and differ from the outgoing (d-1)-spherical waves induced by isotropic point sources
These quasi-$(d-1)$-spherical waves, which are referred to as
% 1.) quasi-cylindrical waves in the two-dimensional Euclidean space
\acp{QCW} in
the two-dimensional Euclidean space, i.e. $d = 2$, reflect
% 2.) anisotropic directivities of the vibrating faces
the anisotropic directivities of
the vibrating faces and, thus, differ from
% 3.) outgoing (d-1)-spherical waves
the outgoing $(d-1)$-spherical waves induced by
isotropic point sources.
% c) LTI measurement process superimposes their pulse echoes such that each incident wave projects a complete SA acquisition sequence on a single pulse-echo measurement
% article:Schiffner2018, Sect. I. Introduction (sec:introduction)
% - In fact, the \ac{LTI} measurement process merely superimposes the weighted and delayed pulse echoes induced by the individual array elements.
The \ac{LTI} measurement process superimposes
% 1.) pulse echoes
their pulse echoes such that
% 2.) incident wave
each incident wave projects
% 3.) complete SA acquisition sequence
a complete \ac{SA} acquisition sequence on
% 4.) single pulse-echo measurement
a single pulse-echo measurement.

%---------------------------------------------------------------------------------------------------------------
% 2.) randomization of the apodization weights and the time delays / random incident waves
%---------------------------------------------------------------------------------------------------------------
% a) highly probable satisfaction of the RIP by certain types of random sensing matrices motivates the randomization of the apodization weights and time delays
% article:Schiffner2018, Sect. II. Compressed Sensing in a Nutshell (sec:compressed_sensing)
% - Fortunately, CERTAIN TYPES OF RANDOM SENSING MATRICES \eqref{eqn:cs_math_prob_general_sensing_matrix} also obey
%   THE \ac{RIP} WITH VERY HIGH PROBABILITY, if
%   THE NUMBER OF OBSERVATIONS IS SUFFICIENTLY LARGE \cite[6]{book:Foucart2013}, \cite{article:TroppPIEEE2010}.
% - Realizations of \ac{IID} random variables governed by certain distributions, e.g.
%   Gaussian or \name{Bernoulli}, as entries \cite[Thm. 5.2]{article:BaraniukCA2008} and
%   randomly and uniformly chosen scaled rows of a \name{Fourier} basis \cite[Thm. 3.3]{article:RudelsonCPAM2008}, for example, require
%   $M \in \bigomega{ s \ln( N / s ) }$ and
%   $M \in \bigomega{ s \ln^{4}( N ) }$ observations, respectively.
% - THESE ORDERS OF GROWTH ARE almost linear in $s$ and, thus, SIGNIFICANTLY BETTER THAN THAT GUARANTEED BY THE WORST-CASE COHERENCE.
The highly probable satisfaction of
% 1.) restricted isometry property (RIP)
the \ac{RIP} by
% 2.) certain types of random sensing matrices
certain types of
random sensing matrices for
% 3.) relatively few observations
relatively few observations
(cf. \cref{sec:compressed_sensing}) motivates
% 4.) randomization of the apodization weights and the time delays
the randomization of
% 5.) apodization weights
the apodization weights and
% 6.) time delays
the time delays.
Unlike
% 1.) steered QPWs
the steered \acp{QPW}, which practically implement
% 2.) steered PWs simplifying image recovery
the steered \acp{PW} in
ultrafast \ac{UI},
% 3.) resulting random waves
the resulting random waves decorrelate
% 4.) pulse echoes of the lossy heterogeneous object's admissible structural building blocks
the pulse echoes of
the lossy heterogeneous object's admissible structural building blocks.
% c) random incident waves strongly compress both the acquisition time and the data volume of a complete SA acquisition sequence
They strongly compress both
% 1.) acquisition time
the acquisition time and
% 2.) data volume
the data volume of
a complete \ac{SA} acquisition sequence and potentially improve
the image recovery.

%%%%%%%%%%%%%%%%%%%%%%%%%%%%%%%%%%%%%%%%%%%%%%%%%%%%%%%%%%%%%%%%%%%%%%%%%%%%%%%%%%%%%%%%%%%%%%%%%%%%%%%%%%%%%%%%
% 1.) superpositions of quasi-(d-1)-spherical waves
%%%%%%%%%%%%%%%%%%%%%%%%%%%%%%%%%%%%%%%%%%%%%%%%%%%%%%%%%%%%%%%%%%%%%%%%%%%%%%%%%%%%%%%%%%%%%%%%%%%%%%%%%%%%%%%%
\subsection{Superpositions of Quasi-$(d-1)$-Spherical Waves}
\label{subsec:syn_p_in_sup_qsw}
%%%%%%%%%%%%%%%%%%%%%%%%%%%%%%%%%%%%%%%%%%%%%%%%%%%%%%%%%%%%%%%%%%%%%%%%%%%%%%%%%%%%%%%%%%%%%%%%%%%%%%%%%%%%%%%%
% graphic: block diagram of the syntheses of the incident waves
%%%%%%%%%%%%%%%%%%%%%%%%%%%%%%%%%%%%%%%%%%%%%%%%%%%%%%%%%%%%%%%%%%%%%%%%%%%%%%%%%%%%%%%%%%%%%%%%%%%%%%%%%%%%%%%%
%
\begin{figure*}[t!]
 \centering%
  \input{syntheses_incident_waves/figures/latex/syn_inc_field_sup_qsw_sgn_proc_chain.tex}
 \caption{}
 \label{fig:V}
\end{figure*}
C
{% a) block diagram of the syntheses of the incident waves
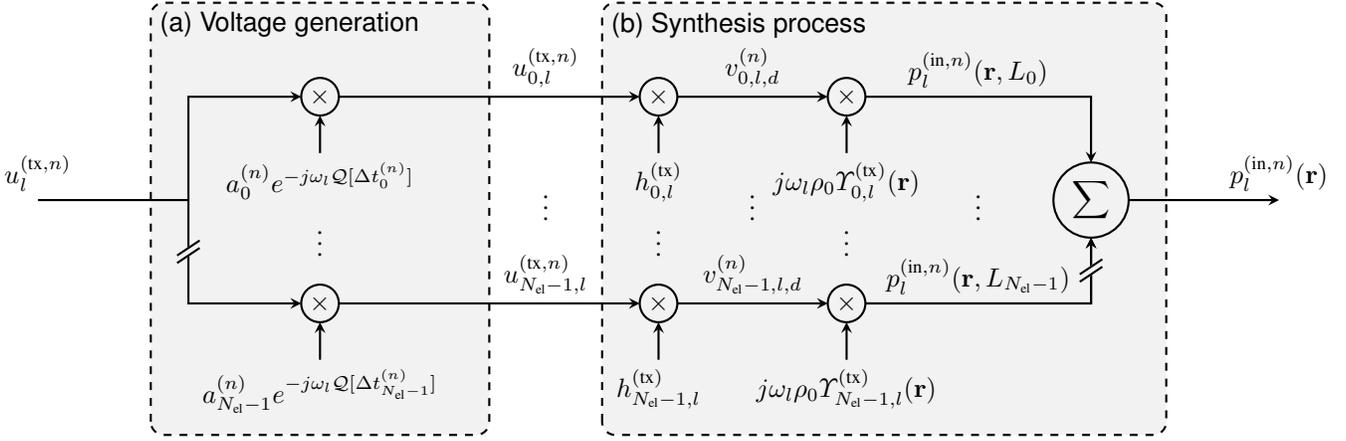
 Block diagram of
 % 1.) syntheses of the incident waves
 the syntheses of
 the incident waves.
 % b) voltage generation maps the reference voltage signals to the excitation voltages
 Given
 % 1.) apodization weights
 the apodization weights
 $a_{m}^{(n)} \in \R$ and
 % 2.) time delays
 the time delays
 $\Delta t_{m}^{(n)} \in \Rnonneg$,
 % 3.) voltage generation
 the voltage generation
 (cf. \subref{fig:syn_sup_qsw_sgn_proc_chain_v_tx}) maps
 % 4.) reference voltage signals exciting all array elements
 the reference voltage signals
 $u_{l}^{(\text{tx}, n)} \in \C$ to
 % 5.) excitation voltages
 the excitation voltages
 \eqref{eqn:syn_p_in_types_v_tx}.
 % c) excitation voltages control the synthesis process that superimposes the quasi-(d-1)-spherical waves emitted by the individual elements of the planar transducer array to form various types of incident waves
 These voltages control
 % 1.) synthesis process
 the synthesis process
 (cf. \subref{fig:syn_sup_qsw_sgn_proc_chain_synthesis}) that superimposes
 % 2.) quasi-(d-1)-spherical waves
 the quasi-$(d-1)$-spherical waves emitted by
 the individual elements of
 the planar transducer array
 \eqref{eqn:syn_sup_qsw_p_in_qsw} to form
 % 3.) incident waves generated by the entire planar transducer array
 various types of
 incident waves
 \eqref{eqn:syn_sup_qsw_p_in}.
}%
{syn_sup_qsw_sgn_proc_chain}

The array elements transduce
% 1.) excitation voltages
their excitation voltages
$u_{m, l}^{(\text{tx}, n)} \in \C$ into
% 2.) homogeneous r_{d}-component of the particle velocity
the homogeneous normal velocities on
% 3.) planar faces L_{m}
their planar faces
$L_{m} \subset \R^{d-1}$
\cite{article:LabyedITUFFC2014,article:NgITUFFC2006}.
%  $v_{m, l, d}^{(n)} \in \C$,
% b) RS diffraction equations uniquely solve the Helmholtz equation for the rigid baffle and represent the individual quasi-(d-1)-spherical waves by the incident acoustic pressure fields
% book:Cobbold2006, Sect. 2.1: Rayleigh-Sommerfeld Diffraction Equations
% - To DETERMINE THE PRESSURE OR VELOCITY FIELD PRODUCED BY
%   A VIBRATING SOURCE EXCITED BY AN ARBITRARY WAVEFORM, we
%   must solve the APPROPRIATE WAVE EQUATION that characterizes the manner in which the wave is propagated and
%   must CONSTRAIN THE SOLUTION TO MEET THE BOUNDARY CONDITIONS defined by the problem. (p. 97)
The \acl{RS} diffraction equations uniquely solve
% 1.) Helmholtz equations for the incident acoustic pressure fields
the \name{Helmholtz} equations
\eqref{eqn:lin_mod_sol_wave_eq_pde_p_in} for
% 2.) rigid baffle
these boundary conditions
%the rigid baffle
(cf. e.g.
\cite[(2.48)]{book:Devaney2012},		% "wavefield", d-dimensional space, dispersive homogeneous medium, temporal frequency domain [CHECKED: CORRECT! (Dirichlet, d=2,3)] (3-34)]
\cite[(13) of §8.11]{book:Born1999}%		% "wavefield", three-dimensional space, lossless homogeneous medium, temporal frequency domain [CHECKED: CORRECT! (12)]
) and represent
% 3.) individual quasi-(d-1)-spherical waves
the individual quasi-$(d-1)$-spherical waves by
% 4.) incident acoustic pressure fields [individual quasi-(d-1)-spherical waves]
the incident acoustic pressure fields
\begin{subequations}
\label{eqn:syn_sup_qsw_p_in_qsw}
\begin{equation}
 %--------------------------------------------------------------------------------------------------------------
 % a) incident acoustic pressure fields [individual quasi-(d-1)-spherical waves]
 %--------------------------------------------------------------------------------------------------------------
  p_{l}^{(\text{in}, n)}( \vect{r}, L_{m} )
  =
  j \omega_{l} \rho_{0}
  h_{m, l}^{(\text{tx})}
  u_{m, l}^{(\text{tx}, n)}
  \varUpsilon_{m, l}^{(\text{tx})}( \vect{r} )
 \label{eqn:syn_sup_qsw_p_in_qsw_exp}
\end{equation}
for
% 6.) all sequential pulse-echo measurements, all relevant discrete frequencies, and all array elements
all $( n, l, m ) \in \setconsnonneg{ N_{\text{in}} - 1 } \times \setsymbol{L}_{ \text{BP} }^{(n)} \times \setconsnonneg{ N_{\text{el}} - 1 }$, where
% 7.) transmitter electromechanical transfer functions
$h_{m, l}^{(\text{tx})} \in \C$ denote
the electromechanical transfer functions, and
% 8.) apodized spatial transmit functions
the apodized spatial transmit functions
\begin{equation}
 %--------------------------------------------------------------------------------------------------------------
 % b) apodized spatial transmit functions
 %--------------------------------------------------------------------------------------------------------------
  \varUpsilon_{m, l}^{(\text{tx})}( \vect{r} )
  =
  - 2
  \int_{ L_{m} }
    \chi_{m, l}^{(\text{tx})}( \vect{r}_{\rho}' )
    g_{l}( \vect{r}_{\rho} - \vect{r}_{\rho}', r_{d} )
  \text{d} \vect{r}_{\rho}'
 \label{eqn:syn_sup_qsw_p_in_qsw_spat_trans}
\end{equation}
\end{subequations}
characterize
% 9.) anisotropic directivities of the vibrating faces
the anisotropic directivities of
the vibrating faces for
% 10.) positive axial coordinate
all $r_{d} > 0$, similar to
% 11.) apodized spatial receive functions
the apodized spatial receive functions
\eqref{eqn:lin_mod_exc_sup_qsw_volt_rx_spat_trans}
(cf. \cref{tab:lin_mod_scan_config_instrum_params}).
%\begin{equation}
 %--------------------------------------------------------------------------------------------------------------
 % homogeneous r_{d}-component of the particle velocity on the planar face L_{m}
 %--------------------------------------------------------------------------------------------------------------
  %v_{m, l, d}^{(n)}
  %=
  %h_{m, l}^{(\text{tx})}
  %u_{m, l}^{(\text{tx}, n)},
 %\label{eqn:lin_mod_scan_config_trans_array_transfer_v_d}
%\end{equation}
% 1.) transmitter apodization functions
%the transmitter apodization functions
%$\chi_{m, l}^{(\text{tx})}: L_{m} \mapsto \C$ into %, which account for
% b) apodized spatial transmit functions relate the geometry of the vibrating faces and the effect of the acoustic lens to the velocity potential induced by a temporally-impulsive normal velocity
%They relate
% 1.) geometry of the vibrating faces
%the geometry of
%the vibrating faces and
% 2.) effect of the acoustic lens
%the effect of
%the acoustic lens to
% 3.) velocity potential induced by a temporally-impulsive normal velocity
%the velocity potential induced by
%a temporally-impulsive normal velocity.
% c) superpositions represent the incident waves emitted by the entire planar transducer array by the incident acoustic pressure fields
Their superpositions represent
% 1.) incident waves emitted by the entire planar transducer array
the incident waves by
% 2.) incident acoustic pressure fields [superpositions of quasi-(d-1)-spherical waves]
the acoustic pressure fields
\begin{equation}
 %--------------------------------------------------------------------------------------------------------------
 % incident acoustic pressure fields [superpositions of quasi-(d-1)-spherical waves]
 %--------------------------------------------------------------------------------------------------------------
  p_{l}^{(\text{in}, n)}( \vect{r} )
  =
  j \omega_{l} \rho_{0}
  \sum_{ m = 0 }^{ N_{\text{el}} - 1 }
    h_{m, l}^{(\text{tx})}
    u_{m, l}^{(\text{tx}, n)}
    \varUpsilon_{m, l}^{(\text{tx})}( \vect{r} )
 \label{eqn:syn_sup_qsw_p_in}
\end{equation}
for
% 3.) all sequential pulse-echo measurements and all relevant discrete frequencies
all $( n, l ) \in \setconsnonneg{ N_{\text{in}} - 1 } \times \setsymbol{L}_{ \text{BP} }^{(n)}$, where
% 4.) excitation voltages
the excitation voltages determine
% 5.) synthesized types of incident waves
the synthesized types of
incident waves
(cf. \cref{fig:syn_sup_qsw_sgn_proc_chain_synthesis}).

%%%%%%%%%%%%%%%%%%%%%%%%%%%%%%%%%%%%%%%%%%%%%%%%%%%%%%%%%%%%%%%%%%%%%%%%%%%%%%%%%%%%%%%%%%%%%%%%%%%%%%%%%%%%%%%%
% 2.) types of incident waves
%%%%%%%%%%%%%%%%%%%%%%%%%%%%%%%%%%%%%%%%%%%%%%%%%%%%%%%%%%%%%%%%%%%%%%%%%%%%%%%%%%%%%%%%%%%%%%%%%%%%%%%%%%%%%%%%
\subsection{Types of Incident Waves}
\label{subsec:syn_p_in_types}
%---------------------------------------------------------------------------------------------------------------
% 1.) generation of the excitation voltages
%---------------------------------------------------------------------------------------------------------------
% a) generation of the excitation voltages typically applies quantized apodization weights and time delays to the reference voltage signals
% article:Schiffner2018, Sect. I. Introduction (sec:introduction)
% - The latter typically applies apodization weights and time delays to a common reference voltage signal exciting the individual array elements.
% - Their specifications provide hundreds of degrees of freedom to synthesize alternative types of incident waves.
% book:Bushberg2011, Chapter 14: Ultrasound / Sect. 14.4: Ultrasound Beam Properties
% - FIGURE 14-15
The generation of
the excitation voltages typically applies
% 1.) quantized apodization weights
quantized apodization weights
$a_{m}^{(n)} \in \R$ and
% 2.) quantized time delays
time delays
$\Delta t_{m}^{(n)} \in \Rnonneg$ to
% 3.) reference voltage signals identically exciting all array elements
the reference voltage signals
$u_{l}^{(\text{tx}, n)} \in \C$, whose electric energies are
constant for
% 4.) all sequential pulse-echo measurements
all $n \in \setconsnonneg{ N_{\text{in}} - 1 }$.
% b) apodization weights of unity absolute values then ensure the energy equivalence of the incident waves and only require the quantization of the time delays
% \norm{ \vect{u}^{(\text{tx}, n)} }{2}^{2} \norm{ \vect{a}^{(n)} }{2}^{2}
Apodization weights of
unity absolute values then ensure
% 1.) energy equivalence of the incident waves
the energy equivalence of
the incident waves and only require
% 2.) quantization of the time delays
the quantization of
the time delays.
% c) excitation voltages for all sequential pulse-echo measurements and all array elements
For
% 1.) clock signal with the frequency f_{\text{clk}}
a clock signal with
the frequency
$f_{\text{clk}} = 1 / T_{\text{clk}} \in \Rplus$ and
% 2.) quantization operator
the quantization operator
\begin{subequations}
\label{eqn:syn_p_in_types_v_tx}
\begin{equation}
 %--------------------------------------------------------------------------------------------------------------
 % a) quantization operator providing the admissible time delays
 %--------------------------------------------------------------------------------------------------------------
  \mathcal{Q} \bigl[ \Delta t_{m}^{(n)} \bigr]
  =
  \dround{ \Delta t_{m}^{(n)} f_{\text{clk}} }{1} T_{\text{clk}},
 \label{eqn:syn_p_in_types_v_tx_quantization}
\end{equation}
% 3.) generated excitation voltages
the generated excitation voltages are
\begin{equation}
 %--------------------------------------------------------------------------------------------------------------
 % b) excitation voltages (single pulse-echo measurement, monofrequent, single array element)
 %--------------------------------------------------------------------------------------------------------------
  u_{m, l}^{(\text{tx}, n)}
  =
  u_{l}^{(\text{tx}, n)}
  a_{m}^{(n)}
  e^{ -j \omega_{l} \mathcal{Q} \left[ \Delta t_{m}^{(n)} \right] }
 \label{eqn:syn_p_in_types_v_tx_expression}
\end{equation}
\end{subequations}
for
% 4.) all sequential pulse-echo measurements, all relevant discrete frequencies, and all array elements
all $( n, l, m ) \in \setconsnonneg{ N_{\text{in}} - 1 } \times \setsymbol{L}_{ \text{BP} }^{(n)} \times \setconsnonneg{ N_{\text{el}} - 1 }$, where
% 5.) asymmetric brackets denote the rounding to the nearest integer
the asymmetric brackets denote
the rounding to
the nearest integer and
% 6.) complex exponential functions induce the phase shifts corresponding to the time delays
the complex exponential functions induce
the phase shifts corresponding to
the time delays
(cf. \cref{fig:syn_sup_qsw_sgn_proc_chain_v_tx}).

%%%%%%%%%%%%%%%%%%%%%%%%%%%%%%%%%%%%%%%%%%%%%%%%%%%%%%%%%%%%%%%%%%%%%%%%%%%%%%%%%%%%%%%%%%%%%%%%%%%%%%%%%%%%%%%%
% 1.) steered quasi-plane waves
%%%%%%%%%%%%%%%%%%%%%%%%%%%%%%%%%%%%%%%%%%%%%%%%%%%%%%%%%%%%%%%%%%%%%%%%%%%%%%%%%%%%%%%%%%%%%%%%%%%%%%%%%%%%%%%%
\subsubsection{Steered Quasi-Plane Waves}
%\label{subsubsec:syn_p_in_types_qpw}
%%%%%%%%%%%%%%%%%%%%%%%%%%%%%%%%%%%%%%%%%%%%%%%%%%%%%%%%%%%%%%%%%%%%%%%%%%%%%%%%%%%%%%%%%%%%%%%%%%%%%%%%%%%%%%%%
% graphic: emission of a steered quasi-plane wave (QPW) by a linear transducer array (two-dimensional Euclidean space)
%%%%%%%%%%%%%%%%%%%%%%%%%%%%%%%%%%%%%%%%%%%%%%%%%%%%%%%%%%%%%%%%%%%%%%%%%%%%%%%%%%%%%%%%%%%%%%%%%%%%%%%%%%%%%%%%
%
\begin{figure*}[t!]
 \centering%
  \input{syntheses_incident_waves/figures/latex/syn_inc_field_sup_qsw_emission.tex}
 \caption{}
 \label{fig:V}
\end{figure*}
C
{%--------------------------------------------------------------------------------------------------------------
 % 1.) general description
 %--------------------------------------------------------------------------------------------------------------
 % a) syntheses of a steered QPW and a superposition of both randomly-apodized and randomly-delayed QCWs in the two-dimensional Euclidean space
 Syntheses of
 % 1.) steered QPW
 a steered \acf{QPW}\acused{QPW}
 (cf. \subref{fig:lin_mod_exc_sup_qsw_emission_qpw}) and
 % 2.) superposition of both randomly-apodized and randomly-delayed QCWs
 a proposed type of
 random wave
 (cf. \subref{fig:lin_mod_exc_sup_qsw_emission_rnd_apo_del}) in
 % 3.) two-dimensional Euclidean space
 the two-dimensional Euclidean space, i.e. $d = 2$.
 % b) gray semicircles represent the individual QCWs emitted by each element of the linear transducer array
 The gray semicircles represent
 the individual \acfp{QCW}\acused{QCW} emitted by
 each element of
 the linear transducer array.
 % c) dashed semicircles indicate negatively-apodized QCWs
 Their dashed variants indicate
 negatively-apodized \acp{QCW}, i.e.
 $a_{m}^{(0)} = - 1$ in
 \eqref{eqn:syn_p_in_types_v_tx_rnd_apo_del}, whereas
 % d) solid semicircles indicate positively-apodized QCWs
 their solid variants indicate
 positively-apodized \acp{QCW}, i.e.
 $a_{m}^{(0)} = 1$ in
 \eqref{eqn:syn_p_in_types_v_tx_qpw} and
 \eqref{eqn:syn_p_in_types_v_tx_rnd_apo_del}.
 % f) distinct radii reflect the time delays
 Their distinct radii reflect
 the time delays in
 \eqref{eqn:syn_p_in_types_v_tx_qpw} and
 \eqref{eqn:syn_p_in_types_v_tx_rnd_apo_del}, whose
 % g) maximum time delays
 maxima equal
 $\Delta t_{\text{max}}^{(0)} = \max_{ m \in \setconsnonneg{ N_{\text{el}} - 1 } }\{ \Delta t_{m}^{(0)} \} = ( N_{\text{el}} - 1 ) \Delta r_{\text{el}, 1} \tabs{ \uvectcomp{ \vartheta }{ 1 }^{(0)} } / c_{\text{ref}}$. % = ( N_{\text{el}} - 1 ) T_{\text{inc}}^{(n)}$.
 %, at
 % 4.) time instant \Delta t_{\text{max}}
 %the time instant
 %$t = \mathcal{Q}[ \Delta t_{\text{max}} ]$.
 %--------------------------------------------------------------------------------------------------------------
 % 2.) steered QPW
 %--------------------------------------------------------------------------------------------------------------
 % a) dashed black line indicates the approximated planar wavefront
 The dashed black line indicates
 the approximated planar wavefront.
 % b) black arrow shows the preferred direction of propagation
 The black arrow shows
 the preferred direction of
 propagation.
 %$\uvect{\vartheta}^{(0)} = \trans{ ( \cos( \vartheta ), \sin( \vartheta ) ) }$ with
 %$\vartheta = 13 \pi / 36 = \SI{65}{\degree}$.
 % c) positive r_{1}-component requires the reference position to equal r_{\text{el}, 0, 1} \uvect{1}
 %Its positive $r_{1}$-component requires
 % 1.) components of the reference positions for the time delays
 %the reference position with
 %the components
 %\eqref{eqn:syn_p_in_types_v_tx_qpw_r_ref} to equal
 %$\vect{r}_{\text{ref}}^{(0)} = r_{\text{el}, 0, 1} \uvect{1}$.
}%
{lin_mod_exc_sup_qsw_emission}

%---------------------------------------------------------------------------------------------------------------
% 1.) apodization weights and time delays for steered QPWs
%---------------------------------------------------------------------------------------------------------------
% a) steered QPWs denote the approximations of steered PWs synthesized by the UI system
Steered \acp{QPW} denote
the approximations of
steered \acp{PW} synthesized by
the \ac{UI} system.
% b) time delays in the excitation voltages depend affine-linearly on the center coordinates of the vibrating faces on each coordinate axis
The time delays in
% 1.) excitation voltages
the excitation voltages
\eqref{eqn:syn_p_in_types_v_tx} depend affine-linearly on
% 2.) center coordinates of the vibrating faces
the center coordinates of
the vibrating faces on
each coordinate axis, whereas
% c) all apodization weights equal unity
all apodization weights equal
unity.
% d) apodization weights and time delays in the excitation voltages
% book:Schmerr2015, Sect. 8.1 Beam Steering in 3-D
% - Consider an element of a 2-D array as shown in Fig. 8.1 where we want to steer the ultrasonic beam of
%   the array in the direction of the unit vector, u. (p. 169)
% - Steering of the beam in this direction can be accomplished by applying a LINEARLY VARYING TIME SHIFT,
%   ∆t= u*x / c, over the face of the array and evaluating that phase at the centroids of the individual elements. (p. 169)
% - Since the delays in Eq. (8.1) contain both positive and negative values, we can simply add a constant delay equal to
%   the magnitude of largest negative value to obtain a proper time delay law, ∆t_{mn}^{d}, given by (8.4). (p. 169)
These specifications yield
\begin{subequations}
\label{eqn:syn_p_in_types_v_tx_qpw}
\begin{align}
 %--------------------------------------------------------------------------------------------------------------
 % a) apodization weights
 %--------------------------------------------------------------------------------------------------------------
  a_{m}^{(n)}
  &=
  1
  & \text{and} & &
 %--------------------------------------------------------------------------------------------------------------
 % b) time delays
 %--------------------------------------------------------------------------------------------------------------
  \Delta t_{m}^{(n)}
  &=
  \frac{
    \dinprod{ \vect{r}_{\text{el}, m} - \vect{r}_{\text{ref}}^{(n)} }{ \uvect{\vartheta}^{(n)} }{1}
  }{
    c_{\text{ref}}
  }
 \label{eqn:syn_p_in_types_v_tx_qpw_apo_del}
\end{align}
for
% 1.) all sequential pulse-echo measurements and all array elements
all $( n, m ) \in \setconsnonneg{ N_{\text{in}} - 1 } \times \setconsnonneg{ N_{\text{el}} - 1 }$, where
% 2.) center coordinates of the vibrating faces
$\vect{r}_{\text{el}, m} \in \mathcal{M}$ denote
the center coordinates of
the vibrating faces,
% 3.) preferred directions of propagation
$\uvect{\vartheta}^{(n)} = \trans{ ( \uvectcomp{ \vartheta }{ 1 }^{(n)}, \dotsc, \uvectcomp{ \vartheta }{ d }^{(n)} ) } \in \uhemisphere{d-1}$ indicate
the preferred directions of
propagation, and
% 4.) reference positions for the time delays and their components
$\vect{r}_{\text{ref}}^{(n)} = \trans{ ( r_{\text{ref}, 1}^{(n)}, \dotsc, r_{\text{ref}, d-1}^{(n)}, 0 ) }$ are
the reference positions with
the components
\begin{equation}
 %--------------------------------------------------------------------------------------------------------------
 % c) components of the reference positions for the time delays
 %--------------------------------------------------------------------------------------------------------------
  r_{\text{ref}, \delta}^{(n)}
  =
  \begin{cases}
    - M_{\text{el}, \delta} \Delta r_{\text{el}, \delta} & \text{for } \uvectcomp{ \vartheta }{ \delta }^{(n)} \geq 0,\\
      M_{\text{el}, \delta} \Delta r_{\text{el}, \delta} & \text{for } \uvectcomp{ \vartheta }{ \delta }^{(n)} < 0,
  \end{cases}
 \label{eqn:syn_p_in_types_v_tx_qpw_r_ref}
\end{equation}
\end{subequations}
for
% 5.) all coordinate axes
all $\delta \in \setcons{ d - 1 }$
(cf. \cref{tab:lin_mod_scan_config_instrum_params}).
% e) components of the reference positions ensure the nonnegativity of the time delays and the causality of the voltage generation
The latter ensure
% 1.) nonnegativity of the time delays
the nonnegativity of
the time delays and, thus,
% 2.) causality of the voltage generation
the causality of
the voltage generation.
% f) finite number of elements and their anisotropic directivities limit the accuracies of the approximations to a bounded area in front of the array
% article:Schiffner2018, Sect. I. Introduction (sec:introduction)
% - The FINITE NUMBER OF EVENLY SPACED ARRAY ELEMENTS and their ANISOTROPIC DIRECTIVITIES, for example, prevent
%   the exact syntheses of
%   [1.)] NON-DIFFRACTING BEAMS, whose spatial extent and amount of transferred energy are unlimited, and
%   [2.)] outgoing $(d-1)$-spherical waves, whose isotropic sources are points.
The finite number of
elements and
their anisotropic directivities limit
the accuracies of
the approximations to
bounded areas in front of
the array
(cf. \cref{fig:lin_mod_exc_sup_qsw_emission_qpw}).

%%%%%%%%%%%%%%%%%%%%%%%%%%%%%%%%%%%%%%%%%%%%%%%%%%%%%%%%%%%%%%%%%%%%%%%%%%%%%%%%%%%%%%%%%%%%%%%%%%%%%%%%%%%%%%%%
% 2.) superpositions of randomly-apodized quasi-(d-1)-spherical waves
%%%%%%%%%%%%%%%%%%%%%%%%%%%%%%%%%%%%%%%%%%%%%%%%%%%%%%%%%%%%%%%%%%%%%%%%%%%%%%%%%%%%%%%%%%%%%%%%%%%%%%%%%%%%%%%%
\subsubsection{Randomly-Apodized Quasi-$(d-1)$-Spherical Waves}
%\label{subsubsec:syn_p_in_types_rnd_apo}
%---------------------------------------------------------------------------------------------------------------
% 1.) superposition of randomly-apodized quasi-(d-1)-spherical waves
%---------------------------------------------------------------------------------------------------------------
% a) first type of random wave uses realizations of i.i.d. random variables as the apodization weights
The first type of
random wave uses
% 1.) realizations of i.i.d. random variables as the apodization weights
realizations of
\ac{IID} random variables $\Lambda_{m}^{(n)}$ as
the apodization weights in
% 2.) excitation voltages
the excitation voltages
\eqref{eqn:syn_p_in_types_v_tx}, whereas
% b) all time delays equal zero
all time delays equal
zero.
% c) apodization weights and time delays in the excitation voltages
These specifications yield
\begin{subequations}
\label{eqn:syn_p_in_types_v_tx_rnd_apo}
\begin{align}
 %--------------------------------------------------------------------------------------------------------------
 % a) apodization weights
 %--------------------------------------------------------------------------------------------------------------
  a_{m}^{(n)}
  &=
  \Lambda_{m}^{(n)}
  & \text{and} & &
 %--------------------------------------------------------------------------------------------------------------
 % b) time delays
 %--------------------------------------------------------------------------------------------------------------
  \Delta t_{m}^{(n)}
  &=
  0
 \label{eqn:syn_p_in_types_v_tx_rnd_apo_apo_del}
\end{align}
for
% 1.) all sequential pulse-echo measurements and all transducer elements
all $( n, m ) \in \setconsnonneg{ N_{\text{in}} - 1 } \times \setconsnonneg{ N_{\text{el}} - 1 }$.
% d) Bernoulli distribution exclusively requires inverting buffer amplifiers and enables a relatively simple hardware realization
The \name{Bernoulli} distribution
\begin{equation}
 %--------------------------------------------------------------------------------------------------------------
 % c) Bernoulli distribution
 %--------------------------------------------------------------------------------------------------------------
  \Lambda_{m}^{(n)}
  =
  \begin{cases}
    1  & \text{with probability } 0.5,\\
    -1 & \text{with probability } 0.5,
  \end{cases}
 \label{eqn:syn_p_in_types_v_tx_rnd_apo_apo_bernoulli}
\end{equation}
\end{subequations}
exclusively requires
% 1.) inverting buffer amplifiers
inverting buffer amplifiers and, thus, enables
% 2.) relatively simple hardware realization
a relatively simple hardware realization.
% e) Bernoulli distribution maintains the acoustic energies transferred by the steered QPWs
%It maintains
%the acoustic energies transferred by
%the steered \acp{QPW}.

%%%%%%%%%%%%%%%%%%%%%%%%%%%%%%%%%%%%%%%%%%%%%%%%%%%%%%%%%%%%%%%%%%%%%%%%%%%%%%%%%%%%%%%%%%%%%%%%%%%%%%%%%%%%%%%%
% 3.) superpositions of randomly-delayed quasi-(d-1)-spherical waves
%%%%%%%%%%%%%%%%%%%%%%%%%%%%%%%%%%%%%%%%%%%%%%%%%%%%%%%%%%%%%%%%%%%%%%%%%%%%%%%%%%%%%%%%%%%%%%%%%%%%%%%%%%%%%%%%
\subsubsection{Randomly-Delayed Quasi-$(d-1)$-Spherical Waves}
\label{subsubsec:syn_p_in_types_rnd_del}
%---------------------------------------------------------------------------------------------------------------
% 1.) apodization weights and time delays for the superpositions of randomly-delayed quasi-(d-1)-spherical waves
%---------------------------------------------------------------------------------------------------------------
% a) second type of random wave assigns random permutations of uniformly spaced time instants to the time delays in the excitation voltages
The second type of
random wave assigns
% 1.) random permutations of uniformly spaced time instants
random permutations of
uniformly spaced time instants to
the time delays in
% 2.) excitation voltages
the excitation voltages
\eqref{eqn:syn_p_in_types_v_tx}, whereas
% b) all apodization weights equal unity
all apodization weights equal
unity.
% c) apodization weights and time delays in the excitation voltages
These specifications yield
\begin{subequations}
\label{eqn:syn_p_in_types_v_tx_rnd_del}
\begin{align}
 %--------------------------------------------------------------------------------------------------------------
 % a) apodization weights
 %--------------------------------------------------------------------------------------------------------------
  a_{m}^{(n)}
  &=
  1
  & \text{and} & &
 %--------------------------------------------------------------------------------------------------------------
 % b) time delays
 %--------------------------------------------------------------------------------------------------------------
  \Delta t_{m}^{(n)}
  &=
  \dpermutel{ \setconsnonneg{ N_{\text{el}} - 1 } }{ m }{ n }{1}
  T_{\text{inc}}^{(n)}
 \label{eqn:syn_p_in_types_v_tx_rnd_del_apo_del}
\end{align}
for
% 1.) all sequential pulse-echo measurements and all transducer elements
all $( n, m ) \in \setconsnonneg{ N_{\text{in}} - 1 } \times \setconsnonneg{ N_{\text{el}} - 1 }$, where
% 2.) elements of index m in random permutations
$\tpermutel{ \setconsnonneg{ N_{\text{el}} - 1 } }{ m }{ n }$ denote
the elements of
index $m$ in
random permutations of
the set
$\setconsnonneg{ N_{\text{el}} - 1 }$ and
% 3.) fixed time periods
$T_{\text{inc}}^{(n)} \in \Rplus$ are
fixed time periods.
% d) fixed time periods significantly influence the properties of the incident waves and those of the observation operators
The latter significantly influence
the properties of
% 1.) incident acoustic pressure fields generated by the entire planar transducer array
the superpositions and, consequently,
those of
% 2.) pulse-echo measurement process
the pulse-echo measurement process
\eqref{eqn:lin_mod_v_rx_born_obs_proc}.

%---------------------------------------------------------------------------------------------------------------
% 2.) specifications of the fixed time periods
%---------------------------------------------------------------------------------------------------------------
% a) superpositions converge to steered QPWs with the preferred directions of propagation \uvect{d}
In
the limiting cases
$T_{\text{inc}}^{(n)} \rightarrow 0+$,
the superpositions converge to
% 1.) steered QPWs
steered \acp{QPW} with
% 2.) preferred directions of propagation
the preferred directions of
propagation
$\uvect{\vartheta}^{(n)} = \uvect{d}$, because
% 3.) apodization weights and time delays in the excitation voltages (superpositions of randomly-delayed quasi-(d-1)-spherical waves)
the apodization weights and
the time delays in
\eqref{eqn:syn_p_in_types_v_tx_rnd_del_apo_del} converge to
% 4.) apodization weights and time delays in the excitation voltages (steered QPWs)
those in
\eqref{eqn:syn_p_in_types_v_tx_qpw}.
% b) fixed time periods smaller than the upper bounds in the specified observation time intervals induce range ambiguities that can be resolved by the proposed method
%\TODO{notwendig? ordentlich!}
% TODO: move to discussion!
%Fixed time periods smaller than
%the round-trip times-of-flight (TOFs)
%the upper bounds in
% 1.) specified observation time intervals for the received RF voltage signals
%the specified observation time intervals for
%the received \ac{RF} voltage signals
%\eqref{eqn:lin_mod_scan_config_volt_rx_obs_interval}, i.e.
%$T_{\text{inc}}^{(n)} < t_{\text{ub}}^{(n)}$, induce
%range ambiguities that
%are undesired in
%the established image recovery methods but
%can be resolved by
%the proposed method.
%Since the choice of
%a fixed time interval
%$T_{\text{inc}}$ equal to or exceeding
%the upper bound in
%% 1.) specified observation time interval for the received RF voltage signals
%the specified observation time interval for
%the received \ac{RF} voltage signals
%\eqref{eqn:lin_mod_scan_config_volt_rx_obs_interval}, i.e.
%$T_{\text{inc}} \geq t_{\text{ub}}^{(n)}$, results in
%% 1.) complete SA acquisition sequence
%the complete \ac{SA} acquisition sequence,
% c) specific fixed time periods result in the syntheses times of the steered QPWs
The specific fixed time periods
\begin{equation}
\begin{split}
 %--------------------------------------------------------------------------------------------------------------
 % c) fixed time periods permuting the time delays for steered QPWs
 %--------------------------------------------------------------------------------------------------------------
  T_{\text{inc}}^{(n)}
  &=
  \frac{ 1 }{ N_{\text{el}} - 1 }
  \underset{ m \in \setconsnonneg{ N_{\text{el}} - 1 } }{ \max }
  \left\{
    \frac{
      \dinprod{ \vect{r}_{\text{el}, m} - \vect{r}_{\text{ref}}^{(n)} }{ \uvect{\vartheta}^{(n)} }{1}
    }{
      c_{\text{ref}}
    }
  \right\}\\
  &=
  \frac{ 1 }{ N_{\text{el}} - 1 }
  \sum_{ \delta = 1 }^{ d - 1 }
    ( N_{\text{el}, \delta} - 1 )
    \frac{
      \Delta r_{\text{el}, \delta}
      \dabs{ \uvectcomp{ \vartheta }{ \delta }^{(n)} }{1}
    }{
      c_{\text{ref}}
    }
\end{split}
\label{eqn:syn_p_in_types_v_tx_rnd_del_interval}
\end{equation}
\end{subequations}
result in
% 1.) syntheses times
the syntheses times of
% 2.) steered QPWs
the steered \acp{QPW}.
% d) specific fixed time periods induce random permutations of the time delays specified for the steered QPWs (two-dimensional Euclidean space)
In
the two-dimensional Euclidean space, i.e. $d = 2$,
they simplify to
$T_{\text{inc}}^{(n)} = \Delta r_{\text{el}, 1} \tabs{ \uvectcomp{ \vartheta }{ 1 }^{(n)} } / c_{\text{ref}}$ and induce
random permutations of
the time delays specified for
% 2.) steered QPWs
the steered \acp{QPW} in
\eqref{eqn:syn_p_in_types_v_tx_qpw}.

%%%%%%%%%%%%%%%%%%%%%%%%%%%%%%%%%%%%%%%%%%%%%%%%%%%%%%%%%%%%%%%%%%%%%%%%%%%%%%%%%%%%%%%%%%%%%%%%%%%%%%%%%%%%%%%%
% 4.) superpositions of randomly-apodized and randomly-delayed quasi-(d-1)-spherical waves
%%%%%%%%%%%%%%%%%%%%%%%%%%%%%%%%%%%%%%%%%%%%%%%%%%%%%%%%%%%%%%%%%%%%%%%%%%%%%%%%%%%%%%%%%%%%%%%%%%%%%%%%%%%%%%%%
\subsubsection{Randomly-Apodized and Randomly-Delayed Quasi-$(d-1)$-Spherical Waves}
%\label{subsubsec:syn_p_in_types_rnd_apo_del}
%---------------------------------------------------------------------------------------------------------------
% 1.) superposition of randomly-apodized and randomly-delayed quasi-(d-1)-spherical waves
%---------------------------------------------------------------------------------------------------------------
% a) third type of random wave combines realizations of i.i.d. random variables as the apodization weights with random permutations of uniformly spaced time instants as the time delays
The third type of
random wave combines
% 1.) realizations of i.i.d. random variables as the apodization weights
realizations of
\ac{IID} random variables $\Lambda_{m}^{(n)}$ as
the apodization weights with
% 2.) random permutations of uniformly spaced time instants as the time delays
random permutations of
uniformly spaced time instants as
the time delays in
% 3.) excitation voltages
the excitation voltages
\eqref{eqn:syn_p_in_types_v_tx}.
% b) apodization weights and time delays in the excitation voltages
These specifications yield
\begin{align}
 %--------------------------------------------------------------------------------------------------------------
 % a) apodization weights
 %--------------------------------------------------------------------------------------------------------------
  a_{m}^{(n)}
  &=
  \Lambda_{m}^{(n)}
  & \text{and} & &
 %--------------------------------------------------------------------------------------------------------------
 % b) time delays
 %--------------------------------------------------------------------------------------------------------------
  \Delta t_{m}^{(n)}
  &=
  \dpermutel{ \setconsnonneg{ N_{\text{el}} - 1 } }{ m }{ n }{1}
  T_{\text{inc}}^{(n)}
 \label{eqn:syn_p_in_types_v_tx_rnd_apo_del}
\end{align}
for
% 1.) all sequential pulse-echo measurements and all transducer elements
all $( n, m ) \in \setconsnonneg{ N_{\text{in}} - 1 } \times \setconsnonneg{ N_{\text{el}} - 1 }$, where
% 2.) Bernoulli distribution
the \name{Bernoulli} distribution
\eqref{eqn:syn_p_in_types_v_tx_rnd_apo_apo_bernoulli} is combined with
% 3.) fixed time periods permuting the time delays for steered QPWs
the fixed time periods
\eqref{eqn:syn_p_in_types_v_tx_rnd_del_interval}.
% c) specifications result in irregular wavefronts that do not resemble any common models
They result in
irregular wavefronts that
do not resemble
any common models
(cf. \cref{fig:lin_mod_exc_sup_qsw_emission_rnd_apo_del}).

%%%%%%%%%%%%%%%%%%%%%%%%%%%%%%%%%%%%%%%%%%%%%%%%%%%%%%%%%%%%%%%%%%%%%%%%%%%%%%%%%%%%%%%%%%%%%%%%%%%%%%%%%%%%%%%%
% 5.) image recovery based on compressed sensing
%%%%%%%%%%%%%%%%%%%%%%%%%%%%%%%%%%%%%%%%%%%%%%%%%%%%%%%%%%%%%%%%%%%%%%%%%%%%%%%%%%%%%%%%%%%%%%%%%%%%%%%%%%%%%%%%
\section{Image Recovery Based on Compressed Sensing}
\label{sec:recovery}
%---------------------------------------------------------------------------------------------------------------
% 1.) overview of the image recovery
%---------------------------------------------------------------------------------------------------------------
% a) proposed method aims at recovering the compressibility fluctuations inside the specified FOV from only a few sequential measurements of the received RF voltage signals
The proposed method aims at
recovering
% 1.) relative spatial fluctuations in the unperturbed compressibility
the compressibility fluctuations
\eqref{eqn:lin_mod_mech_model_tis_simple_rel_fluctuations} inside
% 2.) specified FOV
the specified \ac{FOV}
(cf. \cref{fig:lin_mod_scan_configuration}) from
% 3.) a few sequential measurements of the received RF voltage signals
only a few recordings of
the \ac{RF} voltage signals, which are modeled by
% 4.) Born approximation of the received RF voltage signals
their \name{Born} approximation
\eqref{eqn:lin_mod_v_rx_born}.
% b) numerical solution of this linear ISP necessitates suitable spatial discretizations of the physical models
The numerical solution of
% 1.) linear ISP
this linear \ac{ISP} necessitates
% 2.) suitable spatial discretizations of the physical models
suitable spatial discretizations of
the physical models.
% c) discretizations result in an ill-conditioned and typically underdetermined linear algebraic system that does not support any direct solution
These discretizations, however, result in
% 1.) ill-conditioned linear algebraic system
an ill-conditioned and, for
only a few sequential pulse-echo measurements,
% 2.) underdetermined dense linear algebraic system
typically underdetermined dense linear algebraic system
\eqref{eqn:cs_math_prob_general_obs_error} that does not support
any direct solution.
% d) proposed method circumvents this difficulty by reformulating the discretized linear ISP as an instance of the CS problem
The proposed method circumvents
% 1.) difficulty [ no direct solution ]
this difficulty by reformulating
% 2.) discretized linear ISP
the discretized linear \ac{ISP} as
an instance of
% 3.) CS problem associated with the corrupted observations
the \ac{CS} problem
\eqref{eqn:cs_math_prob_general}.
% e) sparsity-promoting lq-minimization method ensures the stable recovery if the sensing matrix meets one of the sufficient conditions
Postulating
the existence of
% 1.) nearly-sparse representation
a nearly-sparse representation of
% 2.) discretized compressibility fluctuations
the discretized compressibility fluctuations in
% 3.) known orthonormal basis
a known orthonormal basis
\eqref{eqn:def_transform_coefficients},
% 4.) sparsity-promoting lq-minimization method
the sparsity-promoting $\ell_{q}$-minimization method
\eqref{eqn:cs_lq_minimization} ensures
% 5.) stable recovery
its stable recovery if
% 6.) sensing matrix
the sensing matrix
\eqref{eqn:cs_math_prob_general_sensing_matrix} meets
one of
% 7.) sufficient conditions
the sufficient conditions
(cf. \cref{sec:compressed_sensing}).
% f) sufficient conditions require sufficiently uncorrelated pulse echoes of the admissible structural building blocks
%The latter require
%sufficiently uncorrelated pulse echoes of
%the admissible structural building blocks.

%%%%%%%%%%%%%%%%%%%%%%%%%%%%%%%%%%%%%%%%%%%%%%%%%%%%%%%%%%%%%%%%%%%%%%%%%%%%%%%%%%%%%%%%%%%%%%%%%%%%%%%%%%%%%%%%
% 1.) spatial discretizations of the physical models
%%%%%%%%%%%%%%%%%%%%%%%%%%%%%%%%%%%%%%%%%%%%%%%%%%%%%%%%%%%%%%%%%%%%%%%%%%%%%%%%%%%%%%%%%%%%%%%%%%%%%%%%%%%%%%%%
\subsection{Spatial Discretizations of the Physical Models}
%\label{subsec:recovery_discretization}
%%%%%%%%%%%%%%%%%%%%%%%%%%%%%%%%%%%%%%%%%%%%%%%%%%%%%%%%%%%%%%%%%%%%%%%%%%%%%%%%%%%%%%%%%%%%%%%%%%%%%%%%%%%%%%%%
% table: geometric parameters of the spatial discretizations
%%%%%%%%%%%%%%%%%%%%%%%%%%%%%%%%%%%%%%%%%%%%%%%%%%%%%%%%%%%%%%%%%%%%%%%%%%%%%%%%%%%%%%%%%%%%%%%%%%%%%%%%%%%%%%%%
\begin{table*}[tb]
 \centering
 \caption{%
  Geometric parameters of
  the spatial discretizations for
  all $\delta \in \setcons{ d - 1 }$,
  $\zeta \in \setcons{ d }$,
  $m \in \setconsnonneg{ N_{\text{el}} - 1 }$.
 }
 \label{tab:recon_disc_params}
 \small
 \begin{tabular}{%
  @{}%
  >{$}l<{$}%		01.) symbol
  p{0.92\textwidth}%	02.) meaning
  @{}%
 }
 \toprule
  \multicolumn{1}{@{}H}{Symbol} &
  \multicolumn{1}{H@{}}{Meaning}\\
  \cmidrule(r){1-1}\cmidrule(l){2-2}
 \addlinespace
 %--------------------------------------------------------------------------------------------------------------
 % a) spatial discretization of the planar transducer array [\delta \in \setcons{ d - 1 }]
 %--------------------------------------------------------------------------------------------------------------
  % 1.) number of grid points per vibrating face along the r_{\delta}-axis
  N_{\text{mat}, \delta} &
  Number of
  grid points per
  vibrating face along
  the $r_{\delta}$-axis,
  $N_{\text{mat}, \delta} \in \N$\\
  % 2.) constant spacing between the adjacent grid points on each vibrating face along the r_{\delta}-axis [w_{\text{el}, \delta} \in \Rplus]
  \Delta r_{\text{mat}, \delta} &
  Constant spacing between
  the adjacent grid points on
  each vibrating face along
  the $r_{\delta}$-axis,
  $\Delta r_{\text{mat}, \delta} = w_{\text{el}, \delta} / N_{\text{mat}, \delta}$\\
  % 3.) coordinates of the grid points on each vibrating face
  \vect{r}_{\text{mat}, \nu}^{(m)} &
  Coordinates of
  the grid points on
  each vibrating face,\par
  $\mathcal{V}_{m} = \{ \vect{r}_{\text{mat}, \nu}^{(m)} \in \R^{d}: \vect{r}_{\text{mat}, \nu}^{(m)} = \vect{r}_{\text{el}, m} + \sum_{ \delta = 1 }^{ d - 1 } ( \nu_{\delta} - M_{\text{mat}, \delta} ) \Delta r_{\text{mat}, \delta} \uvect{\delta}, \nu_{\delta} \in \setconsnonneg{ N_{\text{mat}, \delta} - 1 }, \nu = \mathcal{I}( \vectsym{\nu}, \vect{N}_{\text{mat}} ) \}$, where\par
  % 3.a) shift of index along the r_{\delta}-axis
  $M_{\text{mat}, \delta} = ( N_{\text{mat}, \delta} - 1 ) / 2$ and
  % 3.b) forward index transform
  $\mathcal{I}( \vectsym{\nu}, \vect{N}_{\text{mat}} ) = \sum_{ \delta = 1 }^{ d - 2 } \nu_{\delta} \prod_{ k = \delta + 1 }^{ d - 1 } N_{\text{mat}, k} + \nu_{d-1}$\\
  % 4.) total number of grid points per vibrating face
  N_{\text{mat}} &
  Total number of
  grid points per
  vibrating face,
  $N_{\text{mat}} = \tabs{ \mathcal{V}_{m} } = \prod_{ \delta = 1 }^{ d - 1 } N_{\text{mat}, \delta}$\\
  % 5.) (d-1)-dimensional surface element
  \Delta A &
  $(d-1)$-dimensional surface element,
  $\Delta A = \prod_{ \delta = 1 }^{ d - 1 } \Delta r_{\text{mat}, \delta}$\\
 %--------------------------------------------------------------------------------------------------------------
 % b) spatial discretization of the field of view [\zeta \in \setcons{ d }]
 %--------------------------------------------------------------------------------------------------------------
  % 1.) number of grid points in the FOV along the r_{\zeta}-axis
  N_{\text{lat}, \zeta} &
  Number of
  grid points in
  the \ac{FOV} along
  the $r_{\zeta}$-axis,
  $N_{\text{lat}, \zeta} \in \N$\\
  % 2.) constant spacing between the adjacent grid points in the FOV along the r_{\zeta}-axis
  \Delta r_{\text{lat}, \zeta} &
  Constant spacing between
  the adjacent grid points in
  the \ac{FOV} along
  the $r_{\zeta}$-axis,
  $\Delta r_{\text{lat}, \zeta} \in \Rplus$\\
  % 3.) offset of the grid points in the FOV
  \vect{r}_{\text{lat}, 0} &
  Offset of
  the grid points in
  the \ac{FOV},
  $\vect{r}_{\text{lat}, 0} \in \R^{d-1} \times \Rplus$\\
  % 4.) coordinates of the grid points in the FOV
  \vect{r}_{\text{lat}, i} &
  Coordinates of
  the grid points in
  the \ac{FOV},\par
  $\mathcal{L} = \{ \vect{r}_{\text{lat}, i} \in \R^{d}: \vect{r}_{\text{lat}, i} = \vect{r}_{\text{lat}, 0} + \sum_{ \zeta = 1 }^{ d } i_{\zeta} \Delta r_{\text{lat}, \zeta} \uvect{\zeta}, i_{\zeta} \in \setconsnonneg{ N_{\text{lat}, \zeta} - 1 }, i = \mathcal{I}( \vect{i}, \vect{N}_{\text{lat}} ) \}$, where\par
  % 4.1) forward index transform
  $\mathcal{I}( \vect{i}, \vect{N}_{\text{lat}} ) = \sum_{ \zeta = 1 }^{ d - 1 } i_{\zeta} \prod_{ k = \zeta + 1 }^{ d } N_{\text{lat}, k} + i_{d}$\\
  % 5.) total number of grid points in the FOV
  N_{\text{lat}} &
  Total number of
  grid points in
  the \ac{FOV},
  $N_{\text{lat}} = \tabs{ \mathcal{L} } = \prod_{ \zeta = 1 }^{ d } N_{\text{lat}, \zeta}$\\
  % 6.) d-dimensional volume element
  \Delta V &
  $d$-dimensional volume element,
  $\Delta V = \prod_{ \zeta = 1 }^{ d } \Delta r_{\text{lat}, \zeta}$\\
 \addlinespace
 \bottomrule
 \end{tabular}
\end{table*}

%---------------------------------------------------------------------------------------------------------------
% 1.) spatial discretizations of the planar transducer array and the FOV
%---------------------------------------------------------------------------------------------------------------
% a) two types of regular grids emulate the geometries and the spatial heterogeneities of the vibrating faces and the FOV containing the lossy heterogeneous object
Two types of
regular grids emulate
% 1.) geometry of each vibrating face
the geometries and
% 2.) spatial heterogeneity of each vibrating face
the spatial heterogeneities of
the vibrating faces and
% 3.) FOV containing the lossy heterogeneous object
the \ac{FOV} containing
the lossy heterogeneous object
(cf. \cref{tab:recon_disc_params}).
% b) expressions for the discretized apodization functions and the relative spatial fluctuations in the unperturbed compressibility based on Dirac delta distributions
The apodization functions
(cf. \cref{tab:lin_mod_scan_config_instrum_params}) and
% 1.) relative spatial fluctuations in the unperturbed compressibility
the compressibility fluctuations
\eqref{eqn:lin_mod_mech_model_tis_simple_rel_fluctuations} are discretized as
\begin{subequations}
\label{eqn:recovery_disc_space_trans_array_spat_trans}
\begin{align}
 %--------------------------------------------------------------------------------------------------------------
 % a) discretized transmitter apodization functions
 %--------------------------------------------------------------------------------------------------------------
  \chi_{m, l}^{(\text{tx})}( \vect{r}_{\rho} )
  &=
  \Delta A
  \sum_{ \nu = 0 }^{ N_{\text{mat}} - 1 }
    \chi_{m, \nu, l}^{(\text{tx})}
    \delta\bigl[ \vect{r}_{\rho} - \vect{r}_{\text{mat}, \nu, \rho}^{(m)} \bigr],
 \label{eqn:recovery_disc_space_trans_array_spat_trans_tx}\\
 %--------------------------------------------------------------------------------------------------------------
 % b) discretized receiver apodization functions
 %--------------------------------------------------------------------------------------------------------------
  \chi_{m, l}^{(\text{rx})}( \vect{r}_{\rho} )
  &=
  \Delta A
  \sum_{ \nu = 0 }^{ N_{\text{mat}} - 1 }
    \chi_{m, \nu, l}^{(\text{rx})}
    \delta\bigl[ \vect{r}_{\rho} - \vect{r}_{\text{mat}, \nu, \rho}^{(m)} \bigr],
 \label{eqn:recovery_disc_space_trans_array_spat_trans_rx}\\
 %--------------------------------------------------------------------------------------------------------------
 % c) discretized relative spatial fluctuations in the unperturbed compressibility
 %--------------------------------------------------------------------------------------------------------------
  \gamma^{(\kappa)}( \vect{r} )
  &=
  \Delta V
  \sum_{ i = 0 }^{ N_{\text{lat}} - 1 }
    \gamma_{i}^{(\kappa)}
    \delta( \vect{r} - \vect{r}_{\text{lat}, i} ),
 \label{eqn:recovery_disc_space_fov_rel_fluctuations_bp_sampled_kappa}
\end{align}
\end{subequations}
where
% 2.) regular samples of the transmitter apodization functions
$\chi_{m, \nu, l}^{(\text{tx})} = \chi_{m, l}^{(\text{tx})}[ \vect{r}_{\text{mat}, \nu, \rho}^{(m)} ]$,
% 3.) regular samples of the receiver apodization functions
$\chi_{m, \nu, l}^{(\text{rx})} = \chi_{m, l}^{(\text{rx})}[ \vect{r}_{\text{mat}, \nu, \rho}^{(m)} ]$, and
% 4.) regular samples of the relative spatial fluctuations in the unperturbed compressibility
$\gamma_{i}^{(\kappa)} = \gamma^{(\kappa)}( \vect{r}_{\text{lat}, i} )$ denote
their regular samples for
% 5.) all array elements, all grid points per vibrating face, and all relevant discrete frequencies
all $( m, \nu, l ) \in \setconsnonneg{ N_{\text{el}} - 1 } \times \setconsnonneg{ N_{\text{mat}} - 1 } \times \setsymbol{L}_{ \text{BP} }^{(n)}$ or
% 6.) all grid points
all $i \in \setconsnonneg{ N_{\text{lat}} - 1 }$.

%%%%%%%%%%%%%%%%%%%%%%%%%%%%%%%%%%%%%%%%%%%%%%%%%%%%%%%%%%%%%%%%%%%%%%%%%%%%%%%%%%%%%%%%%%%%%%%%%%%%%%%%%%%%%%%%
% 2.) computations of the incident acoustic pressure fields
%%%%%%%%%%%%%%%%%%%%%%%%%%%%%%%%%%%%%%%%%%%%%%%%%%%%%%%%%%%%%%%%%%%%%%%%%%%%%%%%%%%%%%%%%%%%%%%%%%%%%%%%%%%%%%%%
\subsection{Computations of the Incident Acoustic Pressure Fields}
%\label{subsec:recovery_p_in}
%---------------------------------------------------------------------------------------------------------------
% 1.) discretized incident acoustic pressure fields [superpositions of quasi-(d-1)-spherical waves]
%---------------------------------------------------------------------------------------------------------------
% a) discretized transmitter apodization functions transform the incident acoustic pressure fields into the discretized incident acoustic pressure fields
The insertions of
% 1.) apodized spatial transmit functions
the apodized spatial transmit functions
\eqref{eqn:syn_sup_qsw_p_in_qsw_spat_trans} and
% 2.) discretized transmitter apodization functions
the discretized transmitter apodization functions
\eqref{eqn:recovery_disc_space_trans_array_spat_trans_tx} into
% 3.) incident acoustic pressure fields [superpositions of quasi-(d-1)-spherical waves]
the incident acoustic pressure fields
\eqref{eqn:syn_sup_qsw_p_in} yield
% 4.) discretized incident acoustic pressure fields [superpositions of quasi-(d-1)-spherical waves]
% geometric requirement: $\vect{r}_{\text{mat}, \nu, \rho}^{(m)} \in L_{m}$ for delta functions to integrate properly
% \vect{r} = \vect{r}_{\text{lat}, i}
the discretizations
\begin{equation}
\begin{split}
 %--------------------------------------------------------------------------------------------------------------
 % discretized incident acoustic pressure fields [superpositions of quasi-(d-1)-spherical waves]
 %--------------------------------------------------------------------------------------------------------------
  p_{l}^{(\text{in}, n)}( \vect{r}_{\text{lat}, i} )
  =
  & - j 2 \omega_{l} \rho_{0} \Delta A
    \sum_{ m = 0 }^{ N_{\text{el}} - 1 }
      h_{m, l}^{(\text{tx})}
      u_{m, l}^{(\text{tx}, n)}\\
  &   \times
      \sum_{ \nu = 0 }^{ N_{\text{mat}} - 1 }
        \chi_{m, \nu, l}^{(\text{tx})}
        g_{l}\bigl[ \vect{r}_{\text{lat}, i} - \vect{r}_{\text{mat}, \nu}^{(m)} \bigr]
\end{split}
\label{eqn:recovery_p_in}
\end{equation}
for
% 5.) all sequential pulse-echo measurements, all relevant discrete frequencies, and all grid points
all $( n, l, i ) \in \setconsnonneg{ N_{\text{in}} - 1 } \times \setsymbol{L}_{ \text{BP} }^{(n)} \times \setconsnonneg{ N_{\text{lat}} - 1 }$, where
% 6.) acoustic energy at a specified grid point
the acoustic energy, which includes
% 7.) all sequential pulse-echo measurements
all sequential pulse-echo measurements and
% 8.) all relevant discrete frequencies
all relevant discrete frequencies, equals
\begin{equation}
 %--------------------------------------------------------------------------------------------------------------
 % incident acoustic energy at a specified grid point (all pulse-echo measurements, multifrequent)
 %--------------------------------------------------------------------------------------------------------------
  E^{(\text{in})}( \vect{r}_{\text{lat}, i} )
  =
  \sum_{ n = 0 }^{ N_{\text{in}} - 1 }
  \sum_{ l \in \setsymbol{L}_{ \text{BP} }^{(n)} }
    \dabs{ p_{l}^{(\text{in}, n)}( \vect{r}_{\text{lat}, i} ) }{1}^{2}
 \label{eqn:recovery_p_in_energy}
\end{equation}
for
% 9.) all grid points
all $i \in \setconsnonneg{ N_{\text{lat}} - 1 }$.

%%%%%%%%%%%%%%%%%%%%%%%%%%%%%%%%%%%%%%%%%%%%%%%%%%%%%%%%%%%%%%%%%%%%%%%%%%%%%%%%%%%%%%%%%%%%%%%%%%%%%%%%%%%%%%%%
% 3.) linear algebraic system induced by the Born approximation
%%%%%%%%%%%%%%%%%%%%%%%%%%%%%%%%%%%%%%%%%%%%%%%%%%%%%%%%%%%%%%%%%%%%%%%%%%%%%%%%%%%%%%%%%%%%%%%%%%%%%%%%%%%%%%%%
\subsection{Linear System Induced by the \name{Born} Approximation}
%\label{subsec:recovery_systems_linear_equations}
%---------------------------------------------------------------------------------------------------------------
% 1.) discretized Born approximation of the recorded RF voltage signals (single pulse-echo measurement, monofrequent, single array element)
%---------------------------------------------------------------------------------------------------------------
% a)
With
the discretized versions of
% 1.) discretized receiver apodization functions
the receiver apodization functions
\eqref{eqn:recovery_disc_space_trans_array_spat_trans_rx} and
% 2.) discretized relative spatial fluctuations in the unperturbed compressibility
the compressibility fluctuations
\eqref{eqn:recovery_disc_space_fov_rel_fluctuations_bp_sampled_kappa},
% 3.) Born approximation of the recorded RF voltage signals
the \name{Born} approximation of
the recorded \ac{RF} voltage signals
\eqref{eqn:lin_mod_v_rx_born} yields
% 4.) discretized Born approximation of the recorded RF voltage signals
the linear algebraic equations
\begin{subequations}
\label{eqn:recovery_sys_lin_eq_v_rx_born}
\begin{equation}
\begin{split}
 %--------------------------------------------------------------------------------------------------------------
 % a) discretized Born approximation of the recorded RF voltage signals (single pulse-echo measurement, monofrequent, single array element)
 %--------------------------------------------------------------------------------------------------------------
  u_{m, l}^{(\text{rx}, n)}
  \approx
  u_{m, l}^{(\text{B}, n)}
  =
  \sum_{ i = 0 }^{ N_{\text{lat}} - 1 }
    \phi_{m, l, i}\bigl[ p_{l}^{(\text{in}, n)} \bigr]
    \gamma_{i}^{(\kappa)}
\end{split}
\label{eqn:recovery_sys_lin_eq_v_rx_born_sum}
\end{equation}
for
% 5.) all sequential pulse-echo measurements, all relevant discrete frequencies, and all array elements
all $( n, l, m ) \in \setconsnonneg{ N_{\text{in}} - 1 } \times \setsymbol{L}_{ \text{BP} }^{(n)} \times \setconsnonneg{ N_{\text{el}} - 1 }$, where
% 6.) entries of the observation process (single pulse-echo measurement, monofrequent, single array element)
the coefficients satisfy
% geometric requirement: $\vect{r}_{\text{mat}, \nu, \rho}^{(m)} \in L_{m}$, $\vect{r}_{\text{lat}, i} \in \Omega$ for delta functions to integrate properly
\begin{equation}
\begin{split}
 %--------------------------------------------------------------------------------------------------------------
 % b) entries of the observation process (single pulse-echo measurement, monofrequent, single array element)
 %--------------------------------------------------------------------------------------------------------------
  \phi_{m, l, i}\bigl[ p_{l}^{(\text{in}, n)} \bigr]
  =
  & 2 {\munderbar{k}_{l}}^{2} \Delta A \Delta V
    h_{m, l}^{(\text{rx})}
    p_{l}^{(\text{in}, n)}( \vect{r}_{\text{lat}, i} )\\
  & \times
    \sum_{ \nu = 0 }^{ N_{\text{mat}} - 1 }
      \chi_{m, \nu, l}^{(\text{rx})}
      g_{l}\bigl[ \vect{r}_{\text{mat}, \nu}^{(m)} - \vect{r}_{\text{lat}, i} \bigr],
\end{split}
\label{eqn:recovery_sys_lin_eq_v_rx_born_coef}
\end{equation}
\end{subequations}
and
% 7.) number of observations (all pulse-echo measurements, multifrequent, all array elements)
the number of
observations equals
\begin{equation}
 %--------------------------------------------------------------------------------------------------------------
 % number of observations (all pulse-echo measurements, multifrequent, all array elements)
 %--------------------------------------------------------------------------------------------------------------
  N_{\text{obs}}
  =
  N_{\text{el}}
  \sum_{ n = 0 }^{ N_{\text{in}} - 1 }
    N_{f, \text{BP}}^{(n)}.
 \label{eqn:recovery_sys_lin_eq_num_obs}
\end{equation}

%---------------------------------------------------------------------------------------------------------------
% 2.) linear algebraic system (all pulse-echo measurements, multifrequent, all array elements)
%---------------------------------------------------------------------------------------------------------------
% a) vertical stacking of all partial linear algebraic systems yields the linear algebraic system
Stacking
the regular samples in
% 1.) discretized relative spatial fluctuations in the unperturbed compressibility
the discretized compressibility fluctuations
\eqref{eqn:recovery_disc_space_fov_rel_fluctuations_bp_sampled_kappa} for
% 2.) all grid points
all grid points into
% 3.) vector of bandpass-filtered relative spatial fluctuations in compressibility
the complex-valued%
\footnote{
  % a) positivity of the relevant discrete frequencies forming the sets results in the recovery of complex-valued compressibility fluctuations that contain only positive spatial frequencies along the r_{d}-axis
  The positivity of
  % 1.) relevant discrete frequencies
  the relevant discrete frequencies forming
  % 2.) sets of relevant discrete frequencies
  the sets
  \eqref{eqn:recon_disc_axis_f_discrete_BP} results in
  the recovery of
  % 3.) vector stacking the regular samples in the discretized relative spatial fluctuations in the unperturbed compressibility
  complex-valued compressibility fluctuations
  \eqref{eqn:recovery_sys_lin_eq_gamma_kappa_bp_vector} that contain only
  % 4.) positive spatial frequencies along the r_{d}-axis
  positive spatial frequencies along
  the $r_{d}$-axis.
}
$N_{\text{lat}} \times 1$ vector
\begin{equation}
 %--------------------------------------------------------------------------------------------------------------
 % vector of bandpass-filtered relative spatial fluctuations in compressibility
 %--------------------------------------------------------------------------------------------------------------
  \vectsym{\gamma}^{(\kappa)}
  =
  \trans{
    \begin{bmatrix}
      \gamma_{ 0 }^{(\kappa)} & \hdots & \gamma_{ N_{\text{lat}} - 1 }^{(\kappa)}
    \end{bmatrix}
  }
 \label{eqn:recovery_sys_lin_eq_gamma_kappa_bp_vector}
\end{equation}
and
% 4.) recorded RF voltage signals and their Born approximation (single pulse-echo measurement, monofrequent, all array elements)
the recorded \ac{RF} voltage signals
\eqref{eqn:recovery_disc_freq_v_rx_Fourier_series_coef} and
% 5.) Born approximation
their \name{Born} approximation
\eqref{eqn:recovery_sys_lin_eq_v_rx_born} into
% 6.) recorded RF voltage signals and their Born approximation (all pulse-echo measurements, multifrequent, all array elements)
the complex-valued
$N_{\text{obs}} \times 1$ vectors
\begin{subequations}
\label{eqn:recovery_sys_lin_eq_v_rx_born_all_f_all_in}
\begin{align}
 %--------------------------------------------------------------------------------------------------------------
 % a) relevant Fourier coefficients of the recorded RF voltage signals (all pulse-echo measurements, multifrequent, all array elements)
 %--------------------------------------------------------------------------------------------------------------
  \vect{u}^{(\text{rx})}
  &=
  \vertcat_{ n = 0 }^{ N_{\text{in}} - 1 }
    \vertcat_{ l \in \setsymbol{L}_{ \text{BP} }^{(n)} }
      \vertcat_{ m = 0 }^{ N_{\text{el}} - 1 }
        u_{ m, l }^{(\text{rx}, n)},
 \label{eqn:recovery_sys_lin_eq_v_rx_born_all_f_all_in_v_rx}\\
 %--------------------------------------------------------------------------------------------------------------
 % b) approximate relevant Fourier coefficients of the recorded RF voltage signals (all pulse-echo measurements, multifrequent, all array elements)
 %--------------------------------------------------------------------------------------------------------------
  \vect{u}^{(\text{B})}
  &=
  \vertcat_{ n = 0 }^{ N_{\text{in}} - 1 }
    \vertcat_{ l \in \setsymbol{L}_{ \text{BP} }^{(n)} }
      \vertcat_{ m = 0 }^{ N_{\text{el}} - 1 }
        u_{ m, l }^{(\text{B}, n)},
 \label{eqn:recovery_sys_lin_eq_v_rx_born_all_f_all_in_v_rx_born}
\end{align}
respectively,
% 7.) observation process (all pulse-echo measurements, multifrequent, all array elements)
the complex-valued
$N_{\text{obs}} \times N_{\text{lat}}$ matrix
\begin{equation}
 %--------------------------------------------------------------------------------------------------------------
 % c) observation process (all pulse-echo measurements, multifrequent, all array elements)
 %--------------------------------------------------------------------------------------------------------------
  \mat{\Phi}\bigl[ p^{(\text{in})} \bigr]
  =
  \vertcat_{ n = 0 }^{ N_{\text{in}} - 1 }
    \vertcat_{ l \in \setsymbol{L}_{ \text{BP} }^{(n)} }
      \vertcat_{ m = 0 }^{ N_{\text{el}} - 1 }
        \horzcat_{ i = 0 }^{ N_{\text{lat}} - 1 }
          \phi_{m, l, i}\bigl[ p_{l}^{(\text{in}, n)} \bigr],
 \label{eqn:recovery_sys_lin_eq_v_rx_born_all_f_all_in_mat}
\end{equation}
represents
the pulse-echo measurement process and defines
% 8.) linear algebraic system (all pulse-echo measurements, multifrequent, all array elements)
the ill-conditioned, and, for
only a few sequential measurements,
% 9.) typically underdetermined dense linear algebraic system
typically underdetermined dense linear algebraic system
\begin{equation}
 %--------------------------------------------------------------------------------------------------------------
 % d) linear algebraic system (all pulse-echo measurements, multifrequent, all array elements)
 %--------------------------------------------------------------------------------------------------------------
  \vect{u}^{(\text{rx})}
  \approx
  \vect{u}^{(\text{B})}
  =
  \mat{\Phi}\bigl[ p^{(\text{in})} \bigr]
  \vectsym{\gamma}^{(\kappa)}.
 \label{eqn:recovery_sys_lin_eq_v_rx_born_all_f_all_in_mat_vec}
\end{equation}
\end{subequations}

%---------------------------------------------------------------------------------------------------------------
% 3.) unwanted properties of the discretized linear ISP and regularization by CS
%---------------------------------------------------------------------------------------------------------------
% a) unwanted properties prevent the direct recovery of the compressibility fluctuations from the recorded RF voltage signals
These unwanted properties, which result from
% 1.) discretization of the Fredholm integral equations of the first kind
% book:Hansen2010, Chapter 1:
%
% book:Hansen1998, Chapter 1: Setting the Stage / Sect. 1.1. Problems with Ill-Conditioned Matrices
% - The numerical treatment of SYSTEMS OF EQUATIONS WITH AN ILL-CONDITIONED COEFFICIENT MATRIX depends on
%   the type of ill-conditioning of A. (p. 2)
% - DISCRETE ILL-POSED PROBLEMS ARISE FROM
%   THE DISCRETIZATION OF ILL-POSED PROBLEMS such as
%   FREDHOLM INTEGRAL EQUATIONS OF THE FIRST KIND. (p. 2)
% - Here all the singular values of A, as well as the SVD components of the solution, on the average,
%   decay gradually to zero, and we say that
%   a DISCRETE PICARD CONDITION (see §4.5) IS SATISFIED. (p. 2)
% - Since there is no gap in the singular value spectrum, there is
%   NO NOTION OF A NUMERICAL RANK for these matrices. (p. 2)
% - For discrete ill-posed problems, the goal is to find a balance between
%   [1.)] the RESIDUAL NORM and
%   [2.)] the SIZE OF THE SOLUTION that matches
%   the errors in the data as well as one's expectations to the computed solution. (pp. 2, 3)
% - Here, "size" should be interpreted in a rather broad sense; e.g., size can be measured by a norm, a seminorm, or a Sobolev norm. (p. 3)
the discretization of
the \name{Fredholm} integral equations of
the first kind
\eqref{eqn:lin_mod_v_rx_born}
\cite[]{book:Hansen2010}
\cite[2, 3]{book:Hansen1998},
% 2.) relatively small number of sequential pulse-echo measurements per image
the relatively small number of
sequential pulse-echo measurements per
image, and
% 3.) slow spatial decays of the outgoing free-space Green's functions
% article:ChewITAP1997: Fast solution methods in electromagnetics
% I. Introduction
% - These numerical techniques involve either solving
%   [1.)] PARTIAL-DIFFERENTIAL EQUATIONS with THE FINITE-DIFFERENCE METHOD (FDM) [6]–[9] or THE FINITE-ELEMENT METHOD (FEM) [10]–[12] WHICH RESULT IN
%     SPARSE MATRICES, or
%   [2.)] INTEGRAL EQUATIONS which are CONVERTED INTO
%     DENSE MATRIX EQUATIONS USING THE METHOD OF MOMENTS (MoM) [1]–[5]. (p. 533)
% III. INTEGRAL EQUATION SOLVERS
% - However, INTEGRAL EQUATION SOLVERS RESULT IN DENSE MATRICES. (p. 535)
% - THE LONG RANGE INTERACTION IN ELECTRODYNAMICS FALLS OFF AS 1/r;
%   THIS DECAY CANNOT BE IGNORED EVEN OVER LARGE DISTANCES.
% article:RokhlinJCP1990: Rapid solution of integral equations of scattering theory in two dimensions
% - ONE OF PRINCIPAL DIFFICULTIES arising in
%   the solution of large-scale scattering problems of integral equations is the fact that
%   THE GREEN'S FUNCTION FOR THE HELMHOLTZ EQUATION DECAYS SLOWLY.
% - drawback: discretization yields DENSE large-scale systems of linear algebraic equations
% - As a result,
%   THE KERNELS OF THE OBTAINED INTEGRAL EQUATIONS ARE NOT SPARSE, AND THEIR DISCRETIZATION LEADS TO
%   DENSE LARGE-SCALE SYSTEMS OF LINEAR ALGEBRAIC EQUATIONS.
the slow spatial decays of
% 4.) outgoing free-space Green's functions (two- and three-dimensional Euclidean spaces)
the outgoing free-space \name{Green}'s functions
\eqref{eqn:app_helmholtz_green_free_space_2_3_dim}
\cite{article:ChewITAP1997,article:RokhlinJCP1990}, prevent
the direct recovery of
% 5.) vector stacking the regular samples in the discretized relative spatial fluctuations in the unperturbed compressibility
the compressibility fluctuations
\eqref{eqn:recovery_sys_lin_eq_gamma_kappa_bp_vector} from
% 6.) vector stacking the relevant Fourier coefficients of the recorded RF voltage signals (all pulse-echo measurements, multifrequent, all array elements)
the recorded \ac{RF} voltage signals
\eqref{eqn:recovery_sys_lin_eq_v_rx_born_all_f_all_in_v_rx}.
% b) reformulation of this discretized linear ISP as an instance of the CS problem circumvents this difficulty
The reformulation of
% 1.) discretized linear ISP
this discretized linear \ac{ISP} as
an instance of
% 2.) CS problem associated with the corrupted observations
the \ac{CS} problem
\eqref{eqn:cs_math_prob_general}, however, circumvents
this difficulty.

%%%%%%%%%%%%%%%%%%%%%%%%%%%%%%%%%%%%%%%%%%%%%%%%%%%%%%%%%%%%%%%%%%%%%%%%%%%%%%%%%%%%%%%%%%%%%%%%%%%%%%%%%%%%%%%%
% 4.) regularization of the discretized linear inverse scattering problem
%%%%%%%%%%%%%%%%%%%%%%%%%%%%%%%%%%%%%%%%%%%%%%%%%%%%%%%%%%%%%%%%%%%%%%%%%%%%%%%%%%%%%%%%%%%%%%%%%%%%%%%%%%%%%%%%
\subsection{Regularization of the Inverse Scattering Problem}
\label{subsec:recovery_regularization}
%---------------------------------------------------------------------------------------------------------------
% 1.) linear algebraic system (all pulse-echo measurements, multifrequent, all array elements, additive errors)
%---------------------------------------------------------------------------------------------------------------
% a) complex-valued vector of transform coefficients constitutes a nearly-sparse representation of the compressibility fluctuations
% article:Schiffner2018, Sect. II. Compressed Sensing in a Nutshell (sec:compressed_sensing)
% - \ac{CS} replaces the identity by an inequality using the known upper bound on the $\ell_{2}$-norm of the additive errors and postulates that
%   a known dictionary of structural building blocks, e.g. an orthonormal basis or a frame, represents
%   the high-dimensional vector almost sparsely \cite{article:TroppPIEEE2010}.
% - Let the column vectors $\vectsym{\psi}_{n} \in \C^{ N }$, $n \in \setcons{ N }$, of
%   the unitary matrix $\mat{\Psi} \in \C^{ N \times N }$, which represents
%   a suitable orthonormal basis of $\C^{ N }$, e.g. the \name{Fourier}, a wavelet, or the canonical basis, define
%   the admissible structural building blocks.
% - The vector of transform coefficients constitutes a nearly-sparse representation of the high-dimensional vector, if
%   the sorted absolute values of its components decay rapidly.
Postulating
the knowledge of
% 1.) suitable orthonormal basis
a suitable orthonormal basis of
$\C^{ N_{\text{lat}} }$, which is represented by
% 2.) complex-valued unitary matrix of structural building blocks
the complex-valued unitary
$N_{\text{lat}} \times N_{\text{lat}}$ matrix
$\mat{\Psi}$ of
structural building blocks
$\vectsym{\psi}_{i}$,
% 3.) complex-valued vector of transform coefficients
the complex-valued
$N_{\text{lat}} \times 1$ vector of
transform coefficients
\begin{equation}
 %--------------------------------------------------------------------------------------------------------------
 % nearly-sparse representation / vector of transform coefficients
 %--------------------------------------------------------------------------------------------------------------
  \vectsym{\theta}^{(\kappa)}
  =
  \herm{ \mat{\Psi} }
  \vectsym{\gamma}^{(\kappa)}
 \label{eqn:recovery_reg_sparse_representation}
\end{equation}
constitutes
% 4.) nearly-sparse representation
a nearly-sparse representation of
% 5.) vector stacking the regular samples in the discretized relative spatial fluctuations in the unperturbed compressibility
the compressibility fluctuations
\eqref{eqn:recovery_sys_lin_eq_gamma_kappa_bp_vector}.
% b) linear algebraic system for the nearly-sparse representation and additive errors
% article:Schiffner2018, Sect. II. Compressed Sensing in a Nutshell (sec:compressed_sensing)
% - Both vectors satisfy the UNDERDETERMINED LINEAR ALGEBRAIC SYSTEM, where
%   the known matrix $\mat{\Phi} \in \C^{ M \times N }$ represents the nonadaptive observation process, and
%   the UNKNOWN VECTOR $\vectsym{\eta} \in \C^{ M }$ denotes ADDITIVE ERRORS OF BOUNDED $\ell_{2}$-NORM $\tnorm{ \vectsym{\eta} }{2} \leq \eta$.
Inserting
% 1.) nearly-sparse representation
this representation, defining
% 2.) sensing matrix (all pulse-echo measurements, multifrequent, all array elements)
the complex-valued
$N_{\text{obs}} \times N_{\text{lat}}$ sensing matrix
\begin{equation}
 %--------------------------------------------------------------------------------------------------------------
 % sensing matrix (all pulse-echo measurements, multifrequent, all array elements)
 %--------------------------------------------------------------------------------------------------------------
  \mat{A}\bigl[ p^{(\text{in})} \bigr]
  =
  \mat{\Phi}\bigl[ p^{(\text{in})} \bigr]
  \mat{\Psi},
 \label{eqn:recovery_reg_sensing_matrix}
\end{equation}
and accounting for
% 3.) additive errors of bounded l2-norm
an unknown complex-valued
$N_{\text{obs}} \times 1$ vector of
additive errors of
bounded $\ell_{2}$-norm
$\tnorm{ \vectsym{\eta} }{2} \leq \eta$,
% 4.) linear algebraic system (all pulse-echo measurements, multifrequent, all array elements)
the linear algebraic system
\eqref{eqn:recovery_sys_lin_eq_v_rx_born_all_f_all_in} becomes
\begin{equation}
 %--------------------------------------------------------------------------------------------------------------
 % linear algebraic system (all pulse-echo measurements, multifrequent, all array elements, additive errors)
 %--------------------------------------------------------------------------------------------------------------
  \vect{u}^{(\text{rx})}
  =
  \underbrace{
    \mat{\Phi}\bigl[ p^{(\text{in})} \bigr]
    \mat{\Psi}
  }_{ = \mat{A}[ p^{(\text{in})} ] }
  \vectsym{\theta}^{(\kappa)}
  +
  \vectsym{\eta}
  =
  \underbrace{
    \mat{A}\bigl[ p^{(\text{in})} \bigr]
    \vectsym{\theta}^{(\kappa)}
  }_{ = \vect{u}^{(\text{B})} }
  +\:
  \vectsym{\eta}.
 \label{eqn:recovery_reg_prob_general_obs_trans_coef_error}
\end{equation}
% c) additive errors reflect the inaccuracies of the discretized physical models and the voltage measurements
The additive errors reflect
the inaccuracies of
% 1.) discretized physical models
the discretized physical models and
% 2.) voltage measurements
the voltage measurements.

%---------------------------------------------------------------------------------------------------------------
% 2.) l2-normalization of the sensing matrix's column vectors
%---------------------------------------------------------------------------------------------------------------
% a) column vectors of the sensing matrix define the pulse echoes of the admissible structural building blocks
The column vectors of
% 1.) sensing matrix (all pulse-echo measurements, multifrequent, all array elements)
the sensing matrix
\eqref{eqn:recovery_reg_sensing_matrix}, i.e.
% 2.) approximate vectors stacking the relevant Fourier coefficients of the recorded RF voltage signals (all pulse-echo measurements, multifrequent, all array elements)
the \name{Born} approximations of
the recorded \ac{RF} voltage signals
\eqref{eqn:recovery_sys_lin_eq_v_rx_born_all_f_all_in_v_rx_born} induced by
the individual components of
% 3.) nearly-sparse representation
the nearly-sparse representation
\eqref{eqn:recovery_reg_sparse_representation}, define
% 4.) pulse echoes
the pulse echoes of
% 5.) admissible structural building blocks
the admissible structural building blocks.
Although
their $\ell_{2}$-normalization by
diagonal weighting matrices minimizes both
% 1.) restricted isometry ratio
the restricted isometry ratio and
% 2.) restricted isometry constant
the restricted isometry constant for
% 3.) 1-sparse representations
$1$-sparse representations
\eqref{eqn:recovery_reg_sparse_representation}
(cf. \cref{sec:compressed_sensing}),
% c) l2-normalization by diagonal weighting matrices potentially amplifies the additive errors
it potentially amplifies
the additive errors.
% d) recorded electric energies characterize the transfer behavior of the sensing matrix
In fact,
% 1.) density of population
the density of
population and
% 2.) dynamic range of the recorded electric energies
the dynamic range of
the recorded electric energies
\begin{equation}
 %--------------------------------------------------------------------------------------------------------------
 % recorded electric energies in the pulse echoes (all pulse-echo measurements, multifrequent, all array elements)
 %--------------------------------------------------------------------------------------------------------------
  E_{i}^{(\text{B})}
  =
  \dnorm{ \vect{a}_{i}\bigl[ p^{(\text{in})} \bigr] }{2}{1}^{2}
 \label{eqn:recovery_reg_v_rx_born_trans_coef_energy}
\end{equation}
for
% 1.) all coefficients
all $i \in \setcons{ N_{\text{lat}} }$, which characterize
% 2.) transfer behavior
the transfer behavior, vary significantly with
the orthonormal basis.
% e) imposition of a hard threshold on the l2-norms mitigates this amplification (pseudo-inverse filter)
The imposition of
% 1.) hard threshold on the l2-norms
a hard threshold on
the $\ell_{2}$-norms, whose value is dictated by
% 2.) signal-to-noise ratio (SNR)
the \ac{SNR} of
% 3.) vector stacking the relevant Fourier coefficients of the recorded RF voltage signals (all pulse-echo measurements, multifrequent, all array elements)
the recorded \ac{RF} voltage signals
\eqref{eqn:recovery_sys_lin_eq_v_rx_born_all_f_all_in_v_rx}, mitigates
this amplification.

%---------------------------------------------------------------------------------------------------------------
% 3.) l2-normalization of the sensing matrix by diagonal weighting matrices
%---------------------------------------------------------------------------------------------------------------
% a) factor \xi \in ( 0; 1 ] specifies the lower bound on the l2-norms of the sensing matrix's column vectors
Let
% 1.) factor \xi \in ( 0; 1 ]
the factor
$\xi \in ( 0; 1 ]$ specify
% 2.) lower bound on the l2-norms of the sensing matrix's column vectors
the lower bound on
the $\ell_{2}$-norms of
the sensing matrix's column vectors
\begin{equation}
 %--------------------------------------------------------------------------------------------------------------
 % lower bound on the l2-norms of the sensing matrix's column vectors
 %--------------------------------------------------------------------------------------------------------------
  a_{\xi, \text{lb}}\bigl[ p^{(\text{in})} \bigr]
  =
  \xi
  \underset{ i \in \setcons{ N_{\text{lat}} } }{ \max }
  \left\{
    \dnorm{ \vect{a}_{i}\bigl[ p^{(\text{in})} \bigr] }{2}{1}
  \right\}.
 \label{eqn:recovery_reg_norm_l2_norms_lb}
\end{equation}
% b) thresholded l2-norms constitute the individual entries of the weighting matrix and its inverse matrix
The thresholded $\ell_{2}$-norms of
these column vectors
\begin{equation}
 %--------------------------------------------------------------------------------------------------------------
 % thresholded l2-norms of the sensing matrix's column vectors
 %--------------------------------------------------------------------------------------------------------------
  a_{ \xi, i }\bigl[ p^{(\text{in})} \bigr]
  =
  \max
  \left\{
    \dnorm{ \vect{a}_{i}\bigl[ p^{(\text{in})} \bigr] }{2}{1},
    a_{\xi, \text{lb}}\bigl[ p^{(\text{in})} \bigr]
  \right\}
 \label{eqn:recovery_reg_norm_l2_norms_thresholded}
\end{equation}
populate
% 1.) diagonal weighting matrix
the real-valued
$N_{\text{lat}} \times N_{\text{lat}}$ weighting matrix
\begin{subequations}
\label{eqn:recovery_reg_norm_weighting_matrices}
\begin{align}
 %--------------------------------------------------------------------------------------------------------------
 % a) diagonal weighting matrix
 %--------------------------------------------------------------------------------------------------------------
  \mat{W}_{\xi}
  &=
  \ddiag{
    \begin{matrix}
      a_{ \xi, 1 }\bigl[ p^{(\text{in})} \bigr] & \hdots & a_{ \xi, N_{\text{lat}} }\bigl[ p^{(\text{in})} \bigr]
    \end{matrix}
  }{2}
 \label{eqn:recovery_reg_norm_weighting_matrix}\\
\intertext{%
and
% 2.) diagonal inverse weighting matrix
its inverse matrix%
}
 %--------------------------------------------------------------------------------------------------------------
 % b) diagonal inverse weighting matrix
 %--------------------------------------------------------------------------------------------------------------
  \mat{W}_{\xi}^{-1}
  &=
  \diag{
    \begin{matrix}
      \frac{ 1 }{ a_{ \xi, 1 }\left[ p^{(\text{in})} \right] } & \hdots & \frac{ 1 }{ a_{ \xi, N_{\text{lat}} }\left[ p^{(\text{in})} \right] }
    \end{matrix}
  },
 \label{eqn:recovery_reg_norm_weighting_matrix_inv}
\end{align}
\end{subequations}
whose
dependences on
% 3.) discretized incident acoustic pressure fields [superpositions of quasi-(d-1)-spherical waves]
the incident acoustic pressure fields
\eqref{eqn:recovery_p_in} are omitted for
the sake of
notational lucidity.
% c) right multiplication of the sensing matrix by the diagonal inverse weighting matrix yields the normalized sensing matrix
The right multiplication of
% 1.) sensing matrix (all pulse-echo measurements, multifrequent, all array elements)
the sensing matrix
\eqref{eqn:recovery_reg_sensing_matrix} by
% 2.) diagonal inverse weighting matrix
the diagonal inverse weighting matrix
\eqref{eqn:recovery_reg_norm_weighting_matrix_inv} yields
% 3.) normalized sensing matrix (all pulse-echo measurements, multifrequent, all array elements)
the complex-valued normalized
$N_{\text{obs}} \times N_{\text{lat}}$ sensing matrix
\begin{equation}
 %--------------------------------------------------------------------------------------------------------------
 % normalized sensing matrix (all pulse-echo measurements, multifrequent, all array elements)
 %--------------------------------------------------------------------------------------------------------------
  \bar{\mat{A}}_{\xi}\bigl[ p^{(\text{in})} \bigr]
  =
  \mat{A}\bigl[ p^{(\text{in})} \bigr]
  \mat{W}_{\xi}^{-1}
  =
  \mat{\Phi}\bigl[ p^{(\text{in})} \bigr]
  \mat{\Psi}
  \mat{W}_{\xi}^{-1},
 \label{eqn:recon_reg_norm_sensing_matrix}
\end{equation}
whose
% 4.) column vectors
column vectors exhibit
unity $\ell_{2}$-norms, if
the $\ell_{2}$-norm of
the corresponding column vector in
% 5.) original sensing matrix (all pulse-echo measurements, multifrequent, all array elements)
the original sensing matrix
\eqref{eqn:recovery_reg_sensing_matrix} is
not smaller than
% 6.) lower bound on the l2-norms of the sensing matrix's column vectors
the specified lower bound
\eqref{eqn:recovery_reg_norm_l2_norms_lb}.
% d) normalized sensing matrix maintains the potential dense population of the sensing matrix
This matrix maintains
% 1.) potential dense population
the potential dense population of
% 2.) sensing matrix (all pulse-echo measurements, multifrequent, all array elements)
the sensing matrix
\eqref{eqn:recovery_reg_sensing_matrix}.

%---------------------------------------------------------------------------------------------------------------
% 4.) normalized CS problem and sparsity-promoting methods for its solution
%---------------------------------------------------------------------------------------------------------------
% a) insertions of the weighting matrices and the normalized sensing matrix into the linear algebraic system yield the equivalent normalized linear algebraic system
With
the normalized versions of
% 1.) unit vector stacking the relevant Fourier coefficients of the recorded RF voltage signals (all pulse-echo measurements, multifrequent, all array elements)
the recorded \ac{RF} voltage signals
$\bar{\vect{u}}^{(\text{rx})} = \vect{u}^{(\text{rx})} / \tnorm{ \vect{u}^{(\text{rx})} }{2}$,
% 2.) normalized additive errors of bounded l2-norm
the additive errors
$\bar{\vectsym{\eta}} = \vectsym{\eta} / \tnorm{ \vect{u}^{(\text{rx})} }{2}$ of
bounded $\ell_{2}$-norm
$\tnorm{ \bar{\vectsym{\eta}} }{2} \leq \bar{\eta} = \eta / \tnorm{ \vect{u}^{(\text{rx})} }{2}$, and
% 3.) normalized nearly-sparse representation
the nearly-sparse representation
\begin{equation}
 %--------------------------------------------------------------------------------------------------------------
 % normalized nearly-sparse representation / nearly-sparse normalized vector of transform coefficients
 %--------------------------------------------------------------------------------------------------------------
  \bar{\vectsym{\theta}}_{\xi}^{(\kappa)}
  =
  \frac{ 1 }{ \dnorm{ \vect{u}^{(\text{rx})} }{2}{1} }
  \mat{W}_{\xi}
  \vectsym{\theta}^{(\kappa)},
 \label{eqn:recon_reg_norm_trans_coef}
\end{equation}
the insertions of
% 4.) weighting matrices
the weighting matrices
\eqref{eqn:recovery_reg_norm_weighting_matrices} and
% 5.) normalized sensing matrix
the normalized sensing matrix
\eqref{eqn:recon_reg_norm_sensing_matrix} into
% 6.) linear algebraic system (all pulse-echo measurements, multifrequent, all array elements, additive errors)
the linear algebraic system
\eqref{eqn:recovery_reg_prob_general_obs_trans_coef_error} yield
% 7.) normalized linear algebraic system (all pulse-echo measurements, multifrequent, all array elements, additive errors)
the equivalent system
\begin{equation}
\begin{split}
 %--------------------------------------------------------------------------------------------------------------
 % normalized linear algebraic system (all pulse-echo measurements, multifrequent, all array elements, additive errors)
 %--------------------------------------------------------------------------------------------------------------
  \bar{\vect{u}}^{(\text{rx})}
  &=
  \frac{ 1 }{ \dnorm{ \vect{u}^{(\text{rx})} }{2}{1} }
  \Bigl[
    \underbrace{
      \mat{A}\bigl[ p^{(\text{in})} \bigr]
      \mat{W}_{\xi}^{-1}
    }_{ = \bar{\mat{A}}_{\xi}[ p^{(\text{in})} ] }
    \mat{W}_{\xi}
    \vectsym{\theta}^{(\kappa)}
    +
    \vectsym{\eta}
  \Bigr]\\
  &=
  \bar{\mat{A}}_{\xi}\bigl[ p^{(\text{in})} \bigr]
  \bar{\vectsym{\theta}}_{\xi}^{(\kappa)}
  +
  \bar{\vectsym{\eta}}.
\end{split}
\label{eqn:recovery_reg_norm_obs_trans_coef_error}
\end{equation}
% b) CS problem associated with the normalized linear algebraic system
The associated normalized \ac{CS} problem
\eqref{eqn:cs_math_prob_general} reads
\begin{equation}
\begin{alignedat}{2}
 %--------------------------------------------------------------------------------------------------------------
 % CS problem associated with the normalized linear algebraic system
 %--------------------------------------------------------------------------------------------------------------
  &
  \text{Recover}
  &
  \text{nearly-sparse }
  \bar{\vectsym{\theta}}_{\xi}^{(\kappa)}
  \in
  \C^{ N_{\text{lat}} }\\
  &
  \text{subject to}
  \quad
  &
  \dnorm{ \bar{\vect{u}}^{(\text{rx})} - \bar{\mat{A}}_{\xi}\bigl[ p^{(\text{in})} \bigr] \bar{\vectsym{\theta}}_{\xi}^{(\kappa)} }{2}{1}
  &\leq
  \bar{\eta}
\end{alignedat}
\label{eqn:recovery_reg_norm_prob_general}
\end{equation}
and
% c) summary of sparsity-promoting methods for the solution of the CS problem
the sparsity-promoting $\ell_{q}$-minimization method for
its stable solution
\eqref{eqn:cs_lq_minimization},
$q \in [ 0; 1 ]$, recovers
% 1.) complex-valued normalized vector of recovered transform coefficients
the complex-valued normalized
$N_{\text{lat}} \times 1$ vector of
transform coefficients
\begin{equation}
\begin{alignedat}{2}
 %--------------------------------------------------------------------------------------------------------------
 % sparsity-promoting lq-minimization method
 %--------------------------------------------------------------------------------------------------------------
  \hspace{-0.75em}
  \hat{\bar{\vectsym{\theta}}}_{\xi}^{(\kappa, q, \eta)}
  &\in
  \underset{ \tilde{\vectsym{\theta}} \in \C^{ N_{\text{lat}} } }{ \arg\min }
  \dnorm{ \tilde{\vectsym{\theta}} }{q}{1}\\
  &
  \mspace{24.5mu}
  \text{subject to}
  &
  \dnorm{ \bar{\vect{u}}^{(\text{rx})} - \bar{\mat{A}}_{\xi}\bigl[ p^{(\text{in})} \bigr] \tilde{\vectsym{\theta}} }{2}{1}
  &\leq
  \bar{\eta}.
\end{alignedat}
\tag{$\recmethodnorm{ q }{ \xi }{ \eta }$}
\label{eqn:recovery_reg_norm_lq_minimization}
\end{equation}
% d) inversions of both the normalization and the basis transform estimate the compressibility fluctuations
The inversions of both
% 1.) inversion of the normalization
the normalization in
\eqref{eqn:recon_reg_norm_trans_coef} and
% 2.) inversion of the basis transform
the basis transform in
\eqref{eqn:recovery_reg_sparse_representation} estimate
% 3.) vector stacking the regular samples in the discretized relative spatial fluctuations in the unperturbed compressibility
the compressibility fluctuations
\eqref{eqn:recovery_sys_lin_eq_gamma_kappa_bp_vector} as
\begin{equation}
 %--------------------------------------------------------------------------------------------------------------
 % estimated vector stacking the regular samples in the discretized relative spatial fluctuations in the unperturbed compressibility
 %--------------------------------------------------------------------------------------------------------------
  \hat{\vectsym{\gamma}}_{\xi}^{(\kappa, q, \eta)}
  =
  \dnorm{ \vect{u}^{(\text{rx})} }{2}{1}
  \mat{\Psi}
  \mat{W}_{\xi}^{-1}
  \hat{\bar{\vectsym{\theta}}}_{\xi}^{(\kappa, q, \eta)}.
 \label{eqn:recovery_reg_norm_lq_minimization_sol_mat_params}
\end{equation}
% e) doubled real parts estimate the physically meaningful real-valued regular samples in the discretized compressibility fluctuations for all grid points
\TODO{shorten! estimate two times...}
Their doubled real parts estimate
% 1.) physically meaningful real-valued regular samples
the physically meaningful real-valued regular samples in
% 2.) discretized relative spatial fluctuations in the unperturbed compressibility
the discretized compressibility fluctuations
\eqref{eqn:recovery_disc_space_fov_rel_fluctuations_bp_sampled_kappa} for
% 3.) all grid points
all grid points.

%%%%%%%%%%%%%%%%%%%%%%%%%%%%%%%%%%%%%%%%%%%%%%%%%%%%%%%%%%%%%%%%%%%%%%%%%%%%%%%%%%%%%%%%%%%%%%%%%%%%%%%%%%%%%%%%
% 6.) implementation
%%%%%%%%%%%%%%%%%%%%%%%%%%%%%%%%%%%%%%%%%%%%%%%%%%%%%%%%%%%%%%%%%%%%%%%%%%%%%%%%%%%%%%%%%%%%%%%%%%%%%%%%%%%%%%%%
\section{Implementation}
\label{sec:implementation}
%---------------------------------------------------------------------------------------------------------------
% 1.) overview of the implementation
%---------------------------------------------------------------------------------------------------------------
% a) Fourier coefficients of the recorded RF voltage signals were determined first
The \name{Fourier} coefficients of
the recorded \ac{RF} voltage signals
\eqref{eqn:recovery_disc_freq_v_rx_Fourier_series_coef} were determined first.
% b) proposed method influenced the statistical properties of the additive errors in the linear algebraic system
The proposed method influenced
% 1.) statistical properties
the statistical properties of
% 2.) additive errors of bounded l2-norm
the additive errors in
% 3.) linear algebraic system (all pulse-echo measurements, multifrequent, all array elements, additive errors)
the linear algebraic system
\eqref{eqn:recovery_reg_prob_general_obs_trans_coef_error}.
% c) expected energies permit an estimate of the l2-norm of the normalized additive errors in the normalized CS problem
In
the absence of
any model inaccuracies,
% 1.) additive errors of bounded l2-norm:
%   \tnorm{ \vectsym{\eta} }{2} \leq \eta
% 2.) normalized additive errors of bounded l2-norm:
%   \bar{\vectsym{\eta}} = \vectsym{\eta} / \tnorm{ \vect{u}^{(\text{rx})} }{2}
% 3.) l2-norm of the normalized additive errors
%   \tnorm{ \bar{\vectsym{\eta}} }{2} = \tnorm{ \vectsym{\eta} }{2} / \tnorm{ \vect{u}^{(\text{rx})} }{2} \leq \bar{\eta} = \eta / \tnorm{ \vect{u}^{(\text{rx})} }{2}
the expected energies of both
% 1.) additive errors of bounded l2-norm
these errors and
% 2.) vector stacking the relevant Fourier coefficients of the recorded RF voltage signals (all pulse-echo measurements, multifrequent, all array elements)
the recorded \ac{RF} voltage signals
\eqref{eqn:recovery_sys_lin_eq_v_rx_born_all_f_all_in_v_rx} permitted
an estimate of
the $\ell_{2}$-norm of
% 3.) normalized additive errors of bounded l2-norm
the additive errors in
% 4.) normalized CS problem
the normalized \ac{CS} problem
\eqref{eqn:recovery_reg_norm_prob_general}.
% d) proposed method additionally requires an efficient implementation of the sparsity-promoting lq-minimization method
% g) auxiliary functions compose the FMM and a fast basis transform
Matrix-free compositions of
% 1.) fast multipole method for the observation process
the \ac{FMM} and
% 2.) fast basis transform for the nearly-sparse representation
a fast basis transform accelerated
% 3.) sparsity-promoting lq-minimization method
the sparsity-promoting $\ell_{q}$-minimization method
\eqref{eqn:recovery_reg_norm_lq_minimization}.
% j) discretized incident acoustic pressure fields have to be evaluated numerically
%Finally,
% 1.) discretized incident acoustic pressure fields [superpositions of quasi-(d-1)-spherical waves]
%the discretized incident acoustic pressure fields
%\eqref{eqn:recovery_p_in} have to be evaluated numerically.

%%%%%%%%%%%%%%%%%%%%%%%%%%%%%%%%%%%%%%%%%%%%%%%%%%%%%%%%%%%%%%%%%%%%%%%%%%%%%%%%%%%%%%%%%%%%%%%%%%%%%%%%%%%%%%%%
% 1.) determination of the relevant Fourier coefficients
%%%%%%%%%%%%%%%%%%%%%%%%%%%%%%%%%%%%%%%%%%%%%%%%%%%%%%%%%%%%%%%%%%%%%%%%%%%%%%%%%%%%%%%%%%%%%%%%%%%%%%%%%%%%%%%%
\subsection{Determination of the Relevant \name{Fourier} Coefficients}
\label{subsec:imp_fourier_coefficients}
%---------------------------------------------------------------------------------------------------------------
% 1.) determination of the relevant Fourier coefficients
%---------------------------------------------------------------------------------------------------------------
% a) concurrent ADC of the RF voltage signals generated by all array elements at the rates f_{\text{s}}^{(n)} quantized the endpoints of each recording time interval
% article:Schiffner2018, Sect. III. Linear Physical Model for the Pulse-Echo Measurement Process / Sect. III.A. Pulse-Echo Measurement Process
% - The \ac{UI} system sequentially performs $N_{\text{in}} \in \N$ independent pulse-echo measurements using a planar transducer array
%   (cf. \cref{fig:lin_mod_scan_configuration,tab:lin_mod_scan_config_instrum_params}).
% - Each measurement begins at the time instant $t = 0$ and triggers
%   the CONCURRENT RECORDING OF
%   THE \ac{RF} VOLTAGE SIGNALS $\tilde{u}_{m}^{(\text{rx}, n)}: \setsymbol{T}_{ \text{rec} }^{(n)} \mapsto \R$ GENERATED BY
%   ALL ARRAY ELEMENTS $m \in \setconsnonneg{ N_{\text{el}} - 1 }$ in the specified time interval
%   [ \setsymbol{T}_{ \text{rec} }^{(n)} = \bigl[ t_{\text{lb}}^{(n)}; t_{\text{ub}}^{(n)} \bigr], ] (eqn:lin_mod_scan_config_volt_rx_obs_interval) where
%   $t_{\text{lb}}^{(n)} \in \Rnonneg$ and $t_{\text{ub}}^{(n)} > t_{\text{lb}}^{(n)}$ denote
%   its lower and upper bounds, respectively.
The concurrent \ac{ADC} of
% 1.) RF voltage signals generated by all array elements
the \ac{RF} voltage signals generated by
all array elements at
% 2.) rates f_{\text{s}}^{(n)}
the rates
$f_{\text{s}}^{(n)} = 1 / T_{\text{s}}^{(n)} > 2 f_{\text{ub}}^{(n)}$ quantized
% 3.) endpoints
the endpoints of
% 4.) specified recording time intervals for the RF voltage signals generated by all array elements
each recording time interval
\eqref{eqn:lin_mod_scan_config_volt_rx_obs_interval} as
% 5.) quantized lower bounds in the specified recording time intervals (t_{\text{lb}}^{(n)} \in \Rnonneg)
$t_{\text{lb}}^{(n)} = q_{\text{lb}}^{(n)} T_{\text{s}}^{(n)}$ and
% 6.) quantized upper bounds in the specified recording time intervals (t_{\text{ub}}^{(n)} > t_{\text{lb}}^{(n)})
$t_{\text{ub}}^{(n)} = q_{\text{ub}}^{(n)} T_{\text{s}}^{(n)}$, where
% 7.) nonnegative quantized bounds
$q_{\text{lb}}^{(n)}, q_{\text{ub}}^{(n)} \in \Nnonneg$ and
% 8.) quantized upper bounds are larger than quantized lower bounds
$q_{\text{ub}}^{(n)} > q_{\text{lb}}^{(n)}$.
% b) concurrent ADC recorded N_{t}^{(n)} real-valued samples per signal and N_{\text{el}} N_{t}^{(n)} samples per pulse-echo measurement
% book:Briggs1995, Chapter 2: The Discrete Fourier Transform / Sect. 2.4.: DFT Approximations to Fourier Series Coefficients [DERIVATION OF THE DFT]
% - As closely as the DFT is related to the Fourier transform, it may be argued that
%   IT HOLDS EVEN MORE KINSHIP TO THE COEFFICIENTS OF THE FOURIER SERIES. (p. 33)
% - It is a simple matter to use the Fourier series to derive the DFT formula, and we will do so shortly. (p. 33)
% - With this prelude to Fourier series, we are now in a position to derive
%   the DFT AS AN APPROXIMATION to the integral that gives
%   the FOURIER SERIES COEFFICIENTS c_{k}. (p. 38)
% - Now we consider APPROXIMATIONS TO THE INTEGRAL
%   [ c_{k} = \frac{ 1 }{ A } \int_{ - A / 2 }^{ A / 2 } f( x ) e^{ -j 2 \pi k x / A } dx. ] (2.17) (p. 38)
% - As before,
%   the INTERVAL OF INTEGRATION is divided into N SUBINTERVALS OF EQUAL LENGTH, and
%   let the GRID SPACING BE \Delta x = A / N. (p. 39)
% - A grid with N + 1 equally spaced points over the interval [ -A/2, A/2 ] is defined by
%   the points x_{n} = n \Delta x for n = -N/2:N/2. (p. 39)
% - Furthermore, we let [ g( x ) = f( x ) e^{ -j 2 \pi k x / A } ] be the integrand in this expression. (p. 39)
% - Applying the TRAPEZOID RULE gives the approximations
%   [ c_{k} = \frac{ 1 }{ A } \int_{ - A / 2 }^{ A / 2 } g( x ) dx \approx \frac{ \Delta x }{ 2 A } [ g( -A/2 ) + 2 \sum_{ n = - N/2 + 1 }^{ N/2 - 1 } g( x_{n} ) + g( A/2 ) ]. ] (p. 39)
It recorded
% 1.) number of recorded real-valued samples per RF voltage signal
$N_{t}^{(n)} = q_{\text{ub}}^{(n)} - q_{\text{lb}}^{(n)}$ real-valued samples per
signal%
\footnote{%
 % a) assuming identical samples at the endpoints of the recording time intervals
 % book:Briggs1995, Chapter 2: The Discrete Fourier Transform / Sect. 2.4.: DFT Approximations to Fourier Series Coefficients [DERIVATION OF THE DFT]
 % - NOW THE QUESTION OF ENDPOINT VALUES ENTERS IN A CRITICAL WAY. (p. 39)
 % - We have already seen that if
 %   the periodic extension of f is DISCONTINUOUS AT THE ENDPOINTS x = \pm A/2, then, when its Fourier series converges,
 %   it converges to the AVERAGE VALUE
 %   [ \frac{ f( -A/2+ ) + f( A/2- ) }{ 2 }. ] (p. 39)
 % - Therefore, it is the AVERAGE VALUE OF f AT THE ENDPOINTS THAT MUST BE USED IN THE TRAPEZOID RULE. (p. 39)
 % - Noting that the kernel e^{ -j 2 \pi k x / A } has the value (-1)^{k} at x = \pm A/2, we see that
 %   function g that must be used for the trapezoid rule is
 %   [ g( x ) = f( x ) e^{ -j 2 \pi k x / A } for x \neq \pm A/2,
 %            = \frac{ (-1)^{k} }{ 2 } [ f( -A/2+ ) + f( A/2- ) ] for x = \pm A/2. ] (pp. 39, 40)
 % - It should be verified that this choice of g, dictated by the convergence properties of the Fourier series, guarantees that
 %   g( -A/2 ) = g( A/2 ). (p. 40)
 Assuming identical samples at
 % 1.) endpoints
 the endpoints of
 % 2.) specified recording time intervals for the RF voltage signals generated by all array elements
 the recording time intervals
 \eqref{eqn:lin_mod_scan_config_volt_rx_obs_interval}, i.e.
 $\tilde{u}_{m}^{(\text{rx}, n)}[ t_{\text{lb}}^{(n)} ] = \tilde{u}_{m}^{(\text{rx}, n)}[ t_{\text{ub}}^{(n)} ]$ for
 % 3.) all sequential pulse-echo measurements and all array elements
 all $( n, m ) \in \setconsnonneg{ N_{\text{in}} - 1 } \times \setconsnonneg{ N_{\text{el}} - 1 }$.
 % b) averages of the left and right limits have to be used at any point of discontinuity
 % book:Briggs1995, Chapter 2: The Discrete Fourier Transform / Sect. 2.4.: DFT Approximations to Fourier Series Coefficients [DERIVATION OF THE DFT]
 % - In a similar way,
 %   an AVERAGE VALUE must be used at ANY GRID POINTS AT WHICH f HAS DISCONTINUITIES. (p. 40)
 % - There are SUBTLETIES CONCERNING THE USE OF AVERAGE VALUES AT THE ENDPOINTS AND DISCONTINUITES, but
 %   the importance of this issue will be emphasized many times in hopes of removing the subtlety! (p. 40)
 In general,
 the averages of
 the left and right limits replace
 any discontinuities
 \cite[40]{book:Briggs1995}.
} and, thus,
% 2.) number of recorded real-valued samples per pulse-echo measurement
$N_{\text{el}} N_{t}^{(n)}$ samples per
pulse-echo measurement for
% 3.) all sequential pulse-echo measurements
all $n \in \setconsnonneg{ N_{\text{in}} - 1 }$.
Normalized $N_{t}^{(n)}$-point \acp{DFT}
%\footnote{%
%average of first and last sample%
%\begin{equation*}
  %u_{m, l}^{(\text{rx}, n)}
  %=
  %\frac{ 1 }{ N_{t}^{(n)} }
  %e^{ -j 2 \pi l q_{\text{lb}}^{(n)} / N_{t}^{(n)} }
  %\sum_{ q = 0 }^{ N_{t}^{(n)} - 1 }
    %\tilde{u}_{m}^{(\text{rx}, n)}[ ( q_{\text{lb}}^{(n)}  + q ) T_{\text{s}}^{(n)} ]
    %e^{ -j 2 \pi l q / N_{t}^{(n)} }
%\end{equation*}
%$\tilde{u}_{m}^{(\text{rx}, n)}( q_{\text{lb}}^{(n)} T_{\text{s}}^{(n)} )$ is the average value
%}
(cf. e.g.
\cite[Sect. 3.3.2]{book:Mallat2009},
\cite[Sect. 2.2.3]{book:Manolakis2005},
\cite[Sect. 6.2]{book:Briggs1995}%
) provided
% 1.) Fourier coefficients of the recorded RF voltage signals
the \name{Fourier} coefficients
\eqref{eqn:recovery_disc_freq_v_rx_Fourier_series_coef} forming
% 2.) vector stacking the relevant Fourier coefficients of the recorded RF voltage signals (all pulse-echo measurements, multifrequent, all array elements)
the vector
\eqref{eqn:recovery_sys_lin_eq_v_rx_born_all_f_all_in_v_rx} for
% 3.) quantized recording times
the quantized recording times
$T_{ \text{rec} }^{(n)} = N_{t}^{(n)} T_{\text{s}}^{(n)}$.
% d) approximate efficiencies of the regular sampling in combination with the subsequent computation of the DFTs
% article:SchiffnerITUFFC2018, Sect. III.D Pulse-Echo Measurement Process (subsec:lin_mod_measurement_process)
% - Multiple time-domain methods for the \ac{ADC} of the received \ac{RF} voltage signals permit the determination of
%   the relevant \name{Fourier} coefficients.
% - These methods differ in their \emph{efficiency}, i.e. the quotient of
%   the data volume occupied by the quantized relevant \name{Fourier} coefficients and
%   the data volume digitized during the pulse-echo measurement.
% 1.) numbers of relevant discrete frequencies (effective time-bandwidth products)
The effective time-bandwidth products
\eqref{eqn:recon_disc_axis_f_discrete_BP_TB_product},
% 2.) quantized recording times
the quantized recording times,
% 3.) lower bounds on the sampling rates
the lower bounds on
the sampling rates, and
% 4.) effective bandwidths
the effective bandwidths approximate
% 5.) efficiencies
the efficiencies of
these procedures as
\begin{equation}
 %--------------------------------------------------------------------------------------------------------------
 % approximate efficiencies of the regular sampling in combination with the subsequent computation of the DFTs
 %--------------------------------------------------------------------------------------------------------------
  \text{Efficiency}^{(n)}
  =
  \frac{
    2 N_{f, \text{BP}}^{(n)}
  }{
    N_{t}^{(n)}
  }
  \approx
  2 T_{\text{s}}^{(n)} B_{ u }^{(n)}
  <
  1 - \frac{ f_{\text{lb}}^{(n)} }{ f_{\text{ub}}^{(n)} }
 \label{eqn:imp_fourier_coef_efficiency}
\end{equation}
for
% 6.) all sequential pulse-echo measurements
all $n \in \setconsnonneg{ N_{\text{in}} - 1 }$, where
% 7.) quantized complex-valued Fourier coefficient
a \name{Fourier} coefficient occupies
twice the data volume of
% 8.) quantized real-valued sample
a signal sample.
% e) upper bounds show that the digitized data volumes exceeded those occupied by the relevant Fourier coefficients
The upper bounds show that
the digitized data volumes exceeded
those occupied by
the relevant \name{Fourier} coefficients.

%%%%%%%%%%%%%%%%%%%%%%%%%%%%%%%%%%%%%%%%%%%%%%%%%%%%%%%%%%%%%%%%%%%%%%%%%%%%%%%%%%%%%%%%%%%%%%%%%%%%%%%%%%%%%%%%
% 2.) additive errors
%%%%%%%%%%%%%%%%%%%%%%%%%%%%%%%%%%%%%%%%%%%%%%%%%%%%%%%%%%%%%%%%%%%%%%%%%%%%%%%%%%%%%%%%%%%%%%%%%%%%%%%%%%%%%%%%
\subsection{Additive Errors}
%\label{subsec:imp_obs_errors}
%---------------------------------------------------------------------------------------------------------------
% 1.) additive errors
%---------------------------------------------------------------------------------------------------------------
% a) additive errors corrupted the recorded samples of all RF voltage signals
% article:Schiffner2018, Sect. VI. Implementation / Sect. VI-A Determination of the Relevant Fourier Coefficients (subsec:imp_fourier_coefficients)
% - It [concurrent ADC] RECORDED
%   $N_{t}^{(n)} = q_{\text{ub}}^{(n)} - q_{\text{lb}}^{(n)}$ REAL-VALUED SAMPLES PER SIGNAL and, thus,
%   $N_{\text{el}} N_{t}^{(n)}$ SAMPLES PER PULSE-ECHO MEASUREMENT for
%   all $n \in \setconsnonneg{ N_{\text{in}} - 1 }$.
Additive errors, which
were statistically modeled as
\ac{GWN}
(cf. e.g. \cite[110]{book:Manolakis2005}) with
% 2.) zero mean
zero mean and
% 3.) variance \sigma_{\eta}^{2}
the variance $\sigma_{\eta}^{2}$, corrupted
% 4.) recorded samples of all RF voltage signals
the recorded samples of
all \ac{RF} voltage signals.
% b) expected energy of the recorded RF voltage signals
% article:Schiffner2018, Sect. V. Image Recovery Based on Compressed Sensing / Sect. V-D Regularization of the Inverse Scattering Problem (subsec:recovery_regularization)
% - Inserting this representation [nearly-sparse],
%   defining the complex-valued $N_{\text{obs}} \times N_{\text{lat}}$ sensing matrix
%   [ \mat{A}\bigl[ p^{(\text{in})} \bigr] = \mat{\Phi}\bigl[ p^{(\text{in})} \bigr] \mat{\Psi}, ] (eqn:recovery_reg_sensing_matrix) and accounting for
%   an UNKNOWN COMPLEX-VALUED $N_{\text{obs}} \times 1$ VECTOR OF ADDITIVE ERRORS OF BOUNDED $\ell_{2}$-NORM
%   $\tnorm{ \vectsym{\eta} }{2} \leq \eta$,
%   the linear algebraic system \eqref{eqn:recovery_sys_lin_eq_v_rx_born_all_f_all_in} becomes
%   [ \vect{u}^{(\text{rx})} = \underbrace{ \mat{\Phi}\bigl[ p^{(\text{in})} \bigr] \mat{\Psi} }_{ = \mat{A}[ p^{(\text{in})} ] } \vectsym{\theta}^{(\kappa)} + \vectsym{\eta} = \underbrace{ \mat{A}\bigl[ p^{(\text{in})} \bigr] \vectsym{\theta}^{(\kappa)} }_{ = \vect{u}^{(\text{B})} } +\: \vectsym{\eta}. ]
%   (eqn:recovery_reg_prob_general_obs_trans_coef_error)
% - The ADDITIVE ERRORS reflect the inaccuracies of the discretized physical models and the voltage measurements.
The expected energy of
% 1.) vector stacking the relevant Fourier coefficients of the recorded RF voltage signals (all pulse-echo measurements, multifrequent, all array elements)
the recorded \ac{RF} voltage signals
\eqref{eqn:recovery_sys_lin_eq_v_rx_born_all_f_all_in_v_rx} in
% 2.) linear algebraic system (all pulse-echo measurements, multifrequent, all array elements, additive errors)
the linear algebraic system
\eqref{eqn:recovery_reg_prob_general_obs_trans_coef_error} was
\begin{equation*}
 %--------------------------------------------------------------------------------------------------------------
 % expected energy of the vector stacking the relevant Fourier coefficients of the recorded RF voltage signals
 %--------------------------------------------------------------------------------------------------------------
  \expect{
    \dnorm{ \vect{u}^{(\text{rx})} }{2}{1}^{2}
  }
  =
  \dnorm{ \vect{u}^{(\text{B})} }{2}{1}^{2}
  +
  \sigma_{\eta}^{2}
  N_{\text{el}}
  \sum_{ n = 0 }^{ N_{\text{in}} - 1 }
    \frac{
      N_{f, \text{BP}}^{(n)}
    }{
      N_{t}^{(n)}
    }.
\end{equation*}
% c) expected energy permitted the l2-norm of the normalized additive errors in the normalized CS problem and the sparsity-promoting lq-minimization method to be estimated
% article:Schiffner2018, Sect. V. Image Recovery Based on Compressed Sensing / Sect. V-D Regularization of the Inverse Scattering Problem (subsec:recovery_regularization)
% - With the NORMALIZED VERSIONS of
%   the recorded \ac{RF} voltage signals $\bar{\vect{u}}^{(\text{rx})} = \vect{u}^{(\text{rx})} / \tnorm{ \vect{u}^{(\text{rx})} }{2}$,
%   the ADDITIVE ERRORS $\bar{\vectsym{\eta}} = \vectsym{\eta} / \tnorm{ \vect{u}^{(\text{rx})} }{2}$ of bounded $\ell_{2}$-norm
%   $\tnorm{ \bar{\vectsym{\eta}} }{2} \leq \bar{\eta} = \eta / \tnorm{ \vect{u}^{(\text{rx})} }{2}$, and
%   the nearly-sparse representation [...], the insertions of
%   the weighting matrices \eqref{eqn:recovery_reg_norm_weighting_matrices} and
%   the normalized sensing matrix \eqref{eqn:recon_reg_norm_sensing_matrix} into
%   the linear algebraic system \eqref{eqn:recovery_reg_prob_general_obs_trans_coef_error} yield
%   the equivalent system [...].
% => \tnorm{ \bar{\vectsym{\eta}} }{2} = \tnorm{ \vectsym{\eta} }{2} / \tnorm{ \vect{u}^{(\text{rx})} }{2} \leq \bar{\eta}
% \bar{\eta} = \tnorm{ \vectsym{\eta} }{2} / \tnorm{ \vect{u}^{(\text{rx})} }{2}
It permitted
% 1.) l2-norm of the normalized additive errors
the $\ell_{2}$-norm of
the normalized additive errors in
% 2.) normalized CS problem
the normalized \ac{CS} problem
\eqref{eqn:recovery_reg_norm_prob_general} and
% 3.) sparsity-promoting lq-minimization method
the sparsity-promoting $\ell_{q}$-minimization method
\eqref{eqn:recovery_reg_norm_lq_minimization} to be estimated as
\begin{equation}
 %--------------------------------------------------------------------------------------------------------------
 % estimated l2-norm of the normalized additive errors in the normalized CS problem
 %--------------------------------------------------------------------------------------------------------------
  \hat{ \bar{\eta} }
  =
  \left[
    1
    +
    \frac{
      \norm{ \vect{u}^{(\text{B})} }{2}^{2}
    }{
      \sigma_{\eta}^{2}
      N_{\text{el}}
      \sum_{ n = 0 }^{ N_{\text{in}} - 1 }
        N_{f, \text{BP}}^{(n)} / N_{t}^{(n)}
    }
  \right]^{ - \frac{1}{2} }.
 \label{eqn:imp_data_acq_rel_obs_error_est}
\end{equation}

%%%%%%%%%%%%%%%%%%%%%%%%%%%%%%%%%%%%%%%%%%%%%%%%%%%%%%%%%%%%%%%%%%%%%%%%%%%%%%%%%%%%%%%%%%%%%%%%%%%%%%%%%%%%%%%%
% 3.) computation of the incident acoustic pressure fields
%%%%%%%%%%%%%%%%%%%%%%%%%%%%%%%%%%%%%%%%%%%%%%%%%%%%%%%%%%%%%%%%%%%%%%%%%%%%%%%%%%%%%%%%%%%%%%%%%%%%%%%%%%%%%%%%
%\subsection{Computation of the Incident Acoustic Pressure Fields}
%\label{subsec:imp_p_in}
%\input{implementation/implementation_p_in.tex}

%%%%%%%%%%%%%%%%%%%%%%%%%%%%%%%%%%%%%%%%%%%%%%%%%%%%%%%%%%%%%%%%%%%%%%%%%%%%%%%%%%%%%%%%%%%%%%%%%%%%%%%%%%%%%%%%
% 4.) sparsity-promoting lq-minimization method
%%%%%%%%%%%%%%%%%%%%%%%%%%%%%%%%%%%%%%%%%%%%%%%%%%%%%%%%%%%%%%%%%%%%%%%%%%%%%%%%%%%%%%%%%%%%%%%%%%%%%%%%%%%%%%%%
\subsection{Sparsity-Promoting $\ell_{q}$-Minimization Method}
\label{subsec:imp_lq_minimization}
%---------------------------------------------------------------------------------------------------------------
% 1.) sparsity-promoting lq-minimization method
%---------------------------------------------------------------------------------------------------------------
% a) SPGL1 implemented the convex sparsity-promoting l1-minimization method
% article:VanDenBergSIAMJSC2009: Probing the Pareto Frontier for Basis Pursuit Solutions
Spectral projected gradient for
$\ell_{1}$-minimization (\acs{SPGL1})\acused{SPGL1}
\cite{article:VanDenBergSIAMJSC2009} implemented
% 1.) convex sparsity-promoting l1-minimization method
the convex $\ell_{1}$-minimization method
\eqreflqmin{eqn:recovery_reg_norm_lq_minimization}{ 1 }.
\name{Foucart}'s algorithm
\cite[Sect. 4]{article:FoucartACHA2009} iteratively applied
% 1.) convex l1-minimization method
this method based on
\ac{SPGL1} to
a sequence of
renormalized \ac{CS} problems to approximate
% 2.) nonconvex sparsity-promoting lq-minimization method, q \in [ 0; 1 )
the nonconvex $\ell_{q}$-minimization method
\eqref{eqn:recovery_reg_norm_lq_minimization} for
the half-open parameter interval
$q \in [ 0; 1 )$.
% c) SPGL1 is iterative and left multiplied a sequence of recursively-generated vectors by the normalized sensing matrix or its adjoint
% article:RokhlinJCP1990: Rapid Solution of Integral Equations of Scattering Theory in Two Dimensions
% - most iterative solvers: APPLICATION OF THE MATRIX TO A SEQUENCE OF RECURSIVELY GENERATED VECTORS
% article:RokhlinJCP1985: Rapid solution of integral equations of classical potential theory
% - MOST ITERATIVE SCHEMES FOR SOLUTION OF LINEAR SYSTEMS resulting from classical potential theory
%   REQUIRE APPLICATION OF THE MATRIX OF THE SYSTEM TO A SEQUENCE OF RECURSIVELY GENERATED VECTORS.
\ac{SPGL1} is
iterative and left multiplied
% 1.) sequence of recursively-generated vectors
a sequence of
recursively-generated vectors by
% 2.) normalized sensing matrix
the potentially densely-populated normalized sensing matrix
\eqref{eqn:recon_reg_norm_sensing_matrix} or
% 3.) adjoint of the normalized sensing matrix
its adjoint.
% d) matrix-free implementation interpreted each type of matrix-vector product as a linear map and dedicated a customized auxiliary function to its numerical evaluation
Its matrix-free implementation interpreted
each type of
matrix-vector product as
a linear map and dedicated
a customized auxiliary function to
its numerical evaluation.
% e) auxiliary functions aimed at circumventing the explicit storage of the associated matrix and accelerating the numerical computations
Both functions aimed at
% 1.) circumventing the explicit storage of the associated matrix in the fast but limited RAM
circumventing
the explicit storage of
the associated matrix in
the fast but limited \ac{RAM} and
% 2.) accelerating the numerical computations
accelerating
the numerical computations.
% f) memory consumption of the normalized sensing matrix and the number of multiplications executed by the associated matrix-vector product pose challenges
In fact,
% 1.) memory consumption of the normalized sensing matrix
the memory consumption of
the normalized sensing matrix
\eqref{eqn:recon_reg_norm_sensing_matrix} and
% 2.) number of multiplications executed by the associated matrix-vector product
the number of
multiplications executed by
the associated matrix-vector product pose
challenges for
modern \ac{UI} systems.
% g) memory consumption of the normalized sensing matrix and the number of multiplications executed by the associated matrix-vector product
They equal
% 1.) memory consumption of the normalized sensing matrix
$M_{\text{conv}} = N_{\text{obs}} N_{\text{lat}} w_{\text{c}}$, where
$w_{\text{c}} \in \Rplus$ denotes
the amount of
memory allocated to
a complex-valued variable, and
% 2.) number of multiplications executed by the associated matrix-vector product
$N_{\text{mul},\text{conv}} = N_{\text{obs}} N_{\text{lat}}$,
respectively.
% h) auxiliary functions composed the FMM, a fast basis transform, and the normalization of the column vectors
For this reason,
both functions composed
% 1.) fast multipole method for the observation process
the \ac{FMM},
% 2.) fast basis transform for the nearly-sparse representation
a fast basis transform, and
% 3.) normalization of the column vectors
\TODO{ordentlich, punkt 3}
the normalization of
the column vectors.
% h) FMM efficiently approximated the action of the observation process or its adjoint on a suitable vector
%The \ac{FMM} efficiently approximated
%the action of
% 1.) observation process (all pulse-echo measurements, multifrequent, all transducer elements)
%the observation process
%\eqref{eqn:recovery_sys_lin_eq_v_rx_born_all_f_all_in_mat} or
% 2.) adjoint of the observation process (all pulse-echo measurements, multifrequent, all transducer elements)
%its adjoint on
%a suitable vector, whereas
% i) fast basis transform efficiently inferred the nearly-sparse representation from the discretized relative spatial fluctuations in compressibility or vice versa
%the fast basis transform efficiently inferred
% 1.) nearly-sparse representation
%the nearly-sparse representation
%\eqref{eqn:recovery_reg_sparse_representation} from
% 2.) vector stacking the regular samples in the discretized relative spatial fluctuations in the unperturbed compressibility
%the compressibility fluctuations
%\eqref{eqn:recovery_sys_lin_eq_gamma_kappa_bp_vector} or
%vice versa.
% j) diagonal inverse weighting matrix readily permitted the efficient normalization of the column vectors by numerically evaluating the associated matrix-vector products
The latter corresponded to
a multiplication by
% 1.) diagonal inverse weighting matrix
the diagonal inverse weighting matrix
\eqref{eqn:recovery_reg_norm_weighting_matrix_inv} and readily permitted
an efficient evaluation.

%%%%%%%%%%%%%%%%%%%%%%%%%%%%%%%%%%%%%%%%%%%%%%%%%%%%%%%%%%%%%%%%%%%%%%%%%%%%%%%%%%%%%%%%%%%%%%%%%%%%%%%%%%%%%%%%
% 5.) fast multipole method for the observation process
%%%%%%%%%%%%%%%%%%%%%%%%%%%%%%%%%%%%%%%%%%%%%%%%%%%%%%%%%%%%%%%%%%%%%%%%%%%%%%%%%%%%%%%%%%%%%%%%%%%%%%%%%%%%%%%%
\subsection{Fast Multipole Method for the Observation Process}
\label{subsec:imp_fmm_obs_process}
The \ac{FMM}
(cf. e.g.
\cite[Chapt. 9]{book:Gibson2014},
\cite{article:CoifmanIAPM1993,article:RokhlinJCP1990}%
) efficiently approximated
the action of
% 1.) observation process (all pulse-echo measurements, multifrequent, all transducer elements)
the observation process
\eqref{eqn:recovery_sys_lin_eq_v_rx_born_all_f_all_in_mat} or
% 2.) adjoint of the observation process (all pulse-echo measurements, multifrequent, all transducer elements)
its adjoint on
a suitable vector.
It substituted
% 1.) outgoing free-space Green's functions (two- and three-dimensional Euclidean spaces)
the outgoing free-space \name{Green}'s functions
\eqref{eqn:app_helmholtz_green_free_space_2_3_dim} in
% 2.) entries of the observation process (single pulse-echo measurement, monofrequent, single transducer element)
the entries of
the observation process
\eqref{eqn:recovery_sys_lin_eq_v_rx_born_coef} by
% 3.) error-regulated truncated multipole expansions
error-regulated truncated multipole expansions if
the grid points
% 4.) discrete position of the point-like relative spatial fluctuation in the unperturbed compressibility
$\vect{r}_{\text{lat}, i} \in \mathcal{L}$ and
% 5.) discrete positions of the mathematical points
$\vect{r}_{\text{mat}, \nu}^{(m)} \in \mathcal{V}_{m}$, satisfied
% 6.) specific geometric relationship
a specific geometric relationship
\cite[Chapt. 9]{book:Gibson2014},
\cite{article:CoifmanIAPM1993}.
This substitution decomposed
% 1.) observation process (all pulse-echo measurements, multifrequent, all transducer elements)
the observation process
\eqref{eqn:recovery_sys_lin_eq_v_rx_born_all_f_all_in_mat} into
the sum
$\mat{\Phi}[ p^{(\text{in})} ] \approx \mat{\Phi}^{(\text{near})}[ p^{(\text{in})} ] + \mat{\Phi}^{(\text{far})}[ p^{(\text{in})} ]$, where
% 1.) point-like relative spatial fluctuations in the unperturbed compressibility located on the lattice points close to the planar transducer array
$\mat{\Phi}^{(\text{near})}[ p^{(\text{in})} ]$ accounted for
the grid points close to
the planar transducer array, and
% 2.) point-like relative spatial fluctuations in compressibility located on the lattice points exceeding a specified distance from the planar transducer array
$\mat{\Phi}^{(\text{far})}[ p^{(\text{in})} ]$ accounted for
those exceeding
a specified distance from
the planar transducer array
\cite[Sect. 9.1]{book:Gibson2014},
\cite[(8) and (23)]{article:CoifmanIAPM1993}.
% d) sparse population of the summand \mat{\Phi}^{(\text{near})}[ p^{(\text{in})} ] enabled both its explicit storage in the RAM and fast numerical evaluations of the associated matrix-vector products
The sparse population of
the summand
$\mat{\Phi}^{(\text{near})}[ p^{(\text{in})} ]$ enabled both
% 1.) explicit storage in the RAM
its explicit storage in
the \ac{RAM} and
% 2.) fast numerical evaluations of the associated matrix-vector products
fast numerical evaluations of
the associated matrix-vector products.
% e) additional blockwise decomposition of the summand \mat{\Phi}^{(\text{far})}[ p^{(\text{in})} ] into the products of only a few unique matrices provides similar benefits
The resulting block structure of
the summand
$\mat{\Phi}^{(\text{far})}[ p^{(\text{in})} ]$, which consisted of
the products of
only a few unique
% 1.) diagonal translation matrices
diagonal translation matrices and
% 2.) densely-populated aggregation and disaggregation matrices
densely-populated aggregation and
disaggregation matrices, provided
similar benefits.
% f) small number of unique matrices enabled their explicit storage in the RAM
% book:Gibson2014, Chapter 9: The Fast Multipole Method
% - As we will see in this chapter,
%   the FMM can be applied to vector Helmholtz problems,
%   [1] ALLOWING FOR A FAST COMPUTATION OF THE MATRIX-VECTOR PRODUCT IN AN ITERATIVE SOLVER, and
%   [2] ELIMINATING THE NEED TO STORE MANY OF THE MOM MATRIX ELEMENTS EXPLICITLY. (p. 330)
% - THIS RESULTS IN
%   [1] A SPEED INCREASE and
%   [2] A REDUCTION IN MEMORY REQUIREMENT, allowing us to solve
%   EXISTING PROBLEMS FASTER, as well as LARGER PROBLEMS THAT COULD NOT BE ATTEMPTED BEFORE. (p. 330)
The small number of
unique matrices enabled
their explicit storage in
the \ac{RAM}, whereas
% g) diagonal population of the translation matrices simultaneously reduced the computational costs
the diagonal population of
the translation matrices concurrently reduced
the computational costs.

% ill-conditioned sensing matrix can be approximated by a low-rank matrix
% TODO: move to description of FMM
\TODO{low-rank approximation of the matrix, accuracy of FMM?}

%---------------------------------------------------------------------------------------------------------------
% 2.) implementation details
%---------------------------------------------------------------------------------------------------------------
% a) two C programs based on CUDA implemented parallelized versions of the FMM for the observation process and its adjoint
Two \name{C} programs based on
\name{CUDA}
(NVIDIA Corp., Santa Clara, CA, USA) implemented
parallelized versions of
the \ac{FMM} for
% 1.) observation process (all pulse-echo measurements, multifrequent, all transducer elements)
the observation process
\eqref{eqn:recovery_sys_lin_eq_v_rx_born_all_f_all_in_mat} and
% 2.) adjoint of the observation process (all pulse-echo measurements, multifrequent, all transducer elements)
its adjoint.
% b) Tesla K40c performs all computations with 32 bit single precision
A \name{Tesla K40c} \ac{GPU} performed
all computations with
% TODO: single precision is redundant: 32 bit = single
$\SI{32}{\bit}$ single precision. %, which
%was found to be sufficiently accurate.
% c) MATLAB interface based on the MEX framework simplified the data transfers and the analysis of the results
A \name{Matlab}
(The MathWorks, Inc., Natick, MA, USA) interface, which used
% 1.) MEX framework
the \name{MEX} framework and
% 2.) double-precision (64 bit = 8 B) floating-point format for real-valued variables
the $\SI{8}{\byte}$ floating-point format for
real-valued variables, i.e.
% 3.) amount of memory allocated to a complex-valued variable
$w_{\text{c}} = \SI{16}{\byte}$, simplified
% 4.) data transfers
the data analyses. %and analyses
% 5.) data analyses

%%%%%%%%%%%%%%%%%%%%%%%%%%%%%%%%%%%%%%%%%%%%%%%%%%%%%%%%%%%%%%%%%%%%%%%%%%%%%%%%%%%%%%%%%%%%%%%%%%%%%%%%%%%%%%%%
% 6.) fast basis transform for the nearly-sparse representation
%%%%%%%%%%%%%%%%%%%%%%%%%%%%%%%%%%%%%%%%%%%%%%%%%%%%%%%%%%%%%%%%%%%%%%%%%%%%%%%%%%%%%%%%%%%%%%%%%%%%%%%%%%%%%%%%
\subsection{Fast Basis Transform for the Nearly-Sparse Representation}
%\label{subsec:imp_fast_basis_transform}
%---------------------------------------------------------------------------------------------------------------
% 1.) fast basis transform for the nearly-sparse representation
%---------------------------------------------------------------------------------------------------------------
% a) fast basis transform implemented the matrix-vector products between the unitary matrix \mat{\Psi} or its adjoint and a suitable vector
% article:Schiffner2018, Sect. V. Image Recovery Based on Compressed Sensing / Sect. V.D. Regularization of the Inverse Scattering Problem
% - Postulating the knowledge of a suitable orthonormal basis of $\C^{ N_{\text{lat}} }$, which is represented by
%   the complex-valued unitary $N_{\text{lat}} \times N_{\text{lat}}$ matrix $\mat{\Psi}$ of structural building blocks $\vectsym{\psi}_{i}$,
%   the complex-valued $N_{\text{lat}} \times 1$ vector of transform coefficients [...] (eqn:recovery_reg_sparse_representation) constitutes
%   a nearly-sparse representation of the compressibility fluctuations \eqref{eqn:recovery_sys_lin_eq_gamma_kappa_bp_vector}.
A fast basis transform implemented
the matrix-vector products between
% 1.) unitary matrix \mat{\Psi}
the unitary matrix $\mat{\Psi}$, which represents
% 2.) suitable orthonormal basis of \C^{ N_{\text{lat}} }
the orthonormal basis specified in
% 3.) nearly-sparse representation
the nearly-sparse representation
\eqref{eqn:recovery_reg_sparse_representation}, or
% 4.) adjoint of the unitary matrix \mat{\Psi}
its adjoint and
% 5.) suitable vector
a suitable vector.
% b) FFT efficiently implemented the matrix-vector products involving the N_{\text{lat}}-dimensional unitary DFT matrix or its adjoint
The \ac{FFT}, for example,
efficiently implemented
the matrix-vector products involving
the unitary \ac{DFT} matrix $\mat{\Psi}$, which represents
the discrete \name{Fourier} basis, or
its adjoint.
% c) various wavelet or wave atom bases provide comparably fast transforms
% book:Mallat2009:
% article:DemanetACHA2007: Wave atoms and sparsity of oscillatory patterns
% 3. Digital wave atoms / 3.1. Implementation of wave atoms: 1D warmup
% - The complexity of each inverse FFT at scale j is O(j2^{j}), and there are O(2{j}) frequency bumps at scale j, indexed by m, so the total complexity is
%   [O( N * log(N) )] with N = 2^{2J}. (p. 377)
% - Likewise, the complexity of the inverse transform is O( N * log(N) ). (p. 378)
Various wavelet
\cite{book:Mallat2009} or
wave atom bases
\cite{article:DemanetACHA2007} provide
comparably-fast transforms that circumvent
% 1.) explicit storage of the matrices
the explicit storage of
the matrices in
% 2.) RAM
the \ac{RAM} and minimize
% 3.) computational costs
the computational costs.

%%%%%%%%%%%%%%%%%%%%%%%%%%%%%%%%%%%%%%%%%%%%%%%%%%%%%%%%%%%%%%%%%%%%%%%%%%%%%%%%%%%%%%%%%%%%%%%%%%%%%%%%%%%%%%%%
% 7.) computation of the weighting matrices
%%%%%%%%%%%%%%%%%%%%%%%%%%%%%%%%%%%%%%%%%%%%%%%%%%%%%%%%%%%%%%%%%%%%%%%%%%%%%%%%%%%%%%%%%%%%%%%%%%%%%%%%%%%%%%%%
%\subsection{Computation of the Weighting Matrices}
%\label{subsec:imp_weighting_mat}
%\input{implementation/implementation_weighting_matrices.tex}

%%%%%%%%%%%%%%%%%%%%%%%%%%%%%%%%%%%%%%%%%%%%%%%%%%%%%%%%%%%%%%%%%%%%%%%%%%%%%%%%%%%%%%%%%%%%%%%%%%%%%%%%%%%%%%%%
% 7.) simulation study
%%%%%%%%%%%%%%%%%%%%%%%%%%%%%%%%%%%%%%%%%%%%%%%%%%%%%%%%%%%%%%%%%%%%%%%%%%%%%%%%%%%%%%%%%%%%%%%%%%%%%%%%%%%%%%%%
\section{Simulation Study}
\label{sec:simulation_study}
The proposed method was validated using
synthetic \ac{RF} voltage signals.
% b) voltage signals were generated by numerical simulations of the pulse-echo measurement process probing two lossy heterogeneous objects
These were generated by
numerical simulations of
a typical pulse-echo measurement process probing
% 1.) two lossy heterogeneous objects
two lossy heterogeneous objects by
% 2.) each type of incident wave
each type of
incident wave.
% c) first object mimicked a typical wire phantom
The first object mimicked
a wire phantom, whereas %, i.e.
%an ensemble of
%isolated thin wires immersed in
%a lossy homogeneous fluid, whereas
% d) second object approximated the structure and the properties of human soft tissues
the second object approximated
the structure and
the properties of
human soft tissues.
% e) discretized compressibility fluctuations permitted sparse representations in the canonical and the Fourier bases
%Their compressibility fluctuations permitted
%sparse representations in
%the canonical and
%the \name{Fourier} bases. %, which defined
% f) bases defined the admissible structural building blocks as
%the admissible structural building blocks as
% 1.) individual samples
%individual samples and
% 2.) complex exponential functions of distinct spatial frequencies
%complex exponential functions of
%distinct spatial frequencies.
% g) additive errors of five distinct energy levels corrupted these synthetic RF voltage signals
Additive errors of
five distinct energy levels corrupted
these synthetic \ac{RF} voltage signals.

\subsection{Parameters}
\label{subsec:sim_study_parameters}
%%%%%%%%%%%%%%%%%%%%%%%%%%%%%%%%%%%%%%%%%%%%%%%%%%%%%%%%%%%%%%%%%%%%%%%%%%%%%%%%%%%%%%%%%%%%%%%%%%%%%%%%%%%%%%%%
% table: values of all simulation parameters (two-dimensional Euclidean space)
%%%%%%%%%%%%%%%%%%%%%%%%%%%%%%%%%%%%%%%%%%%%%%%%%%%%%%%%%%%%%%%%%%%%%%%%%%%%%%%%%%%%%%%%%%%%%%%%%%%%%%%%%%%%%%%%
\begin{table*}[tb]
 \centering
 \caption{%
  Values of
  all simulation parameters for
  the two-dimensional Euclidean space, i.e.
  $d = 2$,
  $\bar{b} = b / ( 2 \pi )^{\zeta}$.
 }
 \renewcommand{\arraystretch}{1}
 \label{tab:sim_study_parameters}
 \small
 \begin{tabular}{%
  @{}%
  l%    01.) Geometric and electromechanical parameters of the instrumentation / Wire phantom
  @{\hspace{0.9em}}%
  l%    02.) Geometric and electromechanical parameters of the instrumentation / Wire phantom
  @{\hspace{0.9em}}%
  l%    03.) Pulse-echo parameters / Tissue-mimicking phantom
  @{\hspace{0.9em}}%
  l%    04.) Synthesis parameters / Tissue-mimicking phantom
  @{\hspace{0.9em}}%
  l%    05.) Geometric parameters of the discretizations / Regularization
  @{}%
 }
 \toprule
  \multicolumn{2}{@{}H}{a) Instrumentation (cf. \cref{tab:lin_mod_scan_config_instrum_params})} &
  \multicolumn{1}{H}{b) Pulse-echo measurements} &
  \multicolumn{1}{H}{c) Wave syntheses} &
  \multicolumn{1}{H@{}}{d) Discretization (cf. \cref{tab:recon_disc_params})}\\
  \cmidrule(r){1-2}\cmidrule(lr){3-3}\cmidrule(lr){4-4}\cmidrule(l){5-5}
 \addlinespace
 %--------------------------------------------------------------------------------------------------------------
  % 1.a.1) total number of identical array elements (geometric parameter of the linear transducer array)
  % 1.a.2) number of vibrating faces along the r_{1}-axis (geometric parameter of the linear transducer array)
  $N_{\text{el}} = N_{\text{el}, 1} = 128$ &
  % 1.b) width of the identical kerfs separating the vibrating faces along the r_{1}-axis (geometric parameter of the linear transducer array)
  $k_{\text{el}, 1} = 0$ & % real value: $k_{\text{el}, 1} = \SI{25}{\micro\meter}$
  % 1.c) number of sequential pulse-echo measurements / sampling rate (pulse-echo parameter)
  $N_{\text{in}} = 1$,
  $f_{\text{s}}^{(0)} = \SI{20}{\mega\hertz}$ & % T_{\text{s}}^{(0)} = \SI{50}{\nano\second}
  % 1.d) reference voltage signal identically exciting all array elements (synthesis parameter)
  $\tilde{u}^{(\text{tx}, 0)}( t ) = \hat{u} \sin( \omega_{\text{c}} t )$ & % $t \in [ 0; T_{\text{c}} ]$
  % 1.e.1) total number of grid points per vibrating face (spatial discretization)
  % 1.e.2) number of grid points per vibrating face along the r_{1}-axis (spatial discretization)
  $N_{\text{mat}} = N_{\text{mat}, 1} = 4$\\
 %--------------------------------------------------------------------------------------------------------------
  % 2.a.1) constant spacing between the centers of the adjacent faces along the r_{1}-axis (element pitch, geometric parameter of the linear transducer array)
  % 2.a.2) identical width of the vibrating faces along the r_{1}-axis (geometric parameter of the linear transducer array)
  $\Delta r_{\text{el}, 1} = w_{\text{el}, 1} = \SI{304.8}{\micro\meter}$ & % real value: $w_{\text{el}, 1} = \SI{279.8}{\micro\meter}$
  % 2.b) transmitter and receiver apodization functions (geometric parameter of the linear transducer array)
  $\chi_{m, l}^{(\text{tx})}( r_{1} ) = \chi_{m, l}^{(\text{rx})}( r_{1} ) = 1$ & % $r_{1} \in L_{m}$
  % 2.c) lower frequency bound (pulse-echo parameter)
  $f_{\text{lb}}^{(0)} = \SI{2.6}{\mega\hertz}$ &
  % 2.d) frequency of the clock signal in the quantization operator providing the admissible time delays (synthesis parameter)
  $f_{\text{clk}} = \SI{80}{\mega\hertz}$ & % \SI{12.5}{\nano\second}
  % 2.e) constant spacing between the adjacent grid points on each vibrating face along the r_{1}-axis (spatial discretization)
  $\Delta r_{\text{mat}, 1} = \SI{76.2}{\micro\meter}$\\
 %--------------------------------------------------------------------------------------------------------------
  % 3.a) center frequency (electromechanical parameter of the instrumentation)
  $f_{\text{c}} = \omega_{\text{c}} / ( 2 \pi ) = \SI{4}{\mega\hertz}$ &
  % 3.b) fractional bandwidth (electromechanical parameter of the instrumentation)
  $B_{h, \text{frac}} = 0.7$ &
  % 3.c) upper frequency bound (pulse-echo parameter)
  $f_{\text{ub}}^{(0)} = \SI{5.4}{\mega\hertz}$ &
  % 3.d) preferred direction of propagation (synthesis parameter)
  \acs{QPW}: $\uvect{\vartheta}^{(0)} = \uvect{2}$ &
  % 3.e) numbers of grid points in the FOV along both coordinate axes (spatial discretization)
  $N_{\text{lat}, 1} = N_{\text{lat}, 2} = \num{512}$\\
 %--------------------------------------------------------------------------------------------------------------
  % 4.a) receiver electromechanical transfer functions (electromechanical parameter of the instrumentation)
  $h_{m, l}^{(\text{rx})} = \SI{1}{ \volt \meter \per \newton }$ &
  % 4.b)
  &
  % 4.c) effective bandwidth of the recorded RF voltage signals (pulse-echo parameter)
  $B_{ u }^{(0)} = \SI{2.8}{\mega\hertz}$ &
  % 4.d) preferred direction of propagation (synthesis parameter)
  $\uvectcomp{ \vartheta }{ 1 }^{(0)} \approx \cos( \SI{77.6}{\degree} )$&
  % 4.e) constant spacings between the adjacent grid points in the FOV along both coordinate axes (spatial discretization)
  $\Delta r_{\text{lat}, 1} = \Delta r_{\text{lat}, 2} = \Delta r_{\text{mat}, 1}$\\
 %--------------------------------------------------------------------------------------------------------------
  % 5.a) cutoff time (electromechanical parameter of the instrumentation)
  $t_{\text{cut}} = 12 \ln( 10 ) / ( \omega_{\text{c}} B_{h, \text{frac}} )$ &
  % 5.b) parameter a in modulated Gaussian pulse (electromechanical parameter of the instrumentation)
  $a = 3 \ln( 10 ) / {t_{\text{cut}}}^{2}$ &
  % 5.c)  
  &
  % 5.d)
  &
  % 5.e) arbitrary offset of the grid points in the FOV (spatial discretization)
  $\vect{r}_{\text{lat}, 0} = \trans{ ( - 511, 1 ) } \Delta r_{\text{lat}, 1} / 2$\\
 %--------------------------------------------------------------------------------------------------------------
  % 6.a) combined transmitter electromechanical transfer functions (electromechanical parameter of the instrumentation)
  \multicolumn{2}{@{}l}{%
    $\tilde{h}_{m}^{(\text{c})}( t ) = e^{ -a ( t - t_{\text{cut}} )^{2} } \cos[ \omega_{\text{c}} ( t - t_{\text{cut}} ) ]$,
    $t \in [ 0; 2 t_{\text{cut}} ]$
  } &
  % 6.c)
  &
  % 6.d)
  &
  % 6.e) total number of grid points in the FOV (spatial discretization)
  $N_{\text{lat}} = \num{262144}$\\
 \addlinespace
  \multicolumn{2}{@{}H}{e) Wire phantom} &
  \multicolumn{2}{H}{f) Tissue-mimicking phantom} &
  \multicolumn{1}{H@{}}{g) Regularization}\\
  \cmidrule(r){1-2}\cmidrule(lr){3-4}\cmidrule(l){5-5}
 \addlinespace
 %--------------------------------------------------------------------------------------------------------------
  % 1.f) number of nonzero components (wire phantom)
  $\tnorm{ \vectsym{\gamma}^{(\kappa)} }{0} = 21$ &
  % 1.g) quantized bounds in the specified common observation time interval for the recorded RF voltage signals (wire phantom)
  $q_{\text{lb}}^{(0)} = \num{1}$, % \SI{50}{\nano\second}
  $q_{\text{ub}}^{(0)} = \num{1648}$ & % \SI{82.4}{\micro\second}
  % 1.h) number of nonzero components (tissue-mimicking phantom)
  $\tnorm{ \vectsym{\theta}^{(\kappa)} }{0} = 10$ &
  % 1.i) quantized bounds in the specified common observation time interval for the recorded RF voltage signals (tissue-mimicking phantom)
  $q_{\text{lb}}^{(0)} = \num{0}$,
  $q_{\text{ub}}^{(0)} = \num{1607}$ & % \SI{80.35}{\micro\second}
  % 1.j) admissible maximum number of iterations in SPGL1 (regularization)
  $N_{\text{iter}} = \num{1000}$\\
 %--------------------------------------------------------------------------------------------------------------
  % 2.f) canonical basis defined the admissible structural building blocks (wire phantom)
  $\mat{\Psi} = \mat{I}$ &
  % 2.g) quantized common observation time / number of quantized real-valued samples per recorded RF voltage signal (wire phantom)
  $T_{ \text{rec} }^{(0)} = 1647 T_{\text{s}}^{(0)}$ &
  % 2.h) discrete Fourier basis defined the admissible structural building blocks (tissue-mimicking phantom)
  $\mat{\Psi} = \mat{\Psi}_{\text{\acs{DFT}}}$ &
  % 2.i) quantized common observation time / number of quantized real-valued samples per recorded RF voltage signal (tissue-mimicking phantom)
  $T_{ \text{rec} }^{(0)} = 1607 T_{\text{s}}^{(0)}$ &
  % 2.j) norm parameters $q \in \{ 0.5; 1 \} were investigated (regularization)
  $q \in \{ 0.5; 1 \}$\\
 %--------------------------------------------------------------------------------------------------------------
  % 3.f) absorption parameters in the complex-valued wavenumber (wire phantom)
  $b = \SI{0.217}{ \deci\bel \mega\hertz\tothe{-2} \per \meter }$ & % $b = \SI{2.17e-3}{ \deci\bel \mega\hertz\tothe{-2} \per \centi\meter }$
  % 3.g) lower and upper bounds defining the set of admissible frequency indices (wire phantom)
  $l_{\text{lb}}^{(0)} = \num{215}$,
  $l_{\text{ub}}^{(0)} = \num{444}$ &
  % 3.h) absorption parameters in the complex-valued wavenumber (tissue-mimicking phantom)
  $b = \SI{0.5}{ \deci\bel \mega\hertz\tothe{-1} \per \centi\meter }$ &
  % 3.i) lower and upper bounds defining the set of admissible frequency indices (tissue-mimicking phantom)
  $l_{\text{lb}}^{(0)} = \num{209}$,
  $l_{\text{ub}}^{(0)} = \num{433}$ &
  % 3.j) number of realizations of the additive errors per admissible reference SNR (regularization)
  $N_{\text{rcn}} = 10$\\  
 %--------------------------------------------------------------------------------------------------------------
  % 4.f) reference angular frequency (wire phantom)
  $\omega_{\text{ref}} = \omega_{\text{c}}$ &
  % 4.g) number of relevant discrete frequencies (wire phantom)
  $N_{f, \text{BP}}^{(0)} = \num{230}$ &
  % 4.h) reference angular frequency (tissue-mimicking phantom)
  $\omega_{\text{ref}} = \omega_{\text{c}}$ &
  % 4.i) number of relevant discrete frequencies (tissue-mimicking phantom)
  $N_{f, \text{BP}}^{(0)} = \num{225}$ &
  % 4.j) normalization parameters generating a sequence of renormalized CS problems in Foucart's algorithm (regularization)
  $\epsilon_{n} = 1 / ( 2 + n )$, $n \in \setconsnonneg{ 4 }$\\
 %--------------------------------------------------------------------------------------------------------------
  % 5.f) reference phase velocity [exponent \zeta = \num{2} prevented dispersion and the phase velocity was constant] (wire phantom)
  $c_{\text{ref}} = \SI{1500}{\meter\per\second}$ &
  % 5.g) total number of observations in the observation process (wire phantom)
  $N_{\text{obs}} = \num{29440}$ &
  % 5.h) reference phase velocity (tissue-mimicking phantom)
  $c_{\text{ref}} = \SI{1540}{\meter\per\second}$ &  
  % 5.i) total number of observations in the observation process (tissue-mimicking phantom)
  $N_{\text{obs}} = \num{28800}$ &
  % 5.j) (regularization)
  \\
 \addlinespace
 \bottomrule
 \end{tabular}
\end{table*}

%%%%%%%%%%%%%%%%%%%%%%%%%%%%%%%%%%%%%%%%%%%%%%%%%%%%%%%%%%%%%%%%%%%%%%%%%%%%%%%%%%%%%%%%%%%%%%%%%%%%%%%%%%%%%%%%
% 1.) instrumentation
%%%%%%%%%%%%%%%%%%%%%%%%%%%%%%%%%%%%%%%%%%%%%%%%%%%%%%%%%%%%%%%%%%%%%%%%%%%%%%%%%%%%%%%%%%%%%%%%%%%%%%%%%%%%%%%%
\subsubsection{Instrumentation}
%\label{subsubsec:sim_study_params_scan_config}
%---------------------------------------------------------------------------------------------------------------
% 1.) geometric and electromechanical parameters of the instrumentation
%---------------------------------------------------------------------------------------------------------------
% a) commercial linear transducer array was emulated in the two-dimensional Euclidean space to reduce the computational costs
A commercial linear transducer array was emulated in
% 1.) two-dimensional Euclidean space
the two-dimensional Euclidean space
(cf. \cref{tab:sim_study_parameters}(a)).
% b) kerfs of width zero simplified the implementation
The kerfs of
width zero simplified
the implementation.
\TODO{unit in table}
The products
% 1.) h_{m, l}^{(\text{c})} = -j \omega_{l} \rho_{0} h_{m, l}^{(\text{tx})}
$h_{m, l}^{(\text{c})} = -j \omega_{l} \rho_{0} h_{m, l}^{(\text{tx})}$ in
% 2.) discretized incident acoustic pressure fields [superpositions of quasi-(d-1)-spherical waves]
the incident acoustic pressure fields
\eqref{eqn:recovery_p_in} corresponded to
% 3.) modulated Gaussian pulse in the time domain
a modulated Gaussian pulse in
the time domain.

%%%%%%%%%%%%%%%%%%%%%%%%%%%%%%%%%%%%%%%%%%%%%%%%%%%%%%%%%%%%%%%%%%%%%%%%%%%%%%%%%%%%%%%%%%%%%%%%%%%%%%%%%%%%%%%%
% 2.) pulse-echo measurement process
%%%%%%%%%%%%%%%%%%%%%%%%%%%%%%%%%%%%%%%%%%%%%%%%%%%%%%%%%%%%%%%%%%%%%%%%%%%%%%%%%%%%%%%%%%%%%%%%%%%%%%%%%%%%%%%%
\subsubsection{Pulse-Echo Measurement Process}
%\label{subsubsec:sim_study_params_measurement_process}
%---------------------------------------------------------------------------------------------------------------
% 1.) pulse-echo measurement process
%---------------------------------------------------------------------------------------------------------------
% a) single pulse-echo measurement was simulated for each type of incident wave
% article:Schiffner2018, Sect. III. Linear Physical Model for the Pulse-Echo Measurement Process / Sect. III.A. Pulse-Echo Measurement Process
% - The \ac{UI} system SEQUENTIALLY PERFORMS $N_{\text{in}} \in \N$ INDEPENDENT PULSE-ECHO MEASUREMENTS using a planar transducer array
%   (cf. \cref{fig:lin_mod_scan_configuration,tab:lin_mod_scan_config_instrum_params}).
A single pulse-echo measurement was simulated for
% 1.) each type of incident wave
each type of
incident wave
(cf. \cref{tab:sim_study_parameters}(b)).
% b) distinct recording time intervals were specified for the RF voltage signals recorded from each object
% article:Schiffner2018, Sect. III. Linear Physical Model for the Pulse-Echo Measurement Process / Sect. III.A. Pulse-Echo Measurement Process
% - Each measurement begins at the time instant $t = 0$ and triggers
%   the CONCURRENT RECORDING OF THE \ac{RF} VOLTAGE SIGNALS $\tilde{u}_{m}^{(\text{rx}, n)}: \setsymbol{T}_{ \text{rec} }^{(n)} \mapsto \R$ GENERATED BY
%   ALL ARRAY ELEMENTS $m \in \setconsnonneg{ N_{\text{el}} - 1 }$ in the SPECIFIED TIME INTERVAL
%   [ \setsymbol{T}_{ \text{rec} }^{(n)} = \bigl[ t_{\text{lb}}^{(n)}; t_{\text{ub}}^{(n)} \bigr], ] (eqn:lin_mod_scan_config_volt_rx_obs_interval) where
%   $t_{\text{lb}}^{(n)} \in \Rnonneg$ and $t_{\text{ub}}^{(n)} > t_{\text{lb}}^{(n)}$ denote
%   its lower and upper bounds, respectively.
% 1.) specified recording time interval for the RF voltage signals generated by all array elements
The object-specific recording time interval
\eqref{eqn:lin_mod_scan_config_volt_rx_obs_interval} was identical for
% 1.) all types of incident waves
all types of
incident waves.
% c) lower and upper frequency bounds were derived from the modulated Gaussian pulse
The lower and
upper frequency bounds were derived from
the modulated Gaussian pulse.
% d) relevant Fourier coefficients were determined at the approximate efficiency of \text{Efficiency}^{(0)} \approx \SI{28}{\percent}
The relevant \name{Fourier} coefficients were determined at
% 1.) approximate efficiency of the regular sampling in combination with the subsequent computation of the DFTs
the approximate efficiency
\eqref{eqn:imp_fourier_coef_efficiency} of
$\text{Efficiency}^{(0)} \approx \SI{28}{\percent}$.

%%%%%%%%%%%%%%%%%%%%%%%%%%%%%%%%%%%%%%%%%%%%%%%%%%%%%%%%%%%%%%%%%%%%%%%%%%%%%%%%%%%%%%%%%%%%%%%%%%%%%%%%%%%%%%%%
% 3.) syntheses of the incident waves
%%%%%%%%%%%%%%%%%%%%%%%%%%%%%%%%%%%%%%%%%%%%%%%%%%%%%%%%%%%%%%%%%%%%%%%%%%%%%%%%%%%%%%%%%%%%%%%%%%%%%%%%%%%%%%%%
\subsubsection{Syntheses of the Incident Waves}
\label{subsubsec:sim_study_params_inc_waves}
%---------------------------------------------------------------------------------------------------------------
% 1.) syntheses of the incident waves
%---------------------------------------------------------------------------------------------------------------
% a) reference voltage signal in the excitation voltages corresponded to a single period of a sinusoid
% article:Schiffner2018, Sect. IV. Syntheses of the Incident Waves / Sect. IV-B Types of Incident Waves (subsec:syn_p_in_types)
% - The generation of the excitation voltages typically applies quantized apodization weights $a_{m}^{(n)} \in \R$ and time delays $\Delta t_{m}^{(n)} \in \Rnonneg$ to
%   the REFERENCE VOLTAGE SIGNALS $u_{l}^{(\text{tx}, n)} \in \C$, whose
%   electric energies are constant for all $n \in \setconsnonneg{ N_{\text{in}} - 1 }$.
% - [...] the generated excitation voltages are
%   [ u_{m, l}^{(\text{tx}, n)} = u_{l}^{(\text{tx}, n)} a_{m}^{(n)} e^{ -j \omega_{l} \mathcal{Q} \left[ \Delta t_{m}^{(n)} \right] } ] (eqn:syn_p_in_types_v_tx_expression)
%   for all $( n, l, m ) \in \setconsnonneg{ N_{\text{in}} - 1 } \times \setsymbol{L}_{ \text{BP} }^{(n)} \times \setconsnonneg{ N_{\text{el}} - 1 }$ [...].
%
% MATLAB:
% excitation = sin( 2 * pi * f_tx * (0:1/f_s:1/f_tx) ) * 0.5e5 * f_s;
% N_samples_A_in_td = numel(excitation) + numel(impulse_response) - 1;
% impulse_response_dft = fft(impulse_response, N_samples_A_in_td);
% A_in_td_dft = impulse_response_dft .* fft(excitation, N_samples_A_in_td);
% A_in_td = ifft(A_in_td_dft) / f_s;
The reference voltage signal in
% 1.) excitation voltages (single pulse-echo measurement, monofrequent, single array element)
the excitation voltages
\eqref{eqn:syn_p_in_types_v_tx} corresponded to
% 2.) single period of a sinusoid
a single period of
a sinusoid, whose amplitude was
% 3.) irrelevant
irrelevant in
% 4.) LTI measurement process
the \ac{LTI} measurement process
(cf. \cref{tab:sim_study_parameters}(c)).
% b) frequency of the clock signal in the quantization operator matched the specifications of a commercial UI system
% article:ChengITUFFC2006: Extended high-frame rate imaging method with limited-diffraction beams
% V. In Vitro and In Vivo Experiments
% - For steered plane wave or conventional delay-and-sum methods,
%   linear time delays are applied to the transducers to steer the beams. (p. 890)
% - The precision of the time delay of the system is 6.25 ns, which is determined by a 160 MHz clock. (p. 890)
The frequency of
% 1.) clock signal
the clock signal in
% 2.) quantization operator providing the admissible time delays
the quantization operator
\eqref{eqn:syn_p_in_types_v_tx_quantization} matched
% 3.) specifications of commercial UI systems
the specifications of
commercial \ac{UI} systems.
% c) steered QPW propagated preferentially along the r_{2}-axis
The steered \ac{QPW} propagated preferentially along
the $r_{2}$-axis.
% d) fixed time period permuting the time delays for a steered QPW
% article:Schiffner2018, Sect. IV-B.3) Superpositions of Randomly-Delayed Quasi-(d-1)-Spherical Waves (subsubsec:syn_p_in_types_rnd_del)
% - In the two-dimensional Euclidean space, i.e. $d = 2$, they simplify to
%   $T_{\text{inc}}^{(n)} = \Delta r_{\text{el}, 1} \tabs{ \uvectcomp{ \vartheta }{ 1 }^{(n)} } / c_{\text{ref}}$ and induce
%   RANDOM PERMUTATIONS OF THE TIME DELAYS SPECIFIED FOR THE STEERED \acp{QPW} in \eqref{eqn:syn_p_in_types_v_tx_qpw}.
%
% => \tabs{ \uvectcomp{ \vartheta }{ 1 }^{(0)} } = T_{\text{inc}}^{(0)} c_{\text{ref}} / \Delta r_{\text{el}, 1}
% => \cos( \vartheta_{0} ) = T_{\text{inc}}^{(0)} c_{\text{ref}} / \Delta r_{\text{el}, 1} [ \cos( \vartheta_{0} ) >= 0 ]
% => \vartheta_{0} = \acos( T_{\text{inc}}^{(0)} c_{\text{ref}} / \Delta r_{\text{el}, 1} )
%
% wire phantom:
% c_{\text{ref}} = \SI{1500}{\meter\per\second}, T_{\text{inc}}^{(0)} = \SI{43.6369}{\nano\second}
% => \vartheta_{0} \approx 77.5992°
% tissue-mimicking phantom:
% c_{\text{ref}} = \SI{1540}{\meter\per\second}, T_{\text{inc}}^{(0)} = \SI{42.5035}{\nano\second}
% => \vartheta_{0} \approx 77.5992°
The fixed time period
\eqref{eqn:syn_p_in_types_v_tx_rnd_del_interval} emerged from
% 1.) preferred direction of propagation
the direction
$\uvect{\vartheta}^{(0)} = \trans{ ( \cos( \vartheta ), \sin( \vartheta ) ) }$.

%%%%%%%%%%%%%%%%%%%%%%%%%%%%%%%%%%%%%%%%%%%%%%%%%%%%%%%%%%%%%%%%%%%%%%%%%%%%%%%%%%%%%%%%%%%%%%%%%%%%%%%%%%%%%%%%
% 4.) spatial discretizations
%%%%%%%%%%%%%%%%%%%%%%%%%%%%%%%%%%%%%%%%%%%%%%%%%%%%%%%%%%%%%%%%%%%%%%%%%%%%%%%%%%%%%%%%%%%%%%%%%%%%%%%%%%%%%%%%
\subsubsection{Spatial Discretizations}
\label{subsubsec:sim_study_params_disc_space}
The constant spacing between
% 1.) adjacent grid points
the adjacent grid points on
% 2.) each vibrating face
each vibrating face along
% 3.) r_{1}-axis
the $r_{1}$-axis ensured
% 4.) approximately four grid points
approximately \num{3.7} points per
% 5.) smallest wavelength
% article:Schiffner2018, Sect. III. Linear Physical Model for the Pulse-Echo Measurement Process / Sect. B. Acoustic Model for Human Soft Tissues
% - Given reference values of
%   the angular frequency $\omega_{\text{ref}} \in \Rplus$ and
%   the associated phase velocity $c_{\text{ref}} \in \Rplus$, the complex-valued wavenumber \cite{article:WatersITUFFC2005,article:SzaboJASA1995}
%   [ \munderbar{k}_{l} = \frac{ \omega_{l} }{ c_{\text{ref}} } + \beta_{\text{E,ref}, l} - j \bar{b} \abs{ \omega_{l} }^{ \zeta } ]
%   [                     \beta_{l} = \omega_{l} / c_{l} ] where
%   the phase term $\beta_{l} \in \R$ sums the real-valued wavenumber $k_{\text{ref}, l} = \omega_{l} / c_{\text{ref}}$ and
%   the excess dispersion term [...] combines power-law absorption with dispersion.
smallest wavelength
(cf. \cref{tab:sim_study_parameters}(d)).
% b) FOV was square shaped and laterally centered in front of the linear transducer array
% independent parameters required for the coordinates of the grid points in the FOV ( d = 2, \delta \in \{ 1, 2 \} ):
% 1.) N_{\text{lat}, 1}, N_{\text{lat}, 2} \in \N
%     [number of grid points in the FOV along the r_{1}- and r_{2}-axes]
% 2.) \Delta r_{\text{lat}, 1}, \Delta r_{\text{lat}, 2} \in \Rplus
%     [constant spacing between the adjacent grid points in the FOV along the r_{1}- and r_{2}-axes]
% 3.) \vect{r}_{\text{lat}, 0} = \trans{ ( r_{\text{lat}, 0, 1}, r_{\text{lat}, 0, 2} ) } \in \R \times \Rplus
%     [arbitrary offset of the grid points in the FOV]
% => \mathcal{L} = \{ \vect{r}_{\text{lat}, i} \in \R^{2}: \vect{r}_{\text{lat}, i} = \vect{r}_{\text{lat}, 0} + \sum_{ \delta = 1 }^{ 2 } i_{\delta} \Delta r_{\text{lat}, \delta} \uvect{\delta}, i_{\delta} \in \setconsnonneg{ N_{\text{lat}, \delta} - 1 }, i = \mathcal{I}\left( \vect{i}, \vect{N}_{\text{lat}} \right) \}
%    [coordinates of the grid points in the FOV]
% => N_{\text{lat}} = \abs{ \mathcal{L} } = \prod_{ \delta = 1 }^{ 2 } N_{\text{lat}, \delta}
%    [total number of grid points in the FOV]
% => \Delta V = \prod_{ \delta = 1 }^{ 2 } \Delta r_{\text{lat}, \delta}
%    [d-dimensional volume element]
The \ac{FOV} was
% 1.) square shaped
square shaped and
% 2.) laterally centered
laterally centered in front of
% 3.) linear transducer array
the linear transducer array.
% c) symmetry enables simple computation of the incident field
The identical spacings between
the adjacent grid points on
each vibrating face and
in the \ac{FOV}, i.e.
$\Delta r_{\text{lat}, 1} = \Delta r_{\text{lat}, 2} = \Delta r_{\text{mat}, 1}$, simplified
the computations of
% 1.) discretized incident acoustic pressure fields [superpositions of quasi-(d-1)-spherical waves]
the incident acoustic pressure fields
\eqref{eqn:recovery_p_in} and
% 2.) implementation of the FMM
the implementation of
the \ac{FMM}.

%%%%%%%%%%%%%%%%%%%%%%%%%%%%%%%%%%%%%%%%%%%%%%%%%%%%%%%%%%%%%%%%%%%%%%%%%%%%%%%%%%%%%%%%%%%%%%%%%%%%%%%%%%%%%%%%
% 5.) wire phantom
%%%%%%%%%%%%%%%%%%%%%%%%%%%%%%%%%%%%%%%%%%%%%%%%%%%%%%%%%%%%%%%%%%%%%%%%%%%%%%%%%%%%%%%%%%%%%%%%%%%%%%%%%%%%%%%%
\subsubsection{Wire Phantom}
\label{subsubsec:sim_study_params_obj_A}
%---------------------------------------------------------------------------------------------------------------
% 1.) wire phantom
%---------------------------------------------------------------------------------------------------------------
% a) wires were represented by 21 identical nonzero components in the vector aggregating the regular samples in the discretized relative spatial fluctuations in the unperturbed compressibility
The wires were represented by
identical nonzero components in
% 1.) vector stacking the regular samples in the discretized relative spatial fluctuations in the unperturbed compressibility
the compressibility fluctuations
\eqref{eqn:recovery_sys_lin_eq_gamma_kappa_bp_vector}
(cf. \cref{tab:sim_study_parameters}(e)).
\TODO{Why does the simulation recover real vectors?!?}
% b) axial distances from the linear transducer array / axial and lateral spacings
Their axial distances from
% 1.) linear transducer array
the linear transducer array ranged from
% 2.) 5 - 37 mm
\SIrange{5}{37}{\milli\meter}, and
% 3.) axial and lateral spacings
their axial and
lateral spacings amounted to
% 4.) 5 mm and 10 mm
approximately \SI{5}{\milli\meter} and
\SI{10}{\milli\meter},
respectively.
% c) canonical basis defined the admissible structural building blocks as individual samples and induced a 21-sparse representation
The canonical basis defined
% 1.) structural building block
the structural building block with
% 2.) index n \in \setcons{ N_{\text{lat}} }
the index
$n \in \setcons{ N_{\text{lat}} }$ as
% 3.) individual sample located at the position \vect{r}_{ \text{lat}, n - 1 }
% index_1 = \ceil{ n / N_{\text{lat}, 2} }
% index_2 = n - ( index_1 - 1 ) * N_{\text{lat}, 2}
% r_{ \text{lat}, n, 1 } = ( index_1 - 513 / 2 ) * \Delta r_{\text{lat}, 1}
% r_{ \text{lat}, n, 2 } = ( index_2 - 0.5 ) * \Delta r_{\text{lat}, 1}
the individual sample located at
the position
$\vect{r}_{ \text{lat}, n - 1 } \in \mathcal{L}$ and induced
% 4.) nearly-sparse representation
a sparse representation
\eqref{eqn:recovery_reg_sparse_representation}.
% d) absorption parameters equaled those of pure water at a temperature of 20 °C
% book:Duck1990, Chapter 4: Acoustic Properties of Tissue at Ultrasonic Frequencies / Sect. 4.1.11: Acoustic velocity through some materials other than tissue
% Sect. 4.1.11.1: Water
% - The acoustic velocity in water is given in Table 4.8, including its temperature dependence. (p. 94)
% - Pure water non-dispersive. (p. 95)
% - Table 4.8: Acoustic velocity and attenuation, and non-linearity parameter B/A for pure water at atmospheric pressure (p. 95)
%   20°C | 1482.3 m/s | 25 * 1e-3 Np / ( m MHz^{2} ) | 4.96 B/A
The absorption parameters in
% 1.) complex-valued wavenumber with respect to k_{\text{ref}}
the wavenumber
\eqref{eqn:lin_mod_mech_model_tis_abs_time_causal_wavenumber_complex_kref} equaled
those of
% 2.) pure water at a temperature of 20 °C
pure water at
a temperature of
$\SI{20}{\celsius}$
\cite[Table 4.8]{book:Duck1990}, where
% 3.) quadratic frequency dependence
the quadratic frequency dependence prevented
% 4.) dispersion
dispersion.
% e) quantized recording time interval and the associated number of relevant discrete frequencies resulted in the number of observations of N_{\text{obs}} / N_{\text{lat}} \approx \SI{11.23}{\percent}
% T_{ \text{rec} }^{(0)} = 1647 T_{\text{s}}^{(0)}
% N_{f, \text{BP}}^{(0)} = 230
% N_{\text{lat}} = 262144
% => N_{\text{obs}} = 29440
% => N_{\text{obs}} / N_{\text{lat}} = 29440 / 262144 \approx 11.2305 %
The quantized recording time interval
\eqref{eqn:lin_mod_scan_config_volt_rx_obs_interval} and
% 1.) number of relevant discrete frequencies (effective time-bandwidth products)
the associated number of
relevant discrete frequencies
\eqref{eqn:recon_disc_axis_f_discrete_BP_TB_product} resulted in
% 2.) number of observations (all pulse-echo measurements, multifrequent, all array elements)
%the number of
%observations
%\eqref{eqn:recovery_sys_lin_eq_num_obs} of
the ratio
% indeterminancy
$N_{\text{obs}} / N_{\text{lat}} \approx \SI{11.23}{\percent}$.

%%%%%%%%%%%%%%%%%%%%%%%%%%%%%%%%%%%%%%%%%%%%%%%%%%%%%%%%%%%%%%%%%%%%%%%%%%%%%%%%%%%%%%%%%%%%%%%%%%%%%%%%%%%%%%%%
% 6.) tissue-mimicking phantom
%%%%%%%%%%%%%%%%%%%%%%%%%%%%%%%%%%%%%%%%%%%%%%%%%%%%%%%%%%%%%%%%%%%%%%%%%%%%%%%%%%%%%%%%%%%%%%%%%%%%%%%%%%%%%%%%
\subsubsection{Tissue-Mimicking Phantom}
\label{subsubsec:sim_study_params_obj_B}
%---------------------------------------------------------------------------------------------------------------
% 1.) tissue-mimicking phantom
%---------------------------------------------------------------------------------------------------------------
% a) discrete Fourier basis defined the structural building blocks as complex exponential functions
The discrete \name{Fourier} basis defined
% 1.) structural building block
the structural building block with
% 2.) index n \in \setcons{ N_{\text{lat}} }
the index
$n \in \setcons{ N_{\text{lat}} }$ as
% 3.) complex exponential function with the normalized discrete lateral and axial frequencies
% index_1 = \ceil{ n / N_{\text{lat}, 2} }
% index_2 = n - ( index_1 - 1 ) * N_{\text{lat}, 2}
% \hat{K}_{ n, 1 } = ( \ceil{ N_{\text{lat}, 1} / 2 } - N_{\text{lat}, 1} + index_1 - 1 ) / N_{\text{lat}, 1} = ( \tceil{ n / 512 } - 257 ) / 512
% \hat{K}_{ n, 2 } = ( index_2 - 1 ) / N_{\text{lat}, 2} = ( n - ( index_1 - 1 ) * N_{\text{lat}, 2} - 1 ) / N_{\text{lat}, 2} = ( n + 511 ) / 512 - \tceil{ n / 512 }
the complex exponential function with
the normalized discrete lateral and
axial frequencies
$\hat{K}_{ n, 1 } = ( \tceil{ n / 512 } - 257 ) / 512$ and
$\hat{K}_{ n, 2 } = ( n + 511 ) / 512 - \tceil{ n / 512 }$,
respectively
(cf. \cref{tab:sim_study_parameters}(f)).
% b) random specification of 10 coefficients with respect to the discrete Fourier basis generated the relative spatial fluctuations in the unperturbed compressibility
% MATLAB:
% theta_kappa_abs_mu = 1e-1;
% theta_kappa = zeros( N_lattice_axis(2), N_lattice_axis(1) );
% theta_kappa( indices_k_kappa ) = theta_kappa_abs_mu .* exp( 2j * pi * rand( 1, N_coefficients_kappa ) );
% gamma_kappa = reshape( psi_fourier( N_lattice_axis, theta_kappa, 2, [] ), [N_lattice_axis(2), N_lattice_axis(1)] );
% gamma_kappa_abs_max = max( abs( gamma_kappa(:) ) );
% theta_kappa = theta_kappa / gamma_kappa_abs_max * 1e-1;
% gamma_kappa = gamma_kappa / gamma_kappa_abs_max * 1e-1;
Nonzero components of
% 1.) identical absolute value
identical absolute value and
% 2.) uniformly distributed phase
uniformly distributed phase in
% 3.) nearly-sparse representation
the sparse representation
\eqref{eqn:recovery_reg_sparse_representation} spawned
% 4.) vector stacking the regular samples in the discretized relative spatial fluctuations in the unperturbed compressibility
dense compressibility fluctuations
\eqref{eqn:recovery_sys_lin_eq_gamma_kappa_bp_vector}.
% c) typical absorption parameters for soft tissues governed the wavenumber
% article:JensenProgBMB2007: Medical ultrasound imaging
% 2. Basic ultrasound
% - Typically, an ATTENUATION OF 0.5 dB/(MHz cm) IS EXPERIENCED IN THE SOFT TISSUES. (p. 154)
% book:Duck1990, Chapter 4: Acoustic Properties of Tissue at Ultrasonic Frequencies / Sect. 4.3: Ultrasonic attenuation: absorption and scatter
% Sect. 4.3.8: Values of acoustic absorption coefficients in tissue
% - Measured values of ABSORPTION COEFFICIENTS FOR ULTRASOUND IN SOFT TISSUE are given in Tables 4.19 and 4.20. (p. 115)
% - Values at particular frequencies are included in Table 4.19, and
%   the POWER-LAW EXPRESSION Equation 4.30 used as the basis for the values given in Table 4.20. (p. 115)
% - Table 4.20: Ultrasound absorption coefficient (ii); \alpha = a f^{b} (p. 117)
Typical absorption parameters for
% 1.) soft tissues
soft tissues
\cite[Table 4.20]{book:Duck1990} governed
% 2.) complex-valued wavenumber with respect to k_{\text{ref}}
the wavenumber
\eqref{eqn:lin_mod_mech_model_tis_abs_time_causal_wavenumber_complex_kref}, where
% 3.) linear frequency dependence
the linear frequency dependence implied
% 4.) anomalous dispersion
anomalous dispersion.
% d) quantized recording time interval and the associated number of relevant discrete frequencies resulted in the ratio N_{\text{obs}} / N_{\text{lat}} \approx \SI{10.99}{\percent}
% T_{ \text{rec} }^{(0)} = 1607 T_{\text{s}}^{(0)}
% N_{f, \text{BP}}^{(0)} = 225
% N_{\text{lat}} = 262144
% => N_{\text{obs}} = 28800
% => N_{\text{obs}} / N_{\text{lat}} = 28800 / 262144 \approx 10.9863 %
The quantized recording time interval
\eqref{eqn:lin_mod_scan_config_volt_rx_obs_interval} and
% 1.) number of relevant discrete frequencies (effective time-bandwidth product)
the associated number of
relevant discrete frequencies
\eqref{eqn:recon_disc_axis_f_discrete_BP_TB_product} resulted in
% 2.) number of observations (all pulse-echo measurements, multifrequent, all array elements)
%the number of
%observations
%\eqref{eqn:recovery_sys_lin_eq_num_obs} of
the ratio
$N_{\text{obs}} / N_{\text{lat}} \approx \SI{10.99}{\percent}$.

%%%%%%%%%%%%%%%%%%%%%%%%%%%%%%%%%%%%%%%%%%%%%%%%%%%%%%%%%%%%%%%%%%%%%%%%%%%%%%%%%%%%%%%%%%%%%%%%%%%%%%%%%%%%%%%%
% 7.) additive errors
%%%%%%%%%%%%%%%%%%%%%%%%%%%%%%%%%%%%%%%%%%%%%%%%%%%%%%%%%%%%%%%%%%%%%%%%%%%%%%%%%%%%%%%%%%%%%%%%%%%%%%%%%%%%%%%%
\subsubsection{Additive Errors}
\label{subsubsec:sim_study_params_obs_errors}
%---------------------------------------------------------------------------------------------------------------
% 1.) additive errors
%---------------------------------------------------------------------------------------------------------------
% a) five variances specified additive errors of distinct energy levels
% article:Schiffner2018, Sect. VI. Implementation / Sect. VI-B. Additive Errors (subsec:imp_obs_errors)
% - ADDITIVE ERRORS, which were statistically modeled as
%   \ac{GWN} (cf. e.g. \cite[110]{book:Manolakis2005}) WITH ZERO MEAN AND THE VARIANCE $\sigma_{\eta}^{2}$, CORRUPTED
%   THE RECORDED SAMPLES OF ALL \ac{RF} VOLTAGE SIGNALS.
The five variances
\begin{equation}
 %--------------------------------------------------------------------------------------------------------------
 % variances of the zero-mean GWN
 %--------------------------------------------------------------------------------------------------------------
  \sigma_{\eta}^{2}
  =
  \frac{
    2
    \norm{ \vect{u}^{(\text{B}, \text{\acs{QPW}})} }{2}^{2}
  }{
    N_{\text{el}}
  }
  10^{ -\frac{ \text{SNR}_{\text{dB}} }{ 10 \si{\deci\bel} } },
 \label{eqn:sim_study_params_obs_errors_variance}
\end{equation}
where
% 1.) energy of the Born approximation induced by the QPW
% time-domain energy: 2 N_{t}^{(0)} \tnorm{ \vect{u}^{(\text{B}, \text{\acs{QPW}})} }{2}^{2}
% time-domain power: 2 \tnorm{ \vect{u}^{(\text{B}, \text{\acs{QPW}})} }{2}^{2} / N_{\text{el}}
$\tnorm{ \vect{u}^{(\text{B}, \text{\acs{QPW}})} }{2}^{2}$ equals
the energy of
% 2.) approximate vector stacking the relevant Fourier coefficients of the recorded RF voltage signals (all pulse-echo measurements, multifrequent, all array elements)
the \name{Born} approximation of
the recorded \ac{RF} voltage signals
\eqref{eqn:recovery_sys_lin_eq_v_rx_born_all_f_all_in_v_rx_born} induced by
% 3.) quasi-plane wave (QPW)
the \ac{QPW}, and
% 4.) reference signal-to-noise ratio (SNR) in \deci\bel
$\text{SNR}_{\text{dB}} \in \{ \SI{3}{\deci\bel}, \SI{6}{\deci\bel}, \SI{10}{\deci\bel}, \SI{20}{\deci\bel}, \SI{30}{\deci\bel} \}$ is
the reference \ac{SNR}, specified
% 5.) additive errors
additive errors of
% 6.) five distinct energy levels
distinct energy levels.

%%%%%%%%%%%%%%%%%%%%%%%%%%%%%%%%%%%%%%%%%%%%%%%%%%%%%%%%%%%%%%%%%%%%%%%%%%%%%%%%%%%%%%%%%%%%%%%%%%%%%%%%%%%%%%%%
% 8.) regularization
%%%%%%%%%%%%%%%%%%%%%%%%%%%%%%%%%%%%%%%%%%%%%%%%%%%%%%%%%%%%%%%%%%%%%%%%%%%%%%%%%%%%%%%%%%%%%%%%%%%%%%%%%%%%%%%%
\subsubsection{Regularization}
\label{subsubsec:sim_study_params_regularization}
%---------------------------------------------------------------------------------------------------------------
% 1.) regularization
%---------------------------------------------------------------------------------------------------------------
% a) approximation of the estimated l2-norm of the normalized additive errors in the normalized CS problem
% 1.) variances of the zero-mean GWN
The variances
\eqref{eqn:sim_study_params_obs_errors_variance},
% 2.) number of relevant discrete frequencies (effective time-bandwidth product)
the effective time-bandwidth product
\eqref{eqn:recon_disc_axis_f_discrete_BP_TB_product}, and
% 3.) quantized recording time
the quantized recording time permitted
the approximation of
% 4.) estimated l2-norm of the normalized additive errors in the normalized CS problem
the estimated $\ell_{2}$-norm of
the normalized additive errors
\eqref{eqn:imp_data_acq_rel_obs_error_est} as
\begin{equation*}
 %--------------------------------------------------------------------------------------------------------------
 % approximation of the estimated l2-norm of the normalized additive errors in the normalized CS problem
 %--------------------------------------------------------------------------------------------------------------
  \hat{ \bar{\eta} }
  \approx
  \left[
    1
    +
    \frac{
      \norm{ \vect{u}^{(\text{B})} }{2}^{2}
      f_{\text{s}}^{(0)}
    }{
      \norm{ \vect{u}^{(\text{B}, \text{\acs{QPW}})} }{2}^{2}
      2 B_{ u }^{(0)}
    }
    10^{ \frac{ \text{SNR}_{\text{dB}} }{ 10 \si{\deci\bel} } }
  \right]^{ - \frac{1}{2} }.
\end{equation*}
% b) empirical threshold factors for the normalization of the sensing matrices
% 1.) QPW
% SNR_cs = [3, 6, 10, 20, 30, inf];		% specify SNR in dB
% norms_cols_thresh = 10.^(-SNR_cs / 20);	% thresholds for normalization according to actual SNR
% 2.) random incident waves
% SNR_cs_act = 10 * log10( data_RF_tgc_cs_power_mean ./ noise_RF_tgc_cs_variance );
% => SNR_cs_act = 10 * log10( norm( data_RF_tgc_cs(:) )^2 ./ norm( data_RF_tgc_qpw(:) )^2 ) + SNR_cs;
% norms_cols_thresh = 10.^(-SNR_cs_act / 20);	% thresholds for normalization according to actual SNR
% => norms_cols_thresh = norm( data_RF_tgc_qpw(:) ) ./ norm( data_RF_tgc_cs(:) ) * 10.^( - SNR_cs / 20 );
\TODO{exception: equal threshold for 3 and 6 dB}
For
each reference \ac{SNR},
% 1.) empirical threshold factors for the normalization of the sensing matrices
the empirical factors
\begin{equation}
 %--------------------------------------------------------------------------------------------------------------
 % empirical threshold factors for the normalization of the sensing matrices
 %--------------------------------------------------------------------------------------------------------------
  \xi
  =
  \frac{
    \norm{ \vect{u}^{(\text{B}, \text{\acs{QPW}})} }{2}
  }{
    \norm{ \vect{u}^{(\text{B})} }{2}
  }
  10^{ - \frac{ \text{SNR}_{\text{dB}} }{ 20 \si{\deci\bel} } }
 \label{eqn:sim_study_params_reg_factor_threshold}
\end{equation}
specified
% 2.) lower bounds on the l2-norms of the sensing matrices' column vectors
the lower bounds on
the $\ell_{2}$-norms of
the column vectors
\eqref{eqn:recovery_reg_norm_l2_norms_lb}.
% c) maximum number of iterations in SPGL1 was N_{\text{iter}}
% article:Schiffner2018, Sect. VI. Implementation / Sect. VI-C. Sparsity-Promoting lq-Minimization Method (subsec:imp_lq_minimization)
% - \acs{SPGL1} is ITERATIVE and left multiplied a sequence of recursively-generated vectors by
%   the potentially densely-populated normalized sensing matrix \eqref{eqn:recon_reg_norm_sensing_matrix} or its adjoint.
The maximum number of
iterations in
\ac{SPGL1} was
$N_{\text{iter}}$
(cf. \cref{tab:sim_study_parameters}(g)).
% d) normalization parameters \epsilon_{n} induced a sequence of five renormalized CS problems in Foucart's algorithm
% article:Schiffner2018, Sect. VI. Implementation / Sect. VI-C. Sparsity-Promoting lq-Minimization Method (subsec:imp_lq_minimization)
% - \name{Foucart}'s algorithm \cite[Sect. 4]{article:FoucartACHA2009} iteratively applied this method based on \ac{SPGL1} to
%   a sequence of RENORMALIZED \ac{CS} PROBLEMS to approximate the nonconvex $\ell_{q}$-minimization method \eqref{eqn:recovery_reg_norm_lq_minimization} for
%   the half-open parameter interval $q \in [ 0; 1 )$.
% - cs_2d_mlfma_options.q = 0.5;
% - cs_2d_mlfma_options.epsilon_n = 1 ./ (1 + (1:5)); [cs_2d_mlfma_options.epsilon_n = 1 ./ (2 + (0:4))]
The normalization parameters
$\epsilon_{n}$ induced
a sequence of
five renormalized \ac{CS} problems in
\name{Foucart}'s algorithm
(cf. \cref{subsec:imp_lq_minimization}).
% e) Foucart's algorithm entailed six l1 minimizations
Since
\ac{SPGL1} provided
the initial guess,
\name{Foucart}'s algorithm entailed
six $\ell_{1}$ minimizations.

%%%%%%%%%%%%%%%%%%%%%%%%%%%%%%%%%%%%%%%%%%%%%%%%%%%%%%%%%%%%%%%%%%%%%%%%%%%%%%%%%%%%%%%%%%%%%%%%%%%%%%%%%%%%%%%%
% 9.) reference sensing matrices
%%%%%%%%%%%%%%%%%%%%%%%%%%%%%%%%%%%%%%%%%%%%%%%%%%%%%%%%%%%%%%%%%%%%%%%%%%%%%%%%%%%%%%%%%%%%%%%%%%%%%%%%%%%%%%%%
\subsubsection{Reference Sensing Matrices}
\label{subsubsec:sim_study_params_ref_sens_mat}
%---------------------------------------------------------------------------------------------------------------
% 1.) reference sensing matrices
%---------------------------------------------------------------------------------------------------------------
% a) two types of sensing matrices served as benchmarks
Two types of
sensing matrices, which emerged from
\ac{GWN}, served as
benchmarks.
% b) first reference observation process and the associated sensing matrix met the RIP with very high probability
% article:Schiffner2018, Sect. II. Compressed Sensing in a Nutshell (sec:compressed_sensing)
% - Fortunately, CERTAIN TYPES OF RANDOM SENSING MATRICES \eqref{eqn:cs_math_prob_general_sensing_matrix} also obey
%   THE \ac{RIP} WITH VERY HIGH PROBABILITY, if
%   THE NUMBER OF OBSERVATIONS IS SUFFICIENTLY LARGE \cite[6]{book:Foucart2013}, \cite{article:TroppPIEEE2010}.
% - REALIZATIONS OF \ac{IID} RANDOM VARIABLES governed by certain distributions, e.g.
%   GAUSSIAN or \name{Bernoulli}, as entries \cite[Thm. 5.2]{article:BaraniukCA2008} and
%   randomly and uniformly chosen scaled rows of a \name{Fourier} basis \cite[Thm. 3.3]{article:RudelsonCPAM2008}, for example, require
%   $M \in \bigomega{ s \ln( N / s ) }$ and
%   $M \in \bigomega{ s \ln^{4}( N ) }$ observations, respectively.
% article:BaraniukCA2008: A Simple Proof of the Restricted Isometry Property for Random Matrices (real-valued CS problem / canonical basis)
% 6 Discussion
% - Furthermore, we prove above that the RIP HOLDS for \Phi(ω) WITH HIGH PROBABILITY when
%   the matrix is drawn according to one of the distributions
%   [ \phi_{ i, j } \sim \gaussian{ 0 }{ 1 / n } ] (4.4),
%   [ \phi_{ i, j } := + 1 / \sqrt{n} with probability 0.5; - 1 / \sqrt{n} with probability 0.5 ] (4.5), or
%   [ \phi_{ i, j } := + \sqrt{ 3 / n } with probability 1/6; 0 with probability 2/3; - \sqrt{ 3 / n } with probability 1/6 ] (4.6) []. (p. 261)
For
a sufficiently large
% 1.) number of observations (all pulse-echo measurements, multifrequent, all array elements)
number of
observations
\eqref{eqn:recovery_sys_lin_eq_num_obs}, both
% 2.) first reference observation process (RIP)
the real-valued random
$N_{\text{obs}} \times N_{\text{lat}}$ observation process
\begin{subequations}
\begin{align}
 %--------------------------------------------------------------------------------------------------------------
 % first reference observation process and its entries (RIP)
 %--------------------------------------------------------------------------------------------------------------
  \mat{\Phi}^{(\text{\acs{RIP}})}
  =
  \vertcat_{ m = 1 }^{ N_{\text{obs}} }
    \horzcat_{ i = 1 }^{ N_{\text{lat}} }
      \phi_{ m, i }^{(\text{\acs{RIP}})},
  & &
  \phi_{ m, i }^{(\text{\acs{RIP}})}
  \underset{ \text{\acs{IID}} }{ \sim }
  \dgaussian{ 0 }{ \frac{ 1 }{ N_{\text{obs}} } }{1}
 \label{eqn:sim_study_params_ref_obs_proc_rip}
\end{align}
and
% 3.) first reference sensing matrix (RIP)
the associated complex-valued
$N_{\text{obs}} \times N_{\text{lat}}$ sensing matrix
\begin{equation}
 %--------------------------------------------------------------------------------------------------------------
 % first reference sensing matrix (RIP)
 %--------------------------------------------------------------------------------------------------------------
  \mat{A}^{(\text{\acs{RIP}})}
  =
  \mat{\Phi}^{(\text{\acs{RIP}})}
  \mat{\Psi}
 \label{eqn:sim_study_params_ref_sens_mat_rip}
\end{equation}
\end{subequations}
met
% 4.) restricted isometry property (RIP)
the \ac{RIP} with
very high probability
(cf. \cref{sec:compressed_sensing}).
% c) specified variance ensured recorded electric energies of unity expectation
The specified variance ensured
% 1.) recorded electric energies in the pulse echoes (all pulse-echo measurements, multifrequent, all array elements)
recorded electric energies
\eqref{eqn:recovery_reg_v_rx_born_trans_coef_energy} of
% 2.) unity expectation
unity expectation.
% d) replacement of the incident acoustic pressure field in the observation process by complex-valued GWN additionally formed the observation process
The replacement of
% 1.) discretized incident acoustic pressure fields [superpositions of quasi-(d-1)-spherical waves]
the incident acoustic pressure field
\eqref{eqn:recovery_p_in} in
% 2.) observation process (all pulse-echo measurements, multifrequent, all array elements)
the observation process
\eqref{eqn:recovery_sys_lin_eq_v_rx_born_all_f_all_in_mat} by
% 3.) complex-valued GWN (realizations of i.i.d. complex-valued Gaussian random variables)
% MATLAB:
% p_incident_theta_act{index_f} = (randn( N_lattice_axis(2), N_lattice_axis(1) ) + 1j * randn( N_lattice_axis(2), N_lattice_axis(1)) ) * 1e-5;
%$\treal{ p_{l}^{(\text{in}, 0)}( \vect{r}_{\text{lat}, i} ) } \sim \gaussian{ 0 }{ 1 }$
%$\timag{ p_{l}^{(\text{in}, 0)}( \vect{r}_{\text{lat}, i} ) } \sim \gaussian{ 0 }{ 1 }$
complex-valued \ac{GWN} additionally formed
% 4.) second reference observation process (GWN)
the complex-valued structured
$N_{\text{obs}} \times N_{\text{lat}}$ observation process
\TODO{complex GWN}
\begin{subequations}
\begin{align}
 %--------------------------------------------------------------------------------------------------------------
 % second reference observation process and its entries (GWN)
 %--------------------------------------------------------------------------------------------------------------
  \mat{\Phi}^{(\text{\acs{GWN}})}
  =
  \mat{\Phi}\bigl[ p^{(\text{in})} \bigr],
  & &
  p_{l}^{(\text{in}, 0)}( \vect{r}_{\text{lat}, i} )
  \underset{ \text{\acs{IID}} }{ \sim }
  \gaussian{ 0 }{ 1 }
 \label{eqn:sim_study_params_ref_obs_proc_gwn}
\end{align}
%for
% 1.) all admissible frequency indices and all grid points
%$( l, i ) \in \setsymbol{L}_{ \text{BP} }^{(0)} \times \setconsnonneg{ N_{\text{lat}} - 1 }$.
and
% 5.) second reference sensing matrix (GWN)
the associated complex-valued
$N_{\text{obs}} \times N_{\text{lat}}$ sensing matrix
\begin{equation}
 %--------------------------------------------------------------------------------------------------------------
 % second reference sensing matrix (GWN)
 %--------------------------------------------------------------------------------------------------------------
  \mat{A}^{(\text{\acs{GWN}})}
  =
  \mat{\Phi}^{(\text{\acs{GWN}})}
  \mat{\Psi}.
 \label{eqn:sim_study_params_ref_sens_mat_gwn}
\end{equation}
\end{subequations}
% e) complex-valued GWN violated the Helmholtz equations
Although
% 1.) complex-valued GWN
the complex-valued \ac{GWN} violated
% 2.) Helmholtz equations for the incident acoustic pressure fields
the \name{Helmholtz} equations
\eqref{eqn:lin_mod_sol_wave_eq_pde_p_in},
% f) replacement correctly respected the monopole scattering and the reception by the instrumentation
this replacement correctly respected
% 2.) monopole scattering
the monopole scattering and
% 3.) reception by the instrumentation
the reception by
the instrumentation.

%%%%%%%%%%%%%%%%%%%%%%%%%%%%%%%%%%%%%%%%%%%%%%%%%%%%%%%%%%%%%%%%%%%%%%%%%%%%%%%%%%%%%%%%%%%%%%%%%%%%%%%%%%%%%%%%
% 2.) methods
%%%%%%%%%%%%%%%%%%%%%%%%%%%%%%%%%%%%%%%%%%%%%%%%%%%%%%%%%%%%%%%%%%%%%%%%%%%%%%%%%%%%%%%%%%%%%%%%%%%%%%%%%%%%%%%%
\subsection{Methods}
%\label{subsec:sim_study_methods}
%%%%%%%%%%%%%%%%%%%%%%%%%%%%%%%%%%%%%%%%%%%%%%%%%%%%%%%%%%%%%%%%%%%%%%%%%%%%%%%%%%%%%%%%%%%%%%%%%%%%%%%%%%%%%%%%
% 1.) incident acoustic pressure fields
%%%%%%%%%%%%%%%%%%%%%%%%%%%%%%%%%%%%%%%%%%%%%%%%%%%%%%%%%%%%%%%%%%%%%%%%%%%%%%%%%%%%%%%%%%%%%%%%%%%%%%%%%%%%%%%%
\subsubsection{Incident Acoustic Pressure Fields}
%\label{subsubsec:sim_study_methods_p_in}
%---------------------------------------------------------------------------------------------------------------
% 1.) incident acoustic pressure fields
%---------------------------------------------------------------------------------------------------------------
% a) acoustic pressure fields were computed for all types of incident waves
% 1.) discretized incident acoustic pressure fields [superpositions of quasi-(d-1)-spherical waves]
The acoustic pressure fields
\eqref{eqn:recovery_p_in} were computed for
% 2.) all types of incident waves
all types of
incident waves.
% b) spatial and spectral dependencies were analyzed for the discrete frequency closest to the center frequency and three closely spaced positions next to the r_{2}-axis
For
% 1.) wire phantom
the wire phantom,
% 2.) spatial and spectral dependencies
their spatial and
spectral dependencies were analyzed for
% 3.) discrete frequency closest to the center frequency
% f_{ l_{\text{c}} } = \SI{3.9951}{\mega\hertz} \approx f_{\text{c}}, with the index $l_{\text{c}} = 329$
the discrete frequency closest to
the center frequency and
% 4.) three closely spaced positions next to the r_{2}-axis
% \Delta r_{\text{lat}, 1} = \Delta r_{\text{lat}, 2} = \Delta r_{\text{mat}, 1} = \SI{76.2}{\micro\meter}
% indices_pos_ref_x = [256, 256, 256];
% indices_pos_ref_z = [301, 316, 331];
% pos_ref_x = -38.1 um / -38.1 um / -38.1 um
% pos_ref_z = 22.8981 mm / 24.0411 mm / 25.1841 mm
%$\vect{r}_{\text{ref}, i} = \trans{ ( - \Delta r_{\text{lat}, 1} / 2, r_{\text{ref}, i, 2} ) }$,
%$i \in \setcons{ 3 }$, with
%$r_{\text{ref}, 1, 2} = 300.5 \Delta r_{1} \approx \SI{22.9}{\milli\meter}$,
%$r_{\text{ref}, 2, 2} = 315.5 \Delta r_{1} \approx \SI{24}{\milli\meter}$, and
%$r_{\text{ref}, 3, 2} = 330.5 \Delta r_{1} \approx \SI{25.2}{\milli\meter}$.
three closely spaced positions next to
the $r_{2}$-axis,
respectively.
% c) least-squares fit of an affine linear model was subtracted from all unwrapped phases to emphasize the differences
The least-squares fit of
% 1.) affine linear model
an affine linear model to
% 2.) unwrapped phase
the unwrapped phase of
% 3.) incident acoustic pressure field associated with the QPW
the acoustic pressure field
\eqref{eqn:recovery_p_in} associated with
the \ac{QPW} at
% 4.) first position $\vect{r}_{\text{ref}, 1}$
the first position was subtracted from
% 5.) all unwrapped phases
all unwrapped phases to emphasize
% 6.) differences
the differences.

%%%%%%%%%%%%%%%%%%%%%%%%%%%%%%%%%%%%%%%%%%%%%%%%%%%%%%%%%%%%%%%%%%%%%%%%%%%%%%%%%%%%%%%%%%%%%%%%%%%%%%%%%%%%%%%%
% 2.) recorded radio frequency voltage signals
%%%%%%%%%%%%%%%%%%%%%%%%%%%%%%%%%%%%%%%%%%%%%%%%%%%%%%%%%%%%%%%%%%%%%%%%%%%%%%%%%%%%%%%%%%%%%%%%%%%%%%%%%%%%%%%%
\subsubsection{Recorded Radio Frequency Voltage Signals}
%\label{subsubsec:sim_study_methods_v_rx}
%---------------------------------------------------------------------------------------------------------------
% 1.) recorded radio frequency voltage signals
%---------------------------------------------------------------------------------------------------------------
% a) ten realizations of the recorded RF voltage signals were derived from their Born approximation for each type of incident wave and each SNR
% d) number of recovery experiments
 %The number of
 %recovery experiments conducted for
 %each reference \ac{SNR} and
 %each type of
 %incident wave amounted to
 %$N_{\text{rcn}} = 10$.
Ten realizations of
%($N_{\text{rcn}} = 10$) of
% 1.) vectors stacking the relevant Fourier coefficients of the recorded RF voltage signals (all pulse-echo measurements, multifrequent, all array elements)
the recorded \ac{RF} voltage signals
\eqref{eqn:recovery_sys_lin_eq_v_rx_born_all_f_all_in_v_rx} were derived from
% 2.) approximate vectors stacking the relevant Fourier coefficients of the recorded RF voltage signals (all pulse-echo measurements, multifrequent, all array elements)
their \name{Born} approximation
\eqref{eqn:recovery_sys_lin_eq_v_rx_born_all_f_all_in_v_rx_born} for
% 3.) each type of incident wave
each type of
incident wave and
% 4.) each SNR
each \ac{SNR} by inserting
% 5.) nearly-sparse representation / vector of transform coefficients
the sparse representation
\eqref{eqn:recovery_reg_sparse_representation} and
% 6.) additive errors of the specified energy levels
the additive errors into
% 7.) linear algebraic system (all pulse-echo measurements, multifrequent, all array elements, additive errors)
the linear algebraic system
\eqref{eqn:recovery_reg_prob_general_obs_trans_coef_error}.

%%%%%%%%%%%%%%%%%%%%%%%%%%%%%%%%%%%%%%%%%%%%%%%%%%%%%%%%%%%%%%%%%%%%%%%%%%%%%%%%%%%%%%%%%%%%%%%%%%%%%%%%%%%%%%%%
% 3.) recorded electric energies
%%%%%%%%%%%%%%%%%%%%%%%%%%%%%%%%%%%%%%%%%%%%%%%%%%%%%%%%%%%%%%%%%%%%%%%%%%%%%%%%%%%%%%%%%%%%%%%%%%%%%%%%%%%%%%%%
\subsubsection{Recorded Electric Energies}
%\label{subsubsec:sim_study_methods_E_rx}
%---------------------------------------------------------------------------------------------------------------
% 1.) recorded electric energies
%---------------------------------------------------------------------------------------------------------------
% a) recorded electric energies were computed for all sensing matrices except for the first reference
% 1.) recorded electric energies in the pulse echoes (all pulse-echo measurements, multifrequent, all array elements)
The recorded electric energies
\eqref{eqn:recovery_reg_v_rx_born_trans_coef_energy} were computed for
% 2.) all sensing matrices (all pulse-echo measurements, multifrequent, all array elements)
all sensing matrices
\eqref{eqn:recovery_reg_sensing_matrix} except for
% 3.) first reference sensing matrix (RIP)
the first reference
\eqref{eqn:sim_study_params_ref_sens_mat_rip}, which approximately induced
% 4.) expected energies of unity
the expected energies of
unity.
% b) visual inspections revealed the transfer behaviors of the sensing matrices for the tissue-mimicking phantom
% TODO: high dynamic ranges
Their visual inspections revealed
% 1.) transfer behaviors of the associated sensing matrices
the transfer behaviors of
% 2.) sensing matrices (all pulse-echo measurements, multifrequent, all transducer elements)
the sensing matrices
\eqref{eqn:recovery_reg_sensing_matrix} for
% 3.) tissue-mimicking phantom
the tissue-mimicking phantom.

%%%%%%%%%%%%%%%%%%%%%%%%%%%%%%%%%%%%%%%%%%%%%%%%%%%%%%%%%%%%%%%%%%%%%%%%%%%%%%%%%%%%%%%%%%%%%%%%%%%%%%%%%%%%%%%%
% 4.) transform point spread functions
%%%%%%%%%%%%%%%%%%%%%%%%%%%%%%%%%%%%%%%%%%%%%%%%%%%%%%%%%%%%%%%%%%%%%%%%%%%%%%%%%%%%%%%%%%%%%%%%%%%%%%%%%%%%%%%%
\subsubsection{Transform Point Spread Functions}
%\label{subsubsec:sim_study_methods_tpsf}
%---------------------------------------------------------------------------------------------------------------
% 1.) transform point spread functions
%---------------------------------------------------------------------------------------------------------------
% a) TPSFs associated with all sensing matrices were evaluated for all ( n_{1}, n_{2} ) \in \setcons{ N_{\text{lat}} } \times \setsymbol{I}
% article:Schiffner2018, Sect. II. Compressed Sensing in a Nutshell (sec:compressed_sensing)
% - It [TPSF] equals the mutual correlation coefficient of the column vectors given by \cite[(23)]{article:ProvostITMI2009}, \cite[(2)]{article:LustigMRM2007}
%   [ \tpsf{ \mat{A} }{ n_{1} }{ n_{2} } = \frac{ \inprod{ \vect{a}_{ n_{1} } }{ \vect{a}_{ n_{2} } } }{ \norm{ \vect{a}_{ n_{1} } }{2} \norm{ \vect{a}_{ n_{2} } }{2} } ] (eqn:cs_math_tpsf) for
%   all $( n_{1}, n_{2} ) \in \setcons{ N }^{2}$.
% - Owing to the HIGH DIMENSIONALITY OF THE SENSING MATRICES \eqref{eqn:cs_math_prob_general_sensing_matrix},
%   PRACTICAL EVALUATIONS OF THE \ac{TPSF} \eqref{eqn:cs_math_tpsf} USUALLY FIX
%   THE SECOND INDEX according to
%   the expected support of the nearly-sparse representation \eqref{eqn:def_transform_coefficients}, i.e. $n_{2} \in \supp( \vectsym{\theta} )$
%   (cf. e.g. \cite[Fig. 1]{article:ProvostITMI2009}, \cite[Figs. 4 and 5]{article:LustigMRM2007}).
The \acp{TPSF}
\eqref{eqn:cs_math_tpsf} associated with
% 1.) all sensing matrices [sensing matrices \eqref{eqn:recovery_reg_sensing_matrix} induced by all incident waves / both reference sensing matrices]
all sensing matrices were evaluated for
% 2.) all pairs of indices
all $( n_{1}, n_{2} ) \in \setcons{ N_{\text{lat}} } \times \setsymbol{I}$, where
% 3.) nine indices n_{2} \in \setcons{ N_{\text{lat}} }
$\setsymbol{I} \subset \setcons{ N_{\text{lat}} }$ fixed
nine indices.
% b) positions \vect{r}_{ \text{lat}, n_{2} - 1 } were approximately uniformly distributed along the diagonal from ( -17.5, 2 ) mm to ( 17.5, 37 ) mm
% MATLAB:
% N_coordinates = 9;
% direction = N_lattice_axis_cs' - ones(2,1);
% tpsf_coordinates = round( ones(N_coordinates, 2) + linspace(0.05, 0.95, N_coordinates)' * direction' );
% cs_2d_mlfma_options.tpsf_indices = [tpsf_coordinates(:,1) - 1, tpsf_coordinates(:,2)] * [N_lattice_axis_cs(2); 1];
%
% tpsf_coordinates_1 = \round{ 1 + ( 0.05 + ( s - 1 ) * 0.9 / 8 ) * 511 }
% tpsf_coordinates_2 = \round{ 1 + ( 0.05 + ( s - 1 ) * 0.9 / 8 ) * 511 }
% tpsf_indices = ( tpsf_coordinates_1 - 1 ) * 512 + tpsf_coordinates_2
%	       = \round{ ( 0.05 + ( s - 1 ) * 0.9 / 8 ) * 511 } * 513 + 1
%	       = 513 \tround{ 25.55 + ( s - 1 ) 57.4875 } + 1
% index_1 = \ceil{ n / N_{\text{lat}, 2} }
% index_2 = n - ( index_1 - 1 ) * N_{\text{lat}, 2}
% r_{ \text{lat}, n - 1, 1 } = ( index_1 - 513 / 2 ) \Delta r_{\text{lat}, 1} = ( \ceil{ n / 512 } - 513 / 2 ) \Delta r_{\text{lat}, 1}
% r_{ \text{lat}, n - 1, 2 } = ( index_2 - 1 / 2 ) \Delta r_{\text{lat}, 2} = ( n - 512 \ceil{ n / 512 } + 511.5 ) \Delta r_{\text{lat}, 2}
% \Delta r_{\text{lat}, 1} = \Delta r_{\text{lat}, 2} = \SI{76.2}{\micro\meter}
For
% 1.) wire phantom
the wire phantom,
% 2.) positions of the individual compressibility fluctuations
the positions
$\vect{r}_{ \text{lat}, n_{2} - 1 }$ were
% 3.) uniformly distributed
approximately uniformly distributed along
% 4.) diagonal from (-17.4879 mm, 2.0193 mm ) to ( 17.4879 mm, 36.9951 mm )
the diagonal from
$\trans{ ( \SI{-17.5}{\milli\meter}, \SI{2}{\milli\meter} ) }$ to
$\trans{ ( \SI{17.5}{\milli\meter}, \SI{37}{\milli\meter} ) }$ and numbered from
\numrange{1}{9} with
increasing axial coordinate. %, i.e.
%$\setsymbol{I} = \{ n_{2} \in \setcons{ N_{\text{lat}} }: n_{2} = 513 \tround{ 25.55 + ( s - 1 ) 57.4875 } + 1, s \in \setcons{ 9 } \}$.
% c) normalized spatial frequencies \hat{\vect{K}}_{ n_{2} } were approximately uniformly distributed along the semicircle with the center \hat{\vect{K}}_{ \text{c} } and the radius \hat{K}_{ \text{r} }
% MATLAB:
% N_tpsf = 9;
% indices_center = [257, 101];
% indices_radius = 101;
% M_indices = (N_tpsf - 1) / 2;
% indices_phi = (-M_indices:M_indices) * pi / N_tpsf + pi / 2;
% tpsf_indices = round( repmat( indices_center, [N_tpsf, 1] ) + indices_radius * [ cos( indices_phi(:) ), sin( indices_phi(:) ) ] );
%
% \hat{K}_{1} = ( \ceil{ N_{\text{lat}, 1} / 2 } - N_{\text{lat}, 1} + index_1 - 1 ) / N_{\text{lat}, 1} = ( index_1 - 257 ) / 512
% \hat{K}_{2} = ( index_2 - 1 ) / N_{\text{lat}, 2} = ( index_2 - 1 ) / 512
%
% indices_phi = ( s - 0.5 ) \pi / 9 (CHECKED!)
% tpsf_indices_1 = 257 + \tround{ 101 \cos[ ( s - 0.5 ) \pi / 9 ] } (CHECKED!)
% tpsf_indices_2 = 101 + \tround{ 101 \sin[ ( s - 0.5 ) \pi / 9 ] } (CHECKED!)
% tpsf_indices = ( tpsf_indices_1 - 1 ) * 512 + tpsf_indices_2
%	       = 131173 + 512 \tround{ 101 \cos[ ( s - 0.5 ) \pi / 9 ] } + \tround{ 101 \sin[ ( s - 0.5 ) \pi / 9 ] } (CHECKED!)
For
% 1.) tissue-mimicking phantom
the tissue-mimicking phantom,
% 2.) normalized spatial frequencies of the complex exponential functions
the normalized spatial frequencies
$\hat{\vect{K}}_{ n_{2} } = \trans{ ( \hat{K}_{ n_{2}, 1 }, \hat{K}_{ n_{2}, 2 } ) }$ were
% 3.) uniformly distributed
approximately uniformly distributed along
% 4.) semicircle
the semicircle with
% 5.) center \hat{\vect{K}}_{ \text{c} } = \trans{ ( 0, 25 ) } / 128
the center
$\hat{\vect{K}}_{ \text{c} } = \trans{ ( 0, 25 ) } / 128$ and
% 6.) radius \hat{K}_{ \text{r} } = 101 / 512
the radius
$\hat{K}_{ \text{r} } = 101 / 512$ and numbered from
\numrange{1}{9} with
increasing polar angle. %, i.e.
%$\setsymbol{I} = \{ n_{2} \in \setcons{ N_{\text{lat}} }: n_{2} = 131173 + 512 \tround{ 101 \cos[ ( s - 0.5 ) \pi / 9 ] } + \tround{ 101 \sin[ ( s - 0.5 ) \pi / 9 ] }, s \in \setcons{ 9 } \}$.
% d) thresholded l2-norms of the column vectors substituted the original l2-norms in the denominators of the TPSFs to avoid numerical inaccuracies
% TODO: does the replacement affect the reference sensing matrices? -> probably not
The thresholded $\ell_{2}$-norms of
the column vectors
\eqref{eqn:recovery_reg_norm_l2_norms_thresholded}, however, substituted
% 1.) original l2-norms
the original $\ell_{2}$-norms in
the denominators of
% 2.) transform point spread functions (TPSFs)
the \acp{TPSF}
\eqref{eqn:cs_math_tpsf} for
% 3.) tissue-mimicking phantom
this phantom to avoid
% 4.) numerical inaccuracies
the numerical inaccuracies caused by
% 5.) high dynamic ranges
their high dynamic ranges.
% e) empirical threshold factors for the reference SNR of \text{SNR}_{\text{dB}} = \SI{10}{\deci\bel} specified their lower bounds
The empirical factors
\eqref{eqn:sim_study_params_reg_factor_threshold} with
% 1.) reference SNR
$\text{SNR}_{\text{dB}} = \SI{10}{\deci\bel}$ specified
% 2.) lower bounds on the l2-norms of the sensing matrices' column vectors
their lower bounds
\eqref{eqn:recovery_reg_norm_l2_norms_lb}.
% f) each computed TPSF was characterized by its FEHM for each index and its empirical CDF
In addition to
a visual inspection,
% 1.) transform point spread function (TPSF)
each computed \ac{TPSF}
\eqref{eqn:cs_math_tpsf} was characterized by
% 2.) full extent at half maximum (FEHM)
its \ac{FEHM} for
% 3.) each index n_{2} \in \setsymbol{I}
each index
$n_{2} \in \setsymbol{I}$ and
% 4.) empirical cumulative distribution function (CDF)
its empirical \ac{CDF}.
% g) empirical CDF excluded all n_{1} = n_{2} and stated the percentages of diverse pulse echoes whose correlation coefficient did not exceed a specified threshold
The latter excluded
% 1.) all identical pairs of indices
all $n_{1} = n_{2}$ and, thus, stated
% 2.) percentages of diverse pulse echoes
the percentages of
diverse pulse echoes whose
% 3.) correlation coefficient
correlation coefficient did not exceed
% 4.) specified threshold
a specified threshold.

%%%%%%%%%%%%%%%%%%%%%%%%%%%%%%%%%%%%%%%%%%%%%%%%%%%%%%%%%%%%%%%%%%%%%%%%%%%%%%%%%%%%%%%%%%%%%%%%%%%%%%%%%%%%%%%%
% 5.) adjoint normalized sensing matrices
%%%%%%%%%%%%%%%%%%%%%%%%%%%%%%%%%%%%%%%%%%%%%%%%%%%%%%%%%%%%%%%%%%%%%%%%%%%%%%%%%%%%%%%%%%%%%%%%%%%%%%%%%%%%%%%%
\subsubsection{Adjoint Normalized Sensing Matrices}
%\label{subsubsec:sim_study_methods_adjoint}
%---------------------------------------------------------------------------------------------------------------
% 1.) adjoint normalized sensing matrices
%---------------------------------------------------------------------------------------------------------------
% a) normalized recorded RF voltage signals generated by all types of incident waves were left multiplied by the associated adjoint normalized sensing matrices
% 1.) normalized linear algebraic system (all pulse-echo measurements, multifrequent, all array elements, additive errors)
The normalized recorded \ac{RF} voltage signals
\eqref{eqn:recovery_reg_norm_obs_trans_coef_error} generated by
% 2.) all types of incident waves
all types of
incident waves were
left multiplied by
% 3.) adjoint normalized sensing matrices (all pulse-echo measurements, multifrequent, all array elements)
the adjoint normalized sensing matrices
\eqref{eqn:recon_reg_norm_sensing_matrix}.
% b) visual inspections of these products revealed the interference effects affecting the recovery
% article:Schiffner2018, Sect. VI. Implementation / Sect. VI-C Sparsity-Promoting lq-Minimization Method
% - \acs{SPGL1} is iterative and left multiplied
%   a sequence of recursively-generated vectors by
%   the potentially densely-populated normalized sensing matrix \eqref{eqn:recon_reg_norm_sensing_matrix} or its adjoint.
The visual inspections of
these products, which underlay
the implementation of
% 1.) sparsity-promoting lq-minimization method
the sparsity-promoting $\ell_{q}$-minimization method
\eqref{eqn:recovery_reg_norm_lq_minimization}, revealed
% 2.) interference effects
important interference effects affecting
the recovery.

%%%%%%%%%%%%%%%%%%%%%%%%%%%%%%%%%%%%%%%%%%%%%%%%%%%%%%%%%%%%%%%%%%%%%%%%%%%%%%%%%%%%%%%%%%%%%%%%%%%%%%%%%%%%%%%%
% 6.) recovery by lq-minimization
%%%%%%%%%%%%%%%%%%%%%%%%%%%%%%%%%%%%%%%%%%%%%%%%%%%%%%%%%%%%%%%%%%%%%%%%%%%%%%%%%%%%%%%%%%%%%%%%%%%%%%%%%%%%%%%%
\subsubsection{Recovery by $\ell_{q}$-Minimization}
%\label{subsubsec:sim_study_methods_lq_minimization}
%---------------------------------------------------------------------------------------------------------------
% 1.) recovery by lq-minimization
%---------------------------------------------------------------------------------------------------------------
% a) forty instances of the normalized CS problem were solved by the sparsity-promoting lq-minimization method
% 10 realizations per SNR per type of wave => 10 * 5 * 4
The $\num{200}$ instances of
% 1.) CS problem associated with the normalized linear algebraic system
the normalized \ac{CS} problem
\eqref{eqn:recovery_reg_norm_prob_general} generated by
% 2.) all types of incident waves
all types of
incident waves and
% 3.) realizations of the vectors stacking the relevant Fourier coefficients of the recorded RF voltage signals (all pulse-echo measurements, multifrequent, all array elements)
all realizations of
the recorded \ac{RF} voltage signals
\eqref{eqn:recovery_sys_lin_eq_v_rx_born_all_f_all_in_v_rx} were solved by
% 4.) sparsity-promoting lq-minimization method
the sparsity-promoting $\ell_{q}$-minimization method
\eqref{eqn:recovery_reg_norm_lq_minimization}.
% b) structural differences were quantified by the mean SSIM indices
% article:WangISPM2009: Mean squared error: Love it or leave it? A new look at Signal Fidelity Measures
% - These local similarities are expressed using simple, easily computed statistics, and combined together to form local SSIM [7]
%   (2) [local SSIM index], where
%   µ_{x} and µ_{y} are (respectively) the local sample means of x and y,
%   σ_{x} and σ_{y} are (respectively) the local sample standard deviations of x and y, and
%   σ_{xy} is the sample cross correlation of x and y after removing their means. (pp. 105, 106)
% article:WangITIP2004: Image quality assessment: From error visibility to structural similarity
% - This results in a specific form of the SSIM index (13). (p. 605)
\TODO{visual inspection}
In addition to
a visual inspection,
structural differences between
% 1.) estimated vectors stacking the regular samples in the discretized relative spatial fluctuations in the unperturbed compressibility
the recovered compressibility fluctuations
\eqref{eqn:recovery_reg_norm_lq_minimization_sol_mat_params} and
% 2.) specified vectors stacking the regular samples in the discretized relative spatial fluctuations in the unperturbed compressibility
their specified version
\eqref{eqn:recovery_sys_lin_eq_gamma_kappa_bp_vector} were quantified by
% 3.) mean SSIM indices
the mean \ac{SSIM} indices
\cite[(2)]{article:WangISPM2009}, whereas
% c) quantitative differences were measured by the relative RMSEs
quantitative differences were measured by
% 1.) relative RMSEs
the relative \acp{RMSE}.
% d) sparsities and speed of convergence were gauged by the numbers of components within the illustrated dynamic range and the numbers of iterations in SPGL1
The sparsity and
% 1.) speed of convergence
speed of
convergence were gauged by
% 2.) numbers of components within the illustrated dynamic range
the numbers of
components within
the illustrated dynamic range and
% 3.) numbers of iterations in SPGL1
the numbers of
iterations in
\ac{SPGL1},
respectively.
% e) incident acoustic energies and the recorded electric energies were related to the sample means of the relative RMSEs caused by the nonconvex l0.5-minimization method
For
each wire,
% 1.) incident acoustic energies at a specified grid point (all pulse-echo measurements, multifrequent)
the incident acoustic energies
\eqref{eqn:recovery_p_in_energy} and
% 2.) recorded electric energies in the pulse echoes (all pulse-echo measurements, multifrequent, all array elements)
the recorded electric energies
\eqref{eqn:recovery_reg_v_rx_born_trans_coef_energy} were related to
% 3.) sample means of the relative RMSEs caused by the nonconvex l0.5-minimization method
the sample means of
the relative \acp{RMSE} caused by
% 4.) nonconvex l0.5-minimization method
the nonconvex $\ell_{0.5}$-minimization method
\eqreflqmin{eqn:recovery_reg_norm_lq_minimization}{ 0.5 }.

%%%%%%%%%%%%%%%%%%%%%%%%%%%%%%%%%%%%%%%%%%%%%%%%%%%%%%%%%%%%%%%%%%%%%%%%%%%%%%%%%%%%%%%%%%%%%%%%%%%%%%%%%%%%%%%%
% 8.) experimental validation
%%%%%%%%%%%%%%%%%%%%%%%%%%%%%%%%%%%%%%%%%%%%%%%%%%%%%%%%%%%%%%%%%%%%%%%%%%%%%%%%%%%%%%%%%%%%%%%%%%%%%%%%%%%%%%%%
%\section{Experimental Validation}
%\label{sec:experimental_validation}
%\input{experimental_validation/experimental_validation.tex}

%%%%%%%%%%%%%%%%%%%%%%%%%%%%%%%%%%%%%%%%%%%%%%%%%%%%%%%%%%%%%%%%%%%%%%%%%%%%%%%%%%%%%%%%%%%%%%%%%%%%%%%%%%%%%%%%
% 9.) results
%%%%%%%%%%%%%%%%%%%%%%%%%%%%%%%%%%%%%%%%%%%%%%%%%%%%%%%%%%%%%%%%%%%%%%%%%%%%%%%%%%%%%%%%%%%%%%%%%%%%%%%%%%%%%%%%
\section{Results}
\label{sec:results}
%%%%%%%%%%%%%%%%%%%%%%%%%%%%%%%%%%%%%%%%%%%%%%%%%%%%%%%%%%%%%%%%%%%%%%%%%%%%%%%%%%%%%%%%%%%%%%%%%%%%%%%%%%%%%%%%
% 1.) wire phantom
%%%%%%%%%%%%%%%%%%%%%%%%%%%%%%%%%%%%%%%%%%%%%%%%%%%%%%%%%%%%%%%%%%%%%%%%%%%%%%%%%%%%%%%%%%%%%%%%%%%%%%%%%%%%%%%%
\subsection{Wire Phantom}
\subsubsection{Incident Acoustic Pressure Fields}
\label{subsubsec:results_phantom_wire_p_in}
%%%%%%%%%%%%%%%%%%%%%%%%%%%%%%%%%%%%%%%%%%%%%%%%%%%%%%%%%%%%%%%%%%%%%%%%%%%%%%%%%%%%%%%%%%%%%%%%%%%%%%%%%%%%%%%%
% images: discretized incident acoustic pressure fields for the wire phantom (QPW, rnd. apo., rnd. del., rnd. apo. del.)
%%%%%%%%%%%%%%%%%%%%%%%%%%%%%%%%%%%%%%%%%%%%%%%%%%%%%%%%%%%%%%%%%%%%%%%%%%%%%%%%%%%%%%%%%%%%%%%%%%%%%%%%%%%%%%%%
%
\begin{figure*}[t!]
 \centering%
  \input{results/object_A/p_in/figures/latex/sim_study_obj_A_setup_p_in_qpw_rnd_apo_rnd_del_rnd_apo_del_images.tex}
 \caption{}
 \label{fig:V}
\end{figure*}
C
{% a) figure illustrates the discretized incident acoustic pressure fields associated with all incident waves
 Incident acoustic pressure fields
 \eqref{eqn:recovery_p_in} associated with
 % 1.) quasi-plane wave (QPW)
 the \acl{QPW}
 (cf.
 \subref{fig:sim_study_obj_A_p_in_images_qpw_ctr} and
 \subref{fig:sim_study_obj_A_p_in_ref_qpw_f}%
 ) and
 the superpositions of
 % 2.) superposition of randomly-apodized QCWs
 randomly-apodized \acfp{QCW}\acused{QCW}
 (cf.
 \subref{fig:sim_study_obj_A_p_in_images_rnd_apo_ctr} and
 \subref{fig:sim_study_obj_A_p_in_ref_rnd_apo_f}%
 ),
 % 3.) superposition of randomly-delayed QCWs
 randomly-delayed \acp{QCW}
 (cf.
 \subref{fig:sim_study_obj_A_p_in_images_rnd_del_ctr} and
 \subref{fig:sim_study_obj_A_p_in_ref_rnd_del_f}%
 ), and both
 % 4.) superposition of both randomly-apodized and randomly-delayed QCWs
 randomly-apodized and
 randomly-delayed \acp{QCW}
 (cf.
 \subref{fig:sim_study_obj_A_p_in_images_rnd_apo_del_ctr} and
 \subref{fig:sim_study_obj_A_p_in_ref_rnd_apo_del_f}%
 ).
 %--------------------------------------------------------------------------------------------------------------
 % top row
 %--------------------------------------------------------------------------------------------------------------
 % a) top row displays these fields at the discrete frequency closest to the center frequency as functions of the position
 The top row
 (cf.
 \subref{fig:sim_study_obj_A_p_in_images_qpw_ctr} to
 \subref{fig:sim_study_obj_A_p_in_images_rnd_apo_del_ctr}%
 ) displays
 these fields at
 % 1.) discrete frequency closest to the center frequency
 % f_{ l_{\text{c}} } = \SI{3.9951}{\mega\hertz} \approx f_{\text{c}}
 the discrete frequency closest to
 the center frequency %, i.e. $f_{ l_{\text{c}} } \approx f_{\text{c}}$ with the index $l_{\text{c}} = 329$,
 as functions of
 % 2.) grid position
 the position.
 % b) large image shows the normalized absolute value in decibel, whereas the inset image shows the phase in radian
 For each type of
 incident wave,
 the large image shows
 % 1.) normalized absolute value in decibel
 the normalized absolute value
 (right colorbar), whereas
 the inset image shows
 % 2.) phase in radian
 the phase inside
 the region indicated by
 the red square 
 (top colorbar).
 % c) reference value for normalization is the maximum absolute value inside the specified FOV
 % \vect{r} = \trans{ ( \SI{-6.1341}{\milli\meter}; \SI{6.8199}{\milli\meter} ) }
 %The maximum absolute value, which
 %was attained by
 %the superposition of both
 %randomly-apodized and
 %randomly-delayed \acp{QCW} at
 %the position
 %$\vect{r} \approx \trans{ ( \SI{-6.13}{\milli\meter}; \SI{6.82}{\milli\meter} ) }$, normalized
 %the absolute values.
 %--------------------------------------------------------------------------------------------------------------
 % bottom row
 %--------------------------------------------------------------------------------------------------------------
 % a) bottom row displays the normalized absolute values and the normalized phase differences at the three positions as functions of the frequency
 The bottom row
 (cf.
 \subref{fig:sim_study_obj_A_p_in_ref_qpw_f} to
 \subref{fig:sim_study_obj_A_p_in_ref_rnd_apo_del_f}%
 ) displays
 % 1.) normalized absolute values in decibel
 the normalized absolute values and
 % 2.) normalized phase differences
 the normalized phase differences at
 the three positions indicated by
 the markers in
 the top row as
 functions of
 the frequency.
 % b) reference values for the normalizations are the maximum absolute value and the minimum phase difference at the three positions
 %The maximum absolute value and
 %the minimum phase difference at
 %these positions, which
 %were attained by
 %the superposition of
 %randomly-delayed \acp{QCW}, served as
 %reference values.
}%
{sim_study_obj_A_p_in_qpw_rnd_apo_rnd_del_rnd_apo_del_images}

%---------------------------------------------------------------------------------------------------------------
% 1.) spatial dependencies closest to the center frequency (QPW, rnd. apo., rnd. del., rnd. apo. del.)
%---------------------------------------------------------------------------------------------------------------
% a) random waves differed significantly from the QPW
The random waves differed significantly from
% 1.) quasi-plane wave (QPW)
the \ac{QPW}
(cf. \cref{fig:sim_study_obj_A_p_in_qpw_rnd_apo_rnd_del_rnd_apo_del_images}).
% b) interference of the QCWs resulted in beamlike fluctuations
The interference of
the \acp{QCW} introduced
beamlike fluctuations into
the absolute values.
% b)
These were
% 1.) relatively subtle and regular for the QPW
relatively subtle and
regular for
the \ac{QPW} but
% 2.) pronounced and erratic for the random waves
pronounced and
irregular for
the random waves.
% b) their [beamlike fluctuations] dynamic ranges increased from only 8.15 dB for the QPW to 67.54 dB for the superposition of randomly-delayed QCWs
% QPW: 8.1532 dB / rnd. apo.: 65.8998 dB / rnd. del.: 67.5402 dB / rnd. apo. del.: 61.9938 dB
Their dynamic ranges increased from
\SI{8.15}{\deci\bel} for
% 1.) quasi-plane wave (QPW)
the \ac{QPW} to
\SI{67.54}{\deci\bel} for
% 2.) superposition of randomly-delayed QCWs
the superposition of
randomly-delayed \acp{QCW}.
% e) approximately constant phase on lines parallel to the r_{1}-axis indicated plane wavefronts
The paths of
constant phase, which were
approximately parallel to
the $r_{1}$-axis for
the \ac{QPW}, turned
irregular with
heterogeneous normal vectors exhibiting
nonzero $r_{2}$-components for
the random waves.
% f) paths of constant phase indicated a transition from plane wavefronts to irregular wavefronts
\TODO{match adjective with Fig. 4}
They indicated
a transition from
plane to
irregular wavefronts.

%---------------------------------------------------------------------------------------------------------------
% 2.) spectral dependencies for three closely spaced positions next to the r_{2}-axis (QPW, rnd. apo., rnd. del., rnd. apo. del.)
%---------------------------------------------------------------------------------------------------------------
% a) behavior at the three indicated positions
At
the three indicated positions,
% 1.) normalized absolute values reflected the approximate Gaussian shape of the electromechanical pulse echo
the absolute values reflected
the approximate Gaussian shape of
the electromechanical pulse echo, and
% 2.) normalized phase differences depended approximately affine-linearly on the frequency
the phase differences depended approximately affine-linearly on
the frequency.
% b) distinct slopes indicated the times-of-flight of the wavefronts from the linear transducer array to each reference position
Their slopes indicated
% 1.) times-of-flights
the diverse \acp{TOF} of
% 2.) wavefronts
the wavefronts from
% 3.) linear transducer array
the linear transducer array to
% 4.) position
each position.
% c) QPW achieved very similar absolute values for all three reference positions, whereas the random waves induced notches and peaks
The \ac{QPW} achieved
very similar absolute values at
all three positions, whereas
% d) random waves induced notches and peaks at various frequencies that erratically modified the approximate Gaussian shape
the random waves induced
notches and peaks at
various frequencies that
erratically modified
the approximate Gaussian shape for
each position.
% d) QPW additionally achieved linear phases
The former additionally achieved
linear phases, whereas
% e.) normalized unwrapped phases approximately maintain the linear dependence on the normalized frequency
the latter introduced
erratic deviations from
this linear frequency dependence.
% e) modification is relatively modest for the superposition of randomly-apodized QCWs
The superposition of
randomly-apodized \acp{QCW} induced
relatively modest modifications, whereas
% f) both superpositions of QCWs using random time delays induced more pronounced modifications
both superpositions of
\acp{QCW} using
random time delays induced
more pronounced modifications.

%%%%%%%%%%%%%%%%%%%%%%%%%%%%%%%%%%%%%%%%%%%%%%%%%%%%%%%%%%%%%%%%%%%%%%%%%%%%%%%%%%%%%%%%%%%%%%%%%%%%%%%%%%%%%%%%
% 2.) Euclidean norms of the column vectors
%%%%%%%%%%%%%%%%%%%%%%%%%%%%%%%%%%%%%%%%%%%%%%%%%%%%%%%%%%%%%%%%%%%%%%%%%%%%%%%%%%%%%%%%%%%%%%%%%%%%%%%%%%%%%%%%
%\subsubsection{Euclidean Norms of the Column Vectors}
%\label{subsubsec:results_phantom_wire_column_norms}
%\input{results/results_phantom_wire_column_norms.tex}

%%%%%%%%%%%%%%%%%%%%%%%%%%%%%%%%%%%%%%%%%%%%%%%%%%%%%%%%%%%%%%%%%%%%%%%%%%%%%%%%%%%%%%%%%%%%%%%%%%%%%%%%%%%%%%%%
% 3.) point spread functions
%%%%%%%%%%%%%%%%%%%%%%%%%%%%%%%%%%%%%%%%%%%%%%%%%%%%%%%%%%%%%%%%%%%%%%%%%%%%%%%%%%%%%%%%%%%%%%%%%%%%%%%%%%%%%%%%
\subsubsection{Point Spread Functions}
%\label{subsubsec:results_phantom_wire_psf}
%%%%%%%%%%%%%%%%%%%%%%%%%%%%%%%%%%%%%%%%%%%%%%%%%%%%%%%%%%%%%%%%%%%%%%%%%%%%%%%%%%%%%%%%%%%%%%%%%%%%%%%%%%%%%%%%
% images: point spread functions (PSFs, third fixed position)
%%%%%%%%%%%%%%%%%%%%%%%%%%%%%%%%%%%%%%%%%%%%%%%%%%%%%%%%%%%%%%%%%%%%%%%%%%%%%%%%%%%%%%%%%%%%%%%%%%%%%%%%%%%%%%%%
%
\begin{figure}[t!]
 \centering%
  \input{results/object_A/kappa_only/sr_tpsf/figures/latex/sim_study_obj_A_sr_1_tpsf_images.tex}
 \caption{}
 \label{fig:V}
\end{figure}
C
{% a) figure illustrates the absolute values of the PSFs associated with the random observation process and the observation processes induced by the GWN and all incident waves
 Absolute values of
 % 1.) point spread functions (PSFs)
 the \aclp{PSF}\acused{PSF}
 \eqref{eqn:cs_math_tpsf} associated with
 % 2.) random observation process (RIP)
 the random observation process
 (cf. \subref{fig:sim_study_obj_A_sr_1_tpsf_images_rip_1}) and
 % 3.) observation processes (all pulse-echo measurements, multifrequent, all array elements)
 the observation processes
 \eqref{eqn:recovery_sys_lin_eq_v_rx_born_all_f_all_in_mat} induced by
 % 3.a) GWN
 the \acl{GWN}\acused{GWN}
 (cf. \subref{fig:sim_study_obj_A_sr_1_tpsf_images_blgwn_1}),
 % 3.b) quasi-plane wave (QPW)
 the \acl{QPW}\acused{QPW}
 (cf. \subref{fig:sim_study_obj_A_sr_1_tpsf_images_qpw_1}), and
 the superpositions of
 % 3.c) superposition of randomly-apodized QCWs 
 randomly-apodized \acfp{QCW}\acused{QCW}
 (cf. \subref{fig:sim_study_obj_A_sr_1_tpsf_images_rnd_apo_1}),
 % 3.d) superposition of randomly-delayed QCWs
 randomly-delayed \acp{QCW}
 (cf. \subref{fig:sim_study_obj_A_sr_1_tpsf_images_rnd_del_1}), and both
 % 3.e) superposition of both randomly-apodized and randomly-delayed QCWs
 randomly-apodized and
 randomly-delayed \acp{QCW}
 (cf. \subref{fig:sim_study_obj_A_sr_1_tpsf_images_rnd_apo_del_1}).
 % b) green crosshairs indicate the third fixed position
 % index_tpsf = 3;
 % n_{2} = 72334
 % index_1 = 142
 % index_2 = 142
 % \vect{r}_{ n_{2} - 1 } \approx \trans{ ( \SI{-8.7249}{\milli\meter}, \SI{10.7823}{\milli\meter} ) }
 The green crosshairs indicate
 the third fixed position.
 %$\vect{r}_{ \text{lat}, n_{2} - 1 } \approx \trans{ ( \SI{-8.72}{\milli\meter}, \SI{10.78}{\milli\meter} ) }$.
 % c) inset images magnify the regions indicated by the white squares
 The inset images magnify
 the regions indicated by
 the white squares.
}%
{sim_study_obj_A_sr_1_tpsf_images}

%---------------------------------------------------------------------------------------------------------------
% 1.) overview of the computed PSFs for the reference observation processes and all incident waves
%---------------------------------------------------------------------------------------------------------------
% a) all computed PSFs correctly attained their maximum absolute values of unity at the fixed positions
Although
% 1.) point spread functions (PSFs)
all computed \acp{PSF}
\eqref{eqn:cs_math_tpsf} correctly attained
% 2.) maximum absolute values of unity
their maximum absolute values of
unity at
% 3.) fixed positions
the fixed positions,
% b) all computed PSFs differed in their behavior for the remaining positions
they differed in
% 1.) behavior
their behavior for
% 2.) remaining positions
the remaining positions
(cf. \cref{fig:sim_study_obj_A_sr_1_tpsf_images}).
% c) both reference observation processes produced random values close to zero that rendered the maxima sharp and isolated
Both reference observation processes produced
% 1.) random values close to zero
random values close to
zero that rendered
% 2.) isolated sharp maximum at the discrete position \vect{r}_{ n_{2} - 1 }
the maxima sharp and
isolated.
% d) random observation process uniformly distributed these values over the FOV
The random observation process
\eqref{eqn:sim_study_params_ref_obs_proc_rip} uniformly distributed
% 1.) random values close to zero
these values over
% 2.) field of view (FOV)
the \ac{FOV}, whereas
% e) structured version formed noticeable gaps that were laterally adjacent to the maxima and shaped hourglasses of larger absolute values
its structured version
\eqref{eqn:sim_study_params_ref_obs_proc_gwn} formed
% 1.) noticeable gaps
noticeable gaps that were
% 2.) laterally adjacent
laterally adjacent to
% 3.) maximum absolute values of unity
the maxima and shaped
% 4.) hourglasses of larger absolute values
hourglasses of
larger absolute values
(cf. inset image).
% f) observation processes induced by all incident waves concentrated relatively large absolute values close to unity in elliptical-shaped regions around the maxima
The observation processes
\eqref{eqn:recovery_sys_lin_eq_v_rx_born_all_f_all_in_mat} induced by
% 1.) all incident waves
all incident waves concentrated
% 2.) relatively large absolute values close to unity
relatively large absolute values close to
unity in
% 3.) elliptical-shaped regions
elliptical-shaped regions around
% 3.) maximum absolute values of unity
the maxima.
% g) lengths of the minor and major axes
% QPW: ( 0.30; 0.76) mm / rnd. apo.: ( 0.152; 0.61 ) mm / rnd. del.: ( 0.152; 0.46 ) mm / rnd. apo. del.: ( 0.152; 0.53 ) mm
The lengths of
the minor and major axes ranged from
\SIrange{0.15}{0.3}{\milli\meter} and from
\SIrange{0.46}{0.76}{\milli\meter},
respectively.
% h) observation processes induced by all incident waves distributed the nonzero values less uniformly and formed sidelobes of various characters
They distributed
% 1.) nonzero absolute values
the nonzero values less uniformly and formed
% 2.) sidelobes of various characters
sidelobes of
various characters.

%---------------------------------------------------------------------------------------------------------------
% 2.) detailed description of the PSFs for all incident waves
%---------------------------------------------------------------------------------------------------------------
% a) observation process induced by the QPW deviated most significantly from both reference observation processes
The observation process
\eqref{eqn:recovery_sys_lin_eq_v_rx_born_all_f_all_in_mat} induced by
% 1.) quasi-plane wave (QPW)
the \ac{QPW} deviated
most significantly from
% 2.) reference observation processes
both references.
% b) observation process induced by the QPW formed the largest elliptical-shaped region and coherent sidelobes of approximately constant absolute values
It formed
% 1.) largest elliptical-shaped region
the largest elliptical-shaped region and
% 2.) coherent sidelobes
coherent sidelobes of
approximately constant absolute values.
% c) observation processes induced by the random waves resembled that induced by the GWN
In contrast,
% 1.) observation processes (all pulse-echo measurements, multifrequent, all array elements)
the observation processes
\eqref{eqn:recovery_sys_lin_eq_v_rx_born_all_f_all_in_mat} induced by
% 2.) random waves
the random waves resembled
% 3.) observation process induced by the GWN
that induced by
the \ac{GWN}
\eqref{eqn:sim_study_params_ref_obs_proc_gwn}.
% d) sizes of the elliptical-shaped regions decreased relative to the QPW
The sizes of
% 1.) elliptical-shaped regions
the elliptical-shaped regions decreased relative to
% 2.) quasi-plane wave (QPW)
the \ac{QPW}.
% e) sidelobes diffused and fluctuated in their values resulting in more uniform distributions
The sidelobes diffused and
fluctuated in
their values, similar to
a speckle pattern, resulting in
more uniform distributions.
% f) both superpositions of QCWs using random time delays distributed the nonzero values slightly more uniformly than the superposition of randomly-apodized QCWs
Both superpositions of
\acp{QCW} using
random time delays distributed
the nonzero values slightly more uniformly than
% 1.) superposition of randomly-apodized QCWs
the superposition of
randomly-apodized \acp{QCW}.
% g) distributed values appeared more random
In addition,
the distributed values appeared more random.

%%%%%%%%%%%%%%%%%%%%%%%%%%%%%%%%%%%%%%%%%%%%%%%%%%%%%%%%%%%%%%%%%%%%%%%%%%%%%%%%%%%%%%%%%%%%%%%%%%%%%%%%%%%%%%%%
% table: full extents at half maximum (all fixed positions)
%%%%%%%%%%%%%%%%%%%%%%%%%%%%%%%%%%%%%%%%%%%%%%%%%%%%%%%%%%%%%%%%%%%%%%%%%%%%%%%%%%%%%%%%%%%%%%%%%%%%%%%%%%%%%%%%
\begin{table*}[tb]
 \centering
 \caption{%
  % a) table summarizes the FEHMs of the PSFs associated with the observation processes induced by all incident waves
  Full extents at
  half maximum of
  % 1.) point spread functions (PSFs)
  the \aclp{PSF}
  \eqref{eqn:cs_math_tpsf} associated with
  % 2.) observation processes (all pulse-echo measurements, multifrequent, all array elements)
  the observation processes
  \eqref{eqn:recovery_sys_lin_eq_v_rx_born_all_f_all_in_mat} induced by
  % 3.) all incident waves
  all incident waves.
  % b) FEHMs were evaluated for nine uniformly distributed positions along the diagonal from (-17.5, 2 ) mm to ( 17.5, 37 ) mm
  They were evaluated for
  % 1.) nine uniformly distributed positions
  nine uniformly distributed positions along
  % 2.) diagonal from (-17.4879 mm, 2.0193 mm ) to ( 17.4879 mm, 36.9951 mm )
  the diagonal from
  $\trans{ ( \SI{-17.5}{\milli\meter}, \SI{2}{\milli\meter} ) }$ to
  $\trans{ ( \SI{17.5}{\milli\meter}, \SI{37}{\milli\meter} ) }$ and numbered from
  \numrange{1}{9} with
  increasing axial coordinate.
 }
 \label{tab:sim_study_obj_A_sr_1_tpsf_fehm}
 \begin{tabular}{%
  @{}%
  l%																01.) type of incident wave
  S[table-format=1.2,table-number-alignment = right,table-auto-round]%								02.) 1st fixed position
  S[table-format=1.2,table-number-alignment = right,table-auto-round]%								03.) 2nd fixed position
  S[table-format=1.2,table-number-alignment = right,table-auto-round]%								04.) 3rd fixed position
  S[table-format=1.2,table-number-alignment = right,table-auto-round]%								05.) 4th fixed position
  S[table-format=1.2,table-number-alignment = right,table-auto-round]%								06.) 5th fixed position
  S[table-format=1.2,table-number-alignment = right,table-auto-round]%								07.) 6th fixed position
  S[table-format=1.2,table-number-alignment = right,table-auto-round]%								08.) 7th fixed position
  S[table-format=1.2,table-number-alignment = right,table-auto-round]%								09.) 8th fixed position
  S[table-format=1.2,table-number-alignment = right,table-auto-round]%								10.) 9th fixed position
  S[table-format=1.2(3),separate-uncertainty,table-align-uncertainty = true,table-number-alignment = right,table-auto-round]%	11.) sample mean & std. dev
  @{}%
 }
 \toprule
  \multicolumn{1}{@{}H}{\multirow{2}{*}{Incident wave}} &
  \multicolumn{10}{H@{}}{Full extent at half maximum (\si{\milli\meter\squared})}\\
  \cmidrule(l){2-11}
  &
  \multicolumn{1}{H}{1} &
  \multicolumn{1}{H}{2} &
  \multicolumn{1}{H}{3} &
  \multicolumn{1}{H}{4} &
  \multicolumn{1}{H}{5} &
  \multicolumn{1}{H}{6} &
  \multicolumn{1}{H}{7} &
  \multicolumn{1}{H}{8} &
  \multicolumn{1}{H}{9} &
  \multicolumn{1}{H@{}}{$\text{Sample mean} \pm \text{std. dev.}$}\\
  \cmidrule(r){1-1}\cmidrule(lr){2-2}\cmidrule(lr){3-3}\cmidrule(lr){4-4}\cmidrule(lr){5-5}
  \cmidrule(lr){6-6}\cmidrule(lr){7-7}\cmidrule(lr){8-8}\cmidrule(lr){9-9}\cmidrule(lr){10-10}\cmidrule(l){11-11}
 \addlinespace
  \acs{QPW}  & 0.19 & 0.18 & 0.21 & 0.24 & 0.28 & 0.32 & 0.38 & 0.42 & 0.51  & 0.30 \pm 0.11\\
Rnd. apo.  & 0.07 & 0.09 & 0.14 & 0.17 & 0.16 & 0.18 & 0.44 & 0.36 & 0.51  & 0.24 \pm 0.16\\
Rnd. del.  & 0.05 & 0.10 & 0.09 & 0.11 & 0.14 & 0.33 & 0.20 & 0.34 & 0.46  & 0.20 \pm 0.14\\
Rnd. apo. del.  & 0.11 & 0.15 & 0.12 & 0.19 & 0.23 & 0.19 & 0.33 & 0.28 & 0.39  & 0.22 \pm 0.10\\

 \addlinespace
 \bottomrule
 \end{tabular}
\end{table*}

%---------------------------------------------------------------------------------------------------------------
% 3.) full extents at half maximum (all fixed positions)
%---------------------------------------------------------------------------------------------------------------
% a) random waves achieved FEHMs that were smaller than or equal to those of the QPW for all fixed positions, except those numbered s \in \{ 6, 7 \}
% 1: QPW (max)              2: QPW (max)             3: QPW (max)         4: QPW (max)              5: QPW (max)              6: rnd. del. (max)   7: rnd. apo. (max)   8: QPW (max)              9: QPW = rnd. apo. (max)
%    rnd. del.:      73.68%    rnd. apo.:     50%       rnd. del.: 57.14%    rnd. del.:      54.17%    rnd. del.:      50%       Rnd. apo.: 45.45%    rnd. del.: 54.55%    rnd. apo. del.: 33.33%    rnd. apo. del.: 23.53%
%    rnd. apo. del.: 42.11%    rnd. apo. del: 16.67%    rnd. apo.: 33.33%    rnd. apo. del.: 20.83%    rnd. apo. del.: 17.86%    QPW:        3.03%    QPW:       13.63%    rnd. apo.:      14.29%    rnd. del.:      9.8%
The random waves achieved
\acp{FEHM} that were
smaller than or
equal to
those of
the \ac{QPW} for
all fixed positions, except
those numbered
$s \in \{ 6, 7 \}$
(cf. \cref{tab:sim_study_obj_A_sr_1_tpsf_fehm}).
% b) FEHMs generally increased with the axial coordinate of these positions
The \acp{FEHM} generally increased with
the axial coordinate of
these positions.
% c) maximum normalized differences ranged from 23.5 % (rnd. apo. del., s = 9) to 73.7 % (rnd. del., s = 1)
The maximum normalized differences ranged from
% 1.) superposition of both randomly-apodized and randomly-delayed QCWs at the ninth fixed position (s = 9)
\SI{23.5}{\percent} for
the superposition of both
randomly-apodized and
randomly-delayed \acp{QCW} at
the ninth fixed position, i.e.
$s = 9$, to
% 2.) superposition of randomly-delayed QCWs at the first fixed position (s = 1)
\SI{73.7}{\percent} for
the superposition of
randomly-delayed \acp{QCW} at
the first fixed position, i.e.
$s = 1$.
% d) mean FEHMs reflected these reductions relative to the QPW
The mean \acp{FEHM} reflected
these reductions relative to
the \ac{QPW}.
% e) superposition of randomly-apodized QCWs produced the largest sample mean and sample standard deviation among the random waves
The superposition of
randomly-apodized \acp{QCW} produced
the largest sample mean and
sample standard deviation among
the random waves.
% f) both reference observation processes consistently achieved the minimum FEHM of a two-dimensional volume element for all fixed positions
Both reference observation processes consistently achieved
% 1.) minimum FEHM
the minimum \ac{FEHM} of
% 2.) two-dimensional volume element
a two-dimensional volume element
$\Delta V \approx \SI{5.81e-3}{\milli\meter\squared}$ for
% 3.) all fixed positions
all fixed positions.

%%%%%%%%%%%%%%%%%%%%%%%%%%%%%%%%%%%%%%%%%%%%%%%%%%%%%%%%%%%%%%%%%%%%%%%%%%%%%%%%%%%%%%%%%%%%%%%%%%%%%%%%%%%%%%%%
% empirical CDFs: point spread functions (PSFs)
%%%%%%%%%%%%%%%%%%%%%%%%%%%%%%%%%%%%%%%%%%%%%%%%%%%%%%%%%%%%%%%%%%%%%%%%%%%%%%%%%%%%%%%%%%%%%%%%%%%%%%%%%%%%%%%%
%
\begin{figure}[t!]
 \centering%
  \input{results/object_A/kappa_only/sr_tpsf/figures/latex/sim_study_obj_A_sr_1_tpsf_ecdfs.tex}
 \caption{}
 \label{fig:V}
\end{figure}
C
{% a) figure illustrates the empirical CDFs of the PSFs associated with both reference observation processes and the observation processes induced by all incident waves
 Empirical \acfp{CDF}\acused{CDF} of
 % 1.) point spread functions (PSFs)
 the \aclp{PSF}
 \eqref{eqn:cs_math_tpsf} associated with
 % 2.) reference observation processes
 both reference observation processes and
 % 3.) observation processes (all pulse-echo measurements, multifrequent, all array elements)
 the observation processes
 \eqref{eqn:recovery_sys_lin_eq_v_rx_born_all_f_all_in_mat} induced by
 % 4.) all incident waves
 all incident waves.
 % b) inset graphic magnifies the region indicated by the red rectangle
 The inset graphic magnifies
 the region indicated by
 the red rectangle.
}%
{sim_study_obj_A_sr_1_tpsf_ecdfs}

%---------------------------------------------------------------------------------------------------------------
% 4.) empirical CDFs (all fixed positions)
%---------------------------------------------------------------------------------------------------------------
% a) empirical CDFs confirmed the beneficial properties of the random waves
The empirical \acp{CDF} confirmed
% 1.) beneficial properties
the beneficial properties of
% 2.) random waves
the random waves
(cf. \cref{fig:sim_study_obj_A_sr_1_tpsf_ecdfs}).
% b) random observation process primarily attained absolute values ranging from -70 to -30.93 dB
% RIP: -30.9349 dB
% RIP: 4.3245 % @ -70 dB
The random observation process
\eqref{eqn:sim_study_params_ref_obs_proc_rip} primarily attained
% 1.) absolute values ranging from -70 to -30.93 dB
absolute values ranging from
\SIrange{-70}{-30.93}{\deci\bel}.
% c) only approximately 4.3 % of the FOV were attributed to smaller absolute values
Only approximately \SI{4.3}{\percent} of
% 1.) field of view (FOV)
the \ac{FOV} were attributed to
% 2.) smaller absolute values
smaller absolute values.
% d) structured random observation process deviated modestly from this behavior
The structured random observation process
\eqref{eqn:sim_study_params_ref_obs_proc_gwn} deviated
modestly from
this behavior.
% e) absolute values ranged from -70 to -16.79 dB
% BLGWN: -16.7930 dB
% BLGWN: 12.0914 % @ -70 dB
The absolute values ranged from
\SIrange{-70}{-16.79}{\deci\bel}.
% f) increased dynamic range reflected the two gaps that were laterally adjacent to the FEHMs
% article:Schiffner2018, Sect. VIII. Results / Sect. VIII-A. Wire Phantom / Sect. VIII-A.2) Point Spread Functions (subsubsec:results_phantom_wire_psf)
% - The random observation process \eqref{eqn:sim_study_params_ref_obs_proc_rip} uniformly distributed these values over the \ac{FOV}, whereas
%   ITS STRUCTURED VERSION \eqref{eqn:sim_study_params_ref_obs_proc_gwn} FORMED NOTICEABLE GAPS that were
%   laterally adjacent to the maxima and shaped hourglasses of larger absolute values (cf. inset image).
This increased dynamic range reflected
% 1.) noticeable gaps
the gaps that were
% 2.) laterally adjacent to the maxima
laterally adjacent to
the maxima
(cf. \cref{fig:sim_study_obj_A_sr_1_tpsf_images_blgwn_1}).
% g) observation processes induced by the random waves deviated in a stronger but similar fashion from both references
The observation processes
\eqref{eqn:recovery_sys_lin_eq_v_rx_born_all_f_all_in_mat} induced by
% 1.) random waves
the random waves deviated in
% 2.) stronger but similar fashion
a stronger but similar fashion from
% 3.) reference observation processes
both references.
% h) absolute values below -70 dB constituted approximately 49.5 to 56.2 % of the FOV and those above this threshold formed the remaining 43.8 to 50.5 %
% rnd. apo.: -0.1124 dB / rnd. del: -0.1164 dB / rnd. apo. del: -0.1349 dB
% rnd. apo.: 56.1824 % @ -70 dB / rnd. del: 49.4858 % @ -70 dB / rnd. apo. del: 49.7709 % @ -70 dB
The absolute values below
\SI{-70}{\deci\bel} constituted
% 1.) approximately 49.5 to 56.2 %
approximately \SIrange{49.5}{56.2}{\percent} of
% 2.) field of view
the \ac{FOV} and
% 3.) absolute values above -70 dB
those above this threshold, which reached up to
\SI{-0.11}{\deci\bel}, formed
% 4.) remaining 43.8 to 50.5 %
the remaining \SIrange{43.8}{50.5}{\percent}.
% i) superposition of randomly-apodized QCWs distributed the latter values over the smallest percentage of the FOV
The superposition of
randomly-apodized \acp{QCW} distributed
% 1.) absolute values above -70 dB
the latter values over
% 2.) smallest percentage
the smallest percentage of
% 3.) field of view
the \ac{FOV}.
% j) observation process induced by the QPW deviated strongest from both references
Clearly,
% 1.) observation process (all pulse-echo measurements, multifrequent, all array elements)
the observation process
\eqref{eqn:recovery_sys_lin_eq_v_rx_born_all_f_all_in_mat} induced by
% 2.) quasi-plane wave (QPW)
the \ac{QPW} deviated
% 3.) strongest
strongest from
% 4.) reference observation processes
both references.
% k) absolute values ranging from -70 to -0.15 dB strongly concentrated on only 20.1 % of the FOV and indicated the distinctive sidelobes
% article:Schiffner2018, Sect. VIII. Results / Sect. VIII-A. Wire Phantom / Sect. VIII-A.2) Point Spread Functions (subsubsec:results_phantom_wire_psf)
% - The observation process \eqref{eqn:recovery_sys_lin_eq_v_rx_born_all_f_all_in_mat} induced by the \ac{QPW} deviated most significantly from
%   both references.
% - It formed the largest elliptical-shaped region and coherent sidelobes of approximately constant absolute values.
% QPW: -0.1508 dB
% QPW: 79.9494 % @ -70 dB
The absolute values ranging from
\SIrange{-70}{-0.15}{\deci\bel} strongly concentrated on
% 1.) only 20.1 %
only \SI{20.1}{\percent} of
% 2.) field of view
the \ac{FOV} and indicated
% 3.) distinctive sidelobes
the distinctive sidelobes
(cf. \cref{fig:sim_study_obj_A_sr_1_tpsf_images_qpw_1}).

%%%%%%%%%%%%%%%%%%%%%%%%%%%%%%%%%%%%%%%%%%%%%%%%%%%%%%%%%%%%%%%%%%%%%%%%%%%%%%%%%%%%%%%%%%%%%%%%%%%%%%%%%%%%%%%%
% 4.) adjoint normalized sensing matrices
%%%%%%%%%%%%%%%%%%%%%%%%%%%%%%%%%%%%%%%%%%%%%%%%%%%%%%%%%%%%%%%%%%%%%%%%%%%%%%%%%%%%%%%%%%%%%%%%%%%%%%%%%%%%%%%%
\subsubsection{Adjoint Normalized Sensing Matrices}
%\label{subsubsec:results_phantom_wire_adjoint}
%%%%%%%%%%%%%%%%%%%%%%%%%%%%%%%%%%%%%%%%%%%%%%%%%%%%%%%%%%%%%%%%%%%%%%%%%%%%%%%%%%%%%%%%%%%%%%%%%%%%%%%%%%%%%%%%
% images: wire phantom (adjoint, single wave emission)
%%%%%%%%%%%%%%%%%%%%%%%%%%%%%%%%%%%%%%%%%%%%%%%%%%%%%%%%%%%%%%%%%%%%%%%%%%%%%%%%%%%%%%%%%%%%%%%%%%%%%%%%%%%%%%%%
%
\begin{figure}[t!]
 \centering%
  \input{results/object_A/kappa_only/adj_1/normalize_true/figures/latex/sim_study_obj_A_adj_1_images.tex}
 \caption{}
 \label{fig:V}
\end{figure}
C
{% a) figure displays the absolute values of the matrix-vector products between the adjoint normalized sensing matrix and the normalized recorded RF voltage signals
 Absolute values of
 the matrix-vector products between
 % 1.) adjoint normalized sensing matrix (all pulse-echo measurements, multifrequent, all array elements)
 the adjoint normalized sensing matrix
 \eqref{eqn:recon_reg_norm_sensing_matrix} and
 % 2.) normalized linear algebraic system (all pulse-echo measurements, multifrequent, all array elements, additive errors)
 the normalized recorded \ac{RF} voltage signals
 \eqref{eqn:recovery_reg_norm_obs_trans_coef_error} for
 % 3.) quasi-plane wave (QPW)
 the \acl{QPW}\acused{QPW}
 (cf. \subref{fig:sim_study_obj_A_adj_1_images_qpw}) and
 the superpositions of
 % 4.) superposition of randomly-apodized QCWs
 randomly-apodized \acfp{QCW}\acused{QCW}
 (cf. \subref{fig:sim_study_obj_A_adj_1_images_rnd_apo}),
 % 5.) superposition of randomly-delayed QCWs
 randomly-delayed \acp{QCW}
 (cf. \subref{fig:sim_study_obj_A_adj_1_images_rnd_del}), and both
 % 6.) superposition of both randomly-apodized and randomly-delayed QCWs
 randomly-apodized and
 randomly-delayed \acp{QCW}
 (cf. \subref{fig:sim_study_obj_A_adj_1_images_rnd_apo_del}).
 % b) green crosshairs indicate the positions of the wires and coincide with local maxima
 The green crosshairs indicate
 the positions of
 the wires and coincide with
 local maxima.
 % c) inset images magnify the regions indicated by the white squares
 The inset images magnify
 the regions indicated by
 the white squares.
 % d) reference SNR amounted to \text{SNR}_{\text{dB}} = \SI{30}{\deci\bel}
 The reference \ac{SNR} amounted to
 $\text{SNR}_{\text{dB}} = \SI{30}{\deci\bel}$.
}%
{sim_study_obj_A_adj_1_images_kap}

%---------------------------------------------------------------------------------------------------------------
% 1.) significance of the adjoint normalized sensing matrices (adjoint, single wave emission)
%---------------------------------------------------------------------------------------------------------------
% a) all incident waves accurately detected the wires
All incident waves accurately detected
the wires
(cf. \cref{fig:sim_study_obj_A_adj_1_images_kap}).
% b) random waves substituted the coherent sidelobes produced by the QPW by noise-like artifacts with a more uniform spatial distribution
The random waves substituted
% 1.) smooth coherent sidelobes
the coherent sidelobes produced by
the \ac{QPW}, whose
% 2.) absolute values did not fluctuate
absolute values did not fluctuate, by
% 3.) noise-like artifacts
noise-like artifacts with
a more uniform spatial distribution.
% c) noise-like artifacts appeared slightly less uniform for the superposition of randomly-apodized QCWs than for both superpositions of QCWs using random time delays
These artifacts appeared
% 1.) slightly less uniform
slightly less uniform for
% 2.) superposition of randomly-apodized QCWs
the superposition of
randomly-apodized \acp{QCW} than for
% 3.) both superpositions of QCWs using random time delays
both superpositions of
\acp{QCW} using
random time delays.

%%%%%%%%%%%%%%%%%%%%%%%%%%%%%%%%%%%%%%%%%%%%%%%%%%%%%%%%%%%%%%%%%%%%%%%%%%%%%%%%%%%%%%%%%%%%%%%%%%%%%%%%%%%%%%%%
% 5.) recovery by l2-minimization
%%%%%%%%%%%%%%%%%%%%%%%%%%%%%%%%%%%%%%%%%%%%%%%%%%%%%%%%%%%%%%%%%%%%%%%%%%%%%%%%%%%%%%%%%%%%%%%%%%%%%%%%%%%%%%%%
%\subsubsection{Recovery by $\ell_{2}$-Minimization}
%\label{subsubsec:results_phantom_wire_l2_minimization}
%\input{results/results_phantom_wire_l2_minimization.tex}

%%%%%%%%%%%%%%%%%%%%%%%%%%%%%%%%%%%%%%%%%%%%%%%%%%%%%%%%%%%%%%%%%%%%%%%%%%%%%%%%%%%%%%%%%%%%%%%%%%%%%%%%%%%%%%%%
% 6.) recovery by lq-minimization
%%%%%%%%%%%%%%%%%%%%%%%%%%%%%%%%%%%%%%%%%%%%%%%%%%%%%%%%%%%%%%%%%%%%%%%%%%%%%%%%%%%%%%%%%%%%%%%%%%%%%%%%%%%%%%%%
\subsubsection{Recovery by $\ell_{q}$-Minimization}
%\label{subsubsec:results_phantom_wire_lq_minimization}
%%%%%%%%%%%%%%%%%%%%%%%%%%%%%%%%%%%%%%%%%%%%%%%%%%%%%%%%%%%%%%%%%%%%%%%%%%%%%%%%%%%%%%%%%%%%%%%%%%%%%%%%%%%%%%%%
% images: wire phantom (lq-minimization, single pulse-echo measurement, ref. SNR: 30 dB)
%%%%%%%%%%%%%%%%%%%%%%%%%%%%%%%%%%%%%%%%%%%%%%%%%%%%%%%%%%%%%%%%%%%%%%%%%%%%%%%%%%%%%%%%%%%%%%%%%%%%%%%%%%%%%%%%
%
\begin{figure*}[t!]
 \centering%
  \input{results/object_A/kappa_only/sr_1/normalize_true/figures/latex/sim_study_obj_A_sr_1_images.tex}
 \caption{}
 \label{fig:V}
\end{figure*}
C
{% a) figure illustrates the absolute values of the recovered relative spatial fluctuations in the unperturbed compressibility
 Absolute values of
 % 1.) recovered relative spatial fluctuations in the unperturbed compressibility
 the recovered compressibility fluctuations
 \eqref{eqn:recovery_reg_norm_lq_minimization_sol_mat_params} for
 % 2.) quasi-plane wave (QPW)
 the \acl{QPW}\acused{QPW}
 (cf.
 \subref{fig:sim_study_obj_A_sr_1_images_spgl1_l1_qpw} and
 \subref{fig:sim_study_obj_A_sr_1_images_spgl1_lq_qpw}%
 ) and
 the superpositions of
 % 3.) superposition of randomly-apodized QCWs
 randomly-apodized \acfp{QCW}\acused{QCW}
 (cf.
 \subref{fig:sim_study_obj_A_sr_1_images_spgl1_l1_rnd_apo} and
 \subref{fig:sim_study_obj_A_sr_1_images_spgl1_lq_rnd_apo}%
 ),
 % 4.) superposition of randomly-delayed QCWs
 randomly-delayed \acp{QCW}
 (cf.
 \subref{fig:sim_study_obj_A_sr_1_images_spgl1_l1_rnd_del} and
 \subref{fig:sim_study_obj_A_sr_1_images_spgl1_lq_rnd_del}%
 ), and both
 % 5.) superposition of both randomly-apodized and randomly-delayed QCWs
 randomly-apodized and
 randomly-delayed \acp{QCW}
 (cf.
 \subref{fig:sim_study_obj_A_sr_1_images_spgl1_l1_rnd_apo_del} and
 \subref{fig:sim_study_obj_A_sr_1_images_spgl1_lq_rnd_apo_del}%
 ).
 % b) top row shows the results of the convex l1-minimization method
 The top row
 (cf.
 \subref{fig:sim_study_obj_A_sr_1_images_spgl1_l1_qpw} to
 \subref{fig:sim_study_obj_A_sr_1_images_spgl1_l1_rnd_apo_del}%
 ) shows
 the results of
 % 1.) convex l1-minimization method
 the convex $\ell_{1}$-minimization method
 \eqreflqmin{eqn:recovery_reg_norm_lq_minimization}{ 1 }, whereas
 % c) bottom row shows the results of the nonconvex l0.5-minimization method
 the bottom row
 (cf.
 \subref{fig:sim_study_obj_A_sr_1_images_spgl1_lq_qpw} to
 \subref{fig:sim_study_obj_A_sr_1_images_spgl1_lq_rnd_apo_del}%
 ) shows
 those of
 % 1.) nonconvex l0.5-minimization method
 the nonconvex $\ell_{0.5}$-minimization method
 \eqreflqmin{eqn:recovery_reg_norm_lq_minimization}{ 0.5 }.
 % d) large images represent the nonzero absolute values by crosshairs of proportional gray values and sizes
 The large images represent
 the nonzero absolute values by
 crosshairs of
 proportional gray values and
 sizes, whereas
 % e) inset images exclusively use gray values to magnify the regions indicated by the white squares
 the inset images exclusively use
 gray values to magnify
 the regions indicated by
 the white squares.
 % f) reference SNR amounted to \text{SNR}_{\text{dB}} = \SI{30}{\deci\bel}
 The reference \ac{SNR} amounted to
 $\text{SNR}_{\text{dB}} = \SI{30}{\deci\bel}$.
}%
{sim_study_obj_A_sr_1_images_kap}

%---------------------------------------------------------------------------------------------------------------
% 1.) visual inspection of the recovered images (lq-minimization, single pulse-echo measurement, ref. SNR: 30 dB)
%---------------------------------------------------------------------------------------------------------------
% a) all incident waves enabled both the accurate detection and the precise localization of the wires
All incident waves enabled both
% 1.) accurate detection
the accurate detection and
% 2.) precise localization
the precise localization of
% 3.) wires
the wires
(cf. \cref{fig:sim_study_obj_A_sr_1_images_kap}).
% b) spatial extents recovered by the convex l1-minimization method were smaller for the random waves than for the QPW
% article:Schiffner2018, Sect. VIII. Results / Sect. VIII-A. Wire Phantom / Sect. VIII-A.2) Point Spread Functions (subsubsec:results_phantom_wire_psf)
% - The random waves achieved \acp{FEHM} that were smaller than or equal to
%   those of the \ac{QPW} for all fixed positions, except those numbered $s \in \{ 6, 7 \}$ (cf. \cref{tab:sim_study_obj_A_sr_1_tpsf_fehm}).
The spatial extents recovered by
% 1.) convex l1-minimization method
the convex $\ell_{1}$-minimization method
\eqreflqmin{eqn:recovery_reg_norm_lq_minimization}{ 1 } were
% 2.) smaller
smaller for
% 3.) random waves
the random waves than for
% 4.) quasi-plane wave (QPW)
the \ac{QPW}.
% c) reductions were more pronounced for smaller axial coordinates and approximately followed the trend of the normalized differences in the FEHMs of the PSFs
% article:Schiffner2018, Sect. VIII. Results / Sect. VIII-A. Wire Phantom / Sect. VIII-A.2) Point Spread Functions (subsubsec:results_phantom_wire_psf)
% - The \acp{FEHM} generally increased with the axial coordinate of these positions.
% - The maximum normalized differences ranged from
%   \SI{23.5}{\percent} for the superposition of both randomly-apodized and randomly-delayed \acp{QCW} at the NINTH FIXED POSITION, i.e. $s = 9$, to
%   \SI{73.7}{\percent} for the superposition of randomly-delayed \acp{QCW} at the FIRST FIXED POSITION, i.e. $s = 1$.
These reductions were more pronounced for
% 1.) smaller axial coordinates
smaller axial coordinates and, thus, approximately followed
% 2.) trend
the trend of
% 3.) normalized differences in the FEHMs
% 1: QPW (max)              2: QPW (max)             3: QPW (max)         4: QPW (max)              5: QPW (max)              6: rnd. del. (max)   7: rnd. apo. (max)   8: QPW (max)              9: QPW = rnd. apo. (max)
%    rnd. del.:      73.68%    rnd. apo.:     50%       rnd. del.: 57.14%    rnd. del.:      54.17%    rnd. del.:      50%       Rnd. apo.: 45.45%    rnd. del.: 54.55%    rnd. apo. del.: 33.33%    rnd. apo. del.: 23.53%
%    rnd. apo. del.: 42.11%    rnd. apo. del: 16.67%    rnd. apo.: 33.33%    rnd. apo. del.: 20.83%    rnd. apo. del.: 17.86%    QPW:        3.03%    QPW:       13.63%    rnd. apo.:      14.29%    rnd. del.:      9.8%
the normalized differences in
the \acp{FEHM} of
% 4.) point spread functions (PSFs)
the \acp{PSF}
\eqref{eqn:cs_math_tpsf}
(cf. \cref{tab:sim_study_obj_A_sr_1_tpsf_fehm}).
% d) random waves caused less artifacts near the wires than the QPW
Moreover,
% 1.) random waves
the random waves caused less
% 2.) artifacts
artifacts near
% 3.) wires
the wires than
% 3.) quasi-plane wave (QPW)
the \ac{QPW}
(cf. inset images).
% e) both advantages agreed with the reduced numbers of components within the illustrated dynamic range relative to the QPW
Both advantages agreed with
% 1.) reduced numbers
the reduced numbers of
% 2.) components within the illustrated dynamic range
components within
the illustrated dynamic range relative to
% 2.) quasi-plane wave (QPW)
the \ac{QPW}.
% f) normalized differences ranged from 58.4 to 66.4 %
% MATLAB:
% indicator_qpw = sim_obj_A_data_spgl1_l1_qpw{5}.gamma_kappa_recon_dB >= -70;
% indicator_rnd_apo = sim_obj_A_data_spgl1_l1_rnd_apo{5}.gamma_kappa_recon_dB >= -70;
% indicator_rnd_del = sim_obj_A_data_spgl1_l1_rnd_del{5}.gamma_kappa_recon_dB >= -70;
% indicator_rnd_apo_del = sim_obj_A_data_spgl1_l1_rnd_apo_del{5}.gamma_kappa_recon_dB >= -70;
% sum( indicator_qpw(:) ) = 1070
% sum( indicator_rnd_apo(:) ) = 445 	=> 58.4112 %
% sum( indicator_rnd_del(:) ) = 360 	=> 66.3551 %
% sum( indicator_rnd_apo_del(:) ) = 417 => 61.0280 %
The normalized differences ranged from
% 1.) 58.4 %
\SI{58.4}{\percent} for
% 2.) superposition of randomly-apodized QCWs
the superposition of
randomly-apodized \acp{QCW} to
% 3.) 66.4 %
\SI{66.4}{\percent} for
% 4.) superposition of randomly-delayed QCWs
the superposition of
randomly-delayed \acp{QCW}.
% g) nonconvex l0.5-minimization method consistently recovered isolated significant components that strongly resembled
In contrast,
% 1.) nonconvex l0.5-minimization method
the nonconvex $\ell_{0.5}$-minimization method
\eqreflqmin{eqn:recovery_reg_norm_lq_minimization}{ 0.5 } consistently recovered
% 2.) isolated components
isolated components that matched
% 3.) specified vector stacking the regular samples in the discretized relative spatial fluctuations in the unperturbed compressibility
the specified compressibility fluctuations
\eqref{eqn:recovery_sys_lin_eq_gamma_kappa_bp_vector}.
% h) numbers of components within the illustrated dynamic range equaled the number of wires
% MATLAB:
% indicator_qpw = sim_obj_A_data_spgl1_lq_qpw{5}.gamma_kappa_recon_dB >= -70;
% indicator_rnd_apo = sim_obj_A_data_spgl1_lq_rnd_apo{5}.gamma_kappa_recon_dB >= -70;
% indicator_rnd_del = sim_obj_A_data_spgl1_lq_rnd_del{5}.gamma_kappa_recon_dB >= -70;
% indicator_rnd_apo_del = sim_obj_A_data_spgl1_lq_rnd_apo_del{5}.gamma_kappa_recon_dB >= -70;
% sum( indicator_qpw(:) ) = 21
% sum( indicator_rnd_apo(:) ) = 36	=> -71.4286 %
% sum( indicator_rnd_del(:) ) = 21	=> 0 %
% sum( indicator_rnd_apo_del(:) ) = 22	=> -4.7619 %
The numbers of
% 1.) components within the illustrated dynamic range
components within
the illustrated dynamic range equaled
% 2.) number of wires
the number of
% 3.) wires
wires.

%%%%%%%%%%%%%%%%%%%%%%%%%%%%%%%%%%%%%%%%%%%%%%%%%%%%%%%%%%%%%%%%%%%%%%%%%%%%%%%%%%%%%%%%%%%%%%%%%%%%%%%%%%%%%%%%
% graphic: statistics of the lq-minimization (mean SSIM indices, rel. RMSEs, and number of iterations vs. SNR)
%%%%%%%%%%%%%%%%%%%%%%%%%%%%%%%%%%%%%%%%%%%%%%%%%%%%%%%%%%%%%%%%%%%%%%%%%%%%%%%%%%%%%%%%%%%%%%%%%%%%%%%%%%%%%%%%
%
\begin{figure*}[t!]
 \centering%
  \input{results/object_A/kappa_only/sr_1/normalize_true/figures/latex/sim_study_obj_A_sr_1_mean_ssim_index_rel_rmse_N_iter_vs_snr.tex}
 \caption{}
 \label{fig:V}
\end{figure*}
C
{% a) figure illustrates the sample means and the sample standard deviations of the mean SSIM indices, the relative RMSEs, and the normalized numbers of iterations
 Sample means and
 sample standard deviations of
 % 1.) mean SSIM indices
 the mean \acf{SSIM}\acused{SSIM} indices and
 % 2.) relative RMSEs
 the relative \acfp{RMSE}\acused{RMSE} achieved by
 % 3.) recovered relative spatial fluctuations in the unperturbed compressibility
 the recovered compressibility fluctuations
 \eqref{eqn:recovery_reg_norm_lq_minimization_sol_mat_params} and
 % 4.) normalized numbers of iterations
 the normalized numbers of
 iterations in
 \ac{SPGL1}.
 % b) assignment of all incident waves and both parameters q to the columns and rows
 The assignment of
 all incident waves and
 both parameters
 $q \in \{ 0.5; 1 \}$ governing
 % 1.) sparsity-promoting lq-minimization method
 the sparsity-promoting $\ell_{q}$-minimization method
 \eqref{eqn:recovery_reg_norm_lq_minimization} to
 the columns and rows in
 this figure equals
 that in
 \cref{fig:sim_study_obj_A_sr_1_images_kap}.
 % c) dashed red lines indicate the reference SNR
 The dashed red lines indicate
 the reference \ac{SNR} selected for
 \cref{fig:sim_study_obj_A_sr_1_images_kap}.
 % d) reference value for normalization
 The maximum sample mean of
 $\num{1348.6}$ normalized
 the numbers of
 iterations.
}%
{sim_study_obj_A_sr_1_ssim_index_rel_rmse_N_iter_vs_snr_kap}

%---------------------------------------------------------------------------------------------------------------
% 2.) statistics of the image recovery (lq-minimization, single pulse-echo measurement, all ref. SNRs)
%---------------------------------------------------------------------------------------------------------------
% a) mean SSIM indices confirmed the excellent structural recovery
% article:Schiffner2018, Sect. VIII. Results / Sect. VIII-A. Wire Phantom / Sect. VIII-A.4) Recovery by lq-Minimization (subsubsec:results_phantom_wire_lq_minimization)
% - All incident waves enabled both the ACCURATE DETECTION and the PRECISE LOCALIZATION of the wires
%   (cf. \cref{fig:sim_study_obj_A_sr_1_images_kap}).
The mean \ac{SSIM} indices confirmed
% 1.) excellent structural recovery
the excellent structural recovery, whereas
% b) relative RMSEs revealed an increased sensitivity of the quantitative recovery using the random waves to the energy of the additive errors
the relative \acp{RMSE} revealed
% 1.) increased sensitivity
an increased sensitivity of
% 2.) quantitative recovery
the quantitative recovery using
% 3.) random waves
the random waves to
% 4.) energy
the energy of
% 5.) additive errors
the additive errors
(cf. \cref{fig:sim_study_obj_A_sr_1_ssim_index_rel_rmse_N_iter_vs_snr_kap}).
% c) all incident waves achieved mean SSIM indices close to unity and comparable trends in both the relative RMSEs and the numbers of iterations [l1-minimization]
All incident waves achieved
% 1.) mean SSIM indices close to unity
% MATLAB:
% SSIM_index_spgl1_l1_qpw(1,:)*1e2	   = 95.6015   94.9718   97.2248   97.7469   98.3861   99.0972
% SSIM_index_spgl1_l1_rnd_apo(1,:)*1e2	   = 88.4886   94.0421   93.8590   98.2373   98.9030   99.2924
% SSIM_index_spgl1_l1_rnd_del(1,:)*1e2	   = 95.2499   95.6733   96.3136   97.2284   98.8742   99.1607
% SSIM_index_spgl1_l1_rnd_apo_del(1,:)*1e2 = 96.1631   96.4825   96.0857   96.6423   98.6328   99.5759
% minimum: 88.4886 (rnd. apo. @ 3 dB)
% maximum: 98.9030 (rnd. apo. @ 30 dB)
mean \ac{SSIM} indices close to
unity and
% 2.) comparable trends
comparable trends in both
% 3.) relative RMSEs
the relative \acp{RMSE} and
% 4.) numbers of iterations
the numbers of
iterations for
% 5.) all reference SNRs
all reference \acp{SNR} and
% 6.) convex l1-minimization method
the convex $\ell_{1}$-minimization method
\eqreflqmin{eqn:recovery_reg_norm_lq_minimization}{ 1 }.
% d) sample means of the relative RMSEs decreased from at most 87.6 % to at least 34.1 % [l1-minimization]
% MATLAB:
% rel_RMSE_spgl1_l1_qpw(1,:)*1e2         = 85.3469   82.1552   58.9715   55.1754   47.6686   39.1278
% rel_RMSE_spgl1_l1_rnd_apo(1,:)*1e2	 = 87.5926   74.9386   73.5188   43.6958   36.4024   36.3388
% rel_RMSE_spgl1_l1_rnd_del(1,:)*1e2	 = 80.0539   69.9126   61.3011   43.5096   34.0714   40.0679
% rel_RMSE_spgl1_l1_rnd_apo_del(1,:)*1e2 = 74.5985   68.7890   61.5668   52.7007   40.0753   31.3463
% minimum: 34.0714 (rnd. del. @ 30 dB)
% maximum: 87.5926 (rnd. apo. @ 3 dB)
The sample means of
% 1.) relative RMSEs
the relative \acp{RMSE} decreased from
% 2.) at most 87.6 %
at most \SI{87.6}{\percent} for
% 3.) superposition of randomly-apodized QCWs
the superposition of
randomly-apodized \acp{QCW} at
% 4.) lowest reference SNR
the lowest reference \ac{SNR} to
% 5.) at least 34.1 %
at least \SI{34.1}{\percent} for
% 6.) superposition of randomly-delayed QCWs
the superposition of
randomly-delayed \acp{QCW} at
% 7.) highest reference SNR
the highest reference \ac{SNR}.
% e) sample means of the normalized numbers of iterations increased from at least 3.4 % to at most 28.3 % [l1-minimization]
% MATLAB:
% N_iter_max = 1348.6
% N_iter_spgl1_l1_qpw(1,:)*1e2/N_iter_max	  = 3.4406    3.8188    8.3494   15.4605   24.1436   39.7449
% N_iter_spgl1_l1_rnd_apo(1,:)*1e2/N_iter_max	  = 7.2149    7.0888    8.4680   12.8355   23.0535   38.1878
% N_iter_spgl1_l1_rnd_del(1,:)*1e2/N_iter_max	  = 4.7827    6.8071    7.7339   14.7560   25.8194   40.1898
% N_iter_spgl1_l1_rnd_apo_del(1,:)*1e2/N_iter_max = 6.5401    7.2297    9.5284   14.3111   28.3034   62.8059
% minimum:  3.4406 (QPW @ 3 dB)
% maximum: 28.3034 (rnd. apo. del. @ 30 dB)
Concurrently,
% 1.) sample means
the sample means of
% 2.) normalized numbers of iterations
the normalized numbers of
iterations increased from
% 3.) at least 3.4 %
at least \SI{3.4}{\percent} for
% 4.) quasi-plane wave (QPW)
the \ac{QPW} to
% 5.) at most 28.3 %
at most \SI{28.3}{\percent} for
% 6.) superposition of both randomly-apodized and randomly-delayed QCWs
the superposition of both
randomly-apodized and
randomly-delayed \acp{QCW}.
% f) sample standard deviations of the relative RMSEs exceeded those of the mean SSIM indices and the normalized numbers of iterations [l1-minimization]
% MATLAB:
% rel_RMSE_spgl1_l1_qpw(2,:)*1e2         = 2.4453    0.4627   14.5006   11.6007   15.9304         0
% rel_RMSE_spgl1_l1_rnd_apo(2,:)*1e2	 = 2.1477    2.0824    3.4309   10.2053    4.9393         0
% rel_RMSE_spgl1_l1_rnd_del(2,:)*1e2	 = 2.1170    4.3834    7.2373   17.8836    8.9579         0
% rel_RMSE_spgl1_l1_rnd_apo_del(2,:)*1e2 = 1.9915    1.1288    5.9226    8.8835    8.3409         0
% minimum: 0.4627 (QPW @ 6 dB)
% maximum: 17.8836 (rnd. del. @ 20 dB)
The sample standard deviations of
% 1.) relative RMSEs
the relative \acp{RMSE} exceeded
% 2.) sample standard deviations
those of
% 3.) mean SSIM indices
% MATLAB:
% SSIM_index_spgl1_l1_qpw(2,:)*1e2         = 2.0646    1.0759    1.5117    1.0335    0.7981         0
% SSIM_index_spgl1_l1_rnd_apo(2,:)*1e2	   = 3.5880    1.2707    1.8291    1.1469    0.2225         0
% SSIM_index_spgl1_l1_rnd_del(2,:)*1e2	   = 1.1922    1.1272    0.8742    2.6744    0.4605         0
% SSIM_index_spgl1_l1_rnd_apo_del(2,:)*1e2 = 1.2267    0.5903    1.7911    1.6325    0.7683         0
% minimum: 0.2225 (rnd. apo. @ 30 dB)
% maximum: 3.5880 (rnd. apo. @ 3 dB)
the mean \ac{SSIM} indices and
% 4.) normalized numbers of iterations
% MATLAB:
% N_iter_max = 1348.6
% N_iter_spgl1_l1_qpw(2,:)*1e2/N_iter_max	  = 0.8340    0.2645    1.8444    1.8935    2.2622         0
% N_iter_spgl1_l1_rnd_apo(2,:)*1e2/N_iter_max	  = 2.6984    0.6147    1.1228    1.2296    2.4624         0
% N_iter_spgl1_l1_rnd_del(2,:)*1e2/N_iter_max	  = 1.1652    1.2819    1.1766    1.8879    3.9669         0
% N_iter_spgl1_l1_rnd_apo_del(2,:)*1e2/N_iter_max = 0.6866    0.9701    1.2873    1.3484    3.1884         0
% minimum: 0.2645 (QPW @ 6 dB)
% maximum: 3.9669 (rnd. del. @ 30 dB)
the normalized numbers of
iterations for
% 5.) reference SNRs of 10, 20, and 30 dB
$\text{SNR}_{\text{dB}} \geq \SI{10}{\deci\bel}$.
% g) nonconvex l0.5-minimization method consistently improved both the mean SSIM indices and the relative RMSEs for all reference SNRs
The nonconvex $\ell_{0.5}$-minimization method
\eqreflqmin{eqn:recovery_reg_norm_lq_minimization}{ 0.5 } consistently improved both
% true for each single recovery!
% 1.) mean SSIM indices
% MATLAB:
% SSIM_index_spgl1_lq_qpw(1,:)*1e2         = 99.7325   99.8129   99.9100   99.9999  100.0000  100.0000
% SSIM_index_spgl1_lq_rnd_apo(1,:)*1e2	   = 98.7629   99.0758   99.3342   99.9890   99.9997  100.0000
% SSIM_index_spgl1_lq_rnd_del(1,:)*1e2	   = 99.4363   99.5682   99.7416   99.9975   99.9999  100.0000
% SSIM_index_spgl1_lq_rnd_apo_del(1,:)*1e2 = 99.3561   99.3499   99.6033   99.9839   99.9967   99.9985
the mean \ac{SSIM} indices and
% 2.) relative RMSEs
% MATLAB:
% rel_RMSE_spgl1_lq_qpw(1,:)*1e2         =  3.3658    1.9743    1.2019    0.5398    0.2027    0.5261
% rel_RMSE_spgl1_lq_rnd_apo(1,:)*1e2	 = 44.8397   34.3090   37.5912    2.2907    0.2474    0.6639
% rel_RMSE_spgl1_lq_rnd_del(1,:)*1e2	 = 31.0277   25.9121   14.8017    0.8747    0.2224    0.6272
% rel_RMSE_spgl1_lq_rnd_apo_del(1,:)*1e2 = 34.9563   31.3846   27.5957    2.6183    0.5178    0.7432
the relative \acp{RMSE} for
% 3.) all reference SNRs
all reference \acp{SNR}.
% h) random waves caused significantly larger relative RMSEs than the QPW at the low reference SNRs [l0.5-minimization]
The random waves, however, caused
% 1.) significantly larger relative RMSEs
significantly larger relative \acp{RMSE} than
% 2.) quasi-plane wave (QPW)
the \ac{QPW} at
% 3.) low reference SNRs of 3, 6, and 10 dB
the low reference \acp{SNR}, i.e.
$\text{SNR}_{\text{dB}} \in \setsymbol{Q} = \{ \SI{3}{\deci\bel}, \SI{6}{\deci\bel}, \SI{10}{\deci\bel} \}$, and
% 4.) superposition of randomly-apodized QCWs
the superposition of
randomly-apodized \acp{QCW} performed
% 5.) worst
worst.
% i) sample means of the normalized numbers of iterations increased [l0.5-minimization]
% MATLAB:
% N_iter_max = 1348.6
%
% N_iter_spgl1_lq_qpw(1,:)*1e2/N_iter_max	  = 21.4445   23.5207   36.3933   52.2913   75.7897   90.6125
% N_iter_spgl1_l1_qpw(1,:)*1e2/N_iter_max	  =  3.4406    3.8188    8.3494   15.4605   24.1436   39.7449
%
% N_iter_spgl1_lq_rnd_apo(1,:)*1e2/N_iter_max	  = 29.1932   33.8203   43.8974   57.6598   88.6846  116.1946
% N_iter_spgl1_l1_rnd_apo(1,:)*1e2/N_iter_max	  =  7.2149    7.0888    8.4680   12.8355   23.0535   38.1878
%
% N_iter_spgl1_lq_rnd_del(1,:)*1e2/N_iter_max	  = 25.6785   33.2938   37.1867   57.5337   89.4335  103.2923
% N_iter_spgl1_l1_rnd_del(1,:)*1e2/N_iter_max	  =  4.7827    6.8071    7.7339   14.7560   25.8194   40.1898
%
% N_iter_spgl1_lq_rnd_apo_del(1,:)*1e2/N_iter_max = 31.2695   34.9844   41.4578   64.7338  100.0000  133.1751
% N_iter_spgl1_l1_rnd_apo_del(1,:)*1e2/N_iter_max =  6.5401    7.2297    9.5284   14.3111   28.3034   62.8059
% 
% differences:
% QPW:			18.0039   19.7019   28.0439   36.8308   51.6462   50.8676
% rnd. apo.:		21.9783   26.7314   35.4293   44.8243   65.6310   78.0068
% rnd. del.:		20.8957   26.4867   29.4528   42.7777   63.6141   63.1025
% rnd. apo. del.:	24.7293   27.7547   31.9294   50.4227   71.6966   70.3693
% minimum: 18.0039 (QPW @ 3 dB)
% maximum: 71.6966 (rnd. apo. del. @ 30 dB)
The sample means of
% 1.) normalized numbers of iterations
the normalized numbers of
iterations increased significantly by
% 2.) at least 18 %
at least \SI{18}{\percent} for
% 3.) quasi-plane wave (QPW)
the \ac{QPW} at
% 4.) lowest reference SNR
the lowest reference \ac{SNR} to
% 5.) at most 71.7 %
at most \SI{71.7}{\percent} for
% 6.) superposition of both randomly-apodized and randomly-delayed QCWs
the superposition of both
randomly-apodized and
randomly-delayed \acp{QCW} at
% 7.) highest reference SNR
the highest reference \ac{SNR}.
% k) sample standard deviations of the relative RMSEs were reduced [l0.5-minimization]
% MATLAB:
% rel_RMSE_spgl1_lq_qpw(2,:)*1e2         = 0.9370    0.5312    0.4390    0.1265    0.0415         0
% rel_RMSE_spgl1_l1_qpw(2,:)*1e2         = 2.4453    0.4627   14.5006   11.6007   15.9304         0
%
% rel_RMSE_spgl1_lq_rnd_apo(2,:)*1e2	 = 3.2399    5.4849    2.5371    1.8768    0.0453         0
% rel_RMSE_spgl1_l1_rnd_apo(2,:)*1e2	 = 2.1477    2.0824    3.4309   10.2053    4.9393         0
%
% rel_RMSE_spgl1_lq_rnd_del(2,:)*1e2	 = 6.5740    4.0448    2.9058    0.4946    0.0586         0
% rel_RMSE_spgl1_l1_rnd_del(2,:)*1e2	 = 2.1170    4.3834    7.2373   17.8836    8.9579         0
%
% rel_RMSE_spgl1_lq_rnd_apo_del(2,:)*1e2 = 6.6002    4.5136    3.5523    2.8522    0.9964         0
% rel_RMSE_spgl1_l1_rnd_apo_del(2,:)*1e2 = 1.9915    1.1288    5.9226    8.8835    8.3409         0
In addition,
% 1.) sample standard deviations
the sample standard deviations of
% 2.) relative RMSEs
the relative \acp{RMSE} were reduced for
% 3.) reference SNRs of 10, 20, and 30 dB
$\text{SNR}_{\text{dB}} \geq \SI{10}{\deci\bel}$.

%%%%%%%%%%%%%%%%%%%%%%%%%%%%%%%%%%%%%%%%%%%%%%%%%%%%%%%%%%%%%%%%%%%%%%%%%%%%%%%%%%%%%%%%%%%%%%%%%%%%%%%%%%%%%%%%
% graphic: incident acoustic energies, recorded electric energies, and sample means of the relative RMSEs
%%%%%%%%%%%%%%%%%%%%%%%%%%%%%%%%%%%%%%%%%%%%%%%%%%%%%%%%%%%%%%%%%%%%%%%%%%%%%%%%%%%%%%%%%%%%%%%%%%%%%%%%%%%%%%%%
%
\begin{figure}[t!]
 \centering%
  \input{results/object_A/kappa_only/sr_1/normalize_true/figures/latex/sim_study_obj_A_sr_1_p_in_Phi_energy_rel_rec_error_SNR_3.tex}
 \caption{}
 \label{fig:V}
\end{figure}
C
{% a) figure illustrates the normalized incident acoustic energies, the normalized first Born approximations of the recorded electric energies, and the mean relative recovery errors
 % 1.) normalized incident acoustic energies (multiple pulse-echo measurements, multifrequent)
 Normalized incident acoustic energies
 \eqref{eqn:recovery_p_in_energy}
 (cf. \subref{fig:sim_study_obj_A_p_in_energy}),
 % 2.) normalized recorded electric energies in the pulse echoes (all pulse-echo measurements, multifrequent, all array elements)
 normalized recorded electric energies
 \eqref{eqn:recovery_reg_v_rx_born_trans_coef_energy}
 (cf. \subref{fig:sim_study_obj_A_v_rx_born_energy}), and
 % 3.) sample means of the relative RMSEs caused by the nonconvex l0.5-minimization method
 sample means of
 the relative \acfp{RMSE}\acused{RMSE} caused by
 % 4.) nonconvex l0.5-minimization method
 the nonconvex $\ell_{0.5}$-minimization method
 \eqreflqmin{eqn:recovery_reg_norm_lq_minimization}{ 0.5 }
 (cf. \subref{fig:sim_study_obj_A_sr_1_mean_rel_errors}) for
 all individual wires.
 % b) maximum energies normalized both types of energy
 The maximum energies normalized %, which were attained by
 % 1.) superposition of both randomly-apodized and randomly-delayed QCWs
 %the superposition of both
 %randomly-apodized and
 %randomly-delayed \aclp{QCW}\acused{QCW} for
 % 2.) wire with the index 3
 %the wire with
 %the index $3$, normalized
 both types of
 energy.
 % c) r_{2}-coordinate of each wire increased monotonically with its index
 The $r_{2}$-coordinate of
 each wire increased monotonically with
 its index.
 % d) reference SNR amounted to \text{SNR}_{\text{dB}} = \SI{3}{\deci\bel}
 The reference \ac{SNR} amounted to
 $\text{SNR}_{\text{dB}} = \SI{3}{\deci\bel}$.
}%
{sim_study_obj_A_sr_1_p_in_Phi_energy_rel_rec_error}

The variations in
% 1.) incident acoustic energies at a specified grid point (all pulse-echo measurements, multifrequent)
the incident acoustic energies
\eqref{eqn:recovery_p_in_energy} across
% 2.) isolated positions
the isolated positions of
% 3.) wires
the wires were negligible for
% 4.) quasi-plane wave (QPW)
the \ac{QPW} but
% 5.) strong and erratic
strong and
erratic for
% 6.) random waves
the random waves
(cf. \cref{fig:sim_study_obj_A_sr_1_p_in_Phi_energy_rel_rec_error}).
% b) dynamic ranges amounted to at least 7.6 dB for the superposition of randomly-delayed QCWs and at most 8.7 dB for the superposition of randomly-apodized QCWs
% 10*log10( max( p_incident_qpw_1.p_incident_pos_kappa_energy ) / min( p_incident_qpw_1.p_incident_pos_kappa_energy ) ) = 0.3104 dB
% 10*log10( max(p_incident_rnd_apo_1.p_incident_pos_kappa_energy) / min(p_incident_rnd_apo_1.p_incident_pos_kappa_energy) ) = 8.6924 dB
% 10*log10( max(p_incident_rnd_del_1.p_incident_pos_kappa_energy) / min(p_incident_rnd_del_1.p_incident_pos_kappa_energy) ) = 7.6395 dB
% 10*log10( max(p_incident_rnd_apo_del_1.p_incident_pos_kappa_energy) / min(p_incident_rnd_apo_del_1.p_incident_pos_kappa_energy) ) = 8.3245 dB
The dynamic ranges amounted to
% 1.) at least 7.6 dB
at least \SI{7.6}{\deci\bel} for
% 2.) superposition of randomly-delayed QCWs
the superposition of
randomly-delayed \acp{QCW} and
% 3.) at most 8.7 dB
at most \SI{8.7}{\deci\bel} for
% 4.) superposition of randomly-apodized QCWs
the superposition of
randomly-apodized \acp{QCW}.
% c) recorded electric energies strongly reflected these variations and generally decreased with increasing axial coordinates of the identical wires
The recorded electric energies
\eqref{eqn:recovery_reg_v_rx_born_trans_coef_energy} strongly reflected
% 1.) variations in the incident acoustic energies
these variations and generally decreased with
% 2.) increasing axial coordinates
increasing axial coordinates of
% 3.) identical wires
the identical wires.
% d) recorded electric energies reflected the SNRs of the corrupted RF voltage signals induced by the individual wires
Fixing
% 1.) energy
the energy of
% 2.) additive errors
the additive errors in
% 3.) linear algebraic system (all pulse-echo measurements, multifrequent, all array elements, additive errors)
the linear algebraic system
\eqref{eqn:recovery_reg_prob_general_obs_trans_coef_error},
% 4.) recorded electric energies in the pulse echoes (all pulse-echo measurements, multifrequent, all array elements)
they additionally reflected
% 5.) signal-to-noise ratios (SNRs)
the \acp{SNR} of
% 6.) vectors stacking the relevant Fourier coefficients of the recorded RF voltage signals (all pulse-echo measurements, multifrequent, all array elements)
the recorded \ac{RF} voltage signals
\eqref{eqn:recovery_sys_lin_eq_v_rx_born_all_f_all_in_v_rx} induced by
% 7.) individual wires
the individual wires.
% e) individual wires insonified by relatively low incident acoustic energies induced corrupted RF voltage signals of worse SNR than those insonified by relatively high incident acoustic energy
Those insonified by
% 1.) incident acoustic energies at a specified grid point (all pulse-echo measurements, multifrequent)
relatively low incident acoustic energies
\eqref{eqn:recovery_p_in_energy}, e.g.
% 2.) wires with the indices 6, 8, and 16
the wires with
the indices $6$, $8$, and $16$ for
% 3.) superposition of randomly-apodized QCWs
the superposition of
randomly-apodized \acp{QCW}
(cf. blue bars in \subref{fig:sim_study_obj_A_p_in_energy}), induced
% 4.) vectors stacking the relevant Fourier coefficients of the recorded RF voltage signals (all pulse-echo measurements, multifrequent, all array elements)
recorded \ac{RF} voltage signals
\eqref{eqn:recovery_sys_lin_eq_v_rx_born_all_f_all_in_v_rx} of
% 5.) worse SNR
worse \ac{SNR}
(cf. blue bars in \subref{fig:sim_study_obj_A_v_rx_born_energy}) than
% 6.) individual wires
those insonified by
% 7.) incident acoustic energies at a specified grid point (all pulse-echo measurements, multifrequent)
relatively high incident acoustic energies
\eqref{eqn:recovery_p_in_energy}, e.g.
% 8.) wires with the indices 3, 11, and 16
the wires with
the indices $3$, $11$, and $16$ for
% 9.) superposition of both randomly-apodized and randomly-delayed QCWs
the superposition of both
randomly-apodized and
randomly-delayed \acp{QCW}
(cf. gray bars in \subref{fig:sim_study_obj_A_p_in_energy} and \subref{fig:sim_study_obj_A_v_rx_born_energy}).
% f) variations in the SNRs induced variations in the mean relative RMSEs caused by the nonconvex l0.5-minimization method
These variations in
% 1.) signal-to-noise ratios (SNRs)
the \acp{SNR} induced
% 2.) variations
variations in
% 3.) mean relative RMSEs
the mean relative \acp{RMSE} caused by
% 4.) nonconvex l0.5-minimization method
the nonconvex $\ell_{0.5}$-minimization method
\eqreflqmin{eqn:recovery_reg_norm_lq_minimization}{ 0.5 }
(cf. \subref{fig:sim_study_obj_A_sr_1_mean_rel_errors}).

%%%%%%%%%%%%%%%%%%%%%%%%%%%%%%%%%%%%%%%%%%%%%%%%%%%%%%%%%%%%%%%%%%%%%%%%%%%%%%%%%%%%%%%%%%%%%%%%%%%%%%%%%%%%%%%%
% 2.) tissue-mimicking phantom
%%%%%%%%%%%%%%%%%%%%%%%%%%%%%%%%%%%%%%%%%%%%%%%%%%%%%%%%%%%%%%%%%%%%%%%%%%%%%%%%%%%%%%%%%%%%%%%%%%%%%%%%%%%%%%%%
\subsection{Tissue-Mimicking Phantom}
%\label{subsec:results_phantom_tissue}
%%%%%%%%%%%%%%%%%%%%%%%%%%%%%%%%%%%%%%%%%%%%%%%%%%%%%%%%%%%%%%%%%%%%%%%%%%%%%%%%%%%%%%%%%%%%%%%%%%%%%%%%%%%%%%%%
% 1.) recorded electric energies
%%%%%%%%%%%%%%%%%%%%%%%%%%%%%%%%%%%%%%%%%%%%%%%%%%%%%%%%%%%%%%%%%%%%%%%%%%%%%%%%%%%%%%%%%%%%%%%%%%%%%%%%%%%%%%%%
\subsubsection{Recorded Electric Energies}
%\label{subsubsec:results_phantom_tissue_energy_rx}
%%%%%%%%%%%%%%%%%%%%%%%%%%%%%%%%%%%%%%%%%%%%%%%%%%%%%%%%%%%%%%%%%%%%%%%%%%%%%%%%%%%%%%%%%%%%%%%%%%%%%%%%%%%%%%%%
% images: recorded electric energies in the pulse echoes (tissue-mimicking phantom, single wave emission)
%%%%%%%%%%%%%%%%%%%%%%%%%%%%%%%%%%%%%%%%%%%%%%%%%%%%%%%%%%%%%%%%%%%%%%%%%%%%%%%%%%%%%%%%%%%%%%%%%%%%%%%%%%%%%%%%
%
\begin{figure}[t!]
 \centering%
  \input{results/object_B/column_norms/figures/latex/sim_study_obj_B_norms_kappa.tex}
 \caption{}
 \label{fig:V}
\end{figure}
C
{% a) figure illustrates the recorded electric energies induced by all incident waves
 Recorded electric energies
 \eqref{eqn:recovery_reg_v_rx_born_trans_coef_energy} induced by
 % 1.) quasi-plane wave (QPW) (dynamic range: -77.7554 dB)
 the \acl{QPW}\acused{QPW}
 (cf. \subref{fig:sim_study_obj_B_norms_kappa_qpw}) and
 the superpositions of
 % 2.) superposition of randomly-apodized QCWs (dynamic range: -72.0162 dB)
 randomly-apodized \acfp{QCW}\acused{QCW}
 (cf. \subref{fig:sim_study_obj_B_norms_kappa_rnd_apo}),
 % 3.) superposition of randomly-delayed QCWs (dynamic range: -70.6122 dB)
 randomly-delayed \acp{QCW}
 (cf. \subref{fig:sim_study_obj_B_norms_kappa_rnd_del}), and both
 % 4.) superposition of both randomly-apodized and randomly-delayed QCWs (dynamic range: -73.9546 dB)
 randomly-apodized and
 randomly-delayed \acp{QCW}
 (cf. \subref{fig:sim_study_obj_B_norms_kappa_rnd_apo_del}).
 % b) green contours indicate the values -6 dB, -10 dB, and -20 dB
 The green contours indicate
 the values
 \SI{-6}{\deci\bel},
 \SI{-10}{\deci\bel}, and
 \SI{-20}{\deci\bel}.
}%
{sim_study_obj_B_norms_kappa}

%---------------------------------------------------------------------------------------------------------------
% 1.) recorded electric energies in the pulse echoes
%---------------------------------------------------------------------------------------------------------------
% a) transfer behaviors of the sensing matrices resembled those of bandpass filters suppressing relatively low and high spatial frequencies
% article:Schiffner2018, Sect. V. Image Recovery Based on Compressed Sensing / Sect. V-D. Regularization of the Inverse Scattering Problem (subsec:recovery_regularization)
% - In fact,
%   the density of population and
%   the DYNAMIC RANGE OF THE RECORDED ELECTRIC ENERGIES
%   [ E_{i}^{(\text{B})} = \dnorm{ \vect{a}_{i}\bigl[ p^{(\text{in})} \bigr] }{2}{1}^{2} ] (eqn:recovery_reg_v_rx_born_trans_coef_energy) for
%   all $i \in \setcons{ N_{\text{lat}} }$, which
%   CHARACTERIZE THE TRANSFER BEHAVIOR, vary significantly with the orthonormal basis.
The transfer behaviors of
% 1.) sensing matrices (all pulse-echo measurements, multifrequent, all array elements)
the sensing matrices
\eqref{eqn:recovery_reg_sensing_matrix} induced by
% 2.) all incident waves
all incident waves resembled those of
% 3.) bandpass filters suppressing relatively low and high spatial frequencies
bandpass filters suppressing
relatively low and high
spatial frequencies
(cf. \cref{fig:sim_study_obj_B_norms_kappa}).
% b) high dynamic ranges indicated the existence of structural building blocks whose pulse echoes contained relatively low electric energies
The high dynamic ranges exceeding
\SI{70}{\deci\bel} indicated
% 1.) existence of structural building blocks
the existence of
structural building blocks whose
% 2.) pulse echoes
pulse echoes contained
% 3.) recorded electric energies in the pulse echoes (all pulse-echo measurements, multifrequent, all array elements)
relatively low electric energies
\eqref{eqn:recovery_reg_v_rx_born_trans_coef_energy}.
% c) QPW induced relatively large electric energies exceeding -20 dB in a sickle-shaped passband inside the interval of normalized spatial frequencies [ -0.24; 0.24 ] x [ 0.15; 0.49 ]
The \ac{QPW} induced
% 1.) relatively large recorded electric energies in the pulse echoes (all pulse-echo measurements, multifrequent, all array elements)
relatively large electric energies
\eqref{eqn:recovery_reg_v_rx_born_trans_coef_energy} exceeding
% 2.) -20 dB
\SI{-20}{\deci\bel} in
% 3.) sickle-shaped passband inside the interval of normalized spatial frequencies [ -0.24; 0.24 ] x [ 0.15; 0.49 ]
a sickle-shaped passband inside
the interval of
normalized spatial frequencies
$\hat{\vect{K}} \in [ \num{-0.24}; \num{0.24} ] \times [ \num{0.15}; \num{0.49} ]$, whereas
% d) random waves induced those in arbelos-shaped passbands inside the intervals of normalized spatial frequencies [ -0.43; 0.43 ] x [ 0; 0.49 ]
the random waves induced those in
% 1.) arbelos-shaped passbands
arbelos-shaped passbands inside
% 2.) intervals of normalized spatial frequencies [ -0.43; 0.43 ] x [ 0; 0.49 ]
the intervals of
normalized spatial frequencies
$\hat{\vect{K}} \in [ \num{-0.43}; \num{0.43} ] \times [ \num{0}; \num{0.49} ]$.
% e) random waves significantly enlarged the passbands
%The latter waves thus significantly enlarged
%the passbands.
% e) all formed passbands strongly agreed with the predictions of the FDT
% TODO: move to discussion spectral supports
All formed passbands, which were significantly enlarged by
the latter waves, strongly agreed with
the predictions of
the \ac{FDT}
(cf. footnote $\num{1}$ in \cref{sec:introduction}).
% f) bounded FOV expanded the predicted passbands by beamlike artifacts
The bounded \ac{FOV}, however, expanded
the predicted passbands by
beamlike artifacts.
% g) beamlike artifacts re-entered the illustrations at the maximum normalized axial frequency of unity
Owing to
% 1.) periodicity of the DFT
the periodicity of
the \ac{DFT},
% 2.) beamlike artifacts
these artifacts re-entered
the illustrations at
% 3.) maximum normalized axial frequency of unity
the maximum normalized axial frequency of
unity, i.e.
$\hat{K}_{2} = 1$.
% h) absence of aliasing confirmed the specifications of sufficiently small constant spacings between the adjacent grid points in the FOV
The absence of
% 1.) aliasing
aliasing confirmed
% 2.) specifications
the specifications of
% 3.) sufficiently small constant spacings between the adjacent grid points in the FOV along both coordinate axes
sufficiently small constant spacings between
the adjacent grid points in
the \ac{FOV}.

%%%%%%%%%%%%%%%%%%%%%%%%%%%%%%%%%%%%%%%%%%%%%%%%%%%%%%%%%%%%%%%%%%%%%%%%%%%%%%%%%%%%%%%%%%%%%%%%%%%%%%%%%%%%%%%%
% 2.) transform point spread functions
%%%%%%%%%%%%%%%%%%%%%%%%%%%%%%%%%%%%%%%%%%%%%%%%%%%%%%%%%%%%%%%%%%%%%%%%%%%%%%%%%%%%%%%%%%%%%%%%%%%%%%%%%%%%%%%%
\subsubsection{Transform Point Spread Functions}
%\label{subsubsec:results_phantom_tissue_tpsf}
%%%%%%%%%%%%%%%%%%%%%%%%%%%%%%%%%%%%%%%%%%%%%%%%%%%%%%%%%%%%%%%%%%%%%%%%%%%%%%%%%%%%%%%%%%%%%%%%%%%%%%%%%%%%%%%%
% images: transform point spread functions (TPSFs, seventh reference spatial frequency)
%%%%%%%%%%%%%%%%%%%%%%%%%%%%%%%%%%%%%%%%%%%%%%%%%%%%%%%%%%%%%%%%%%%%%%%%%%%%%%%%%%%%%%%%%%%%%%%%%%%%%%%%%%%%%%%%
%
\begin{figure}[t!]
 \centering%
  \input{results/object_B/kappa_only/sr_tpsf/figures/latex/sim_study_obj_B_sr_1_tpsf_images.tex}
 \caption{}
 \label{fig:V}
\end{figure}
C
{% a) figure illustrates the absolute values of the TPSFs associated with the random sensing matrix and the sensing matrices induced by the GWN and all incident waves
 Absolute values of
 % 1.) transform point spread functions (TPSFs)
 the \aclp{TPSF}\acused{TPSF}
 \eqref{eqn:cs_math_tpsf} associated with
 % 2.) random sensing matrix (RIP)
 the random sensing matrix
 (cf. \subref{fig:sim_study_obj_B_sr_1_tpsf_images_rip_1}) and
 % 3.) sensing matrices (all pulse-echo measurements, multifrequent, all array elements)
 the sensing matrices
 \eqref{eqn:recovery_reg_sensing_matrix} induced by
 % 3.a) GWN
 the \acl{GWN}\acused{GWN}
 (cf. \subref{fig:sim_study_obj_B_sr_1_tpsf_images_blgwn_1}),
 % 3.b) quasi-plane wave (QPW)
 the \acl{QPW}\acused{QPW}
 (cf. \subref{fig:sim_study_obj_B_sr_1_tpsf_images_qpw_1}), and
 the superpositions of
 % 3.c) superposition of randomly-apodized QCWs
 randomly-apodized \acfp{QCW}\acused{QCW}
 (cf. \subref{fig:sim_study_obj_B_sr_1_tpsf_images_rnd_apo_1}),
 % 3.d) superposition of randomly-delayed QCWs
 randomly-delayed \acp{QCW}
 (cf. \subref{fig:sim_study_obj_B_sr_1_tpsf_images_rnd_del_1}), and both
 % 3.e) superposition of both randomly-apodized and randomly-delayed QCWs
 randomly-apodized and
 randomly-delayed \acp{QCW}
 (cf. \subref{fig:sim_study_obj_B_sr_1_tpsf_images_rnd_apo_del_1}).
 % b) green crosshairs indicate the seventh fixed normalized spatial frequency
 % index_tpsf = 7;
 % n_{2} = 97970
 % \hat{\vect{K}}_{ n_{2} } \approx \trans{ ( \num{-0.1270}, \num{0.3457} ) }
 The green crosshairs indicate
 the seventh fixed normalized spatial frequency.
 %$\hat{\vect{K}}_{ n_{2} } \approx \trans{ ( \num{-0.13}, \num{0.35} ) }$.
 % c) inset images magnify the regions indicated by the white squares
 The inset images magnify
 the regions indicated by
 the white squares.
}%
{sim_study_obj_B_sr_1_tpsf_images}

%---------------------------------------------------------------------------------------------------------------
% 1.) visual inspection of all computed TPSFs
%---------------------------------------------------------------------------------------------------------------
% a) all computed TPSFs correctly attained their maximum absolute values of unity at the fixed spatial frequencies
Although
% 1.) transform point spread functions (TPSFs)
all computed \acp{TPSF}
\eqref{eqn:cs_math_tpsf} correctly attained
% 2.) maximum absolute values of unity
their maximum absolute values of
unity at
% 3.) fixed spatial frequencies
the fixed spatial frequencies,
% b) all computed TPSFs differed in their behavior for the remaining spatial frequencies
they differed in
% 1.) behavior
their behavior for
% 2.) remaining spatial frequencies
the remaining spatial frequencies
(cf. \cref{fig:sim_study_obj_B_sr_1_tpsf_images}).
% c) both reference sensing matrices produced similar results as the random observation process
% article:Schiffner2018, Sect. VIII. Results / Sect. VIII-A. Wire Phantom / Sect. VIII-A.2) Point Spread Functions (subsubsec:results_phantom_wire_psf)
% - Both reference observation processes produced
%   RANDOM VALUES CLOSE TO ZERO THAT RENDERED THE MAXIMA SHARP AND ISOLATED.
% - The RANDOM OBSERVATION PROCESS \eqref{eqn:sim_study_params_ref_obs_proc_rip} UNIFORMLY DISTRIBUTED THESE VALUES over
%   the \ac{FOV}, whereas [...].
Both reference sensing matrices produced
% 1.) similar results
similar results as
% 2.) first reference observation process and its entries (RIP)
the random observation process
\eqref{eqn:sim_study_params_ref_obs_proc_rip}, i.e.
% 3.) uniformly distributed random values close to zero
uniformly distributed random values close to
zero that rendered
% 4.) sharp isolated maximum at the discrete position \vect{r}_{ n_{2} - 1 }
the maxima sharp and
isolated
(cf. \cref{fig:sim_study_obj_A_sr_1_tpsf_images_rip_1}).
% d) structured random sensing matrix significantly elongated these maxima along the \hat{K}_{2}-axis in addition to modest lateral extensions
The structured random sensing matrix
\eqref{eqn:sim_study_params_ref_sens_mat_gwn}, however, significantly elongated
% 1.) sharp isolated maxima
these maxima along
% 2.) \hat{K}_{2}-axis
the $\hat{K}_{2}$-axis in addition to
% 3.) modest lateral extensions
modest lateral extensions
(cf. inset image).
% e) sensing matrices induced by the random waves produced similar noise-like artifacts inside their passbands
The sensing matrices
\eqref{eqn:recovery_reg_sensing_matrix} induced by
% 1.) random waves
the random waves approximately maintained these
% 2.) extended maxima along the \hat{K}_{2}-axis with modest lateral extensions
maxima but confined
% 3.) similar noise-like artifacts
similar noise-like artifacts to
% 4.) passbands
their passbands.
% f) sensing matrix induced by the QPW formed smooth coherent sidelobes and indicated the presence of unspecified spatial frequencies by secondary isolated maxima
In contrast,
% 1.) sensing matrix (all pulse-echo measurements, multifrequent, all array elements)
the sensing matrix
\eqref{eqn:recovery_reg_sensing_matrix} induced by
% 2.) quasi-plane wave (QPW)
the \ac{QPW} formed
% 3.) smooth coherent sidelobes (lack noise-like features)
smooth coherent sidelobes, whose
% 4.) absolute values
absolute values lacked
% 5.) noise-like features
noise-like features, and indicated
% 6.) presence
the presence of
% 7.) unspecified spatial frequencies
unspecified spatial frequencies by
% 8.) secondary isolated maxima
secondary isolated maxima, e.g.
% 9.) absolute value of approximately -2.4 dB
% K_{max} = \trans{ ( \num{-0.127}, \num{0.3457} ) }, K_{secmax} = \trans{ ( \num{0.123}, \num{0.3496} ) } @ -2.394 dB
an absolute value of
approximately \SI{-2.4}{\deci\bel} at
% 10.) normalized spatial frequency \hat{\vect{K}} \approx \trans{ ( 0.12, 0.35 ) }
the normalized spatial frequency
$\hat{\vect{K}} \approx \trans{ ( \num{0.12}, \num{0.35} ) }$.

%%%%%%%%%%%%%%%%%%%%%%%%%%%%%%%%%%%%%%%%%%%%%%%%%%%%%%%%%%%%%%%%%%%%%%%%%%%%%%%%%%%%%%%%%%%%%%%%%%%%%%%%%%%%%%%%
% table: full extents at half maximum (all fixed spatial frequencies)
%%%%%%%%%%%%%%%%%%%%%%%%%%%%%%%%%%%%%%%%%%%%%%%%%%%%%%%%%%%%%%%%%%%%%%%%%%%%%%%%%%%%%%%%%%%%%%%%%%%%%%%%%%%%%%%%
\begin{table*}[tb]
 \centering
 \caption{%
  % a) table summarizes the FEHMs of the TPSFs associated with the sensing matrices induced by all incident waves
  Full extents at
  half maximum of
  % 1.) transform point spread functions (TPSFs)
  the \aclp{TPSF}
  \eqref{eqn:cs_math_tpsf} associated with
  % 2.) sensing matrices (all pulse-echo measurements, multifrequent, all array elements)
  the sensing matrices
  \eqref{eqn:recovery_reg_sensing_matrix} induced by
  % 3.) all incident waves
  all incident waves.
  % b) FEHMs were evaluated for nine uniformly distributed normalized spatial frequencies along the semicircle with the center \hat{\vect{K}}_{ \text{c} } = \trans{ ( 0, 25 ) } / 128 and the radius \hat{K}_{ \text{r} } = 101 / 512
  They were evaluated for
  % 1.) nine uniformly distributed normalized spatial frequencies
  nine uniformly distributed normalized spatial frequencies along
  % 2.) semicircle
  the semicircle with
  % 3.) center \hat{\vect{K}}_{ \text{c} } = \trans{ ( 0, 25 ) } / 128
  the center
  $\hat{\vect{K}}_{ \text{c} } = \trans{ ( 0, 25 ) } / 128$ and
  % 4.) radius \hat{K}_{ \text{r} } = 101 / 512
  the radius
  $\hat{K}_{ \text{r} } = 101 / 512$ and numbered from
  \numrange{1}{9} with
  increasing polar angle.
 }
 \label{tab:sim_study_obj_B_sr_1_tpsf_fehm}
 \begin{tabular}{%
  @{}%
  l%																01.) type of incident wave
  S[table-format=2.2,table-number-alignment = right,table-auto-round]%								02.) 1st fixed spatial frequency
  S[table-format=2.2,table-number-alignment = right,table-auto-round]%								03.) 2nd fixed spatial frequency
  S[table-format=2.2,table-number-alignment = right,table-auto-round]%								04.) 3rd fixed spatial frequency
  S[table-format=2.2,table-number-alignment = right,table-auto-round]%								05.) 4th fixed spatial frequency
  S[table-format=2.2,table-number-alignment = right,table-auto-round]%								06.) 5th fixed spatial frequency
  S[table-format=2.2,table-number-alignment = right,table-auto-round]%								07.) 6th fixed spatial frequency
  S[table-format=2.2,table-number-alignment = right,table-auto-round]%								08.) 7th fixed spatial frequency
  S[table-format=2.2,table-number-alignment = right,table-auto-round]%								09.) 8th fixed spatial frequency
  S[table-format=2.2,table-number-alignment = right,table-auto-round]%								10.) 9th fixed spatial frequency
  S[table-format=2.2(4),separate-uncertainty,table-align-uncertainty = true,table-number-alignment = right,table-auto-round]%	11.) sample mean & std. dev
  @{}%
 }
 \toprule
  \multicolumn{1}{@{}H}{\multirow{2}{*}{Incident wave}} &
  \multicolumn{10}{H@{}}{Full extent at half maximum ($10^{-6}$)}\\
  \cmidrule(l){2-11}
  &
  \multicolumn{1}{H}{1} &
  \multicolumn{1}{H}{2} &
  \multicolumn{1}{H}{3} &
  \multicolumn{1}{H}{4} &
  \multicolumn{1}{H}{5} &
  \multicolumn{1}{H}{6} &
  \multicolumn{1}{H}{7} &
  \multicolumn{1}{H}{8} &
  \multicolumn{1}{H}{9} &
  \multicolumn{1}{H@{}}{$\text{Sample mean} \pm \text{std. dev.}$}\\
  \cmidrule(r){1-1}\cmidrule(lr){2-2}\cmidrule(lr){3-3}\cmidrule(lr){4-4}\cmidrule(lr){5-5}
  \cmidrule(lr){6-6}\cmidrule(lr){7-7}\cmidrule(lr){8-8}\cmidrule(lr){9-9}\cmidrule(lr){10-10}\cmidrule(l){11-11}
 \addlinespace
  \acs{QPW}  & 64.85 & 26.70 & 22.89 & 30.52 & 3.81 & 30.52 & 22.89 & 26.70 & 64.85  & 32.64 \pm 19.92\\
Rnd. apo.  & 38.15 & 11.44 & 11.44 & 11.44 & 11.44 & 11.44 & 11.44 & 11.44 & 38.15  & 17.38 \pm 11.77\\
Rnd. del.  & 38.15 & 11.44 & 11.44 & 11.44 & 11.44 & 11.44 & 11.44 & 15.26 & 38.15  & 17.80 \pm 11.60\\
Rnd. apo. del.  & 38.15 & 11.44 & 11.44 & 11.44 & 11.44 & 11.44 & 11.44 & 11.44 & 38.15  & 17.38 \pm 11.77\\

 \addlinespace
 \bottomrule
 \end{tabular}
\end{table*}

%---------------------------------------------------------------------------------------------------------------
% 2.) full extents at half maximum (all fixed spatial frequencies)
%---------------------------------------------------------------------------------------------------------------
% a) random waves achieved smaller FEHMs than the QPW for all fixed spatial frequencies, except that numbered s = 5
% 1: QPW, 2: QPW, 3: QPW, 4: QPW, 5: rnd. ~, 6: QPW, 7: QPW, 8: QPW, 9: QPW
% 41.17%  57.15%  50.02%  62.52%  −200.26%   62.52%  50.02%  57.15%  41.17%
The random waves achieved
% 1.) smaller FEHMs
smaller \acp{FEHM} than
% 2.) quasi-plane wave (QPW)
the \ac{QPW} for
% 3.) all fixed spatial frequencies
all fixed spatial frequencies, except
% 4.) fixed spatial frequency number 5
that numbered
$s = 5$
(cf. \cref{tab:sim_study_obj_B_sr_1_tpsf_fehm}).
% b) random waves achieved identical FEHMs ranging from 11.44e-6 to 38.15e-6
These were identical for
% 1.) each fixed spatial frequency
each fixed spatial frequency, except
that numbered
$s = 8$, where
% c) superposition of randomly-delayed QCWs produced a slightly larger FEHM than the two remaining random waves
the superposition of
randomly-delayed \acp{QCW} produced
% 1.) slightly larger FEHM
a slightly larger \ac{FEHM} than
% 2.) other two random waves
the other two random waves.
% d) secondary maxima approximately doubled these FEHMs for all fixed spatial frequencies, except that numbered s = 5
% article:Schiffner2018, Sect. VIII. Results / Sect. VIII-B. Tissue-Mimicking Phantom / Sect. VIII-B.2) Transform Point Spread Functions (subsubsec:results_phantom_tissue_tpsf)
% - In contrast, the sensing matrix \eqref{eqn:recovery_reg_sensing_matrix} induced by the \ac{QPW} formed
%   smooth coherent sidelobes, whose absolute values lacked noise-like features, and
%   a SECONDARY ISOLATED MAXIMUM with
%   an absolute value of approximately \SI{-2.4}{\deci\bel} at
%   the normalized spatial frequency \hat{\vect{K}} \approx \trans{ ( 0.12, 0.35 ) }.
% - The latter erroneously indicated the presence of unspecified lateral frequencies and misguided
%   the sparsity-promoting $\ell_{q}$-minimization method \eqref{eqn:recovery_reg_norm_lq_minimization} for sufficiently large additive errors.
The secondary maxima, which were
erroneously formed by
% 1.) quasi-plane wave (QPW)
the \ac{QPW}
(cf. \cref{fig:sim_study_obj_B_sr_1_tpsf_images_qpw_1}), approximately doubled
% 2.) full extents at half maximum (FEHMs)
these \acp{FEHM} for
% 3.) all fixed spatial frequencies
all fixed spatial frequencies, except
that numbered
$s = 5$.
% e) maximum normalized differences ranged from 41.2 % (s \in \{ 1, 9 \}) to 62.5 % (s \in \{ 4, 6 \})
In fact,
% 1.) maximum normalized differences
the maximum normalized differences ranged from
% 2.) fixed spatial frequencies s \in \{ 1, 9 \}
\SI{41.2}{\percent} at
the fixed spatial frequencies
$s \in \{ 1, 9 \}$ to
% 2.) fixed spatial frequencies s \in \{ 4, 6 \}
\SI{62.5}{\percent} at
the fixed spatial frequencies
$s \in \{ 4, 6 \}$.
% f) FEHMs generally increase with the deviation of the polar angle from the axis of axial frequencies for all incident waves
% TODO: wirklich?
%They generally increase with
%the deviation of
%the polar angle from
%the axis of
%axial frequencies for
%all incident waves.
% g) mean FEHMs reflected these increases relative to the random waves
The mean \acp{FEHM} reflected
these increases relative to
the random waves.
% h) QPW outperformed the random waves by a normalized difference of 200 %
For $s = 5$,
% 1.) fixed normalized spatial frequency \hat{\vect{K}}_{ n_{2} } = \trans{ ( 0, 201 ) } / 512
the fixed normalized spatial frequency
$\hat{\vect{K}}_{ n_{2} } = \trans{ ( 0, 201 ) } / 512$ matched
% 2.) preferred direction of propagation
the preferred direction of
propagation of
% 3.) quasi-plane wave (QPW)
the \ac{QPW} that outperformed
% 4.) random waves
the random waves by
% 5.) normalized difference exceeding 200 %
a normalized difference exceeding
\SI{200}{\percent}.
% i) both reference sensing matrices consistently achieved the minimum FEHM of a two-dimensional normalized spatial frequency element for all fixed spatial frequencies
Both reference sensing matrices consistently achieved
% 1.) minimum FEHM
the minimum \ac{FEHM} of
% 2.) two-dimensional normalized spatial frequency element
a two-dimensional normalized spatial frequency element
$\Delta \hat{K} \approx \num{3.81e-6}$ for
% 3.) all fixed spatial frequencies
all fixed spatial frequencies.

%%%%%%%%%%%%%%%%%%%%%%%%%%%%%%%%%%%%%%%%%%%%%%%%%%%%%%%%%%%%%%%%%%%%%%%%%%%%%%%%%%%%%%%%%%%%%%%%%%%%%%%%%%%%%%%%
% empirical CDFs: transform point spread functions (TPSFs)
%%%%%%%%%%%%%%%%%%%%%%%%%%%%%%%%%%%%%%%%%%%%%%%%%%%%%%%%%%%%%%%%%%%%%%%%%%%%%%%%%%%%%%%%%%%%%%%%%%%%%%%%%%%%%%%%
%
\begin{figure}[t!]
 \centering%
  \input{results/object_B/kappa_only/sr_tpsf/figures/latex/sim_study_obj_B_sr_1_tpsf_ecdfs.tex}
 \caption{}
 \label{fig:V}
\end{figure}
C
{% a) figure illustrates the empirical CDFs of the TPSFs associated with both reference sensing matrices and the sensing matrices induced by all incident waves
 Empirical \acfp{CDF}\acused{CDF} of
 % 1.) transform point spread functions (TPSFs)
 the \aclp{TPSF}
 \eqref{eqn:cs_math_tpsf} associated with
 % 2.) reference sensing matrices
 both reference sensing matrices and
 % 3.) sensing matrices (all pulse-echo measurements, multifrequent, all array elements)
 the sensing matrices
 \eqref{eqn:recovery_reg_sensing_matrix} induced by
 % 4.) all incident waves
 all incident waves.
 % b) inset graphic magnifies the region indicated by the red rectangle
 The inset graphic magnifies
 the region indicated by
 the red rectangle.
}%
{sim_study_obj_B_sr_1_tpsf_ecdfs}

%---------------------------------------------------------------------------------------------------------------
% 3.) empirical CDFs (all fixed spatial frequencies)
%---------------------------------------------------------------------------------------------------------------
% a) empirical CDFs confirmed the beneficial properties of the random waves
The empirical \acp{CDF} confirmed
% 1.) beneficial properties
the beneficial properties of
% 2.) random waves
the random waves
(cf. \cref{fig:sim_study_obj_B_sr_1_tpsf_ecdfs}).
% b) random sensing matrix almost exclusively attained absolute values ranging from -70 to -33.2 dB
% RIP: -33.2029 dB
% RIP: 0.2835 % @ -70 dB
The random sensing matrix
\eqref{eqn:sim_study_params_ref_sens_mat_rip} almost exclusively attained
% 1.) absolute values ranging from -70 to -33.2 dB
absolute values ranging from
\SIrange{-70}{-33.2}{\deci\bel}.
% c) only approximately 0.3 % of the admissible spatial frequencies were attributed to smaller absolute values
Only approximately \SI{0.3}{\percent} of
% 1.) admissible spatial frequencies
the admissible spatial frequencies were attributed to
% 2.) smaller absolute values
smaller absolute values.
% d) structured random sensing matrix deviated marginally from this behavior
The structured random sensing matrix
\eqref{eqn:sim_study_params_ref_sens_mat_gwn} deviated
marginally from
this behavior.
% e) absolute values ranged from -70 dB to -7.78 dB
% BLGWN: -7.7835 dB
% BLGWN: 0.2585 % @ -70 dB
The absolute values, however, ranged from
\SIrange{-70}{-7.78}{\deci\bel}.
% f) increased dynamic range reflected the extended maxima
% article:Schiffner2018, Sect. VIII. Results / Sect. VIII-B. Tissue-Mimicking Phantom / Sect. VIII-B.2) Transform Point Spread Functions (subsubsec:results_phantom_tissue_tpsf)
% - The structured random sensing matrix \eqref{eqn:sim_study_params_ref_sens_mat_gwn}, however, significantly elongated
%   these maxima along the $\hat{K}_{2}$-axis in addition to modest lateral extensions (cf. inset image).
This increased dynamic range reflected
% 1.) extended maxima
the extended maxima
(cf. \cref{fig:sim_study_obj_B_sr_1_tpsf_images_blgwn_1}).
% g) sensing matrices induced by the random waves deviated in a stronger but almost identical fashion from both references
The sensing matrices
\eqref{eqn:recovery_reg_sensing_matrix} induced by
% 1.) random waves
the random waves deviated in
% 2.) stronger but almost identical fashion
a stronger but almost identical fashion from
% 3.) reference sensing matrices
both references.
% h) absolute values below -70 dB constituted approximately 20 % of the admissible spatial frequencies and those above this threshold form the remaining 80 %
% rnd. apo.: -0.4746 dB / rnd. del: -0.5178 dB / rnd. apo. del: -0.4665 dB
% rnd. apo.: 19.9779 % @ -70 dB / rnd. del: 21.1052 % @ -70 dB / rnd. apo. del: 21.5781 % @ -70 dB
The absolute values below
\SI{-70}{\deci\bel} constituted
% 1.) approximately 20 %
approximately \SI{20}{\percent} of
% 2.) admissible spatial frequencies
the admissible spatial frequencies and
% 3.) absolute values above -70 dB
those above this threshold, which reached up to
\SI{-0.47}{\deci\bel}, formed
% 4.) remaining 80 %
the remaining \SI{80}{\percent}.
% i) sensing matrix induced by the QPW deviated strongest from both references
Clearly,
% 1.) sensing matrix (all pulse-echo measurements, multifrequent, all array elements)
the sensing matrix
\eqref{eqn:recovery_reg_sensing_matrix} induced by
% 2.) quasi-plane wave (QPW)
the \ac{QPW} deviated
% 3.) strongest
strongest from
% 4.) reference sensing matrices
both references.
% j) absolute values ranging from -70 to -0.19 dB strongly concentrated on only 11 % of the admissible spatial frequencies and reflected both the distinctive sidelobes and the secondary maxima
% article:Schiffner2018, Sect. VIII. Results / Sect. VIII-B. Tissue-Mimicking Phantom / Sect. VIII-B.2) Transform Point Spread Functions (subsubsec:results_phantom_tissue_tpsf)
% - In contrast, the sensing matrix \eqref{eqn:recovery_reg_sensing_matrix} induced by the \ac{QPW} formed
%   smooth coherent sidelobes, whose absolute values lacked noise-like features, and
%   a SECONDARY ISOLATED MAXIMUM with
%   an absolute value of approximately \SI{-2.4}{\deci\bel} at
%   the normalized spatial frequency \hat{\vect{K}} \approx \trans{ ( 0.12, 0.35 ) }.
% QPW: -0.1869 dB
% QPW: 89.0438 % @ -70 dB
The absolute values ranging from
\SIrange{-70}{-0.19}{\deci\bel} strongly concentrated on
% 1.) only 11 %
only \SI{11}{\percent} of
% 2.) admissible spatial frequencies
the admissible spatial frequencies and reflected both
% 3.) distinctive sidelobes
the distinctive sidelobes and
% 4.) secondary maxima
the secondary maxima
(cf. \cref{fig:sim_study_obj_B_sr_1_tpsf_images_qpw_1}).

%%%%%%%%%%%%%%%%%%%%%%%%%%%%%%%%%%%%%%%%%%%%%%%%%%%%%%%%%%%%%%%%%%%%%%%%%%%%%%%%%%%%%%%%%%%%%%%%%%%%%%%%%%%%%%%%
% 3.) adjoint normalized sensing matrices
%%%%%%%%%%%%%%%%%%%%%%%%%%%%%%%%%%%%%%%%%%%%%%%%%%%%%%%%%%%%%%%%%%%%%%%%%%%%%%%%%%%%%%%%%%%%%%%%%%%%%%%%%%%%%%%%
\subsubsection{Adjoint Normalized Sensing Matrices}
%\label{subsubsec:results_phantom_tissue_adjoint}
%%%%%%%%%%%%%%%%%%%%%%%%%%%%%%%%%%%%%%%%%%%%%%%%%%%%%%%%%%%%%%%%%%%%%%%%%%%%%%%%%%%%%%%%%%%%%%%%%%%%%%%%%%%%%%%%
% spectra: tissue-mimicking phantom (adjoint, single wave emission)
%%%%%%%%%%%%%%%%%%%%%%%%%%%%%%%%%%%%%%%%%%%%%%%%%%%%%%%%%%%%%%%%%%%%%%%%%%%%%%%%%%%%%%%%%%%%%%%%%%%%%%%%%%%%%%%%
%
\begin{figure}[t!]
 \centering%
  \input{results/object_B/kappa_only/adj_1/normalize_true/figures/latex/sim_study_obj_B_adj_1_dft.tex}
 \caption{}
 \label{fig:V}
\end{figure}
C
{% a) figure displays the absolute values of the matrix-vector products between the adjoint normalized sensing matrix and the normalized recorded RF voltage signals
 Absolute values of
 the matrix-vector products between
 % 1.) adjoint normalized sensing matrix (all pulse-echo measurements, multifrequent, all array elements)
 the adjoint normalized sensing matrix
 \eqref{eqn:recon_reg_norm_sensing_matrix} and
 % 2.) normalized linear algebraic system (all pulse-echo measurements, multifrequent, all array elements, additive errors)
 the normalized recorded \ac{RF} voltage signals
 \eqref{eqn:recovery_reg_norm_obs_trans_coef_error} for
 % 3.) quasi-plane wave (QPW)
 the \acl{QPW}\acused{QPW}
 (cf. \subref{fig:sim_study_obj_B_adj_1_dft_qpw}) and
 the superpositions of
 % 4.) superposition of randomly-apodized QCWs
 randomly-apodized \acfp{QCW}\acused{QCW}
 (cf. \subref{fig:sim_study_obj_B_adj_1_dft_rnd_apo}),
 % 5.) superposition of randomly-delayed QCWs
 randomly-delayed \acp{QCW}
 (cf. \subref{fig:sim_study_obj_B_adj_1_dft_rnd_del}), and both
 % 6.) superposition of both randomly-apodized and randomly-delayed QCWs
 randomly-apodized and
 randomly-delayed \acp{QCW}
 (cf. \subref{fig:sim_study_obj_B_adj_1_dft_rnd_apo_del}).
 % b) green crosshairs indicate the specified spatial frequencies and mostly coincide with local maxima
 % mostly: QPW at \hat{\vect{K}} = \trans{ ( -0.2324, 0.2539 ) } -> no local maximum
 % rnd. apo. at \hat{\vect{K}} = \trans{ ( -0.1445, 0.2148 ) }, \trans{ ( 0.1465, 0.207 ) } -> no local maxima
 % rnd. del. at \hat{\vect{K}} = \trans{ ( -0.1445, 0.2148 ) } -> no local maximum
 % rnd. apo. del. at \hat{\vect{K}} = \trans{ ( 0.1465, 0.207 ) } -> no local maximum
 The green crosshairs indicate
 the specified spatial frequencies and mostly coincide with
 local maxima.
 % c) inset images magnify the regions indicated by the white squares
 The inset images magnify
 the regions indicated by
 the white squares.
 % d) reference SNR amounted to \text{SNR}_{\text{dB}} = \SI{10}{\deci\bel}
 The reference \ac{SNR} amounted to
 $\text{SNR}_{\text{dB}} = \SI{10}{\deci\bel}$.
}%
{sim_study_obj_B_adj_1_dft_kap}

%---------------------------------------------------------------------------------------------------------------
% 1.) significance of the adjoint normalized sensing matrices (adjoint, single wave emission)
%---------------------------------------------------------------------------------------------------------------
% a) all incident waves accurately detected at least 80 % of the specified spatial frequencies
All incident waves accurately detected
% 1.) at least 80 %
at least
\SI{80}{\percent} of
% 2.) specified spatial frequencies
the specified spatial frequencies
(cf. \cref{fig:sim_study_obj_B_adj_1_dft_kap}).
% b) QPW produced smooth coherent sidelobes and erroneously indicated the presence of multiple unspecified spatial frequencies by isolated local maxima
% article:Schiffner2018, Sect. VIII. Results / Sect. VIII-B. Tissue-Mimicking Phantom / Sect. VIII-B.2) Transform Point Spread Functions (subsubsec:results_phantom_tissue_tpsf)
% - In contrast, the sensing matrix \eqref{eqn:recovery_reg_sensing_matrix} induced by the \ac{QPW} formed
%   SMOOTH COHERENT SIDELOBES, whose absolute values lacked noise-like features, and
%   a SECONDARY ISOLATED MAXIMUM with
%   an absolute value of approximately \SI{-2.4}{\deci\bel} at
%   the normalized spatial frequency \hat{\vect{K}} \approx \trans{ ( 0.12, 0.35 ) }.
% - The latter ERRONEOUSLY INDICATED THE PRESENCE OF UNSPECIFIED LATERAL FREQUENCIES and MISGUIDED
%   THE SPARSITY-PROMOTING $\ell_{q}$-MINIMIZATION METHOD \eqref{eqn:recovery_reg_norm_lq_minimization} for SUFFICIENTLY LARGE ADDITIVE ERRORS.
The \ac{QPW}, which missed
% 1.) normalized spatial frequency \hat{\vect{K}} \approx \trans{ ( -0.23, 0.25 ) }
the normalized spatial frequency
$\hat{\vect{K}} \approx \trans{ ( \num{-0.23}, \num{0.25} ) }$, produced
% 2.) smooth coherent sidelobes (lack noise-like features)
smooth coherent sidelobes and erroneously indicated
% 3.) presence of multiple unspecified spatial frequencies
the presence of
multiple unspecified spatial frequencies by
% 4.) isolated local maxima
isolated local maxima.
% c) examples for local maxima
A pronounced local maximum was located at
% 1.) K_{1} = \trans{ ( -96, 126 ) } / 512 = \trans{ ( -0.1875, 0.2461 ) }
$\hat{\vect{K}}_{1} \approx \trans{ ( -0.19, 0.25 ) }$ and
% 2.) multiple smaller local maxima
% K_{2} = ( -0.1133, 0.4375 ), K_{3} = ( -0.0293, 0.4648 ), K_{4} = ( 0.01758, 0.4648 ), K_{5} = ( 0.07031, 0.4551 ), K_{6} = ( 0.1055, 0.2715 )
multiple smaller local maxima were located at
$\hat{\vect{K}}_{2} \approx \trans{ ( -0.11, 0.44 ) }$,
$\hat{\vect{K}}_{3} \approx \trans{ ( -0.03, 0.46 ) }$,
$\hat{\vect{K}}_{4} \approx \trans{ ( 0.02, 0.46 ) }$,
$\hat{\vect{K}}_{5} \approx \trans{ ( 0.07, 0.46 ) }$, and
$\hat{\vect{K}}_{6} \approx \trans{ ( 0.11, 0.27 ) }$.
% d) deviating local maxima erroneously indicated the presence of additional spatial frequencies in the object and misguided the sparsity-promoting lq-minimization method
These misguided
% 1.) sparsity-promoting lq-minimization method
the sparsity-promoting $\ell_{q}$-minimization method
\eqref{eqn:recovery_reg_norm_lq_minimization} for
% 2.) sufficiently large additive errors
sufficiently large additive errors.
% e) random waves substituted these coherent sidelobes and the isolated local maxima by a uniform distribution of noise-like features in the passbands
In contrast,
% 1.)
the random waves substituted both
% 2.) smooth coherent sidelobes (lack noise-like features)
these sidelobes and
% 3.) undesired local maxima
the undesired local maxima by
% 4.) similar noise-like artifacts
similar noise-like artifacts inside
% 4.) passbands
their passbands.
% f) similar noise-like artifacts facilitated the identification of the specified spatial frequencies
These facilitated
% 1.) identification
the identification of
% 2.) specified spatial frequencies
the specified spatial frequencies.

%%%%%%%%%%%%%%%%%%%%%%%%%%%%%%%%%%%%%%%%%%%%%%%%%%%%%%%%%%%%%%%%%%%%%%%%%%%%%%%%%%%%%%%%%%%%%%%%%%%%%%%%%%%%%%%%
% 4.) recovery by l2-minimization
%%%%%%%%%%%%%%%%%%%%%%%%%%%%%%%%%%%%%%%%%%%%%%%%%%%%%%%%%%%%%%%%%%%%%%%%%%%%%%%%%%%%%%%%%%%%%%%%%%%%%%%%%%%%%%%%
%\subsubsection{Recovery by $\ell_{2}$-Minimization}
%\label{subsubsec:results_phantom_tissue_l2_minimization}
%\input{results/results_phantom_tissue_l2_minimization.tex}

%%%%%%%%%%%%%%%%%%%%%%%%%%%%%%%%%%%%%%%%%%%%%%%%%%%%%%%%%%%%%%%%%%%%%%%%%%%%%%%%%%%%%%%%%%%%%%%%%%%%%%%%%%%%%%%%
% 5.) recovery by lq-minimization
%%%%%%%%%%%%%%%%%%%%%%%%%%%%%%%%%%%%%%%%%%%%%%%%%%%%%%%%%%%%%%%%%%%%%%%%%%%%%%%%%%%%%%%%%%%%%%%%%%%%%%%%%%%%%%%%
\subsubsection{Recovery by $\ell_{q}$-Minimization}
%\label{subsubsec:results_phantom_tissue_lq_minimization}
%%%%%%%%%%%%%%%%%%%%%%%%%%%%%%%%%%%%%%%%%%%%%%%%%%%%%%%%%%%%%%%%%%%%%%%%%%%%%%%%%%%%%%%%%%%%%%%%%%%%%%%%%%%%%%%%
% spectra and images: tissue-mimicking phantom (lq-minimization, single pulse-echo measurement, ref. SNR: 10 dB)
%%%%%%%%%%%%%%%%%%%%%%%%%%%%%%%%%%%%%%%%%%%%%%%%%%%%%%%%%%%%%%%%%%%%%%%%%%%%%%%%%%%%%%%%%%%%%%%%%%%%%%%%%%%%%%%%
%
\begin{figure*}[t!]
 \centering%
  \input{results/object_B/kappa_only/sr_1/normalize_true/figures/latex/sim_study_obj_B_sr_1_dft_images.tex}
 \caption{}
 \label{fig:V}
\end{figure*}
C
{% a) figure illustrates the absolute values of the recovered relative spatial fluctuations in the unperturbed compressibility and their nearly-sparse representations
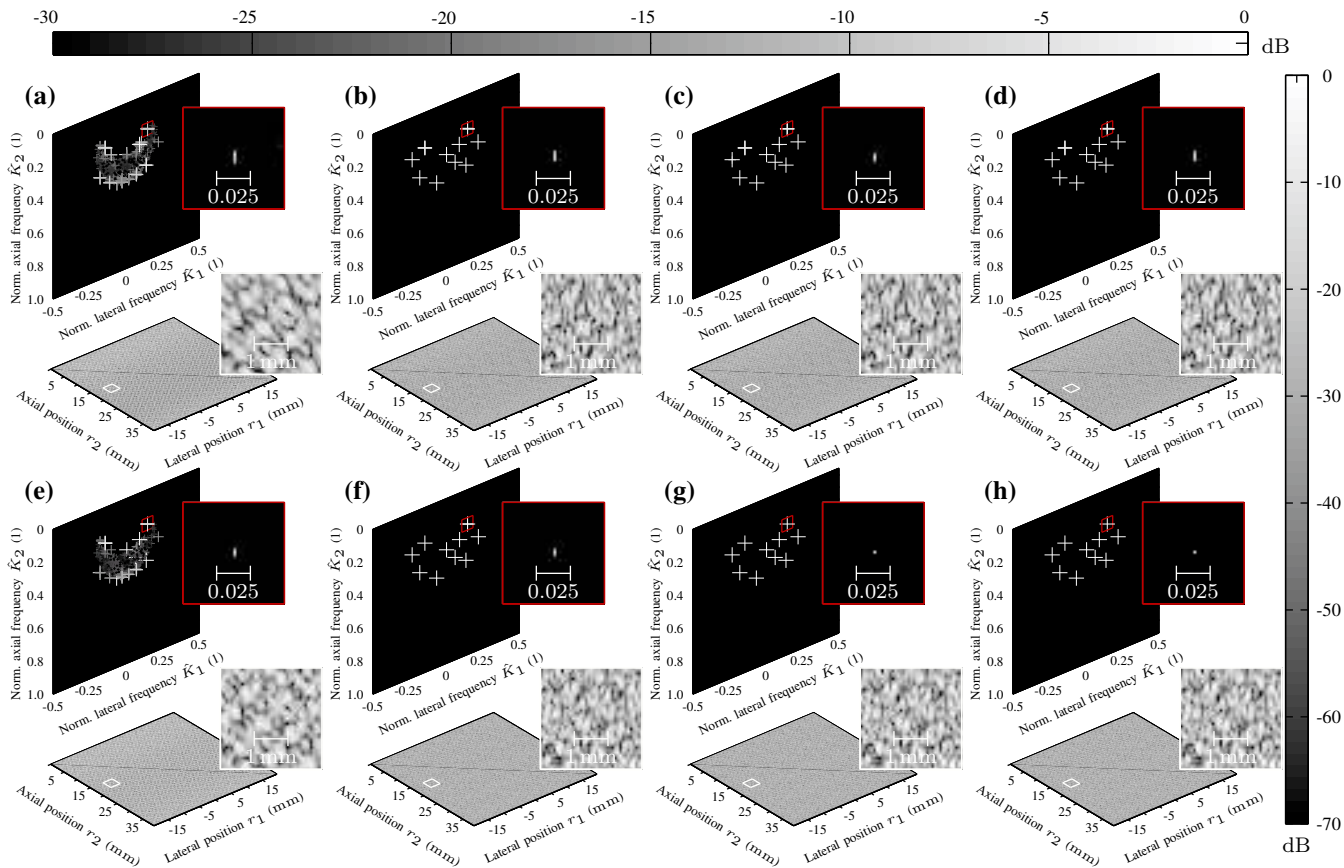
 Absolute values of
 % 1.) recovered relative spatial fluctuations in the unperturbed compressibility
 the recovered compressibility fluctuations
 \eqref{eqn:recovery_reg_norm_lq_minimization_sol_mat_params}
 (top colorbar; reclined) and
 % 2.) nearly-sparse representations
 their nearly-sparse representations
 \eqref{eqn:recovery_reg_sparse_representation}
 (right colorbar; upright) for
 % 3.) quasi-plane wave (QPW)
 the \acl{QPW}\acused{QPW}
 (cf.
 \subref{fig:sim_study_obj_B_sr_1_dft_images_spgl1_l1_qpw} and
 \subref{fig:sim_study_obj_B_sr_1_dft_images_spgl1_lq_qpw}%
 ) and
 the superpositions of
 % 4.) superposition of randomly-apodized QCWs
 randomly-apodized \acfp{QCW}\acused{QCW}
 (cf.
 \subref{fig:sim_study_obj_B_sr_1_dft_images_spgl1_l1_rnd_apo} and
 \subref{fig:sim_study_obj_B_sr_1_dft_images_spgl1_lq_rnd_apo}%
 ),
 % 5.) superposition of randomly-delayed QCWs
 randomly-delayed \acp{QCW}
 (cf.
 \subref{fig:sim_study_obj_B_sr_1_dft_images_spgl1_l1_rnd_del} and
 \subref{fig:sim_study_obj_B_sr_1_dft_images_spgl1_lq_rnd_del}%
 ), and both
 % 6.) superposition of both randomly-apodized and randomly-delayed QCWs
 randomly-apodized and
 randomly-delayed \acp{QCW}
 (cf.
 \subref{fig:sim_study_obj_B_sr_1_dft_images_spgl1_l1_rnd_apo_del} and
 \subref{fig:sim_study_obj_B_sr_1_dft_images_spgl1_lq_rnd_apo_del}%
 ).
 % b) top row shows the results of the convex l1-minimization method
 The top row
 (cf.
 \subref{fig:sim_study_obj_B_sr_1_dft_images_spgl1_l1_qpw} to
 \subref{fig:sim_study_obj_B_sr_1_dft_images_spgl1_l1_rnd_apo_del}%
 ) shows
 the results of
 % 1.) convex l1-minimization method
 the convex $\ell_{1}$-minimization method
 \eqreflqmin{eqn:recovery_reg_norm_lq_minimization}{ 1 }, whereas
 % c) bottom row shows the results of the nonconvex l0.5-minimization method
 the bottom row
 (cf.
 \subref{fig:sim_study_obj_B_sr_1_dft_images_spgl1_lq_qpw} to
 \subref{fig:sim_study_obj_B_sr_1_dft_images_spgl1_lq_rnd_apo_del}%
 ) shows
 those of
 % 1.) nonconvex l0.5-minimization method
 the nonconvex $\ell_{0.5}$-minimization method
 \eqreflqmin{eqn:recovery_reg_norm_lq_minimization}{ 0.5 }.
 % d) large images represent the nonzero absolute values by crosshairs of corresponding gray values and sizes
 The large upright images represent
 the nonzero absolute values by
 crosshairs of
 corresponding gray values and
 sizes, whereas
 % e) inset images exclusively use gray values to magnify the regions indicated by the white squares
 the remaining images exclusively use
 gray values.
 % f) inset images magnify the regions indicated by the red and white squares
 The inset images magnify
 the regions indicated by
 the red and white squares.
 % g) reference SNR amounted to \text{SNR}_{\text{dB}} = \SI{10}{\deci\bel}
 The reference \ac{SNR} amounted to
 $\text{SNR}_{\text{dB}} = \SI{10}{\deci\bel}$.
}%
{sim_study_obj_B_sr_1_dft_images_kap}

%---------------------------------------------------------------------------------------------------------------
% 1.) visual inspection of the recovered images (lq-minimization, single pulse-echo measurement, ref. SNR: 10 dB)
%---------------------------------------------------------------------------------------------------------------
% a) random waves enabled both the accurate detection and the precise determination of the specified spatial frequencies
The random waves enabled both
% 1.) accurate detection of the specified spatial frequencies
the accurate detection and
% 2.) precise determination of the specified spatial frequencies
the precise determination of
the specified spatial frequencies, whereas
% b) QPW underperformed
the \ac{QPW} underperformed
(cf. \cref{fig:sim_study_obj_B_sr_1_dft_images_kap}).
The latter specifically failed at detecting
% 1.) normalized spatial frequency near the edge of the passband
the normalized spatial frequency
$\hat{\vect{K}} \approx \trans{ ( -0.23, 0.25 ) }$ near
the edge of
the passband and erroneously indicated
% 2.) multiple unspecified spatial frequencies
multiple unspecified spatial frequencies.
% d) unspecified spatial frequencies coincided with the coherent sidelobes and the undesired local maxima
% article:Schiffner2018, Sect. VIII.B.3) Conjugate Transpose Normalized Sensing Matrices (subsubsec:results_phantom_tissue_adjoint)
% - 
% article:Schiffner2018, Sect. VIII.B.2) Transform Point Spread Functions (subsubsec:results_phantom_tissue_tpsf)
% - In contrast, the sensing matrix \eqref{eqn:recovery_reg_sensing_matrix} induced by the \ac{QPW} formed
%   SMOOTH COHERENT SIDELOBES, whose absolute values lacked noise-like features, and
%   a SECONDARY ISOLATED MAXIMUM with an absolute value of approximately \SI{-2.4}{\deci\bel} at
%   the normalized spatial frequency $\hat{\vect{K}} \approx \trans{ ( \num{0.12}, \num{0.35} ) }$.
% - This SECONDARY MAXIMUM ERRONEOUSLY INDICATED THE PRESENCE OF AN ADDITIONAL LATERAL FREQUENCY and misguided
%   the sparsity-promoting $\ell_{q}$-minimization method \eqref{eqn:recovery_reg_norm_lq_minimization} for additive errors of
%   relatively high energies.
These coincided with
% 1.) coherent sidelobes
the coherent sidelobes and
% 2.) undesired local maxima
the undesired local maxima at
% K_{2} = ( -0.1133, 0.4375 ), K_{3} = ( -0.0293, 0.4648 ), K_{4} = ( 0.01758, 0.4648 ), K_{5} = ( 0.07031, 0.4551 ), K_{6} = ( 0.1055, 0.2715 )
$\hat{\vect{K}}_{2}$ to
$\hat{\vect{K}}_{6}$
(cf. \cref{fig:sim_study_obj_B_adj_1_dft_qpw}).
% e) axial precision achieved by the random waves surpassed that of the QPW
% article:Schiffner2018, Sect. VIII.B.2) Transform Point Spread Functions (subsubsec:results_phantom_tissue_tpsf)
% - They [FEHMs] generally increase with the deviation of the polar angle from the axis of axial frequencies for
%   all incident waves.
Using
% 1.) convex l1-minimization method
the convex $\ell_{1}$-minimization method
\eqreflqmin{eqn:recovery_reg_norm_lq_minimization}{ 1 },
% 2.) axial extents recovered by the random waves
the axial extents recovered by
the random waves fell below
% 3.) axial extents recovered by the QPW
those recovered by
the \ac{QPW}
(cf. inset images).
% f) advantage was reflected by the normalized differences in the numbers of discrete spatial frequencies indicated within the illustrated dynamic range, which exceeded 98 %
% MATLAB:
% indicator_qpw = sim_obj_B_data_spgl1_l1_qpw{ 3 }.theta_kappa_recon_dB >= -70;
% indicator_rnd_apo = sim_obj_B_data_spgl1_l1_rnd_apo{ 3 }.theta_kappa_recon_dB >= -70;
% indicator_rnd_del = sim_obj_B_data_spgl1_l1_rnd_del{ 3 }.theta_kappa_recon_dB >= -70;
% indicator_rnd_apo_del = sim_obj_B_data_spgl1_l1_rnd_apo_del{ 3 }.theta_kappa_recon_dB >= -70;
% sum( indicator_qpw(:) ) = 1154
% sum( indicator_rnd_apo(:) ) = 15 	=> 98.7002 %
% sum( indicator_rnd_del(:) ) = 19 	=> 98.3536 %
% sum( indicator_rnd_apo_del(:) ) = 16	=> 98.6135 %
This advantage was reflected by
% 1.) normalized differences
the normalized differences in
% 2.) numbers of discrete spatial frequencies indicated within the illustrated dynamic range
the numbers of
discrete spatial frequencies indicated within
the illustrated dynamic range, which exceeded
\SI{98}{\percent}.
% g) nonconvex l0.5-minimization method consistently reduced both the axial extents and the number of unspecified spatial frequencies erroneously indicated by the QPW
The nonconvex $\ell_{0.5}$-minimization method
\eqreflqmin{eqn:recovery_reg_norm_lq_minimization}{ 0.5 } consistently reduced both
% 1.) axial extents
the axial extents recovered by
% 2.) all incident waves
all incident waves and
% 3.) number of unspecified spatial frequencies erroneously indicated by the QPW
the number of
unspecified spatial frequencies erroneously indicated by
% 4.) quasi-plane wave (QPW)
the \ac{QPW}.
% h) numbers of detected spatial frequencies within the illustrated dynamic range equaled the number of specified spatial frequencies
% MATLAB:
% indicator_qpw = sim_obj_B_data_spgl1_lq_qpw{ 3 }.theta_kappa_recon_dB >= -70;
% indicator_rnd_apo = sim_obj_B_data_spgl1_lq_rnd_apo{ 3 }.theta_kappa_recon_dB >= -70;
% indicator_rnd_del = sim_obj_B_data_spgl1_lq_rnd_del{ 3 }.theta_kappa_recon_dB >= -70;
% indicator_rnd_apo_del = sim_obj_B_data_spgl1_lq_rnd_apo_del{ 3 }.theta_kappa_recon_dB >= -70;
% sum( indicator_qpw(:) ) = 627
% sum( indicator_rnd_apo(:) ) = 12	=> 98.0861 %
% sum( indicator_rnd_del(:) ) = 10	=> 98.4051 %
% sum( indicator_rnd_apo_del(:) ) = 10	=> 98.4051 %
In fact,
% 1.) numbers of discrete spatial frequencies indicated within the illustrated dynamic range
the numbers of
discrete spatial frequencies indicated within
the illustrated dynamic range equaled
% 2.) number of specified spatial frequencies
the number of
specified spatial frequencies for both
% 3.) superpositions of QCWs using random time delays
superpositions of
\acp{QCW} using
random time delays, whereas
% i) superposition of randomly-apodized QCWs caused a negligible increase
the superposition of
randomly-apodized \acp{QCW} only caused
a negligible increase.
% j) recovered spatial frequencies strongly resembled the specified sparse representation
% TODO: seltsamer Satz...
The recovered spatial frequencies strongly resembled
% 1.) specified sparse representation
the specified sparse representation
\eqref{eqn:recovery_reg_sparse_representation}.
% k) QPW still overestimated the number of specified spatial frequencies
The \ac{QPW}, however, still overestimated
% 1.) number of specified spatial frequencies
the number of
specified spatial frequencies.
% l) aforementioned findings resulted in significantly different periodic patterns in the recovered relative spatial fluctuations in the unperturbed compressibility
The aforementioned findings resulted in
significantly different periodic patterns in
% 1.) recovered relative spatial fluctuations in the unperturbed compressibility
the recovered compressibility fluctuations
\eqref{eqn:recovery_reg_norm_lq_minimization_sol_mat_params}
(cf. reclined images and their insets).

%%%%%%%%%%%%%%%%%%%%%%%%%%%%%%%%%%%%%%%%%%%%%%%%%%%%%%%%%%%%%%%%%%%%%%%%%%%%%%%%%%%%%%%%%%%%%%%%%%%%%%%%%%%%%%%%
% graphic: statistics of the lq-minimization (mean SSIM indices, rel. RMSEs, and number of iterations vs. SNR)
%%%%%%%%%%%%%%%%%%%%%%%%%%%%%%%%%%%%%%%%%%%%%%%%%%%%%%%%%%%%%%%%%%%%%%%%%%%%%%%%%%%%%%%%%%%%%%%%%%%%%%%%%%%%%%%%
%
\begin{figure*}[t!]
 \centering%
  \input{results/object_B/kappa_only/sr_1/normalize_true/figures/latex/sim_study_obj_B_sr_1_mean_ssim_index_rel_rmse_N_iter_vs_snr.tex}
 \caption{}
 \label{fig:V}
\end{figure*}
C
{% a) figure illustrates the sample means and the sample standard deviations of the mean SSIM indices, the relative RMSEs, and the normalized numbers of iterations
 Sample means and
 sample standard deviations of
 % 1.) mean SSIM indices
 the mean \acf{SSIM}\acused{SSIM} indices and
 % 2.) relative RMSEs
 the relative \acfp{RMSE}\acused{RMSE} achieved by
 % 3.) recovered relative spatial fluctuations in the unperturbed compressibility
 the recovered compressibility fluctuations
 \eqref{eqn:recovery_reg_norm_lq_minimization_sol_mat_params} and
 % 4.) normalized numbers of iterations
 the normalized numbers of
 iterations in
 \ac{SPGL1}. 
 % b) assignment of all incident waves and both parameters q to the columns and rows
 The assignment of
 all incident waves and
 both parameters
 $q \in \{ 0.5; 1 \}$ governing
 % 1.) sparsity-promoting lq-minimization method
 the sparsity-promoting $\ell_{q}$-minimization method
 \eqref{eqn:recovery_reg_norm_lq_minimization} to
 the columns and rows in
 this figure equals
 that in
 \cref{fig:sim_study_obj_B_sr_1_dft_images_kap}.
 % c) dashed red lines indicate the reference SNR
 The dashed red lines indicate
 the reference \ac{SNR} selected for
 \cref{fig:sim_study_obj_B_sr_1_dft_images_kap}.
 % d) reference value for normalization
 The maximum sample mean of
 $\num{1127.6}$ normalized
 the numbers of
 iterations.
}%
{sim_study_obj_B_sr_1_ssim_index_rel_rmse_N_iter_vs_snr_kap}

%---------------------------------------------------------------------------------------------------------------
% 2.) statistics of the image recovery (lq-minimization, single pulse-echo measurement, all ref. SNRs)
%---------------------------------------------------------------------------------------------------------------
% a) all incident waves  mean SSIM indices close to unity and comparable trends in the relative RMSEs
Both
% 1.) mean SSIM indices
the mean \ac{SSIM} indices and
% 2.) relative RMSEs
the relative \acp{RMSE} were consistent for
% 3.) random waves
the random waves, but indicated
% 4.) increased sensitivity
an increased sensitivity of
the recovery based on
% 5.) quasi-plane wave (QPW)
the \ac{QPW} to
% 6.) energy
the energy of
% 7.) additive errors
the additive errors
(cf. \cref{fig:sim_study_obj_B_sr_1_ssim_index_rel_rmse_N_iter_vs_snr_kap}).
% b) random waves produced comparable trends in the mean SSIM indices and the relative RMSEs for all reference SNRs and the convex l1-minimization method
Using
% 1.) convex l1-minimization method
the convex $\ell_{1}$-minimization method
\eqreflqmin{eqn:recovery_reg_norm_lq_minimization}{ 1 },
% 2.) random waves
the random waves achieved
% 3.) mean SSIM indices
% MATLAB:
% SSIM_index_spgl1_l1_qpw(1,:)*1e2	   = 36.6000   41.2488   46.4364   90.4708   99.9325   99.9368
% SSIM_index_spgl1_l1_rnd_apo(1,:)*1e2	   = 80.6549   81.5283   83.8071   86.1310   86.2028   86.1152
% SSIM_index_spgl1_l1_rnd_del(1,:)*1e2	   = 80.2673   82.3483   84.2717   86.3884   90.6166   78.3656
% SSIM_index_spgl1_l1_rnd_apo_del(1,:)*1e2 = 80.2278   81.4986   85.3900   85.2345   86.2158   80.0985
% minimum: 80.2278 (rnd. apo. del. @ 3 dB)
% maximum: 90.6166 (rnd. del. @ 30 dB)
%
% differences to QPW:
% rnd. apo.:		44.0549   40.2794   37.3707   -4.3397  -13.7297  -13.8217
% rnd. del.:		43.6673   41.0994   37.8353   -4.0823   -9.3159  -21.5713
% rnd. apo. del.:	43.6279   40.2498   38.9537   -5.2362  -13.7167  -19.8383
% minimum: -13.7297 (rnd. apo. @ 30 dB)
% maximum:  44.0549 (rnd. apo. @ 3 dB)
mean \ac{SSIM} indices increasing from
\SIrange{80.2}{90.6}{\percent},
% 4.) relative RMSEs
% MATLAB:
% rel_RMSE_spgl1_l1_qpw(1,:)*1e2	 = 74.9216   69.2613   63.4289   20.9912    1.3762    2.0671
% rel_RMSE_spgl1_l1_rnd_apo(1,:)*1e2	 = 32.8928   31.8203   29.1923   26.2924   26.1477   26.3357
% rel_RMSE_spgl1_l1_rnd_del(1,:)*1e2	 = 33.2994   31.0622   28.7238   26.0569   20.5849   36.2633
% rel_RMSE_spgl1_l1_rnd_apo_del(1,:)*1e2 = 33.2734   31.8067   26.8410   27.4907   26.2436   32.9772
% minimum: 20.5849 (rnd. del. @ 30 dB)
% maximum: 33.2994 (rnd. del. @ 3 dB)
%
% differences to QPW:
% rnd. apo.:		42.0289   37.4409   34.2366   -5.3012  -24.7715  -24.2686
% rnd. del.:		41.6222   38.1990   34.7052   -5.0657  -19.2087  -34.1962
% rnd. apo. del.:	41.6482   37.4546   36.5880   -6.4995  -24.8674  -30.9101
% minimum: -24.8674 (rnd. apo. del. @ 30 dB)
% maximum:  42.0289 (rnd. apo. @ 3 dB)
relative \acp{RMSE} decreasing from
\SIrange{33.3}{20.6}{\percent}, and
% 5.) comparable trends
comparable trends in
% 6.) normalized numbers of iterations
the normalized numbers of
iterations for
% 7.) all reference SNRs
all reference \acp{SNR}.
% c) sample means of the normalized numbers of iterations increased from at least 4.64 % to at most 38.17 % [l1-minimization]
% MATLAB:
% N_iter_max = 1127.6
% N_iter_spgl1_l1_qpw(1,:)*1e2/N_iter_max	  = 6.4473    7.8131    9.3650   11.3427   13.1607   20.9294
% N_iter_spgl1_l1_rnd_apo(1,:)*1e2/N_iter_max	  = 5.3210    5.9507    7.3430   11.3249   27.1639   44.3420
% N_iter_spgl1_l1_rnd_del(1,:)*1e2/N_iter_max	  = 5.4363    5.6669    6.6690    9.9858   22.8716   88.6839
% N_iter_spgl1_l1_rnd_apo_del(1,:)*1e2/N_iter_max = 5.2412    5.7201    6.9262    9.6932   25.2128   88.6839
% minimum:  5.2412 (rnd. apo. del. @ 3 dB)
% maximum: 27.1639 (rnd. apo. @ 30 dB)
%
% differences to QPW:
% rnd. apo.:		1.1263    1.8624    2.0220    0.0177  -14.0032  -23.4126
% rnd. del.:		1.0110    2.1462    2.6960    1.3569   -9.7109  -67.7545
% rnd. apo. del.:	1.2061    2.0929    2.4388    1.6495  -12.0521  -67.7545
% minimum: -14.0032 (rnd. apo. @ 30 dB)
% maximum:   2.6960 (rnd. del. @ 10 dB)
The sample means of
% 1.) normalized numbers of iterations
the latter increased from
% 2.) at least 5.2 %
at least \SI{5.2}{\percent} for
% 3.) superposition of both randomly-apodized and randomly-delayed QCWs
the superposition of both
randomly-apodized and
randomly-delayed \acp{QCW} at
% 4.) lowest reference SNR
the lowest reference \ac{SNR} to
% 5.) at most 27.2 %
at most \SI{27.2}{\percent} for
% 6.) superposition of randomly-apodized QCWs
the superposition of
randomly-apodized \acp{QCW} at
% 7.) superposition of randomly-apodized QCWs
the highest reference \ac{SNR}.
% d) QPW produced significantly worse mean SSIM indices and relative RMSEs for the low reference SNRs  [l1-minimization]
In contrast,
% 1.) quasi-plane wave (QPW)
the \ac{QPW} produced
significantly worse
% 2.) mean SSIM indices
mean \ac{SSIM} indices and
% 3.) relative RMSEs
relative \acp{RMSE} for
% 4.) low reference SNRs of 3, 6, and 10 dB
the low reference \acp{SNR}, i.e.
$\text{SNR}_{\text{dB}} \in \setsymbol{Q}$.
% e) sample means of the mean SSIM indices increased from 36.6 to 46.4 %  [l1-minimization]
The sample means of
% 1.) mean SSIM indices
% MATLAB:
% SSIM_index_spgl1_l1_qpw(1,:)*1e2 = 36.6000   41.2488   46.4364   90.4708   99.9325   99.9368
the former metric increased from
\SIrange{36.6}{46.4}{\percent}, whereas
% f) sample means of the relative RMSEs decreased from 74.9 to 63.4 %  [l1-minimization]
% MATLAB:
% rel_RMSE_spgl1_l1_qpw(1,:)*1e2 = 74.9216   69.2613   63.4289   20.9912    1.3762    2.0671
those of
% 2.) relative RMSEs
the latter metric decreased from
\SIrange{74.9}{63.4}{\percent}.
% g) both metrics improved significantly and attained values up to 99.9 and 1.4 %  [l1-minimization]
For
% 1.) high reference SNRs of 20 and 30 dB
the high reference \acp{SNR}, i.e.
$\text{SNR}_{\text{dB}} \in \{ \SI{20}{\deci\bel}, \SI{30}{\deci\bel} \}$, however,
% 2.) mean SSIM indices and relative RMSEs
both metrics improved significantly and attained
values up to
% 3.) 99.9 %
\SI{99.9}{\percent} and down to
% 4.) 1.4 %
\SI{1.4}{\percent},
respectively.
% h) sample means of the normalized numbers of iterations increased from 7.27 % to 23.6 %  [l1-minimization]
% MATLAB:
% N_iter_max = 1127.6
% N_iter_spgl1_l1_qpw(1,:)*1e2/N_iter_max = 6.4473    7.8131    9.3650   11.3427   13.1607   20.9294
The sample means of
% 1.) normalized numbers of iterations
the normalized numbers of
iterations increased from
% 2.) 6.4 %
\SI{6.4}{\percent} at
% 3.) lowest reference SNR
the lowest reference \ac{SNR} to
% 4.) 13.1 %
\SI{13.1}{\percent} at
% 5.) highest reference SNR
the highest reference \ac{SNR}.
Using
% 1.) nonconvex l0.5-minimization method
the nonconvex $\ell_{0.5}$-minimization method
\eqreflqmin{eqn:recovery_reg_norm_lq_minimization}{ 0.5 },
% 2.) random waves
the random waves consistently achieved
% 3.) mean SSIM indices close to unity
% MATLAB:
% SSIM_index_spgl1_lq_qpw(1,:)*1e2	   = 43.6973   49.3404   56.1582   93.0138   99.8934   99.8323
% SSIM_index_spgl1_lq_rnd_apo(1,:)*1e2	   = 99.1838   99.7243   99.8458   99.9954   99.9958   99.9934
% SSIM_index_spgl1_lq_rnd_del(1,:)*1e2	   = 98.7991   99.7009   99.8962   99.9935   99.9957   99.9928
% SSIM_index_spgl1_lq_rnd_apo_del(1,:)*1e2 = 98.6458   99.6973   99.8945   99.9930   99.9962   99.9940
% minimum: 98.6458 (rnd. apo. del. @ 3 dB)
% maximum: 99.9962 (rnd. apo. del. @ 30 dB)
%
% differences to QPW:
% rnd. apo.:		55.4865   50.3839   43.6877    6.9816    0.1024    0.1611
% rnd. del.:		55.1018   50.3605   43.7381    6.9797    0.1023    0.1605
% rnd. apo. del.:	54.9485   50.3569   43.7363    6.9792    0.1029    0.1618
% minimum:  0.1023 (rnd. del. @ 30 dB)
% maximum: 55.4865 (rnd. apo. @ 3 dB)
mean \ac{SSIM} indices close to
unity and
% 4.) relative RMSEs below 6.9 %
% MATLAB:
% rel_RMSE_spgl1_lq_qpw(1,:)*1e2         = 65.8391   60.6768   54.3489   16.3019    1.4052    2.8359
% rel_RMSE_spgl1_lq_rnd_apo(1,:)*1e2	 =  5.3555    3.0675    2.0639    0.4661    0.3106    0.6444
% rel_RMSE_spgl1_lq_rnd_del(1,:)*1e2	 =  6.5540    3.2644    1.7489    0.5210    0.3171    0.6548
% rel_RMSE_spgl1_lq_rnd_apo_del(1,:)*1e2 =  6.8596    3.2637    1.7899    0.5512    0.2990    0.6206
% minimum: 0.2990 (rnd. apo. del. @ 30 dB)
% maximum: 6.8596 (rnd. apo. del. @ 3 dB)
%
% differences to QPW:
% rnd. apo.:		60.4836   57.6093   52.2851   15.8357    1.0946    2.1916
% rnd. del.:		59.2850   57.4124   52.6000   15.7809    1.0882    2.1811
% rnd. apo. del.:	58.9795   57.4131   52.5590   15.7507    1.1062    2.2153
% minimum:  1.0882 (rnd. del. @ 30 dB)
% maximum: 60.4836 (rnd. apo. @ 3 dB)
relative \acp{RMSE} below
\SI{6.9}{\percent} for
% 5.) all reference SNRs
all reference \acp{SNR}.
% i) sample means of the normalized numbers of iterations increased from at least 23 to at most 72.8 % [l0.5-minimization]
% MATLAB:
% N_iter_max = 1127.6
%
% N_iter_spgl1_lq_qpw(1,:)*1e2/N_iter_max	  = 36.4580   45.4949   54.5938   62.7527   66.1848   87.7971
% N_iter_spgl1_l1_qpw(1,:)*1e2/N_iter_max	  =  6.4473    7.8131    9.3650   11.3427   13.1607   20.9294
%
% N_iter_spgl1_lq_rnd_apo(1,:)*1e2/N_iter_max	  = 29.4342   29.1327   33.0259   45.8762  100.0000  111.9191
% N_iter_spgl1_l1_rnd_apo(1,:)*1e2/N_iter_max	  =  5.3210    5.9507    7.3430   11.3249   27.1639   44.3420
%
% N_iter_spgl1_lq_rnd_del(1,:)*1e2/N_iter_max	  = 29.5495   28.6360   31.7311   42.7279   88.5154  167.0805
% N_iter_spgl1_l1_rnd_del(1,:)*1e2/N_iter_max	  =  5.4363    5.6669    6.6690    9.9858   22.8716   88.6839
%
% N_iter_spgl1_lq_rnd_apo_del(1,:)*1e2/N_iter_max = 30.0550   29.5938   33.0702   41.0518   94.5371  157.5913
% N_iter_spgl1_l1_rnd_apo_del(1,:)*1e2/N_iter_max =  5.2412    5.7201    6.9262    9.6932   25.2128   88.6839
%
% differences:
% QPW:			30.0106   37.6818   45.2288   51.4101   53.0241   66.8677
% rnd. apo.:		24.1132   23.1820   25.6829   34.5513   72.8361   67.5772
% rnd. del.:		24.1132   22.9691   25.0621   32.7421   65.6438   78.3966
% rnd. apo. del.:	24.8138   23.8737   26.1440   31.3586   69.3242   68.9074
% minimum: 22.9691 (rnd. del. @ 6 dB)
% maximum: 72.8361 (rnd. apo. @ 30 dB)
The sample means of
% 1.) normalized numbers of iterations
the normalized numbers of
iterations increased significantly by
% 2.) at least 23 %
at least \SI{23}{\percent} for
% 3.) superposition of randomly-delayed QCWs
the superposition of
randomly-delayed \acp{QCW} at
% 4.) 6 dB
$\text{SNR}_{\text{dB}} = \SI{6}{\deci\bel}$ to
% 5.) at most 72.8 %
at most \SI{72.8}{\percent} for
% 6.) superposition of randomly-apodized QCWs
the superposition of
randomly-apodized \acp{QCW} at
% 7.) highest reference SNR
the highest reference \ac{SNR}.
% j) QPW produced only slightly better mean SSIM indices and relative RMSEs than for the convex l1-minimization method [l0.5-minimization]
In contrast,
% 1.) quasi-plane wave (QPW)
the \ac{QPW} produced
% 2.) mean SSIM indices
% MATLAB:
% SSIM_index_spgl1_lq_qpw(1,:)*1e2 = 43.6973   49.3404   56.1582   93.0138   99.8934   99.8323
% SSIM_index_spgl1_l1_qpw(1,:)*1e2 = 36.6000   41.2488   46.4364   90.4708   99.9325   99.9368
only slightly better mean \ac{SSIM} indices and
% 3.) relative RMSEs
% MATLAB:
% rel_RMSE_spgl1_lq_qpw(1,:)*1e2 = 65.8391   60.6768   54.3489   16.3019    1.4052    2.8359
% rel_RMSE_spgl1_l1_qpw(1,:)*1e2 = 74.9216   69.2613   63.4289   20.9912    1.3762    2.0671
relative \acp{RMSE} than for
% 4.) convex l1-minimization method
the convex $\ell_{1}$-minimization method
\eqreflqmin{eqn:recovery_reg_norm_lq_minimization}{ 1 }.
% k) sample means of the normalized numbers of iterations drastically exceeded those for the random waves by up to 22.9 % [l0.5-minimization]
% MATLAB:
% N_iter_max = 1127.6
% N_iter_spgl1_lq_qpw(1,:)*1e2/N_iter_max	  = 36.4580   45.4949   54.5938   62.7527   66.1848   87.7971
% N_iter_spgl1_lq_rnd_apo(1,:)*1e2/N_iter_max	  = 29.4342   29.1327   33.0259   45.8762  100.0000  111.9191
% N_iter_spgl1_lq_rnd_del(1,:)*1e2/N_iter_max	  = 29.5495   28.6360   31.7311   42.7279   88.5154  167.0805
% N_iter_spgl1_lq_rnd_apo_del(1,:)*1e2/N_iter_max = 30.0550   29.5938   33.0702   41.0518   94.5371  157.5913
%
% differences to QPW:
% rnd. apo.:		7.0238   16.3622   21.5679   16.8766  -33.8152  -24.1220
% rnd. del.:		6.9085   16.8588   22.8627   20.0248  -22.3306  -79.2834
% rnd. apo. del.:	6.4030   15.9010   21.5236   21.7010  -28.3523  -69.7943
% minimum:  6.4030 (rnd. apo. del. @ 3 dB)
% maximum: 22.8627 (rnd. del. @ 10 dB)
% 
For
% 1.) all reference SNRs
all reference \acp{SNR} except
$\text{SNR}_{\text{dB}} = \SI{30}{\deci\bel}$,
the sample means of
% 2.) normalized numbers of iterations
the normalized numbers of
iterations drastically exceeded those for
% 3.) random waves
the random waves by up to
% 4.) 22.9 %
\SI{22.9}{\percent}.

%%%%%%%%%%%%%%%%%%%%%%%%%%%%%%%%%%%%%%%%%%%%%%%%%%%%%%%%%%%%%%%%%%%%%%%%%%%%%%%%%%%%%%%%%%%%%%%%%%%%%%%%%%%%%%%%
% 6.) effects of the normalization
%%%%%%%%%%%%%%%%%%%%%%%%%%%%%%%%%%%%%%%%%%%%%%%%%%%%%%%%%%%%%%%%%%%%%%%%%%%%%%%%%%%%%%%%%%%%%%%%%%%%%%%%%%%%%%%%
%\subsubsection{Effects of the Normalization}
%\label{subsubsec:results_phantom_tissue_normalization}
%\input{results/results_phantom_tissue_normalization.tex}

%%%%%%%%%%%%%%%%%%%%%%%%%%%%%%%%%%%%%%%%%%%%%%%%%%%%%%%%%%%%%%%%%%%%%%%%%%%%%%%%%%%%%%%%%%%%%%%%%%%%%%%%%%%%%%%%
% 3.) memory consumption and computational costs
%%%%%%%%%%%%%%%%%%%%%%%%%%%%%%%%%%%%%%%%%%%%%%%%%%%%%%%%%%%%%%%%%%%%%%%%%%%%%%%%%%%%%%%%%%%%%%%%%%%%%%%%%%%%%%%%
\subsection{Memory Consumption and Computational Costs}
The approximate decomposition of
% 1.) observation process (all pulse-echo measurements, multifrequent, all array elements)
the observation process
\eqref{eqn:recovery_sys_lin_eq_v_rx_born_all_f_all_in_mat} by
% 2.) fast multipole method (FMM)
the \ac{FMM} reduced
% 3.) memory consumption
the memory consumption, which theoretically amounted to
% 4.) memory consumption of the observation process for the wire phantom
$M_{\text{conv}} = \SI{115}{\gibi\byte}$ for
% 5.) wire phantom
the wire phantom and
% 6.) memory consumption of the observation process for the tissue-mimicking phantom
$M_{\text{conv}} = \SI{112.5}{\gibi\byte}$ for
% 7.) tissue-mimicking phantom
the tissue-mimicking phantom, to
$M_{\text{\acs{FMM}}} \approx \SI{2.24}{\percent} M_{\text{conv}} \approx \SI{2.58}{\gibi\byte}$ and
% 9.) memory consumption of the FMM for the tissue-mimicking phantom
$M_{\text{\acs{FMM}}} \approx \SI{2.21}{\percent} M_{\text{conv}} \approx \SI{2.49}{\gibi\byte}$,
respectively.
% b) approximate decomposition of the observation process by the FMM concurrently reduced the number of multiplications to 7.25e9 and 6.97e9
It concurrently reduced
% 1.) number of multiplications
the number of
multiplications, which theoretically amounted to
% 2.) number of multiplications executed by the matrix-vector product for the wire phantom
$N_{\text{mul},\text{conv}} \approx \num{7.72e9}$ for
% 3.) wire phantom
the wire phantom and
% 4.) number of multiplications executed by the matrix-vector product for the tissue-mimicking phantom
$N_{\text{mul},\text{conv}} \approx \num{7.55e9}$ for
% 5.) tissue-mimicking phantom
the tissue-mimicking phantom, to
% 6.) number of multiplications executed by the FMM for the wire phantom
$N_{\text{mul},\text{\acs{FMM}}} \approx \SI{93.91}{\percent} N_{\text{mul},\text{conv}} \approx \num{7.25e9}$ and
% 7.) number of multiplications executed by the FMM for the tissue-mimicking phantom
$N_{\text{mul},\text{\acs{FMM}}} \approx \SI{92.32}{\percent} N_{\text{mul},\text{conv}} \approx \num{6.97e9}$,
respectively.

%%%%%%%%%%%%%%%%%%%%%%%%%%%%%%%%%%%%%%%%%%%%%%%%%%%%%%%%%%%%%%%%%%%%%%%%%%%%%%%%%%%%%%%%%%%%%%%%%%%%%%%%%%%%%%%%
% 10.) discussion
%%%%%%%%%%%%%%%%%%%%%%%%%%%%%%%%%%%%%%%%%%%%%%%%%%%%%%%%%%%%%%%%%%%%%%%%%%%%%%%%%%%%%%%%%%%%%%%%%%%%%%%%%%%%%%%%
\section{Discussion}
\label{sec:discussion}
\subsection{Random Waves Decorrelate the Pulse Echoes of the Admissible Structural Building Blocks}
%\label{subsec:discussion_tpsf_measures}
%---------------------------------------------------------------------------------------------------------------
% 1.) isolated sharp maxima and resolutions (reference sensing matrices)
%---------------------------------------------------------------------------------------------------------------
% a) isolated maxima of minimum FEHMs indicate optimal spatial and spectral resolutions
% article:Schiffner2018, Sect. VIII. Results / Sect. VIII-B. Tissue-Mimicking Phantom / Sect. VIII-B.2) Transform Point Spread Functions (subsubsec:results_phantom_tissue_tpsf)
% - Both REFERENCE SENSING MATRICES produced similar results as
%   the RANDOM OBSERVATION PROCESS \eqref{eqn:sim_study_params_ref_obs_proc_rip}, i.e.
%   UNIFORMLY DISTRIBUTED RANDOM VALUES CLOSE TO ZERO THAT RENDERED THE MAXIMA SHARP AND ISOLATED
%   (cf. \cref{fig:sim_study_obj_A_sr_1_tpsf_images_rip_1}).
% article:Schiffner2018, Sect. VIII. Results / Sect. VIII-A. Wire Phantom / Sect. VIII-A.2) Point Spread Functions (subsubsec:results_phantom_wire_psf)
% - Both REFERENCE OBSERVATION PROCESSES produced
%   RANDOM VALUES CLOSE TO ZERO THAT RENDERED THE MAXIMA SHARP AND ISOLATED.
% - The random observation process \eqref{eqn:sim_study_params_ref_obs_proc_rip} UNIFORMLY DISTRIBUTED these values over
%   the \ac{FOV}, whereas its structured version \eqref{eqn:sim_study_params_ref_obs_proc_gwn} formed
%   NOTICEABLE GAPS that were laterally adjacent to the maxima and shaped hourglasses of larger absolute values
%   (cf. inset image).
The isolated maxima of
% 1.) minimum FEHMs
% article:Schiffner2018, Sect. VIII. Results / Sect. VIII-B. Tissue-Mimicking Phantom / Sect. VIII-B.2) Transform Point Spread Functions (subsubsec:results_phantom_tissue_tpsf)
% - Both REFERENCE SENSING MATRICES CONSISTENTLY ACHIEVED
%   THE MINIMUM \ac{FEHM} of a TWO-DIMENSIONAL NORMALIZED SPATIAL FREQUENCY ELEMENT $\Delta \hat{K} \approx \num{3.81e-6}$ for all fixed spatial frequencies.
% article:Schiffner2018, Sect. VIII. Results / Sect. VIII-A. Wire Phantom / Sect. VIII-A.2) Point Spread Functions (subsubsec:results_phantom_wire_psf)
% - Both REFERENCE OBSERVATION PROCESSES CONSISTENTLY ACHIEVED
%   THE MINIMUM \ac{FEHM} of a TWO-DIMENSIONAL VOLUME ELEMENT $\Delta V \approx \SI{5.81e-3}{\milli\meter\squared}$ for all fixed positions.
minimum \acp{FEHM}, which were embedded in
% 2.) random values close to zero
random values close to zero and characterized
% 3.) transform point spread function (TPSF)
the \acp{TPSF}
\eqref{eqn:cs_math_tpsf} associated with
% 4.) reference sensing matrices
all reference sensing matrices
(cf. \cref{%
  fig:sim_study_obj_A_sr_1_tpsf_images_rip_1,%
  fig:sim_study_obj_A_sr_1_tpsf_images_blgwn_1,%
  fig:sim_study_obj_B_sr_1_tpsf_images_rip_1,%
  fig:sim_study_obj_B_sr_1_tpsf_images_blgwn_1%
}), indicate
% 5.) optimal spatial and spectral resolutions
optimal spatial and
spectral resolutions.
% b) isolated maxima of minimum FEHMs resolved the adjacent grid points in the FOV for the wire phantom and the adjacent discrete spatial frequencies for the tissue-mimicking phantom
They resolved
% 1.) adjacent grid points in the FOV
the adjacent grid points in
the \ac{FOV} for
% 2.) wire phantom
the wire phantom and
% 3.) adjacent discrete spatial frequencies
the adjacent discrete spatial frequencies for
% 4.) tissue-mimicking phantom
the tissue-mimicking phantom.
% c) RIP for 4-sparse representations ensured the stable recovery of all 2-sparse representations
% article:Schiffner2018, Sect. VII. Simulation Study / Sect. VII-A. Parameters / Sect. VII-A.9) Reference Sensing Matrices
% - For a sufficiently large number of observations \eqref{eqn:recovery_sys_lin_eq_num_obs}, both
%   the real-valued random $N_{\text{obs}} \times N_{\text{lat}}$ observation process
%   [ ... ] (eqn:sim_study_params_ref_obs_proc_rip) and
%   the ASSOCIATED COMPLEX-VALUED $N_{\text{obs}} \times N_{\text{lat}}$ SENSING MATRIX [ ... ] (eqn:sim_study_params_ref_sens_mat_rip) MET
%   the \ac{RIP} WITH VERY HIGH PROBABILITY (cf. \cref{sec:compressed_sensing}).
In fact,
% 1.) restricted isometry property (RIP)
the \ac{RIP} for
% 2.) 4-sparse representations
$4$-sparse representations
\eqref{eqn:recovery_reg_sparse_representation}, which was met by
% 3.) first reference sensing matrices (RIP)
the random sensing matrices
\eqref{eqn:sim_study_params_ref_sens_mat_rip} with
% 4.) very high probability
very high probability, ensured
% 5.) stable recovery
the stable recovery of
% 6.) all 2-sparse representations
all $2$-sparse representations
\eqref{eqn:recovery_reg_sparse_representation}, including those whose
% 7.) nonzero components
nonzero components populate
% 8.) adjacent grid points or discrete spatial frequencies
adjacent grid points or
discrete spatial frequencies.

%---------------------------------------------------------------------------------------------------------------
% 2.) distribution of the random values close to zero (reference sensing matrices)
%---------------------------------------------------------------------------------------------------------------
% a) incoherent aliasing prevented their sparse approximations and enabled their removal by the sparsity-promoting lq-minimization method
% article:Schiffner2018, Sect. II. Compressed Sensing in a Nutshell (sec:compressed_sensing)
% - For $n_{1} \neq n_{2}$, however, both column vectors typically differ, and the absolute value of the \ac{TPSF} \eqref{eqn:cs_math_tpsf} is desired to be
%   AS CLOSE TO ZERO AS POSSIBLE WITH NOISE-LIKE FEATURES, e.g. a UNIFORM DISTRIBUTION OF THE FLUCTUATING ENERGY OVER THE INDICES
%   \cite{article:ProvostITMI2009,article:LustigMRM2007}.
%   => (1) CLOSE TO ZERO AS POSSIBLE
%   => (2) NOISE-LIKE FEATURES: UNIFORM DISTRIBUTION, FLUCTUATING ENERGY
% - The SMALL ABSOLUTE VALUES indicate that
%   the observation process reliably DISTINGUISHES THE ADMISSIBLE STRUCTURAL BUILDING BLOCKS, and
%   the NOISE-LIKE FEATURES, which are referred to as INCOHERENT ALIASING, do not misguide
%   the sparsity-promoting $\ell_{q}$-minimization method \eqref{eqn:cs_lq_minimization}.
The noise-like properties of
% 1.) random values close to zero
random values close to zero enabled
% 2.) removal
their removal by
% 3.) sparsity-promoting lq-minimization method
the sparsity-promoting $\ell_{q}$-minimization method
\eqref{eqn:recovery_reg_norm_lq_minimization}.
% b) gaps resulted from the single reception angle characterizing the pulse-echo setup
% article:Schiffner2018, Sect. VIII. Results / Sect. VIII-A. Wire Phantom / Sect. VIII-A.2) Point Spread Functions (subsubsec:results_phantom_wire_psf)
% - The random observation process \eqref{eqn:sim_study_params_ref_obs_proc_rip} UNIFORMLY DISTRIBUTED these values over
%   the \ac{FOV}, whereas its structured version \eqref{eqn:sim_study_params_ref_obs_proc_gwn} formed
%   NOTICEABLE GAPS THAT WERE LATERALLY ADJACENT TO THE MAXIMA and shaped
%   HOURGLASSES OF LARGER ABSOLUTE VALUES (cf. inset image).
The gaps formed by
% 1.) second reference observation process and its entries (GWN)
the structured random observation process
\eqref{eqn:sim_study_params_ref_obs_proc_gwn} for
% 2.) wire phantom
the wire phantom
(cf. \cref{fig:sim_study_obj_A_sr_1_tpsf_images_blgwn_1}) arose from
% 3.) single reception angle
the single reception angle in
% 4.) pulse-echo setup
the pulse-echo setup, i.e.
% 5.) fixed position
the fixed position of
% 6.) linear transducer array
the linear transducer array on
% 6.) single edge of the FOV
a single edge of
the \ac{FOV}.
% c) difference indicates potential benefits of the random waves for spatially extended basis functions
% article:Schiffner2018, Sect. VIII. Results / Sect. VIII-B. Tissue-Mimicking Phantom / Sect. VIII-B.2) Transform Point Spread Functions (subsubsec:results_phantom_tissue_tpsf)
% - The STRUCTURED RANDOM SENSING MATRIX \eqref{eqn:sim_study_params_ref_sens_mat_gwn}, however, significantly elongated
%   THESE MAXIMA ALONG THE $\hat{K}_{2}$-axis IN ADDITION TO MODEST LATERAL EXTENSIONS (cf. inset image).
The elimination of
these gaps by
% 1.) second reference sensing matrix (GWN)
the structured random sensing matrix
\eqref{eqn:sim_study_params_ref_sens_mat_gwn} for
% 2.) tissue-mimicking phantom
the tissue-mimicking phantom
(cf. \cref{fig:sim_study_obj_B_sr_1_tpsf_images_blgwn_1}) hints at
% 3.) potential benefits
potential benefits provided by
% 4.) random waves
the random waves for
% 5.) specified structural building blocks
the specified structural building blocks, i.e.
% 6.) complex exponential functions of distinct spatial frequencies
the complex exponential functions.

%---------------------------------------------------------------------------------------------------------------
% 3.) all incident waves degrade the optimal spatial and spectral resolutions
%---------------------------------------------------------------------------------------------------------------
% a) increased FEHMs of the TPSFs associated with the sensing matrices induced by all incident waves indicate degraded spatial and spectral resolutions
The increased \acp{FEHM} of
% 1.) transform point spread functions (TPSFs)
the \acp{TPSF}
\eqref{eqn:cs_math_tpsf} associated with
% 2.) sensing matrices (all pulse-echo measurements, multifrequent, all array elements)
the sensing matrices
\eqref{eqn:recovery_reg_sensing_matrix} induced by
% 3.) all incident waves
all incident waves
(cf. \cref{%
  tab:sim_study_obj_A_sr_1_tpsf_fehm,%
  tab:sim_study_obj_B_sr_1_tpsf_fehm%
}), which resulted from
% 4.) elliptical-shaped regions of absolute values close to unity around the maxima
% article:Schiffner2018, Sect. VIII. Results / Sect. VIII-A. Wire Phantom / Sect. VIII-A.2) Point Spread Functions (subsubsec:results_phantom_wire_psf)
% - The observation processes \eqref{eqn:recovery_sys_lin_eq_v_rx_born_all_f_all_in_mat} induced by all incident waves concentrated
%   RELATIVELY LARGE ABSOLUTE VALUES CLOSE TO UNITY IN ELLIPTICAL-SHAPED REGIONS AROUND THE MAXIMA.
% - The lengths of the minor and major axes ranged from \SIrange{0.15}{0.3}{\milli\meter} and from \SIrange{0.46}{0.76}{\milli\meter}, respectively.
% - They DISTRIBUTED THE NONZERO VALUES LESS UNIFORMLY and FORMED SIDELOBES OF VARIOUS CHARACTERS.
the elliptical-shaped regions of
absolute values close to
unity around
the maxima for
% 5.) wire phantom
the wire phantom
(cf. \cref{%
  fig:sim_study_obj_A_sr_1_tpsf_images_qpw_1,%
  fig:sim_study_obj_A_sr_1_tpsf_images_rnd_apo_1,%
  fig:sim_study_obj_A_sr_1_tpsf_images_rnd_del_1,%
  fig:sim_study_obj_A_sr_1_tpsf_images_rnd_apo_del_1%
}) and
% 6.) extended maxima
% article:Schiffner2018, Sect. VIII. Results / Sect. VIII-B. Tissue-Mimicking Phantom / Sect. VIII-B.2) Transform Point Spread Functions (subsubsec:results_phantom_tissue_tpsf)
% - The sensing matrices \eqref{eqn:recovery_reg_sensing_matrix} induced by the random waves APPROXIMATELY MAINTAINED
%   THESE MAXIMA but CONFINED SIMILAR NOISE-LIKE ARTIFACTS TO THEIR PASSBANDS.
% - In contrast, the sensing matrix \eqref{eqn:recovery_reg_sensing_matrix} induced by the \ac{QPW} formed
%   [1.)] SMOOTH COHERENT SIDELOBES, whose absolute values lacked noise-like features, and indicated
%   [2.)] the PRESENCE OF UNSPECIFIED SPATIAL FREQUENCIES BY SECONDARY ISOLATED MAXIMA, e.g.
%   an absolute value of approximately \SI{-2.4}{\deci\bel} at
%   the normalized spatial frequency $\hat{\vect{K}} \approx \trans{ ( \num{0.12}, \num{0.35} ) }$.
the extended maxima for
% 7.) tissue-mimicking phantom
the tissue-mimicking phantom
(cf. \cref{%
  fig:sim_study_obj_B_sr_1_tpsf_images_qpw_1,%
  fig:sim_study_obj_B_sr_1_tpsf_images_rnd_apo_1,%
  fig:sim_study_obj_B_sr_1_tpsf_images_rnd_del_1,%
  fig:sim_study_obj_B_sr_1_tpsf_images_rnd_apo_del_1%
}), indicate
% 8.) spatial and spectral resolutions
degraded spatial and
spectral resolutions.
% b) incident acoustic pressure fields strongly correlated the pulse echoes of the adjacent grid points and the adjacent discrete spatial frequencies
% article:Schiffner2018, Sect. VII. Simulation Study / Sect. VII-A. Parameters / Sect. VII-A.9) Reference Sensing Matrices (subsubsec:sim_study_params_ref_sens_mat)
% - The replacement of the incident acoustic pressure field \eqref{eqn:recovery_p_in} in the observation process \eqref{eqn:recovery_sys_lin_eq_v_rx_born_all_f_all_in_mat} by
%   COMPLEX-VALUED \ac{GWN} additionally formed the complex-valued structured $N_{\text{obs}} \times N_{\text{lat}}$ observation process
%   [ \mat{\Phi}^{(\text{\acs{GWN}})} = \mat{\Phi}\bigl[ p^{(\text{in})} \bigr], p_{l}^{(\text{in}, 0)}( \vect{r}_{\text{lat}, i} ) \underset{ \text{\acs{IID}} }{ \sim } \gaussian{ 0 }{ 1 } ]
%   (eqn:sim_study_params_ref_obs_proc_gwn) and the associated complex-valued $N_{\text{obs}} \times N_{\text{lat}}$ sensing matrix
%   [ \mat{A}^{(\text{\acs{GWN}})} = \mat{\Phi}^{(\text{\acs{GWN}})} \mat{\Psi}. ] (eqn:sim_study_params_ref_sens_mat_gwn)
In contrast to
% 1.) complex-valued GWN
the \ac{GWN} forming
% 2.) second reference sensing matrix (GWN)
the structured random sensing matrix
\eqref{eqn:sim_study_params_ref_sens_mat_gwn},
% 3.) incident acoustic pressure fields
the incident acoustic pressure fields met
% 4.) Helmholtz equations for the incident acoustic pressure fields
the \name{Helmholtz} equations
\eqref{eqn:lin_mod_sol_wave_eq_pde_p_in} and, thus, strongly correlated
% 5.) strongly correlated pulse echoes
the pulse echoes of
% 6.) adjacent grid points in the FOV
the adjacent grid points and
% 7.) adjacent discrete spatial frequencies
the adjacent discrete spatial frequencies.
% c) FEHMs confirm smaller regions of highly-correlated observations, and the empirical CDFs attest statistical properties more similar to both reference observation processes
In
their attempt to replicate
the desirable properties of
the \ac{GWN}, however,
% 1.) random waves
the random waves significantly outperformed
% 2.) quasi-plane wave (QPW)
the \ac{QPW}.

In fact,
% 1.) noise-like artifacts
the noise-like artifacts
(cf. \cref{%
  fig:sim_study_obj_A_sr_1_tpsf_images_rnd_apo_1,%
  fig:sim_study_obj_A_sr_1_tpsf_images_rnd_del_1,%
  fig:sim_study_obj_A_sr_1_tpsf_images_rnd_apo_del_1,%
  fig:sim_study_obj_B_sr_1_tpsf_images_rnd_apo_1,%
  fig:sim_study_obj_B_sr_1_tpsf_images_rnd_del_1,%
  fig:sim_study_obj_B_sr_1_tpsf_images_rnd_apo_del_1%
}) with
\TODO{exceeding -70 dB}
% 2.) nonzero values
more uniform distributions of
the nonzero values, which populated
% 3.) 43.8 to 50.5 % of the FOV
% article:Schiffner2018, Sect. VIII. Results / Sect. VIII-A. Wire Phantom / Sect. VIII-A.2) Point Spread Functions (subsubsec:results_phantom_wire_psf)
% - The absolute values below \SI{-70}{\deci\bel} constituted approximately \SIrange{49.5}{56.2}{\percent} of
%   the \ac{FOV} and those above this threshold, which reached up to \SI{-0.11}{\deci\bel}, formed the remaining \SIrange{43.8}{50.5}{\percent}.
\SIrange{43.8}{50.5}{\percent} of
the \ac{FOV} for
% 4.) wire phantom
the wire phantom
(cf. \cref{fig:sim_study_obj_A_sr_1_tpsf_ecdfs}) and
% 5.) 80 % of the admissible spatial frequencies
% article:Schiffner2018, Sect. VIII. Results / Sect. VIII-B. Tissue-Mimicking Phantom / Sect. VIII-B.2) Transform Point Spread Functions (subsubsec:results_phantom_tissue_tpsf)
% - The absolute values below \SI{-70}{\deci\bel} constituted approximately \SI{20}{\percent} of
%   the admissible spatial frequencies and those above this threshold, which reached up to \SI{-0.47}{\deci\bel}, formed the remaining \SI{80}{\percent}.
\SI{80}{\percent} of
the admissible spatial frequencies for
% 6.) tissue-mimicking phantom
the tissue-mimicking phantom
(cf. \cref{fig:sim_study_obj_B_sr_1_tpsf_ecdfs}), resembled
% 7.) noise-like artifacts
those induced by
% 8.) second reference sensing matrix (GWN)
the structured random sensing matrices
\eqref{eqn:sim_study_params_ref_sens_mat_gwn}
(cf. \cref{%
  fig:sim_study_obj_A_sr_1_tpsf_images_blgwn_1,%
  fig:sim_study_obj_B_sr_1_tpsf_images_blgwn_1%
}).
% b) noise-like artifacts prevented sparse approximations and enabled their removal by the sparsity-promoting lq-minimization method
Unlike
% 1.) coherent sidelobes
the coherent sidelobes and
% 2.) secondary maxima
the secondary maxima induced by
% 3.) quasi-plane wave (QPW)
the \ac{QPW}
(cf. \cref{%
  fig:sim_study_obj_A_sr_1_tpsf_images_qpw_1,%
  fig:sim_study_obj_B_sr_1_tpsf_images_qpw_1%
}), they prevented
% 1.) sparse approximations
sparse approximations and, thus, enabled
% 2.) removal
their removal by
% 3.) sparsity-promoting lq-minimization method
the sparsity-promoting $\ell_{q}$-minimization method
\eqref{eqn:recovery_reg_norm_lq_minimization}.
% c) reductions in the FEHMs indicate improved spatial and spectral resolutions
% article:Schiffner2018, Sect. VIII. Results / Sect. VIII-B. Tissue-Mimicking Phantom / Sect. VIII-B.2) Transform Point Spread Functions (subsubsec:results_phantom_tissue_tpsf)
% - In fact, the maximum normalized differences ranged from
%   \SI{41.17}{\percent} at the fixed spatial frequencies $s \in \{ 1, 9 \}$ to
%   \SI{62.52}{\percent} at the fixed spatial frequencies $s \in \{ 4, 6 \}$.
% article:Schiffner2018, Sect. VIII. Results / Sect. VIII-A. Wire Phantom / Sect. VIII-A.2) Point Spread Functions (subsubsec:results_phantom_wire_psf)
% - The MAXIMUM NORMALIZED DIFFERENCES ranged from
%   \SI{23.53}{\percent} for the superposition of both randomly-apodized and randomly-delayed \acp{QCW} at
%   the ninth fixed position, i.e. $s = 9$, to
%   \SI{73.68}{\percent} for the superposition of randomly-delayed \acp{QCW} at
%   the first fixed position, i.e. $s = 1$.
The reductions in
% 1.) full extents at half maximum (FEHMs)
the \acp{FEHM}, which ranged from
% 1.) 23.53 to 73.68 %
\SIrange{23.53}{73.68}{\percent} for
% 2.) wire phantom
the wire phantom
(cf. \cref{tab:sim_study_obj_A_sr_1_tpsf_fehm}) and from
% 3.) 41.17 to 62.52 %
\SIrange{41.17}{62.52}{\percent} for
% 4.) tissue-mimicking phantom
the tissue-mimicking phantom
(cf. \cref{tab:sim_study_obj_B_sr_1_tpsf_fehm}) with
% 5.) few exceptions
few exceptions, indicate
% 6.) spatial and spectral resolutions
improved spatial and
spectral resolutions.
% d) slightly increased mean FEHM and the deviation in the empirical CDF produced by the superposition of randomly-apodized QCWs indicate potential benefits for the inclusion of random time delays for the wire phantom
% article:Schiffner2018, Sect. VIII. Results / Sect. VIII-A. Wire Phantom / Sect. VIII-A.2) Point Spread Functions (subsubsec:results_phantom_wire_psf)
% - The SUPERPOSITION OF RANDOMLY-APODIZED \acp{QCW} produced
%   the LARGEST SAMPLE MEAN AND SAMPLE STANDARD DEVIATION among the random waves.
The slightly increased mean \ac{FEHM} and
% 1.) deviation
the deviation in
% 2.) empirical CDF
% article:Schiffner2018, Sect. VIII. Results / Sect. VIII-A. Wire Phantom / Sect. VIII-A.2) Point Spread Functions (subsubsec:results_phantom_wire_psf)
% - The SUPERPOSITION OF RANDOMLY-APODIZED \acp{QCW} distributed
%   the latter values [absolute values above -70 dB] over the smallest percentage of the \ac{FOV}.
the empirical \ac{CDF} produced by
% 3.) superposition of randomly-apodized QCWs
the superposition of
randomly-apodized \acp{QCW} for
% 4.) wire phantom
the wire phantom, however, indicate
% 5.) potential benefits
potential benefits provided by
% 6.) random time delays
the random time delays.

%%%%%%%%%%%%%%%%%%%%%%%%%%%%%%%%%%%%%%%%%%%%%%%%%%%%%%%%%%%%%%%%%%%%%%%%%%%%%%%%%%%%%%%%%%%%%%%%%%%%%%%%%%%%%%%%
% 2.) lq-minimization recovers the images quantitatively
%%%%%%%%%%%%%%%%%%%%%%%%%%%%%%%%%%%%%%%%%%%%%%%%%%%%%%%%%%%%%%%%%%%%%%%%%%%%%%%%%%%%%%%%%%%%%%%%%%%%%%%%%%%%%%%%
\subsection{Image Recovery by $\ell_{q}$-Minimization is Quantitative}
%\subsection{$\ell_{q}$-Minimization Recovers the Images Quantitatively}
\label{subsec:discussion_benefits_lq_minimization}
%---------------------------------------------------------------------------------------------------------------
% 1.) significance of the left multiplications by the adjoint normalized sensing matrices
%---------------------------------------------------------------------------------------------------------------
% a) left multiplications of the normalized recorded RF voltage signals by the adjoint normalized sensing matrices qualitatively recovered the normalized nearly-sparse representations
% article:Schiffner2018, Sect. VI. Implementation / Sect. VI-C. Sparsity-Promoting lq-Minimization Method
% - \ac{SPGL1} is iterative and LEFT MULTIPLIED a sequence of recursively-generated vectors by
%   the potentially densely-populated NORMALIZED SENSING MATRIX \eqref{eqn:recon_reg_norm_sensing_matrix} OR ITS ADJOINT.
The left multiplications of
% 1.) normalized linear algebraic system (all pulse-echo measurements, multifrequent, all array elements, additive errors)
the normalized recorded \ac{RF} voltage signals
\eqref{eqn:recovery_reg_norm_obs_trans_coef_error} by
% 2.) adjoint normalized sensing matrices (all pulse-echo measurements, multifrequent, all array elements)
the adjoint normalized sensing matrices
\eqref{eqn:recon_reg_norm_sensing_matrix} initialized
% 3.) SPGL1
\ac{SPGL1} and qualitatively recovered
% 4.) normalized nearly-sparse representations
the normalized nearly-sparse representations
\eqref{eqn:recon_reg_norm_trans_coef}
(cf. \cref{%
  fig:sim_study_obj_A_adj_1_images_kap,%
  fig:sim_study_obj_B_adj_1_dft_kap%
}).
% b) left multiplications linearly combined the TPSFs for all n_{1} \in \supp[ \vectsym{\theta}^{(\kappa)} ] and added errors
They linearly combined
% 1.) transform point spread functions (TPSFs)
the \acp{TPSF}
\eqref{eqn:cs_math_tpsf} for
% 2.) all supporting indices
all $n_{1} \in \supp[ \vectsym{\theta}^{(\kappa)} ]$ and added
% 3.) projections of the additive errors
errors, as shown in
\eqref{eqn:app_adjoint_tpsf}
(cf. Appendix \ref{app:adjoint}).
% c) left multiplications assigned linear combinations of the zero-lag cross-correlations to the components and significantly enhanced the popular DAS method
Equivalently,
they assigned
% 1.) linear combinations
linear combinations of
% 2.) zero-lag time-domain cross-correlations
the zero-lag cross-correlations
\eqref{eqn:app_adjoint_xcorr_td} between
% 3.) recorded RF voltage signals
the recorded \ac{RF} voltage signals
\eqref{eqn:recovery_disc_freq_v_rx_Fourier_series} and
% 4.) pulse echoes
the pulse echoes of
% 5.) admissible structural building blocks
the admissible structural building blocks to
% 6.) components
the components and, thus, significantly enhanced
% 7.) popular DAS method
the popular \ac{DAS} method.
\TODO{no temporal discretization!}
In fact,
% 1.) popular DAS method
the latter combines
% 2.) canonical basis
the canonical basis with
% 3.) approximation
the approximation of
% 4.) pulse echoes
all pulse echoes by
% 5.) delayed Dirac delta distributions
delayed \name{Dirac} delta distributions
\cite[Sect. IV.A.2]{article:DavidJASA2015}.

%---------------------------------------------------------------------------------------------------------------
% 2.) proposed method enabled the quantitative recovery / convex l1-minimization
%---------------------------------------------------------------------------------------------------------------
% a) proposed method enabled the quantitative recovery of the specified compressibility fluctuations via the sparsity-promoting lq-minimization method
% article:Schiffner2018, Sect. V. Image Recovery Based on Compressed Sensing
% - The proposed method circumvents this difficulty [ no direct solution ] by
%   REFORMULATING THE DISCRETIZED LINEAR \ac{ISP} AS AN INSTANCE OF
%   THE \ac{CS} PROBLEM \eqref{eqn:cs_math_prob_general}.
% - Postulating the existence of a nearly-sparse representation of the discretized compressibility fluctuations in
%   a known orthonormal basis \eqref{eqn:def_transform_coefficients},
%   the sparsity-promoting $\ell_{q}$-minimization method \eqref{eqn:cs_lq_minimization} ensures its stable recovery if
%   the sensing matrix \eqref{eqn:cs_math_prob_general_sensing_matrix} meets one of the sufficient conditions
%   (cf. \cref{sec:compressed_sensing}).
The proposed method, in contrast, enabled
% 1.) quantitative recovery
the quantitative recovery of
% 2.) specified vectors stacking the regular samples in the discretized relative spatial fluctuations in the unperturbed compressibility
the specified compressibility fluctuations
\eqref{eqn:recovery_sys_lin_eq_gamma_kappa_bp_vector} via
% 3.) sparsity-promoting lq-minimization method
the sparsity-promoting $\ell_{q}$-minimization method
\eqref{eqn:recovery_reg_norm_lq_minimization}.
% b) coherent sidelobes and the secondary maxima in the TPSFs induced by the QPW caused artifacts
% article:LustigMRM2007: Sparse MRI: The application of compressed sensing for rapid MR imaging
% THEORY / A Simple, Intuitive Example of Compressed Sensing
% - To get intuition for the importance of incoherence and the feasibility of CS in MRI, consider the example in Fig. 2. (p. 1183)
% - An intuitive plausible recovery procedure is illustrated in Fig. 2e – h. (p. 1184)
% - It is based on
%   [1.)] THRESHOLDING,
%   [2.)] RECOVERING THE STRONG COMPONENTS, and
%   [3.)] CALCULATING THE INTERFERENCE CAUSED BY THEM AND SUBTRACTING IT. (p. 1184)
% - Subtracting the interference of the strong components
%   [1.)] reduces the total interference level and
%   [2.)] enables recovery of weaker, previously submerged components. (p. 1184)
% - By iteratively repeating this procedure, one can recover the rest of the signal components. (p. 1184)
% - A recovery procedure along these lines was proposed by Donoho et al.
%   (Sparse Solution of Underdetermined Linear Equations by Stagewise Orthogonal Matching Pursuit, 2006, Stanford University, Statistics Department, technical report #2006-02) as
%   a fast approximate algorithm for CS reconstruction. (p. 1184)
% - A similar approach of recovery of MR images was proposed in Ref. (22) (p. 1184)
Since
% 1.) identification
the identification of
% 2.) significant components
the significant components in
% 3.) normalized nearly-sparse representations
the normalized nearly-sparse representations
\eqref{eqn:recon_reg_norm_trans_coef} essentially thresholded
% 4.) aforementioned products
the aforementioned products
\cite[Fig. 2]{article:LustigMRM2007}, however,
% 5.) coherent sidelobes
the coherent sidelobes and
% 6.) secondary maxima
the secondary maxima in
% 7.) transform point spread functions (TPSFs)
the \acp{TPSF}
\eqref{eqn:cs_math_tpsf} induced by
% 8.) quasi-plane wave (QPW)
the \ac{QPW}
(cf. \cref{%
  fig:sim_study_obj_A_sr_1_tpsf_images_qpw_1,%
  fig:sim_study_obj_B_sr_1_tpsf_images_qpw_1%
}) caused
% 9.) artifacts
artifacts for
% 10.) sufficiently large additive errors
sufficiently large additive errors
(cf. \cref{%
  fig:sim_study_obj_A_sr_1_images_spgl1_l1_qpw,%
  fig:sim_study_obj_B_sr_1_dft_images_spgl1_l1_qpw,%
  fig:sim_study_obj_B_sr_1_dft_images_spgl1_lq_qpw%
}).
% c) FEHMs increased the numbers of components within the illustrated dynamic range
The \acp{FEHM} for
% 1.) wire phantom
the wire phantom, which significantly exceeded
% 2.) size of a volume element
the size of
a volume element
(cf. \cref{tab:sim_study_obj_A_sr_1_tpsf_fehm}), increased
% 3.) numbers of components within the illustrated dynamic range
the numbers of
components within
the illustrated dynamic range and, thus, the relative \acp{RMSE} for
% 4.) convex l1-minimization method
the convex $\ell_{1}$-minimization method
\eqreflqmin{eqn:recovery_reg_norm_lq_minimization}{ 1 }
(cf. \cref{%
  fig:sim_study_obj_A_sr_1_quality_vs_snr_spgl1_l1_qpw,%
  fig:sim_study_obj_A_sr_1_quality_vs_snr_spgl1_l1_rnd_apo,%
  fig:sim_study_obj_A_sr_1_quality_vs_snr_spgl1_l1_rnd_del,%
  fig:sim_study_obj_A_sr_1_quality_vs_snr_spgl1_l1_rnd_apo_del%
}).
\TODO{special case: QPW for 30 dB}
% TODO: perfect recovery of wire phantom by QPW for SNR 30 dB ?
% d) secondary maxima prevented this reduction for the QPW
Although
% 1.) tissue-mimicking phantom
the tissue-mimicking phantom reduced
% 2.) ratio FEHMs vs.normalized spatial frequency element
this ratio
(cf. \cref{tab:sim_study_obj_B_sr_1_tpsf_fehm}) and, thus, the relative \acp{RMSE} for
% 3.) random waves
the random waves
(cf. \cref{%
  fig:sim_study_obj_B_sr_1_quality_vs_snr_spgl1_l1_rnd_apo,%
  fig:sim_study_obj_B_sr_1_quality_vs_snr_spgl1_l1_rnd_del,%
  fig:sim_study_obj_B_sr_1_quality_vs_snr_spgl1_l1_rnd_apo_del%
}),
% 4.) secondary maxima
the secondary maxima prevented
this reduction for
% 5.) quasi-plane wave (QPW)
the \ac{QPW}
(cf. \cref{fig:sim_study_obj_B_sr_1_quality_vs_snr_spgl1_l1_qpw}).
% e) mean SSIM indices confirmed the excellent recovery of the object's structure exemplified in previous figure
The mean \ac{SSIM} indices, however, confirmed
% 1.) excellent structural recovery
the excellent structural recovery of both
% 2.) both phantoms
phantoms by
% 3.) random waves
the random waves.

%---------------------------------------------------------------------------------------------------------------
% 3.) superiority of the nonconvex l0.5-minimization method to the convex l1-minimization method
%---------------------------------------------------------------------------------------------------------------
% a) consistent improvements of all quality metrics by the nonconvex l0.5-minimization method indicate its superiority to the convex l1-minimization method for the specified compressibility fluctuations
The consistent improvements of
% 1.) mean SSIM indices, relative RMSEs, and number of components within the illustrated dynamic range
all quality metrics by
% 2.) nonconvex l0.5-minimization method
the nonconvex $\ell_{0.5}$-minimization method
\eqreflqmin{eqn:recovery_reg_norm_lq_minimization}{ 0.5 }
(cf. \cref{%
  fig:sim_study_obj_A_sr_1_ssim_index_rel_rmse_N_iter_vs_snr_kap,%
  fig:sim_study_obj_B_sr_1_ssim_index_rel_rmse_N_iter_vs_snr_kap%
}) indicate
% 3.) superiority
its superiority to
% 4.) convex l1-minimization method
the convex $\ell_{1}$-minimization method
\eqreflqmin{eqn:recovery_reg_norm_lq_minimization}{ 1 } for
% 5.) specified vectors stacking the regular samples in the discretized relative spatial fluctuations in the unperturbed compressibility
the specified compressibility fluctuations
\eqref{eqn:recovery_sys_lin_eq_gamma_kappa_bp_vector}.
% b) larger numbers of iterations increase the computational costs and the recovery times
% article:Schiffner2018, Sect. VII. Simulation Study / Sect. VII-A. Parameters / Sect. VII-A.8) Regularization (subsubsec:sim_study_params_regularization)
% - Since \ac{SPGL1} provided the initial guess,
%   \name{Foucart}'s ALGORITHM ENTAILED SIX $\ell_{1}$ MINIMIZATIONS.
% article:FoucartACHA2009: Sparsest solutions of underdetermined linear systems via lq-minimization for 0 < q \leq 1
% 5. Numerical experiments
% - This improvement is obtained at a default cost of 4 times a 10-iteration reweighted l1-minimization, i.e.
%   at a cost of 40 times an l1-minimization. (p. 406)
% article:ChartrandISPL2007: Exact Reconstruction of Sparse Signals via Nonconvex Minimization
% IV. CONCLUSIONS
% - The REQUIRED RECONSTRUCTION TIME IS GENERALLY LONGER THAN WITH p = 1 but much less than with p = 0. (p. 710)
The larger numbers of
iterations, which arose from
% 1.) six executions of SPGL1
the six executions of
\ac{SPGL1}, however, increase
% 2.) computational costs
the computational costs and
% 3.) recovery times
the recovery times.
These findings agree with
% 1.) findings
those derived from
% 2.) numerical experiments
numerical experiments in
% 3.) scientific literature
the literature
(cf. e.g.
\cite[Sect. 5]{article:FoucartACHA2009},
\cite[Sect. 3]{proc:ChartrandICASSP2008},
\cite[Sect. III]{article:ChartrandISPL2007}%
).
In fact,
% 1.) minimization
the minimization of
% 2.) lq-quasinorm, q \in ( 0; 1 )
the $\ell_{q}$-quasinorm, $q \in ( 0; 1 )$, in
% 3.) sparsity-promoting lq-minimization method
the sparsity-promoting method
\eqref{eqn:cs_lq_minimization} exactly recovered
% 4.) nearly-sparse representations
sparse representations
\eqref{eqn:def_transform_coefficients} from
% 5.) significantly smaller number
a significantly smaller number of
% 6.) error-free observations
error-free observations
\cite{proc:ChartrandICASSP2008,article:ChartrandISPL2007}.
Sufficient conditions, which relax
% 1.) upper bounds
the upper bounds on
% 2.) restricted isometry ratio
% article:FoucartACHA2009: Sparsest solutions of underdetermined linear systems via lq-minimization for 0 < q \leq 1
% 2. Exact recovery via lq-minimization
% - Our results are to be stated in terms of a quantity invariant under the change A ← cA, namely
%   [ \gamma_{2s} := {alpha_{2s}}^{-2} {\beta_{2s}}^{2} \geq 1 ]. (p. 396)
the restricted isometry ratio or
% 3.) restricted isometry constant
% article:ChartrandISPL2007: Exact Reconstruction of Sparse Signals via Nonconvex Minimization
constant for
% 4.) smaller parameters q
smaller parameters $q$, ensure
% 5.) stable recovery
\TODO{really stable?}
the stable recovery and justify
% 6.) exact recovery
these findings
\cite[Thm. 3.1]{article:FoucartACHA2009},
\cite[Thm. 1]{article:ChartrandISPL2007}%
\footnote{
  % article:FoucartACHA2009: Sparsest solutions of underdetermined linear systems via lq-minimization for 0 < q \leq 1
  % 1. Introduction
  % - Chartrand [7] studied it
  %   [minimize_{ \vect{z} \in \R^{N} } \norm{ \vect{z} }{q} subject to \mat{A} \vect{z} = \vect{y}, 0 < q \leq 1, (P_{q})] in terms of
  %   Restricted Isometry Constants. (p. 396)
  %   [7] article:ChartrandISPL2007
  % - He stated that s-sparse vectors can be exactly recovered by solving (P_{q}) under the assumption that
  %   δ_{ as } + b δ_{ ( a + 1 ) s } < b − 1 holds for some b > 1 and a := b^{q / ( 2 − q )}. (p. 396)
  % - He then claimed that EXACT RECOVERY OF s-SPARSE VECTORS CAN BE OBTAINED FROM THE SOLUTION OF (P_{q}) FOR SOME q > 0 small enough,
  %   provided that δ_{2s+1} < 1. (p. 396)
  % - There was a MINOR IMPRECISION IN HIS ARGUMENTS, as he neglected the fact that as must be an integer when he chose
  %   the number a under the requirement 1 < a < 1 + 1 / s. (p. 396)
  % - A correct justification would be to define a := 1 + 1 / s, so that
  %   the sufficient condition δ_{as} + b δ_{( a + 1 ) s} < b − 1, where b := a^{( 2 − q ) / q} > 1, becomes feasible for q > 0 small
  %   enough as long as δ_{2s+1} < 1. (p. 396)
  \name{Foucart} \emph{et al.} \cite{article:FoucartACHA2009} corrected
  a minor imprecision in
  the argument.
}.

%---------------------------------------------------------------------------------------------------------------
% 4.) convergence of the proposed implementation for the nonconvex lq-minimization method
%---------------------------------------------------------------------------------------------------------------
% a) intractability of the nonconvex lq-minimization method necessitates the approximation of the global minima by local minima using suitable initializations
% article:Schiffner2018, Sect. II. Compressed Sensing in a Nutshell
% - The parameter $q = 1$ induces the convex $\ell_{1}$-minimization method, whose
%   implementation permits computationally efficient algorithms, whereas
%   the half-open parameter interval $q \in [ 0; 1 )$ induces
%   the nonconvex $\ell_{q}$-minimization method, whose
%   global intractability necessitates local approximations.
The intractability of
% 1.) nonconvex sparsity-promoting lq-minimization method, q \in ( 0; 1 )
the nonconvex $\ell_{q}$-minimization method
\eqref{eqn:recovery_reg_norm_lq_minimization}, $q \in [ 0; 1 )$, necessitates
% 2.) approximation
the approximation of
% 3.) global minima
the global minima by
% 4.) local minima
local minima using
% 5.) suitable initializations
suitable initializations.
% b) although the results indicate the convergence to the global minimum
% article:FoucartACHA2009: Sparsest solutions of underdetermined linear systems via lq-minimization for 0 < q \leq 1
% 4. Description of the algorithm
% - Unfortunately,
%   the CONVERGENCE OF THE WHOLE SEQUENCE (z_{n}) COULD NOT BE ESTABLISHED RIGOROUSLY. (p. 401)
% - However, several points beside the numerical experiments of Section 5 hint at
%   its CONVERGENCE TO THE ORIGINAL s-SPARSE VECTOR x. (p. 401)
% proc:ChartrandICASSP2008: Iteratively reweighted algorithms for compressive sensing
% 2. ALGORITHMS FOR NONCONVEX COMPRESSIVE SENSING
% - However, the fact that
%   in practice we are able to recover signals exactly, combined with
%   theoretical results [1, 18] that give circumstances in which (2) has a unique, global minimizer that is exactly u∗ = x, strongly suggests that
%   THE COMPUTED LOCAL MINIMIZERS ARE ACTUALLY GLOBAL,
%   at least UNDER A BROAD SET OF CIRCUMSTANCES. (p. 3870)
% article:ChartrandISPL2007: Exact Reconstruction of Sparse Signals via Nonconvex Minimization
% II. RESTRICTED ISOMETRY CONSTANTS
% - While WE PROVIDE NO GUARANTEES, the numerical results below strongly suggest that
%   THE LEAST-SQUARES SOLUTION IS OFTEN OR EVEN ALWAYS SUFFICIENTLY CLOSE, IF THERE ARE ENOUGH MEASUREMENTS. (p. 708)
Although
% 1.) proposed implementation
the proposed implementation does not guarantee
% 2.) recovery
the recovery of
% 3.) global minimum
the global minimum
\cite[Sect. 4]{article:FoucartACHA2009},
% 4.) observed improvements
the observed improvements indicate
% 5.) convergence
its convergence to
% 6.) specified nearly-sparse representations
the specified sparse representations
\eqref{eqn:recovery_reg_sparse_representation}.

%---------------------------------------------------------------------------------------------------------------
% 5.) optimal specification of the parameter q
%---------------------------------------------------------------------------------------------------------------
% a) simulation study focused on the two parameters q \in \{ 0.5; 1 \} in the sparsity-promoting lq-minimization method
The simulation study focused on
the two parameters $q \in \{ 0.5; 1 \}$.
% f) choice of q and number of iterations
% g) specification of the parameter q = 0.5 in the simulation study was arbitrary
% TODO: no significant improvement for q < 0.5
%\TODO{why $q = 0.5$? no improvement for lower}
% b) Foucart et al. demonstrated the benefits of retaining the sparsest result provided by a finite set of discrete parameters at the expense of higher computational costs
% article:FoucartACHA2009: Sparsest solutions of underdetermined linear systems via lq-minimization for 0 < q \leq 1
% 5. Numerical experiments
% - Thus, as the approximation of the original problem (P0),
%   ONE WOULD INTUITIVELY EXPECT THAT, among the approximations of the problems (Pq),
%   THE REWEIGHTED l1-MINIMIZATION IS THE BEST OPTION TO RECOVER SPARSE VECTORS. (p. 404)
% - THIS IS NOT THE CASE, though, and THERE APPEARS TO BE SOME ADVANTAGES IN LETTING THE PARAMETER q VARY,
%   as demonstrated by the numerical experiments below. (p. 404)
% - Fig. 2 also shows that the choice q = 0 is NOT UNEQUIVOCALLY THE BEST CHOICE FOR A SINGLE q. (p. 406)
% - Based on these considerations, our preferred values for the parameters are
%   [...] exponents: q ∈ { 0, 0.05, 0.1, 0.2 }. (p. 406)
% - Let us point out that our lq-algorithm allows several choices for q, including q = 0, and that
%   the SPARSEST OUTPUT PRODUCED FROM THESE CHOICES IS EVENTUALLY RETAINED. (p. 406)
% - In this way, it is no surprise that the lq-method performs at least as well as
%   the reweighted l1-minimization. (p. 406)
% - It is surprising, however, that it does perform better, even by a small margin. (p. 406)
\name{Foucart} \emph{et al.} \cite[Sect. 5]{article:FoucartACHA2009} demonstrated
the benefits of retaining
the sparsest result provided by
a finite set of
discrete parameters
$q \in [ 0; 1 ]$ at the expense of
higher computational costs.
% c) reweighted l1 minimization is a special instance of Foucart's algorithm for q = 0
% article:FoucartACHA2009: Sparsest solutions of underdetermined linear systems via lq-minimization for 0 < q \leq 1
% 5. Numerical experiments
% - We point out that the REWEIGHTED l1-MINIMIZATION discussed in [6], which
%   came to our attention while we were testing this scheme, is
%   the SPECIAL INSTANCE OF THE ALGORITHM (18) with
%   q = 0, \epsilon_{n} = \epsilon, z_{0} = minimizer of (P_{1}). (p. 404)
%   [6] article:CandesJFAA2008: Enhancing Sparsity by Reweighted l1 Minimization
The reweighted $\ell_{1}$ minimization
\cite{article:CandesJFAA2008} is
a special instance of
% 2.) Foucart's algorithm
\name{Foucart}'s algorithm for
$q = 0$ that does not unequivocally achieve
the best results.
% d) Achim et al. proposed a method to infer the optimal parameter q from the characteristic exponent of a symmetric \alpha-stable distribution
% article:AchimITCI2015: Reconstruction of Ultrasound RF Echoes Modeled as Stable Random Variables
% - ultrasound RF echoes are best characterized statistically by alpha-stable distributions.
% - Together, these two facts form the basis of an lp minimization approach that employs
%   the iteratively reweighted least squares (IRLS) algorithm, but in which
%   the parameter p is judiciously chosen, by relating it to
%   the characteristic exponent of the underlying alpha-stable distributed data.
\name{Achim} \emph{et al.} \cite{article:AchimITCI2015} proposed
a method to infer
the optimal parameter $q$ from
the characteristic exponent of
a \acl{SaS} distribution modeling
% 1.) temporal samples of the beamformed RF voltage signals
the temporal samples of
an individual beamformed \ac{RF} voltage signal or
% 2.) DFTs of the temporal samples
its \acp{DFT}.
% c) suitable statistical models for the nearly-sparse representation in the normalized CS problem permit similar methods
The author speculates that
suitable statistical models for
% 1.) normalized nearly-sparse representation / nearly-sparse normalized vector of transform coefficients
the normalized nearly-sparse representation
\eqref{eqn:recon_reg_norm_trans_coef} could enable
% 2.) similar methods
similar methods.

%%%%%%%%%%%%%%%%%%%%%%%%%%%%%%%%%%%%%%%%%%%%%%%%%%%%%%%%%%%%%%%%%%%%%%%%%%%%%%%%%%%%%%%%%%%%%%%%%%%%%%%%%%%%%%%%
% 3.) spatially extended structural building blocks increase the robustness against the additive errors
%%%%%%%%%%%%%%%%%%%%%%%%%%%%%%%%%%%%%%%%%%%%%%%%%%%%%%%%%%%%%%%%%%%%%%%%%%%%%%%%%%%%%%%%%%%%%%%%%%%%%%%%%%%%%%%%
\subsection{Spatially Extended Structural Building Blocks Increase the Robustness Against the Additive Errors}
\label{subsec:discussion_robustness_additive_errors}
For
% 1.) wire phantom
the wire phantom,
% 2.) substantial relative RMSEs
the substantial relative \acp{RMSE} at
% 3.) low reference SNRs of 3, 6, and 10 dB
the low reference \acp{SNR}
(cf. \cref{%
  fig:sim_study_obj_A_sr_1_quality_vs_snr_spgl1_lq_rnd_apo,%
  fig:sim_study_obj_A_sr_1_quality_vs_snr_spgl1_lq_rnd_del,%
  fig:sim_study_obj_A_sr_1_quality_vs_snr_spgl1_lq_rnd_apo_del%
}) revealed
% 4.) undesired sensitivity
an undesired sensitivity of
% 5.) estimated vector stacking the regular samples in the discretized relative spatial fluctuations in the unperturbed compressibility
the compressibility fluctuations
\eqref{eqn:recovery_reg_norm_lq_minimization_sol_mat_params} recovered by
% 6.) random waves
the random waves to
% 7.) energy
the energy of
% 8.) additive errors
the additive errors.
% b) strong variations in the incident acoustic energies at these positions and the resulting variations in the sample means of the relative RMSEs suggest that [...]
The variations in
% 1.) incident acoustic energy at a specified grid point (all pulse-echo measurements, multifrequent)
the incident acoustic energies
\eqref{eqn:recovery_p_in_energy} across
% 2.) isolated positions
the isolated positions of
% 3.) wires
the wires and
% 4.) variations
\TODO{chance of low energy and bad SNR}
% TODO: variations in the SNR of the Recorded electric energies
the resulting variations in
% 5.) mean relative RMSEs
the mean relative \acp{RMSE}
(cf. \cref{%
  fig:sim_study_obj_A_p_in_energy,%
  fig:sim_study_obj_A_sr_1_mean_rel_errors%
}) suggest that
% 6.) relocation of the wires to positions of high incident acoustic energies
(i) the energy-guided relocation of
the wires,
% 7.) different realizations of the random waves
(ii) different realizations of
the random waves, or
% 8.) multiple sequential pulse-echo measurements
(iii) multiple sequential pulse-echo measurements overcome
% 9.) sensitivity
this sensitivity.
% c) fixed experimental setup excludes option (i), and only option (iii) ensures consistent results for various configurations
A fixed experimental setup, however, excludes
% 1.) option (i)
option (i), and
% 2.) option (iii)
only option (iii) ensures
% 3.) consistent results
consistent results for
% 4.) various configurations
various configurations.
% d) variations further suggest that spatially extended structural building blocks reduce this sensitivity
The variations further suggest that
% 1.) spatially extended structural building blocks
(iv) spatially extended structural building blocks reduce
% 2.) sensitivity
this sensitivity.

%---------------------------------------------------------------------------------------------------------------
% 2.) enlarged passbands and increased robustness against the unknown additive errors
%---------------------------------------------------------------------------------------------------------------
% a) negligible relative RMSEs at the low reference SNRs indicate the robustness of the compressibility fluctuations recovered by the random waves against the additive errors
% article:Schiffner2018, Sect. VIII. Results / Sect. VIII-B. Tissue-Mimicking Phantom / Sect. VIII-B.4) Recovery by lq-Minimization (subsubsec:results_phantom_tissue_lq_minimization)
% - Using the nonconvex $\ell_{0.5}$-minimization method \eqreflqmin{eqn:recovery_reg_norm_lq_minimization}{ 0.5 },
%   the RANDOM WAVES consistently achieved
%   [1.)] MEAN \ac{SSIM} INDICES CLOSE TO UNITY and
%   [2.)] RELATIVE \acp{RMSE} BELOW \SI{6.86}{\percent} for all investigated reference \acp{SNR}.
% - The sample means of the normalized numbers of iterations increased from
%   at least \SI{27.19}{\percent} at the lowest reference \ac{SNR} to
%   at most \SI{82.13}{\percent} at the highest reference \ac{SNR} for
%   the superposition of randomly-apodized \acp{QCW}.
% - In contrast, the \ac{QPW} PRODUCED ONLY SLIGHTLY BETTER MEAN \ac{SSIM} INDICES and RELATIVE \acp{RMSE} than for
%   the convex $\ell_{1}$-minimization method \eqreflqmin{eqn:recovery_reg_norm_lq_minimization}{ 1 }.
Indeed, for
% 1.) tissue-mimicking phantom
the tissue-mimicking phantom,
% 2.) negligible relative RMSEs
the negligible relative \acp{RMSE} at
% 3.) low reference SNRs of 3, 6, and 10 dB
the low reference \acp{SNR}
(cf. \cref{%
  fig:sim_study_obj_B_sr_1_quality_vs_snr_spgl1_lq_rnd_apo,%
  fig:sim_study_obj_B_sr_1_quality_vs_snr_spgl1_lq_rnd_del,%
  fig:sim_study_obj_B_sr_1_quality_vs_snr_spgl1_lq_rnd_apo_del%
}) indicate
% 4.) robustness
the robustness of
% 5.) recovered relative spatial fluctuations in the unperturbed compressibility
the compressibility fluctuations
\eqref{eqn:recovery_reg_norm_lq_minimization_sol_mat_params} recovered by
% 6.) random waves
the random waves against
% 7.) additive errors
the additive errors.
% b) complex exponential functions scattered the incident acoustic pressure fields at all grid points in the FOV and completely reradiated their spatial variations
In contrast to
% 1.) punctiform wires
the wires,
% 2.) spatially extended complex exponential functions
the spatially extended complex exponential functions scattered
% 3.) discretized incident acoustic pressure fields [superpositions of quasi-(d-1)-spherical waves]
the incident acoustic pressure fields
\eqref{eqn:recovery_p_in} at
% 4.) all grid points in the FOV
all grid points in
the \ac{FOV} and, thus, completely reradiated
% 5.) spatial variations
their spatial variations.
% c) significantly enlarged passbands of the sensing matrices induced by the random waves prove that their scattered waves interfere less destructively on the faces of the array elements than those caused by the QPW
% article:Schiffner2018, Sect. VIII. Results / Sect. VIII-B. Tissue-Mimicking Phantom / Sect. VIII-B.1) Recorded Electric Energies (subsubsec:results_phantom_tissue_energy_rx)
% - The TRANSFER BEHAVIORS OF THE SENSING MATRICES \eqref{eqn:recovery_reg_sensing_matrix} induced by all incident waves resembled those of
%   BANDPASS FILTERS SUPPRESSING RELATIVELY LOW AND HIGH SPATIAL FREQUENCIES (cf. \cref{fig:sim_study_obj_B_norms_kappa}).
% - The \ac{QPW} induced relatively large electric energies \eqref{eqn:recovery_reg_v_rx_born_trans_coef_energy} exceeding \SI{-20}{\deci\bel} in
%   a SICKLE-SHAPED PASSBAND inside the interval of normalized spatial frequencies
%   $\hat{\vect{K}} \in [ \num{-0.24}; \num{0.24} ] \times [ \num{0.15}; \num{0.49} ]$, whereas
%   the RANDOM WAVES induced those in ARBELOS-SHAPED PASSBANDS inside the intervals of normalized spatial frequencies
%   $\hat{\vect{K}} \in [ \num{-0.43}; \num{0.43} ] \times [ \num{0}; \num{0.49} ]$.
% - All formed passbands, which were SIGNIFICANTLY ENLARGED BY THE LATTER WAVES, strongly agreed with
%   the predictions of the \ac{FDT} (cf. footnote $\num{1}$ in \cref{sec:introduction}).
The significantly enlarged passbands of
% 1.) sensing matrices (all pulse-echo measurements, multifrequent, all array elements)
the sensing matrices
\eqref{eqn:recovery_reg_sensing_matrix} induced by
% 2.) random waves
the random waves
(cf. \cref{%
  fig:sim_study_obj_B_norms_kappa_rnd_apo,%
  fig:sim_study_obj_B_norms_kappa_rnd_del,%
  fig:sim_study_obj_B_norms_kappa_rnd_apo_del%
}) prove that
% 3.) scattered waves
their scattered waves interfere
% 4.) less destructively
less destructively on
% 5.) faces of the array elements
the faces of
the array elements than
% 5.) scattered waves
those caused by
% 6.) quasi-plane wave (QPW)
the \ac{QPW}.
% d) significantly enlarged passbands explain both the reductions of the mean FEHMs relative to the QPW for the wire phantom and the robustness against the additive errors
In addition to
% 1.) robustness against the additive errors
the robustness, they explain
% 2.) reductions of the mean FEHMs relative to the QPW
the reductions of
% 3.) mean FEHMs
the mean \acp{FEHM} relative to
% 4.) quasi-plane wave (QPW)
the \ac{QPW} for
% 5.) wire phantom
the wire phantom
(cf. \cref{tab:sim_study_obj_A_sr_1_tpsf_fehm}).
% e) strong agreement with the predictions of the FDT further indicates the correctness of the numerical simulations
% article:Schiffner2018, Sect. VIII. Results / Sect. VIII-B. Tissue-Mimicking Phantom / Sect. VIII-B.1) Recorded Electric Energies (subsubsec:results_phantom_tissue_energy_rx)
% - All formed passbands, [...], STRONGLY AGREED WITH THE PREDICTIONS OF THE \ac{FDT} (cf. footnote $\num{1}$ in \cref{sec:introduction}).
Their strong agreement with
% 1.) predictions
the predictions of
% 2.) FDT
the \ac{FDT} further indicates
% 3.) correctness
the correctness of
% 4.) numerical simulations
the numerical simulations.

%%%%%%%%%%%%%%%%%%%%%%%%%%%%%%%%%%%%%%%%%%%%%%%%%%%%%%%%%%%%%%%%%%%%%%%%%%%%%%%%%%%%%%%%%%%%%%%%%%%%%%%%%%%%%%%%
% 4.) worst-case coherences provide impractical upper bounds on the numbers of nonzero components
%%%%%%%%%%%%%%%%%%%%%%%%%%%%%%%%%%%%%%%%%%%%%%%%%%%%%%%%%%%%%%%%%%%%%%%%%%%%%%%%%%%%%%%%%%%%%%%%%%%%%%%%%%%%%%%%
\subsection{Worst-Case Coherences Provide Impractical Upper Bounds on the Numbers of Nonzero Components}
%\label{subsec:discussion_coherence_measures}
%%%%%%%%%%%%%%%%%%%%%%%%%%%%%%%%%%%%%%%%%%%%%%%%%%%%%%%%%%%%%%%%%%%%%%%%%%%%%%%%%%%%%%%%%%%%%%%%%%%%%%%%%%%%%%%%
% table: upper bounds on the numbers of nonzero components and underlying worst-case coherences
%%%%%%%%%%%%%%%%%%%%%%%%%%%%%%%%%%%%%%%%%%%%%%%%%%%%%%%%%%%%%%%%%%%%%%%%%%%%%%%%%%%%%%%%%%%%%%%%%%%%%%%%%%%%%%%%
\begin{table*}[tb]
 \centering
 \caption{%
  % a) table summarizes the upper bounds on the numbers of nonzero components
  Upper bounds on
  % 1.) numbers of nonzero components in the sparse representations
  the numbers of
  nonzero components
  \eqref{eqn:disc_coherence_sparsity_ub} for
  % 2.) reference sensing matrices
  both reference sensing matrices and
  % 3.) sensing matrices (all pulse-echo measurements, multifrequent, all array elements)
  the sensing matrices
  \eqref{eqn:recovery_reg_sensing_matrix} induced by
  % 4.) all incident waves
  all incident waves.
  % b) best possible values arose from the Welch lower bounds on the worst-case coherences
  Their best possible values arose from
  % 1.) Welch lower bounds on the worst-case coherences of the sensing matrices
  the \name{Welch} lower bounds on
  the worst-case coherences
  \eqref{eqn:disc_coherence_lb_welch}, whereas
  % c) remaining values arose from the lower bounds on the worst-case coherences of the sensing matrices
  their remaining values arose from
  % 1.) lower bounds on the worst-case coherences of the sensing matrices
  the lower bounds on
  the worst-case coherences
  \eqref{eqn:disc_coherence_lb} provided by
  the arguments of
  the maxima in
  the empirical \acp{CDF} of
  the \acp{TPSF}
  \eqref{eqn:cs_math_tpsf}.
 }
 \label{tab:results_sr_1_tpsf_coherence}
 \begin{tabular}{%
  @{}%
  l%										01.) object
  S[table-format = 2,table-number-alignment = right,table-auto-round]%		02.) Welch lower bound
  @{ }>{[}l<{]}%                                                            	03.) Welch lower bound
  S[table-format = 1,table-number-alignment = right,table-auto-round]%		04.) RIP
  @{ }>{[}l<{]}%                                                            	05.) RIP
  S[table-format = 1,table-number-alignment = right,table-auto-round]%		06.) GWN
  @{ }>{[}l<{]}%                                                            	07.) GWN
  S[table-format = 1,table-number-alignment = right,table-auto-round]%		08.) QPW
  @{ }>{[}l<{]}%                                                            	09.) QPW
  S[table-format = 1,table-number-alignment = right,table-auto-round]%		10.) rnd. apo.
  @{ }>{[}l<{]}%                                                            	11.) rnd. apo.
  S[table-format = 1,table-number-alignment = right,table-auto-round]%		12.) rnd. del.
  @{ }>{[}l<{]}%                                                            	13.) rnd. del.
  S[table-format = 1,table-number-alignment = right,table-auto-round]%		14.) rnd. apo. del.
  @{ }>{[}l<{]}%                                                            	15.) rnd. apo. del.
  @{}%
 }
 \toprule
  \multicolumn{1}{@{}H}{\multirow{2}{*}{Object}} &
  \multicolumn{14}{H@{}}{Upper bound on the number of nonzero components (1) [lower bound on the worst-case coherence (\si{\percent})]}\\
  \cmidrule(lr){2-15}
  &
  \multicolumn{2}{H}{\name{Welch} lower bound} &
  \multicolumn{2}{H}{\acs{RIP}} &
  \multicolumn{2}{H}{\acs{GWN}} &
  \multicolumn{2}{H}{\acs{QPW}} &
  \multicolumn{2}{H}{Rnd. apo.} &
  \multicolumn{2}{H}{Rnd. del.} &
  \multicolumn{2}{H@{}}{Rnd. apo. del.}\\
  \cmidrule(r){1-1}\cmidrule(lr){2-3}\cmidrule(lr){4-5}\cmidrule(lr){6-7}\cmidrule(lr){8-9}
  \cmidrule(lr){10-11}\cmidrule(lr){12-13}\cmidrule(l){14-15}
 \addlinespace
  Wire phantom & 91 & 0.55 & 18 & 2.84 & 3 & 14.47 & 1 & 98.28 & 1 & 98.71 & 1 & 98.67 & 1 & 98.46\\
  Tissue-mimicking phantom & 90 & 0.56 & 23 & 2.19 & 1 & 40.82 & 1 & 97.87 & 1 & 94.68 & 1 & 94.21 & 1 & 94.77\\
 \addlinespace
 \bottomrule
 \end{tabular}
\end{table*}

%---------------------------------------------------------------------------------------------------------------
% 1.) upper bounds on the numbers of nonzero components and underlying worst-case coherences
%---------------------------------------------------------------------------------------------------------------
% a) arguments of the maxima in the empirical CDFs of the TPSFs bounded from below the worst-case coherences of the sensing matrices
The arguments of
the maxima in
the empirical \acp{CDF} of
the \acp{TPSF}
\eqref{eqn:cs_math_tpsf}
(cf. \cref{%
  fig:sim_study_obj_A_sr_1_tpsf_ecdfs,%
  fig:sim_study_obj_B_sr_1_tpsf_ecdfs%
}) bounded from
below
% 1.) worst-case coherences of the sensing matrices (all pulse-echo measurements, multifrequent, all array elements)
% book:Foucart2013, Chapter 5: Coherence / Sect. 5.1: Definitions and Basic Properties
% - Definition 5.1.
%   Let \mat{A} \in \C^{ m \times N } be a matrix with l2-NORMALIZED COLUMNS \vect{a}_{1}, ..., \vect{a}_{N}, i.e.,
%   \norm{ \vect{a}_{i} }{2} = 1 for all i \in \setcons{N}.
%   The coherence \mu = \mu( \mat{A} ) of the matrix \mat{A} is defined as
%   [ \mu := \underset{ 1 \leq i \neq j \leq N }{ \max } \abs{ \inprod{ \vect{a}_{i} }{ \vect{a}_{j} } } ]. (5.1) (p. 111)
the worst-case coherences of
the sensing matrices
\eqref{eqn:recovery_reg_sensing_matrix}, which are defined as
\cite[Def. 5.1]{book:Foucart2013}
\begin{equation}
 %--------------------------------------------------------------------------------------------------------------
 % worst-case coherences of the sensing matrices (all pulse-echo measurements, multifrequent, all array elements)
 %--------------------------------------------------------------------------------------------------------------
  \coher\bigl( \mat{A}\bigl[ p^{(\text{in})} \bigr] \bigr)
  =
  \underset{ n_{1} \neq n_{2} }{ \max }
  \Bigl\{
    \dabs{ \dtpsf{ \mat{A}\bigl[ p^{(\text{in})} \bigr] }{1}( n_{1}, n_{2} ) }{1}
  \Bigr\}
 \label{eqn:disc_coherence}
\end{equation}
and obey
% 2.) Welch lower bounds on the worst-case coherences of the sensing matrices
% book:Foucart2013, Chapter 5: Coherence / Sect. 5.
% - Unsurprisingly, a system of l2-normalized vectors is called
%   an EQUIANGULAR TIGHT FRAME if it is both an EQUIANGULAR SYSTEM AND A TIGHT FRAME. (p. 114)
% - Such systems are the ones achieving the lower bound given below and known as the WELCH BOUND. (p. 114)
% - Theorem 5.7.
%   The coherence of a matrix \mat{A} \in \K^{ m \times N } with l2-NORMALIZED COLUMNS satisfies
%   [ mu \geq \sqrt{ \frac{ N - m }{ m ( N - 1 ) } } ]. (5.4)
%   Equality holds if and only if the columns \vect{a}_{1}, ..., \vect{a}_{N} of the matrix \mat{A} form
%   an EQUIANGULAR TIGHT FRAME. (p. 114)
the \name{Welch} lower bounds
\cite[Thm. 5.7]{book:Foucart2013}
\begin{equation}
 %--------------------------------------------------------------------------------------------------------------
 % Welch lower bounds on the worst-case coherences of the sensing matrices
 %--------------------------------------------------------------------------------------------------------------
  \coherlbwelch\bigl( \mat{A}\bigl[ p^{(\text{in})} \bigr] \bigr)
  =
  \sqrt{ \frac{ N_{\text{lat}} - N_{\text{obs}} }{ N_{\text{obs}} ( N_{\text{lat}} - 1 ) } },
 \label{eqn:disc_coherence_lb_welch}
\end{equation}
according to
the inequality
% 3.) lower bounds on the worst-case coherences of the sensing matrices
\begin{equation}
 %--------------------------------------------------------------------------------------------------------------
 % lower bounds on the worst-case coherences of the sensing matrices
 %--------------------------------------------------------------------------------------------------------------
  \coherlbwelch\bigl( \mat{A}\bigl[ p^{(\text{in})} \bigr] \bigr)
  \leq
  \coherlb\bigl( \mat{A}\bigl[ p^{(\text{in})} \bigr] \bigr)
  \leq
  \coher\bigl( \mat{A}\bigl[ p^{(\text{in})} \bigr] \bigr).
 \label{eqn:disc_coherence_lb}
\end{equation}
These measures, in turn, loosely bounded from above
% 1.) restricted isometry constants (RICs)
the \acp{RIC} of
% 2.) normalized sensing matrices (all pulse-echo measurements, multifrequent, all array elements)
the normalized sensing matrices
\eqref{eqn:recon_reg_norm_sensing_matrix} for
% 3.) 2s-sparse representations
$2s$ nonzero components
\cite[Prop. 6.2]{book:Foucart2013}, i.e.
\begin{equation}
 %--------------------------------------------------------------------------------------------------------------
 % loose upper bounds on the RICs provided by the worst-case coherences
 %--------------------------------------------------------------------------------------------------------------
  \resisoconst{ 2s }\bigl( \bar{\mat{A}}_{\xi}\bigl[ p^{(\text{in})} \bigr] \bigr)
  \leq
  ( 2s - 1 )
  \coher\bigl( \mat{A}\bigl[ p^{(\text{in})} \bigr] \bigr)
  <
  \resisoconstub{ 2s }
 \label{eqn:disc_coherence_resisoconst_ub}
\end{equation}
for
% 1.) all factors for the normalization of the sensing matrices
all factors
$\xi \leq \underset{ i \in \setcons{ N_{\text{lat}} } }{ \min }\{ \tnorm{ \vect{a}_{i}[ p^{(\text{in})} ] }{2} \} / \underset{ i \in \setcons{ N_{\text{lat}} } }{ \max }\{ \tnorm{ \vect{a}_{i}[ p^{(\text{in})} ] }{2} \}$, where
% 2.) specified upper bound on the RIC
% article:Schiffner2018, Sect. II: Compressed Sensing (sec:compressed_sensing)
% - Multiple sufficient conditions on the sensing matrix \eqref{eqn:cs_math_prob_general_sensing_matrix} ensure
%   the stable recovery of the nearly-sparse representation \eqref{eqn:def_transform_coefficients} in
%   the \ac{CS} problem \eqref{eqn:cs_math_prob_general} by the sparsity-promoting $\ell_{q}$-minimization method \eqref{eqn:cs_lq_minimization}.
% - These conditions impose specific upper bounds on various characteristic measures quantifying the suitability of
%   the sensing matrix \eqref{eqn:cs_math_prob_general_sensing_matrix}, e.g.
%   [1.)] the null space constants \cite[Def. 4.21]{book:Foucart2013}, \cite[Def. 1.2]{book:Eldar2012},
%   [2.)] the restricted isometry ratio \cite{article:FoucartACHA2009}, and
%   [3.)] the \ac{RIC} \cite{article:FoucartACHA2010,article:CandesCRAS2008,article:CandesSPM2008}.
$\resisoconstub{ 2s }$ denotes
the specific upper bound imposed by
a suitable sufficient condition ensuring
the stable recovery.
% c) complete relaxation of this upper bound and the insertions of the lower bounds on the worst-case coherences yield the upper bounds on the numbers of nonzero components in the sparse representations
% book:Foucart2013, Chapter 6: Restricted Isometry Property / Sect. 6.1: Definitions and Basic Properties
% - We make a FEW REMARKS before establishing the equivalence of these two definitions. (p. 134)
% - The second one is that, although \resisoconst{ s } ≥ 1 is not forbidden,
%   THE RELEVANT SITUATION OCCURS FOR \resisoconst{ s } < 1. (p. 134)
% - Indeed, (6.2) says that EACH COLUMN SUBMATRIX \mat{A}_{S}, S \subset \setcons{N} with \card{S} \leq s,
%   HAS ALL ITS SINGULAR VALUES IN THE INTERVAL [ 1 − \resisoconst{ s }, 1 + \resisoconst{ s } ] and
%   IS THEREFORE INJECTIVE WHEN \resisoconst{ s } < 1. (p. 134)
% - In fact, \resisoconst{ 2s } < 1 IS MORE RELEVANT, since the inequality (6.1) yields
%   \norm{ \mat{A} ( \vect{x} − \vect{x}' ) }{2}^{2} > 0 for
%   all distinct s-sparse vectors \vect{x}, \vect{x}' \in \C^{N}; hence,
%   DISTINCT s-SPARSE VECTORS HAVE DISTINCT MEASUREMENT VECTORS. (p. 134)
The complete relaxation of
% 1.) upper bound on the RIC
this upper bound, i.e.
$\resisoconstub{ 2s } = 1$
\cite[134]{book:Foucart2013}, and
the successive insertions of
% 2.) lower bounds on the worst-case coherences of the sensing matrices
both lower bounds on
the worst-case coherences
\eqref{eqn:disc_coherence_lb} yield
the upper bounds on
% 3.) numbers of nonzero components in the sparse representations
the numbers of
nonzero components
\begin{equation}
 %--------------------------------------------------------------------------------------------------------------
 % upper bounds on the numbers of nonzero components in the sparse representations
 %--------------------------------------------------------------------------------------------------------------
  s
  <
  \frac{ 1 }{ 2 }
  \biggl[
    \frac{ 1 }{ \coherlb\bigl( \mat{A}\bigl[ p^{(\text{in})} \bigr] \bigr) }
    +
    1
  \biggr]
  \leq
  \frac{ 1 }{ 2 }
  \biggl[
    \frac{ 1 }{ \coherlbwelch\bigl( \mat{A}\bigl[ p^{(\text{in})} \bigr] \bigr) }
    +
    1
  \biggr].
 \label{eqn:disc_coherence_sparsity_ub}
\end{equation}

%---------------------------------------------------------------------------------------------------------------
% 2.) upper bounds on the numbers of nonzero components did not provide any practical guarantees
%---------------------------------------------------------------------------------------------------------------
% a) upper bounds [number of nonzero components] did not provide any practical guarantees and seemed to contradict the numerical simulations
These upper bounds, however, did not provide
any practical guarantees for
% 1.) normalized sensing matrices (all pulse-echo measurements, multifrequent, all array elements)
the normalized sensing matrices
\eqref{eqn:recon_reg_norm_sensing_matrix} induced by
% 2.) all incident waves
all incident waves and, thus, seemed to contradict
% 3.) numerical simulations
the numerical simulations
(cf. \cref{tab:results_sr_1_tpsf_coherence}).
% b) both reference sensing matrices only ensured the recovery of at most 22 nonzero components
% TODO: really? bound on RIC is 1! => injectivity only
Moreover,
both reference sensing matrices only ensured
the recovery of
% s = 22 because s < 23
at most $22$ nonzero components, and
% c) neither of the investigated sensing matrices achieved the Welch lower bounds on the worst-case coherences
% article:KutyniokGAMM2013: Theory and applications of compressed sensing
% Sect. 3: Conditions for Sparse Recovery / Sect. 3.2: Sufficient Conditions / Sect. 3.2.1: Mutual Coherence
% - The LOWER BOUND presented in the next result, also KNOWN AS THE WELCH BOUND, is more interesting. (p. 88)
% - Lemma 3.7:
%	- Let A be an m×n matrix. Then we have [\mu(\mat{A}) \in [ \sqrt{ \frac{ n - m }{ m (n - 1) } } ; 1 ]]. (p. 88)
neither of
% 1.) sensing matrices (all pulse-echo measurements, multifrequent, all array elements)
the investigated sensing matrices
\eqref{eqn:recovery_reg_sensing_matrix} achieved
% 2.) Welch lower bounds on the worst-case coherences of the sensing matrices
the \name{Welch} lower bounds
\eqref{eqn:disc_coherence_lb_welch}.
% d) multiple independent studies confirmed the worst-case coherences close to unity in pulse-echo UI
\TODO{literature}
Multiple authors confirmed
% 1.) impractical upper bounds
% article:TillmannITIT2014: The Computational Complexity of the Restricted Isometry Property, the Nullspace Property, and Related Concepts in Compressed Sensing
% V. CONCLUDING REMARKS
% - Instead of the INTRACTABLE RIP, NSP, or spark,
%   the weaker but efficiently computable MUTUAL COHERENCE [24] is sometimes used. (p. 1257)
% - It can be shown that
%   THE MUTUAL COHERENCE YIELDS BOUNDS ON THE RIC, NSC, and the SPARK; see, for instance, [27], [58]. (p. 1257)
% - Thus,
%   IMPOSING CERTAIN CONDITIONS INVOLVING THE MUTUAL COHERENCE OF A MATRIX CAN YIELD UNIQUENESS AND RECOVERABILITY
%   (by basis pursuit or other heuristics), see, e.g., [58]. (p. 1257)
% - However, the SPARSITY LEVELS for which the MUTUAL COHERENCE can guarantee recoverability are
%   quite often TOO SMALL TO BE OF PRACTICAL USE. (p. 1257)
this impracticality
\cite{article:TillmannITIT2014} and
% 2.) worst-case coherences close to unity underlying these upper bounds in pulse-echo UI
the worst-case coherences close to unity in
pulse-echo \ac{UI}
\cite[Fig. 6]{article:BessonITUFFC2018},
%\cite{article:BessonITUFFC2016},
\cite[Table 1]{proc:BessonICIP2016},
\cite[Fig. 9]{article:DavidJASA2015}.

%---------------------------------------------------------------------------------------------------------------
% 3.) reasons for the impracticality of the upper bounds on the numbers of nonzero components
%---------------------------------------------------------------------------------------------------------------
% a) impracticality of the upper bounds arose from both the looseness of the upper bounds and the limited spatial and spectral resolutions
The impracticality of
% 1.) upper bounds on the numbers of nonzero components in the sparse representations
the upper bounds
\eqref{eqn:disc_coherence_sparsity_ub} arose from both
% 2.) looseness of the upper bounds on the RICs provided by the worst-case coherences
the looseness of
the upper bounds on
the \acp{RIC} provided by
the worst-case coherences
\eqref{eqn:disc_coherence_resisoconst_ub} and
% 3.) limited spatial and spectral resolutions
the limited spatial and
spectral resolutions.
% b) limited spatial and spectral resolutions prevented the recovery of sparse representations whose nonzero components populated adjacent grid points or discrete spatial frequencies
The latter prevented
the recovery of
% 1.) sparse representations
sparse representations
\eqref{eqn:recovery_reg_sparse_representation} whose
% 2.) nonzero components
nonzero components populated
% 3.) adjacent grid points or discrete spatial frequencies
adjacent grid points or
discrete spatial frequencies.
% c) upper bounds s < 2 prohibited such configurations and reflected this limitation
The upper bounds $s < 2$ prohibited
such configurations and reflected
this limitation.
% d) upper bounds s < 2 did not contradict the numerical simulations because specific configurations meeting the resolution requirements were recoverable
They did not contradict
the numerical simulations because
specific configurations meeting
the resolution requirements were
recoverable.
% e) increase of the constant spacing between the adjacent grid points along each coordinate axis effectively reduces the worst-case coherences and increases the upper bounds
The increase of
% 1.) constant spacings between the adjacent grid points along each coordinate axis
the constant spacings between
the adjacent grid points in
the \ac{FOV} trivially decorrelates
% 2.) pulse echoes
their pulse echoes and, thus, effectively reduces
% 3.) worst-case coherences of the sensing matrices (all pulse-echo measurements, multifrequent, all array elements)
the worst-case coherences
\eqref{eqn:disc_coherence} and increases
% 4.) upper bounds on the numbers of nonzero components in the sparse representations
the upper bounds
\eqref{eqn:disc_coherence_sparsity_ub}.
% e) faithful discrete representations of the continuous physical models impose upper bounds on these spacings and enforce a trade-off
Faithful discrete representations of
the continuous physical models, however, impose
upper bounds on
these spacings and enforce
a trade-off.
% f) findings emphasize the importance of alternative concepts to assess the suitability of the sensing matrices
% article:TillmannITIT2014: The Computational Complexity of the Restricted Isometry Property, the Nullspace Property, and Related Concepts in Compressed Sensing
% V. CONCLUDING REMARKS
% - This [too small sparsity levels to be of practical use] emphasizes the IMPORTANCE OF OTHER CONCEPTS. (p. 1257)
The impracticality of
% 1.) upper bounds on the numbers of nonzero components in the sparse representations
the upper bounds
\eqref{eqn:disc_coherence_sparsity_ub} emphasizes
the importance of
alternative concepts to assess
the suitability of
% 2.) normalized sensing matrices (all pulse-echo measurements, multifrequent, all array elements)
the normalized sensing matrices
\eqref{eqn:recon_reg_norm_sensing_matrix}, e.g.
% 2.) investigation of the TPSFs
% article:Schiffner2018, Sect. II: Compressed Sensing (sec:compressed_sensing)
% - The \ac{TPSF} frequently quantifies the coherence of the sensing matrices \eqref{eqn:cs_math_prob_general_sensing_matrix} in
%   medical imaging technologies (cf. e.g. \cite{article:ProvostITMI2009,article:LustigMRM2007}).
the presented and
frequently-used investigation of
the \acp{TPSF}
(cf. e.g.
\cite{article:ProvostITMI2009,article:LustigMRM2007}%
) and
% 3.) quantitative recovery experiments
quantitative recovery experiments.
% f) investigation of the TPSFs and recovery experiments empirically revealed the advantages of the random waves over the QPW
These empirically revealed
the advantages of
% 1.) random waves
the random waves over
% 2.) quasi-plane wave (QPW)
the \ac{QPW}.

%%%%%%%%%%%%%%%%%%%%%%%%%%%%%%%%%%%%%%%%%%%%%%%%%%%%%%%%%%%%%%%%%%%%%%%%%%%%%%%%%%%%%%%%%%%%%%%%%%%%%%%%%%%%%%%%
% 5.) multiple sequential pulse-echo measurements per image do not resolve the effective underdeterminedness
%%%%%%%%%%%%%%%%%%%%%%%%%%%%%%%%%%%%%%%%%%%%%%%%%%%%%%%%%%%%%%%%%%%%%%%%%%%%%%%%%%%%%%%%%%%%%%%%%%%%%%%%%%%%%%%%
\subsection{Multiple Sequential Pulse-Echo Measurements per Image do not Resolve the Effective Underdeterminedness}
\label{subsec:discussion_ill_conditioning}
%---------------------------------------------------------------------------------------------------------------
% 1.) multiple sequential pulse-echo measurements per image
%---------------------------------------------------------------------------------------------------------------
% a) number of sequential pulse-echo measurements per image simultaneously controls both the acquisition time and the number of observations
The number of
sequential pulse-echo measurements per
image simultaneously controls both
% 1.) acquisition time
the acquisition time and
% 2.) number of observations (all pulse-echo measurements, multifrequent, all transducer elements)
the number of
observations
\eqref{eqn:recovery_sys_lin_eq_num_obs}.
% b) its sufficient increase formally resolves the typical underdeterminedness of the linear algebraic system in ultrafast UI
% article:Schiffner2018, Sect. V.D. Regularization of the Discretized Linear Inverse Scattering Problem (subsec:recovery_regularization)
% - The linear algebraic system \eqref{eqn:recovery_sys_lin_eq_v_rx_born_all_f_all_in} is
%   [1.)] ill-conditioned and, for only a few sequential pulse-echo measurements,
%   [2.)] TYPICALLY UNDERDETERMINED.
% - The UNDERDETERMINEDNESS RESULTS FROM THE RELATIVELY SMALL NUMBER OF SEQUENTIAL PULSE-ECHO MEASUREMENTS PER IMAGE in
%   ultrafast \ac{UI}.
% - It necessitates the identification of the true vector stacking the regular samples in the discretized compressibility fluctuations
%   \eqref{eqn:recovery_sys_lin_eq_gamma_kappa_bp_vector} among infinitely many admissible vectors.
Its sufficient increase formally resolves
% 1.) typical underdeterminedness
the typical underdeterminedness of
% 2.) linear algebraic system (all pulse-echo measurements, multifrequent, all transducer elements)
the linear algebraic system
\eqref{eqn:recovery_sys_lin_eq_v_rx_born_all_f_all_in} in
% 3.) ultrafast UI
ultrafast \ac{UI}.
% c) wire and the tissue-mimicking phantoms required the minimal numbers 9 and 10 for this purpose
% total number of lattice points: N_{\text{lat}} = \num{262144}
% wire phantom:             N_{\text{lat}} / N_{\text{obs}} = 8.904347826 ( N_{\text{obs}} = \num{29440} )
% tissue-mimicking phantom: N_{\text{lat}} / N_{\text{obs}} = 9.102222222 ( N_{\text{obs}} = \num{28800} )
% 1.) wire phantom
The wire and
% 2.) tissue-mimicking phantom
the tissue-mimicking phantoms, for instance, required
the minimal numbers
$N_{\text{in}} \geq 9$ and
$N_{\text{in}} \geq 10$,
respectively, for
this purpose.
Even
% 1.) observation process (all pulse-echo measurements, multifrequent, all transducer elements)
an observation process
\eqref{eqn:recovery_sys_lin_eq_v_rx_born_all_f_all_in_mat} of
% 2.) full mathematical rank
full mathematical rank, however, remains
% 3.) ill-conditioned
ill-conditioned and, thus, exhibits
% 4.) nontrivial quasi-nullspace
a nontrivial quasi-nullspace, i.e.
a set of
% 5.) nonzero vectors stacking the regular samples in the discretized relative spatial fluctuations in the unperturbed compressibility
nonzero compressibility fluctuations
\eqref{eqn:recovery_sys_lin_eq_gamma_kappa_bp_vector} whose pulse echoes contain
% 6.) relatively low electric energies
relatively low electric energies
\cite[4]{book:Hansen2010},
\cite[1, 3]{book:Hansen1998}.
% e) vectors represent lossy heterogeneous objects that are almost invisible to the pulse-echo measurement process
These vectors represent
lossy heterogeneous objects that are
almost invisible to
the pulse-echo measurement process
\cite[Sects. 6.9.1 and 6.9.4]{book:Devaney2012}.
% TODO: enable discretization -> invisible differences to continuous object; rows vs cols
% f) their existence renders the linear algebraic system effectively underdetermined
% book:Hansen2010, Chapter 1: Introduction and Motivation
% - Hence
%   WE CAN ADD A LARGE AMOUNT OF THIS VECTOR TO THE SOLUTION VECTOR WITHOUT CHANGING THE RESIDUAL VERY MUCH;
%   THE SYSTEM BEHAVES ALMOST LIKE AN UNDERDETERMINED SYSTEM. (p. 4)
% - By supplying the correct additional information we can compute a good approximate solution. (p. 4)
% - The main difficulty is how to choose the parameter δ when we have little knowledge about the exact solution. (p. 4)
Their existence renders
% 1.) linear algebraic system (all pulse-echo measurements, multifrequent, all transducer elements)
the linear algebraic system
\eqref{eqn:recovery_sys_lin_eq_v_rx_born_all_f_all_in} effectively underdetermined because
the additions of
% 2.) linear combinations
their linear combinations to
% 3.) exact solution
any exact solution only induce
% 4.) negligible residuals
negligible residuals
\cite[4]{book:Hansen2010},
\cite[1, 3]{book:Hansen1998}.
% g) complex exponential functions of relatively low and high spatial frequencies exemplified such vectors
% article:Schiffner2018, Sect. VIII. Results / Sect. VIII-B. Tissue-Mimicking Phantom / Sect. VIII-B.1) Recorded Electric Energies (subsubsec:results_phantom_tissue_energy_rx)
% - The TRANSFER BEHAVIORS OF THE SENSING MATRICES \eqref{eqn:recovery_reg_sensing_matrix} induced by all incident waves resembled those of
%   BANDPASS FILTERS SUPPRESSING RELATIVELY LOW AND HIGH SPATIAL FREQUENCIES (cf. \cref{fig:sim_study_obj_B_norms_kappa}).
% - The HIGH DYNAMIC RANGES exceeding \SI{70}{\deci\bel} indicated the existence of structural building blocks whose
%   pulse echoes contained RELATIVELY LOW ELECTRIC ENERGIES \eqref{eqn:recovery_reg_v_rx_born_trans_coef_energy}.
The complex exponential functions of
relatively low and
high spatial frequencies exemplified
such vectors for
% 1.) observation processes (all pulse-echo measurements, multifrequent, all array elements)
the observation processes
\eqref{eqn:recovery_sys_lin_eq_v_rx_born_all_f_all_in_mat} induced by
% 2.) all incident waves
all waves incident on
% 3.) tissue-mimicking phantom
the tissue-mimicking phantom
(cf. \cref{fig:sim_study_obj_B_norms_kappa}).
% TODO: complex exponential functions potentially falsify any recovered vector
% h) maximal passbands are invariant to the number of sequential pulse-echo measurements per image
The maximal passbands, which are limited by
% 1.) electromechanical transfer behavior of the instrumentation
the electromechanical transfer behavior of
the instrumentation and approximately achieved by
% 2.) random waves
the random waves, are invariant to
% 3.) number of sequential pulse-echo measurements per image
the number of
sequential pulse-echo measurements per image.
% i) increase of the latter [number of sequential pulse-echo measurements per image] does not eliminate the need for regularization
Although
the increase of
% 1.) number of sequential pulse-echo measurements per image
the latter does not eliminate
% 2.) need for regularization
the need for
regularization,
% j) increase of the latter enlarges the suboptimal passbands achieved by the steered QPWs, improves the robustness of the recovered compressibility fluctuations against the additive errors
% article:Schiffner2018, Sect. IX. Discussion / Sect. IX-C. Spatially Extended Structural Building Blocks Increase the Robustness Against the Additive Errors
% - The variations in the incident acoustic energies \eqref{eqn:recovery_p_in_energy} across the isolated positions of the wires and
%   the resulting variations in the mean relative \acp{RMSE} (cf. \cref{fig:sim_study_obj_A_p_in_energy,fig:sim_study_obj_A_sr_1_mean_rel_errors}) suggest that
%   (i) the energy-guided relocation of the wires,
%   (ii) different realizations of the random waves, or
%   (iii) MULTIPLE SEQUENTIAL PULSE-ECHO MEASUREMENTS OVERCOME
%   THIS SENSITIVITY.
% - A fixed experimental setup, however, excludes option (i), and only option (iii) ensures consistent results for various configurations.
it (i) enlarges
% 1.) suboptimal passbands achieved by the steered QPWs
the suboptimal passbands achieved by
the steered \acp{QPW} for
% 2.) preferred directions of propagation
various preferred directions of
propagation and
(ii) improves
the robustness of
% 3.) recovered relative spatial fluctuations in the unperturbed compressibility
the recovered compressibility fluctuations
\eqref{eqn:recovery_reg_norm_lq_minimization_sol_mat_params} against
% 4.) additive errors
the additive errors
(cf. \cref{subsec:discussion_robustness_additive_errors}).

%%%%%%%%%%%%%%%%%%%%%%%%%%%%%%%%%%%%%%%%%%%%%%%%%%%%%%%%%%%%%%%%%%%%%%%%%%%%%%%%%%%%%%%%%%%%%%%%%%%%%%%%%%%%%%%%
% 6.) influence of the random apodization weights and time delays
%%%%%%%%%%%%%%%%%%%%%%%%%%%%%%%%%%%%%%%%%%%%%%%%%%%%%%%%%%%%%%%%%%%%%%%%%%%%%%%%%%%%%%%%%%%%%%%%%%%%%%%%%%%%%%%%
%\subsection{Influence of the Random Apodization Weights and Time Delays}
%\label{subsec:disc_parameters}
%\input{discussion/discussion_parameters.tex}

%%%%%%%%%%%%%%%%%%%%%%%%%%%%%%%%%%%%%%%%%%%%%%%%%%%%%%%%%%%%%%%%%%%%%%%%%%%%%%%%%%%%%%%%%%%%%%%%%%%%%%%%%%%%%%%%
% 7.) significance of the matrix-vector products
%%%%%%%%%%%%%%%%%%%%%%%%%%%%%%%%%%%%%%%%%%%%%%%%%%%%%%%%%%%%%%%%%%%%%%%%%%%%%%%%%%%%%%%%%%%%%%%%%%%%%%%%%%%%%%%%
\subsection{Significance of the Fast Multipole Method for the Sparsity-Promoting $\ell_{q}$-Minimization Method}
\label{subsec:disc_matrix_vector_product}
%---------------------------------------------------------------------------------------------------------------
% 1.) significance of the FMM for both auxiliary functions in the proposed matrix-free implementation
%---------------------------------------------------------------------------------------------------------------
% a) numerical evaluations of tens to hundreds of matrix-vector products emphasize the significance of both auxiliary functions in the proposed matrix-free implementation
% article:Schiffner2018, Sect. VI: Implementation / Sect. VI-C: Sparsity-Promoting lq-Minimization Method (subsec:imp_lq_minimization)
% - \ac{SPGL1} is ITERATIVE and LEFT MULTIPLIED A SEQUENCE OF RECURSIVELY-GENERATED VECTORS by
%   the potentially densely-populated normalized sensing matrix \eqref{eqn:recon_reg_norm_sensing_matrix} or
%   its adjoint.
% - Its MATRIX-FREE IMPLEMENTATION interpreted each type of matrix-vector product as a linear map and dedicated
%   a CUSTOMIZED AUXILIARY FUNCTION to its numerical evaluation.
\TODO{relationship between number of iterations and matrix-vector products?}
The numerical evaluations of
% 1.) tens
tens to
% 2.) hundreds
hundreds of
% 3.) matrix-vector products
matrix-vector products involving
% 4.) normalized sensing matrix (all pulse-echo measurements, multifrequent, all array elements)
the normalized sensing matrix
\eqref{eqn:recon_reg_norm_sensing_matrix} or
% 5.) adjoint of the normalized sensing matrix (all pulse-echo measurements, multifrequent, all array elements)
its adjoint by
% 6.) sparsity-promoting lq-minimization method
the sparsity-promoting $\ell_{q}$-minimization method
\eqref{eqn:recovery_reg_norm_lq_minimization}
(cf. \cref{fig:sim_study_obj_A_sr_1_ssim_index_rel_rmse_N_iter_vs_snr_kap,fig:sim_study_obj_B_sr_1_ssim_index_rel_rmse_N_iter_vs_snr_kap}) emphasize
% 7.) significance
the significance of
% 8.) customized auxiliary function for the numerical evaluation of each type of matrix-vector product
both auxiliary functions in
% 9.) proposed matrix-free implementation
the proposed matrix-free implementation.
% b) normalized reductions in the memory consumption and the number of multiplications provided by the FMM confirm its effectiveness in achieving both aims of the auxiliary functions
% article:Schiffner2018, Sect. VIII: Results / Sect. VIII-C: Memory Consumption and Computational Costs (subsec:results_memory_flops)
% - The approximate decomposition of the observation process \eqref{eqn:recovery_sys_lin_eq_v_rx_born_all_f_all_in_mat} by
%   the \ac{FMM} REDUCED THE MEMORY CONSUMPTION, which theoretically amounted to
%   $M_{\text{conv}} = \SI{115}{\gibi\byte}$ for the wire phantom and
%   $M_{\text{conv}} = \SI{112.5}{\gibi\byte}$ for the tissue-mimicking phantom, to
%   $M_{\text{\acs{FMM}}} \approx \SI{2.24}{\percent} M_{\text{conv}} \approx \SI{2.58}{\gibi\byte}$ and
%   $M_{\text{\acs{FMM}}} \approx \SI{2.21}{\percent} M_{\text{conv}} \approx \SI{2.49}{\gibi\byte}$, respectively.
% - It concurrently REDUCED THE NUMBER OF MULTIPLICATIONS, which theoretically amounted to
%   $N_{\text{mul},\text{conv}} \approx \num{7.72e9}$ for the wire phantom and
%   $N_{\text{mul},\text{conv}} \approx \num{7.55e9}$ for the tissue-mimicking phantom, to
%   $N_{\text{mul},\text{\acs{FMM}}} \approx \SI{93.91}{\percent} N_{\text{mul},\text{conv}} \approx \num{7.25e9}$ and
%   $N_{\text{mul},\text{\acs{FMM}}} \approx \SI{92.32}{\percent} N_{\text{mul},\text{conv}} \approx \num{6.97e9}$, respectively.
The normalized reductions in
% 1.) memory consumption
the memory consumption and
% 2.) number of multiplications executed by the associated matrix-vector product
the number of
multiplications achieved by
% 3.) fast multipole method (FMM)
the \ac{FMM}, which amounted up to
% 4.) 97.8 %
\SI{97.8}{\percent} and
% 5.) 7.7 %
\SI{7.7}{\percent},
respectively, confirm
% 6.) effectiveness
its effectiveness in achieving
% 7.) aims
% article:Schiffner2018, Sect. VI: Implementation / Sect. VI-C: Sparsity-Promoting lq-Minimization Method (subsec:imp_lq_minimization)
% - Both functions aimed at
%   [1.)] circumventing the explicit storage of the associated matrix in the fast but limited \ac{RAM} and
%   [2.)] accelerating the numerical computations.
both aims of
% 8.) auxiliary functions
these functions.
% c) modern UI systems typically lack the amounts of fast RAM required for the explicit storage of the observation process
% article:Schiffner2018, Sect. VI: Implementation / Sect. VI-C: Sparsity-Promoting lq-Minimization Method (subsec:imp_lq_minimization)
% - In fact,
%   the memory consumption of the normalized sensing matrix \eqref{eqn:recon_reg_norm_sensing_matrix} and
%   the number of multiplications executed by the associated matrix-vector product pose
%   CHALLENGES FOR MODERN \ac{UI} SYSTEMS.
Although
% 1.) modern UI systems
modern \ac{UI} systems typically lack
% 2.) amounts of fast RAM
the amounts of
fast \ac{RAM} required for
% 3.) explicit storage
the explicit storage of
% 4.) observation process (all pulse-echo measurements, multifrequent, all array elements)
the observation process
\eqref{eqn:recovery_sys_lin_eq_v_rx_born_all_f_all_in_mat},
% d) modern UI systems can readily store its [observation process] approximate decomposition provided by the FMM
they can readily store
% 1.) approximate decomposition of the observation process
its approximate decomposition provided by
% 2.) fast multipole method (FMM)
the \ac{FMM}.
% e) detailed analysis of the FMM for the excitation by steered PWs reported even higher normalized reductions in the memory consumption and the number of multiplications of 99.75 % and 76 %, respectively
% proc:SchiffnerIUS2014: Pulse-Echo Ultrasound Imaging Combining Compressed Sensing and the Fast Multipole Method
% Abstract
% - For an EXAMPLE OF TYPICAL SIZE and in comparison to the conventional approach, we showed that
%   the FMM REQUIRES
%   [1.)] LESS THAN 0.25 % OF THE MEMORY AND
%   [2.)] LESS THAN 24 % OF THE NUMBER OF COMPLEX-VALUED MULTIPLICATIONS. (p. 2205)
A detailed analysis of
% 1.) fast multipole method (FMM)
the \ac{FMM} for
% 2.) excitation
the excitation by
% 3.) steered PWs
steered \acp{PW} reported
% 4.) higher normalized reductions
even higher normalized reductions in
% 5.) memory consumption
the memory consumption and
% 6.) number of multiplications executed by the associated matrix-vector product
the number of
multiplications of
$\SI{99.75}{\percent}$ and
$\SI{76}{\percent}$,
respectively
\cite{proc:SchiffnerIUS2014}.
% f) author hypothesizes that further optimizations enable even higher reductions
% article:Schiffner2018, Sect. VI: Implementation / Sect. VI-D: Fast Multipole Method for the Observation Process (subsec:imp_fmm_obs_process)
% - It [FMM] substituted the OUTGOING FREE-SPACE \name{Green}'s FUNCTIONS \eqref{eqn:app_helmholtz_green_free_space_2_3_dim} in
%   the entries of the observation process \eqref{eqn:recovery_sys_lin_eq_v_rx_born_coef} by
%   ERROR-REGULATED TRUNCATED MULTIPOLE EXPANSIONS if the grid points
%   $\vect{r}_{\text{lat}, i} \in \mathcal{L}$ and $\vect{r}_{\text{mat}, \nu}^{(m)} \in \mathcal{V}_{m}$, satisfied
%   a specific geometric relationship \cite[Chapt. 9]{book:Gibson2014}, \cite{article:CoifmanIAPM1993}.
The author hypothesizes that
% 1.) additional optimizations
additional optimizations in
% 2.) discretization
the discretization of
% 3.) truncated multipole expansions
the truncated multipole expansions enable
% 4.) further improvements
further improvements.

The availability of
% 1.) fast multipole method (FMM)
the \ac{FMM}, which results from
% 2.) usage
the usage of
% 3.) outgoing free-space Green's functions (two- and three-dimensional Euclidean spaces)
the outgoing free-space \name{Green}'s functions
\eqref{eqn:app_helmholtz_green_free_space_2_3_dim}, is
% 4.) crucial advantage
a crucial advantage of
% 5.) analytical derivation
the analytical derivation of
% 6.) observation process (all pulse-echo measurements, multifrequent, all array elements)
the observation process
\eqref{eqn:recovery_sys_lin_eq_v_rx_born_all_f_all_in_mat} over
% 7.) numerical constructions
the numerical constructions using
% 8.) free or commercial simulation software
simulation software or
% 9.) experimental measurements
experimental measurements proposed in
\cite{article:Ghanbarzadeh-DagheyanSensors2018,article:BerthonPMB2018}.
% b) FMM provides explicit analytical expressions for the sparse approximation of the observation process
The \ac{FMM} provides
% 1.) explicit analytical expressions
explicit analytical expressions for
% 2.) sparse approximation
the sparse approximation of
% 3.) observation process (all pulse-echo measurements, multifrequent, all array elements)
the observation process
\eqref{eqn:recovery_sys_lin_eq_v_rx_born_all_f_all_in_mat}, whereas
% c) algebraic methods rely on the numerical values of its entries or its action on a suitable vector
algebraic methods, e.g.
% 1.) singular value decomposition
% article:ChaillatJCP2012: FaIMS: A fast algorithm for the inverse medium problem with multiple frequencies and multiple sources for the scalar Helmholtz equation
% - algorithm to compute an approximate singular value decomposition (SVD) of M or least-squares operators (approximate SVD of the Born operator)
% - can be used to accelerate matrix-vector multiplications and to precondition iterative solvers
% - we consider small perturbations of the background medium and, by invoking the Born approximation, we obtain a linear least-squares problem
% - Finally one could form M and use a dense factorization algorithm, say, use an classical SVD factorization [15].
%   A dense SVD is prohibitively expensive because its work complexity is OðminðNsNxNd;NÞ? ?maxðNsNxNd;NÞÞ.
% - compute approximate SVDs of small submatrices by applying the randomized SVD
%	-> each submatrix is approximated by a low-rank matrix
% - recursive SVD to combine the approximate SVDs of the submatrices
% - complexity is orders of magnitude smaller than the standard SVD factorization
% - matrix-free
%,SchotlandJOSA2001
% \cite[Chapt. 9]{book:Gibson2014}: LU factorization (exponential in the number of unknowns) -> non-iterative
the \acl{SVD}
\cite{book:Hansen2010,book:Hansen1998} and
its enhancements
\cite{article:ChaillatJCP2012},
% 2.) QR-factorization: -> see Chaillat
% article:ChewITAP1997: Fast solution methods in electromagnetics
% III. INTEGRAL EQUATION SOLVERS
% - If the matrix equation is then solved by
%   LU decomposition (Gaussian elimination) or
%   alternatively by an iterative technique such as the CG or related methods [19], [20],
%   the computational labor may be excessive. (p. 535)
%the QR decomposition,
% 2.) adaptive cross approximation (ACA)
% article:Hesford2010: The fast multipole method and Fourier convolution for the solution of acoustic scattering on regular volumetric grids
% - Like methods such as the ADAPTIVE CROSS APPROXIMATION [15–17], efficient solutions are obtained by eliminating approximately redundant information from
%   interactions between sufficiently separated groups.
% - The ADAPTIVE CROSS APPROXIMATION uses an ALGEBRAIC METHOD to construct
%   PRODUCTS OF LOW-RANK MATRICES that cast all pairwise interactions in terms of fewer, dominant interactions.
% article:ZhaoITEC2005: The adaptive cross approximation algorithm for accelerated method of moments computations of EMC problems
the \acl{ACA}
\cite{article:ZhaoITEC2005}, or
% 3.) various transforms
% article:DemanetFCompMath2010: Scattering in Flatland: Efficient Representations via Wave Atoms
% - lossy numerical compression strategy for the boundary integral equation of acoustic scattering in two dimensions
% - system of equations has oscillatory kernels that are represented in a basis of wave atoms, and compressed by thresholding the small coefficients to zero
% - Numerical experiments support the estimate and show that this wave atom representation may be of interest for applications where the same scattering problem needs to
%   be solved for many boundary conditions, for example, the computation of radar cross sections.
% - efficient representation of the operator as a sparse matrix in a system of wave atoms
% - Most of the approaches on sparsifying (1) in well-chosen bases require the construction of the full integral operator
%	-> computational difficulty for large k values
% article:DemanetACHA2007: Wave atoms and sparsity of oscillatory patterns
% article:ChewITAP1997: Fast solution methods in electromagnetics
% III. INTEGRAL EQUATION SOLVERS
% - WAVELET TRANSFORMS [56]–[61] have also been used to yield SPARSE MATRICES that can be solved rapidly. (p. 535)
various transforms
\cite{article:DemanetFCompMath2010,article:DemanetACHA2007,article:ChewITAP1997}, rely on
% 4.) numerical values
the numerical values of
% 5.) entries of the observation process (single pulse-echo measurement, monofrequent, single array element)
its entries
\eqref{eqn:recovery_sys_lin_eq_v_rx_born_coef} or
% 6.) action
its action on
% 7.) suitable vector
a suitable vector.
% d) constancy of the observation process in multiple instances of the normalized CS problem reduces the relevance of the computational overhead
The constancy of
% 1.) observation process (all pulse-echo measurements, multifrequent, all array elements)
the observation process
\eqref{eqn:recovery_sys_lin_eq_v_rx_born_all_f_all_in_mat} in
% 2.) multiple instances
multiple instances of
% 3.) normalized CS problem
the normalized \ac{CS} problem
\eqref{eqn:recovery_reg_norm_prob_general}, which frequently occurs in
% 4.) practice
practice, reduces
% 5.) relevance
the relevance of
% 6.) computational overhead
the increased complexity.

\subsection{Reduction of the Number of Observations Enables Sub-\name{Nyquist} Sampling Rates}
%\label{subsec:disc_sub_nyquist_sampling}
%---------------------------------------------------------------------------------------------------------------
% 1.) reduction of the number of observations and its consequences
%---------------------------------------------------------------------------------------------------------------
% a) number of observations defines the minimal data volume required for the determination of the relevant Fourier coefficients
The number of
observations
\eqref{eqn:recovery_sys_lin_eq_num_obs} defines
% 1.) minimal data volume
the minimal data volume required for
% 2.) determination
the determination of
% 3.) relevant Fourier coefficients
the relevant \name{Fourier} coefficients
\eqref{eqn:recovery_disc_freq_v_rx_Fourier_series_coef}. % forming
% 2.) vector stacking the relevant Fourier coefficients of the recorded RF voltage signals (all pulse-echo measurements, multifrequent, all array elements)
%the vector
%\eqref{eqn:recovery_sys_lin_eq_v_rx_born_all_f_all_in_v_rx}.
% b) it [total number of observations] simultaneously controls both the memory consumption and the computational costs raised by the numerical evaluations of the matrix-vector products
It simultaneously controls both
% 1.) memory consumption
the memory consumption and
% 2.) computational costs
the computational costs raised by
the numerical evaluations of
the matrix-vector products involving
% 3.) normalized sensing matrix (all pulse-echo measurements, multifrequent, all array elements)
the normalized sensing matrix
\eqref{eqn:recon_reg_norm_sensing_matrix} or
% 4.) adjoint of the normalized sensing matrix (all pulse-echo measurements, multifrequent, all array elements)
its adjoint.
% c) methods for the reduction of the number of observations
% article:Schiffner2018, Sect. V-C. Systems of Linear Algebraic Equations Obtained from the First Born Approximation (subsec:recovery_systems_linear_equations)
% - The NUMBER OF OBSERVATIONS \eqref{eqn:recovery_sys_lin_eq_num_obs} exclusively depends on
%   [1.)] the total number of array elements $N_{\text{el}}$,
%   [2.)] the numbers of relevant discrete frequencies \eqref{eqn:recon_disc_axis_f_discrete_BP_TB_product}, and
%   [3.)] the number of sequential pulse-echo measurements $N_{\text{in}}$.
% article:Schiffner2018, Sect. III-D. Pulse-Echo Measurement Process
% - These methods differ in their \emph{efficiency}, i.e. the quotient of
%   the data volume occupied by the quantized relevant \name{Fourier} coefficients and
%   the DATA VOLUME DIGITIZED DURING THE PULSE-ECHO MEASUREMENT.
Besides minimizing
% 1.) number of sequential pulse-echo measurements per image
the number of
sequential pulse-echo measurements per
image,
(i) the deactivation of
selected receiving array elements
\cite[Sect. III-C.1)]{article:BessonITUFFC2018},
\cite[Sects. 3.2 and 4]{proc:BessonICIP2016},
\cite[Sect. V-B]{article:DavidJASA2015},
% 3.) random temporal mixing of the recorded RF voltage signals
% article:BessonITUFFC2018: Ultrafast Ultrasound Imaging as an Inverse Problem: Matrix-Free Sparse Image Reconstruction
% III. SPARSITY-DRIVEN IMAGE RECONSTRUCTION METHODS / C. Compressed Beamforming / 2) Mixing of the Raw Data
% - In order to overcome this drawback [increase of the coherence induced by simple undersampling schemes], TWO STRATEGIES, denoted as
%   [1.)] “CHANNEL MIXING” (CMIX) and
%   [2.)] “CHANNEL AND TIME MIXING” (CTMIX), are proposed. (p. 343)
% - CMIX consists of a RANDOM SUMMATION OF THE SIGNALS coming from the different transducers elements at a given time instant in order to create a mixed output. (p. 343)
% - CTMIX extends the principle of CMIX by considering mixing across both transducer elements and D_{t} TIME SAMPLES to limit the complexity. (p. 343)
(ii) the random temporal mixing of
the recorded \ac{RF} voltage signals
\cite[Sect. III-C.2)]{article:BessonITUFFC2018},
% 4.) random Gaussian projections
% article:ZhangUlt2013: A measurement-domain adaptive beamforming approach for ultrasound instrument based on distributed compressed sensing: Initial development
(iii) random temporal projections
\cite{article:ZhangUlt2013}, and
% 5.) customized alternative discretizations of the frequency axis
% article:ProvostITMI2009: The application of compressed sensing for photo-acoustic tomography
% VI. COMPRESSED SENSING / B. PA Forward Operator as a CS-Matrix / Proof of Concept on Small Phantoms
% - However, one needs to model the transducer response g_{n} appearing in K_{(m,n)(i,j)}. (p. 590)
% - As a first approximation, we used the SIMPLEST BAND-PASS FILTER:
%   frequencies at each angles have been sampled in a rectangular window between 0.2 and 1.2 MHz. (p. 590)
% - To completely define K_{(m,n)(i,j)}, we considered
%   a fixed number of tomographic angles indexed by m with
%   40 RANDOMLY CHOSEN k_{n} / 2 \pi c’s inside the [0.2, 1.2] MHz window for each and
%   the numerical derivative as a sparse basis. (p. 590)
% VII. RESULTS / A. Simulations
% - As before, we generated measurements y_{(m,n)} from a phantom x_{(i,j)} using K_{(m,n)(i,j)} and changed four parameters:
%   [1.)] the number of tomographic angles,
%   [2.)] the frequency window width,
%   [3.)] the SAMPLING OF THIS FREQUENCY WINDOW, and
%   [4.)] the basis used for reconstruction. (p. 591)
% VII. RESULTS / B. Experiments
% - Fig. 6(a) shows reconstruction using 56 angles taken outside a dead angle of 45° and
%   128 frequency samples RANDOMLY CHOSEN inside the [0.5, 6] MHz window. (p. 593)
% - Fig. 6(b) shows reconstruction using 50 angles taken outside a dead angle of 35° and
%   128 frequency samples RANDOMLY CHOSEN inside the [0.5, 6] MHz window. (p. 593)
(iv) customized alternative discretizations of
the frequency axis
\cite[Sects. VI-B and VII]{article:ProvostITMI2009} enable
% 6.) significant reduction
its significant reduction.
% d) customized alternative discretizations specifically reduce the cardinalities of the sets of relevant discrete frequencies below the effective time-bandwidth products
The latter specifically reduce
% 1.) cardinalities
the cardinalities of
% 2.) sets of relevant discrete frequencies
the sets of
relevant discrete frequencies
\eqref{eqn:recon_disc_axis_f_discrete_BP} below
% 3.) numbers of relevant discrete frequencies (effective time-bandwidth products)
the effective time-bandwidth products
\eqref{eqn:recon_disc_axis_f_discrete_BP_TB_product}.
% e) approaches potentially discard essential observations discriminating the pulse echoes of the structural building blocks and typically increase their correlations
% article:BessonITUFFC2018: Ultrafast Ultrasound Imaging as an Inverse Problem: Matrix-Free Sparse Image Reconstruction
% III. SPARSITY-DRIVEN IMAGE RECONSTRUCTION METHODS / C. Compressed Beamforming / 1) Selection of Transducer Elements
% - However, the corresponding measurement model H_{d} [subset of transducer elements] suffers from a HIGH COHERENCE [10], [12].
%   [10] article:DavidJASA2015: Time domain compressive beam forming of ultrasound signals
%   [12] proc:BessonICIP2016: Compressed delay-and-sum beamforming for ultrafast ultrasound imaging
% III. SPARSITY-DRIVEN IMAGE RECONSTRUCTION METHODS / C. Compressed Beamforming / 2) Mixing of the Raw Data
% - Taking into account the INCREASE OF THE COHERENCE induced by these simple undersampling schemes [element deactivation],
%   one may think about DESIGNING A STRATEGY such that
%   the MUTUAL COHERENCE OF THE MEASUREMENT OPERATOR IS OPTIMIZED. (p. 343)
These approaches, however, potentially discard
% 1.) essential observations
essential observations discriminating
% 2.) pulse echoes
the pulse echoes of
% 3.) structural building blocks
the structural building blocks and typically increase
% 4.) correlations
their correlations.

%---------------------------------------------------------------------------------------------------------------
% 2.) optimal selections of the active array elements / treated discrete frequencies
%---------------------------------------------------------------------------------------------------------------
% a) optimal methods minimize these correlations and capture Fourier coefficients of relatively high SNRs
% article:ChernyakovaITUFFC2014: Fourier-Domain Beamforming: The Path to Compressed Ultrasound Imaging
% IV. Rate Reduction by Beamforming in Frequency / B. Reduced Rate Sampling
% - We now address the following question:
%   HOW DO WE OBTAIN THE REQUIRED SET β [set of Fourier coefficients] CORRESPONDING TO THE EFFECTIVE BAND-PASS BANDWIDTH, USING
%   B [cardinality of β] LOW-RATE SAMPLES OF EACH ONE OF THE RECEIVED SIGNALS? (p. 1259)
% 3.) FURTHER REDUCTION IN RATE can be achieved if we want to obtain only a PARTIAL FREQUENCY BEAM'S DATA. (p. 1259)
%	- Explicitly, assume that now we are interested in μ_{BF} ⊂ β_{BF} of size M_{BF} of Fourier coefficients of the beam. (p. 1259)
% 	- THE CHALLENGE NOW IS TO RECOVER THE BEAM FROM SUCH PARTIAL FREQUENCY DATA, because
%	  a simple inverse Fourier transform is insufficient in this case. (p. 1259)
% 	- The choice of μ [set of selected FCs of RF signals], and consequently the analog kernel, is dictated by
%	  the TRANSMITTED PULSE SHAPE. (p. 1260)
Optimal methods therefore (i) minimize
% 1.) correlations [ pulse echoes of the structural building blocks ]
these correlations and (ii) capture
% 2.) Fourier coefficients of the recorded RF voltage signals
\name{Fourier} coefficients
\eqref{eqn:recovery_disc_freq_v_rx_Fourier_series_coef} of
% 3.) relatively high SNR
relatively high \acp{SNR}.
% b) suppression of the admissible noise-like incoherent aliasing of sufficiently small energy enables sub-Nyquist spatiotemporal sampling rates
% article:Schiffner2018, Sect. II: Compressed Sensing in a Nutshell (sec:compressed_sensing)
% - For $n_{1} \neq n_{2}$, however, both column vectors typically differ, and
%   the ABSOLUTE VALUE OF THE \ac{TPSF} \eqref{eqn:cs_math_tpsf} IDEALLY APPROACHES ZERO WITH NOISE-LIKE STATISTICS
%   \cite{article:ProvostITMI2009,article:LustigMRM2007}.
% - These properties, which are referred to as INCOHERENT ALIASING, indicate
%   the reliable discrimination of the admissible structural building blocks by the observation process and guide
%   the sparsity-promoting $\ell_{q}$-minimization method \eqref{eqn:cs_lq_minimization}.
The suppression of
% 1.) incoherent aliasing
the incoherent aliasing by
% 2.) sparsity-promoting lq-minimization method
the sparsity-promoting $\ell_{q}$-minimization method
\eqref{eqn:recovery_reg_norm_lq_minimization} then enables
% 3.) sub-Nyquist spatiotemporal sampling rates
sub-\name{Nyquist} spatiotemporal sampling rates.
In
the simulation study, for instance,
the approximate Gaussian distribution of
the energy in
the \name{Fourier} coefficients
\eqref{eqn:recovery_disc_freq_v_rx_Fourier_series_coef} over
the frequency axis
(cf. \cref{tab:sim_study_parameters}) permits
the selection of
only a few consecutive discrete frequencies around
the center frequency
\cite[Sect. III-B]{article:BurshteinITUFFC2016},
\cite[Sect. IV-B]{article:ChernyakovaITUFFC2014}.
% d) uniform distribution would permit randomly and uniformly distributed discrete frequencies
% article:ChernyakovaITUFFC2014: Fourier-Domain Beamforming: The Path to Compressed Ultrasound Imaging
% IV. Rate Reduction by Beamforming in Frequency / B. Reduced Rate Sampling
%   [2.)] FLAT SPECTRUM:
%   - On the other hand, if the SPECTRUM OF THE TRANSMITTED PULSE IS FLAT, which is the case for LINEAR FREQUENCY-MODULATED CHIRPS [27], [28], then
%     the performance of CS recovery improves when μ is COMPRISED OF ELEMENTS OF β CHOSEN UNIFORMLY AT RANDOM. (p. 1260)
%   - The RESULTING SAMPLING OPERATION CAN BE IMPLEMENTED USING THE TECHNIQUES PROPOSED IN [3] and [11]. (p. 1260)
%     [3] article:TurITSP2011, [11] article:Baransky2012
A uniform distribution of
this energy, in contrast, would permit
the selection of
randomly and uniformly distributed discrete frequencies
\cite[Sect. IV-B]{article:ChernyakovaITUFFC2014}.

%---------------------------------------------------------------------------------------------------------------
% 3.) comparison to compressed beamforming
%---------------------------------------------------------------------------------------------------------------
% a) proposed method resembles the compressed beamforming methods
% article:ChernyakovaITUFFC2018: Fourier-Domain Beamforming and Structure-Based Reconstruction for Plane-Wave Imaging
% Abstract
% - To further reduce the rate [sampling and processing] we exploit
%   the STRUCTURE OF THE BEAMFORMED SIGNAL and use
%   COMPRESSED SENSING METHODS to RECOVER THE BEAMFORMED SIGNAL from its PARTIAL FREQUENCY DATA obtained at a SUB-NYQUIST RATE. (p. )
% proc:SchiffnerIUS2016a: A low-rate parallel Fourier domain beamforming method for ultrafast pulse-echo imaging
% article:BurshteinITUFFC2016: Sub-Nyquist Sampling and Fourier Domain Beamforming in Volumetric Ultrasound Imaging
% article:ChernyakovaITUFFC2014: Fourier-Domain Beamforming: The Path to Compressed Ultrasound Imaging
% V. Further Reduction Through Compressed Sensing
% - We now address RECONSTRUCTION OF THE BEAMFORMED SIGNAL FROM PARTIAL FREQUENCY DATA. (p. 1260)
% - Explicitly, we aim to reconstruct the beamformed signal from its M_{BF} Fourier coefficients, denoted by μ_{BF}. (p. 1260)
% - To this end, we use CS techniques while EXPLOITING THE FRI STRUCTURE OF THE BEAMFORMED SIGNAL. (p. 1260)
% article:WagnerITSP2012: Compressed Beamforming in Ultrasound Imaging
Using
sub-\name{Nyquist} temporal sampling rates,
% 1.) proposed method
the proposed method resembles
% 2.) compressed beamforming methods
the compressed beamforming methods
\cite{article:ChernyakovaITUFFC2018,proc:SchiffnerIUS2016a,article:BurshteinITUFFC2016,article:ChernyakovaITUFFC2014,article:WagnerITSP2012}.
In fact,
both types of
methods leverage
% 1.) sparsity-promoting lq-minimization method
a sparsity-promoting $\ell_{q}$-minimization method
\eqref{eqn:cs_lq_minimization} to recover
% 2.) specific signals
specific signals from
% 3.) Fourier coefficients of the recorded RF voltage signals
only a few \name{Fourier} coefficients
\eqref{eqn:recovery_disc_freq_v_rx_Fourier_series_coef} associated with
the selected discrete frequencies.
% c) proposed method recovers d-dimensional acoustic material parameters based on realistic physical models
The former, however, recovers
$d$-dimensional acoustic material parameters based on
realistic physical models.
% d) proposed method minimizes the number of sequential pulse-echo measurements per image and maximizes the frame rate
It minimizes
the number of
sequential pulse-echo measurements per
image and, thus, maximizes
the frame rate.
The latter methods, in contrast, use
% 1.) popular DAS method
% article:Schiffner2018, Sect. I: Introduction (sec:introduction)
% - The popular \ac{DAS} method, for example, focuses
%   the echo signals on specified points in the \ac{FOV} to quantify their echogeneity.
% - Emitting steered PWs \cite{article:ProvostPMB2014,article:MaceITUFFC2013,article:MontaldoITUFFC2009}, whose
%   spatial extent and energy content are unlimited, or
%   outgoing $\{ 1, 2 \}$-spherical waves \cite{article:ProvostPMB2014,article:PapadacciITUFFC2014,article:JensenUlt2006}, whose
%   isotropic sources are points, it adds
%   the signal samples at the round-trip \acp{TOF} \cite[(2), (6)]{article:MontaldoITUFFC2009}, \cite[(4), (5)]{article:JensenUlt2006}.
the popular \ac{DAS} method to focus
% 2.) recorded RF voltage signals
the recorded \ac{RF} voltage signals.
% f) compressed beamforming methods model the one-dimensional focused signals composing the image as quantized finite streams of known pulses
They model
% 1.) one-dimensional focused signals
the one-dimensional focused signals composing
% 2.) image
the image as
% 3.) finite streams of known pulses
quantized finite streams of
known pulses%
\footnote{
  Finite streams of
  known pulses are
  special instances of
  \acl{FRI} signals, i.e.
  analog signals defined by
  a finite number of
  parameters in
  an underlying signal model per
  unit of
  time
  (cf. e.g.
  \cite{article:BluISPM2008} and
  \cite{article:VetterliITSP2002}%
  ).
} and individually recover
% 4.) hundreds
hundreds of
% 5.) nearly-sparse parameter vectors
nearly-sparse parameter vectors defining
% 6.) streams
these streams.
% g) compressed beamforming methods primarily reduce the sampling rates and only the most recent evolution permits comparably high frame rates
% article:ChernyakovaITUFFC2018: Fourier-Domain Beamforming and Structure-Based Reconstruction for Plane-Wave Imaging
% proc:SchiffnerIUS2016a: A low-rate parallel Fourier domain beamforming method for ultrafast pulse-echo imaging
They primarily reduce
the temporal sampling rates, and only
their most recent versions additionally support
high frame rates
\cite{article:ChernyakovaITUFFC2018,proc:SchiffnerIUS2016a}.
% h) superior physical model driving the proposed method further increases the frame rate and significantly improves the image quality
% article:ChernyakovaITUFFC2014: Fourier-Domain Beamforming: The Path to Compressed Ultrasound Imaging
% VII. Discussion and Conclusions
% - The FRI MODEL IMPLIES AN ASSUMPTION THAT THE TRANSMITTED PULSE SHAPE REMAINS UNCHANGED during its propagation through the tissue. (p. 1266)
% - This is, of course, a SIMPLIFIED MODEL OF ULTRASOUND PROPAGATION, because frequency-dependent attenuation [46, ch. 5] is not taken into account. (p. 1266)
% - The RESULTS REPORTED IN THIS WORK CAN BE POTENTIALLY IMPROVED WITH AN APPROPRIATE GENERALIZATION OF THE FRI MODEL. (p. 1266)
The superior physical model driving
the proposed method, however, further increases
% 1.) frame rate
the frame rate and significantly improves
% 2.) image quality
the image quality.

%%%%%%%%%%%%%%%%%%%%%%%%%%%%%%%%%%%%%%%%%%%%%%%%%%%%%%%%%%%%%%%%%%%%%%%%%%%%%%%%%%%%%%%%%%%%%%%%%%%%%%%%%%%%%%%%
% 9.) experimental validations
%%%%%%%%%%%%%%%%%%%%%%%%%%%%%%%%%%%%%%%%%%%%%%%%%%%%%%%%%%%%%%%%%%%%%%%%%%%%%%%%%%%%%%%%%%%%%%%%%%%%%%%%%%%%%%%%
\subsection{Experimental Validations}
%\label{subsec:disc_experimental_validation}
%---------------------------------------------------------------------------------------------------------------
% 1.) experimental validations of the proposed method using steered QPWs
%---------------------------------------------------------------------------------------------------------------
% a) experimental validation of the proposed method using the PICMUS data and N_{\text{in}} \in \{ 1, 3 \} sequential pulse-echo measurements based on steered QPWs
An experimental validation of
% 1.) proposed method
the proposed method using
% 2.) PICMUS data
the \ac{PICMUS} data and
% 3.) N_{\text{in}} \in \{ 1, 3 \} sequential pulse-echo measurements
$N_{\text{in}} \in \{ 1, 3 \}$ sequential pulse-echo measurements based on
% 4.) steered QPWs
steered \acp{QPW} was presented in
\cite{proc:SchiffnerIUS2016b}.
% b) additional experimental validations of the proposed method at increasing stages of evolution
Comparisons to
% 1.) established image recovery methods
established image recovery methods, including
% 2.) synthetic aperture (SA)
\ac{SA},
% 3.) delay-and-sum (DAS)
\ac{DAS},
% 4.) filtered backpropagation (FBP)
\acl{FBP}, and
% 5.) progressive scanning by focused beams
the progressive scanning by
focused beams, at
increasing stages of
evolution were shown in
\cite{proc:SchiffnerIUS2013a,proc:SchiffnerIUS2012,article:SchiffnerBMT2012,proc:SchiffnerIUS2011}.

%%%%%%%%%%%%%%%%%%%%%%%%%%%%%%%%%%%%%%%%%%%%%%%%%%%%%%%%%%%%%%%%%%%%%%%%%%%%%%%%%%%%%%%%%%%%%%%%%%%%%%%%%%%%%%%%
% 10.) limitations
%%%%%%%%%%%%%%%%%%%%%%%%%%%%%%%%%%%%%%%%%%%%%%%%%%%%%%%%%%%%%%%%%%%%%%%%%%%%%%%%%%%%%%%%%%%%%%%%%%%%%%%%%%%%%%%%
\subsection{Limitations}
%\label{subsec:disc_limitations}
%---------------------------------------------------------------------------------------------------------------
% 1.) important investigations left for future research
%---------------------------------------------------------------------------------------------------------------
% a) multiple important investigations had to be left for future research
Multiple important investigations had to be left for
future research.
These mainly include
statistical analyses of
% 1.) admissible numbers of significant components
the admissible numbers of
significant components in
% 2.) nearly-sparse representation
the nearly-sparse representation
\eqref{eqn:recovery_reg_sparse_representation} for various
% 3.) numbers of sequential pulse-echo measurements per image
numbers of
sequential pulse-echo measurements per image,
% 4.) parameters q governing the sparsity-promoting lq-minimization method
parameters $q$ governing
the sparsity-promoting $\ell_{q}$-minimization method
\eqref{eqn:recovery_reg_norm_lq_minimization},
% 5.) energy levels of the additive errors
energy levels of
the additive errors,
% 6.) realizations of the random waves
realizations of
the random waves, and
% 7.) orthonormal bases
orthonormal bases
(cf. e.g.
  \cite[Figs. 1 and 2]{proc:ChartrandICASSP2008},
  \cite[Figs. 3 and 4]{article:ChartrandISPL2007}%
).
% c) large numbers of parameter combinations and the high computational costs currently prevent such combinatorial analyses for normalized CS problems of the investigated sizes
The large numbers of
parameter combinations and
% 1.) high computational costs induced by the proposed implementation of the sparsity-promoting lq-minimization method
the high computational costs induced by
the proposed implementation of
% 2.) sparsity-promoting lq-minimization method
the sparsity-promoting $\ell_{q}$-minimization method
\eqref{eqn:recovery_reg_norm_lq_minimization} currently prevent
such combinatorial analyses for
% 3.) normalized CS problems
normalized \ac{CS} problems
\eqref{eqn:recovery_reg_norm_prob_general} of
the investigated sizes.
% d) presented study focused on simple types of random waves that can readily be synthesized by programmable UI systems
Furthermore,
the presented study focused on
% 1.) simple types of random waves
simple types of
random waves that
can readily be synthesized by
% 2.) programmable UI systems
programmable \ac{UI} systems.
% e) more advanced syntheses increase the complexity of the transmit hardware
Although
more advanced syntheses, e.g.
\TODO{use correct description}
% 1.) element specific coded excitation voltages
element specific coded excitation voltages, increase
the complexity of
the transmit hardware,
% f) more advanced syntheses potentially further decorrelate the pulse echoes and deserve additional studies
they potentially further decorrelate
the pulse echoes and, thus, deserve
additional studies.
\TODO{article:IlovitshNatComBio2018, structured illumination}
% g) numerical simulations of the pulse-echo measurement process in the two-dimensional Euclidean space exclude variations along the elevational direction
Moreover,
the numerical simulations of
% 1.) pulse-echo measurement process in the two-dimensional Euclidean space
the pulse-echo measurement process in
the two-dimensional Euclidean space, i.e. $d = 2$, exclude
% 2.) variations along the elevational direction
variations along
the elevational direction.
% b) two-dimensional Euclidean space prevented the simulation of the vibrating faces' height and the elevational focus induced by the acoustic lens
%Although
%this space prevented
%the simulation of
% 1.) width of the vibrating faces along the r_{2}-axis
%the vibrating faces' height and
% 2.) elevational focus induced by the acoustic lens
%the elevational focus induced by
%the acoustic lens,
% h) numerical simulations [two-dimensional] underestimate the diffraction-induced decay of the ultrasonic waves and predict less realistic results
They underestimate
% 1.) diffraction-induced decay of the ultrasonic waves
the diffraction-induced decay of
the ultrasonic waves, which is asymptotically proportional to
$\norm{ \vect{r} }{2}^{ - ( d - 1 ) / 2 }$ for
% 2.) outgoing free-space Green's functions (two- and three-dimensional Euclidean spaces)
the outgoing free-space \name{Green}'s functions
\eqref{eqn:app_helmholtz_green_free_space_2_3_dim}, and predict
% 3.) less realistic results
less realistic results.
% i) rectilinear boundary conditions prohibit the usage of curved transducer arrays
% TODO: curved arrays incompatible with Rayleigh-Sommerfeld diffraction integrals, 
The rectilinear boundary conditions prohibit
the usage of
% 1.) curved transducer arrays
curved transducer arrays, and
% j) nonlinear wave propagation
% article:NgITUFFC2006: Modeling ultrasound imaging as a linear, shift-variant system
% I. Introduction
% - Modern clinical practice sometimes exploits higher-order harmonics generated by nonlinearities during transmission. (p. 549)
% - We note a comment in [3] that, although the forward propagation of waves in such a case is nonlinear,
%   A LINEAR MODEL WOULD STILL HOLD FOR THE BACK PROPAGATION provided that the scattering is weak
%   (which is usually the case in soft tissue). (p. 549)
the linear wave propagation neglects
finite amplitude effects.
% j) first Born approximation excludes multiple scattering
% article:Schiffner2018,
% - The \name{Born} approximation \eqref{eqn:lin_mod_pert_born_p_B_1} represents single scattering \cite[709]{book:Born1999}, i.e.
%   the approximated scattered acoustic pressure field is the exclusive response to
%   the incident acoustic pressure field and does not interact with
%   the compressibility fluctuations \eqref{eqn:lin_mod_mech_model_tis_simple_rel_fluctuations}.
% - This map dominates practical image recovery methods in \ac{UI} and allows the systematic derivation of analytical inversion schemes.
Eventually,
the \name{Born} approximation, which dominates
% 1.) practical image recovery methods in UI
practical image recovery methods in
\ac{UI}, excludes
% 2.) multiple scattering
multiple scattering and, thus, any
% 3.) phase aberrations
phase aberrations.
% k) presented results outline general potential pitfalls and benefits of random incident waves
Despite
these limitations and
open investigations, however,
the presented results reliably outline
the potential drawbacks and
the benefits of
random incident waves in
fast compressed pulse-echo \ac{UI}.

\section{Conclusion and Outlook}
\label{sec:conclusion_outlook}
%---------------------------------------------------------------------------------------------------------------
% 1.) summary of the major findings (take-home message)
%---------------------------------------------------------------------------------------------------------------
% a) proposed syntheses of random waves applied random apodization weights, time delays, or combinations thereof to reference excitation voltages
% article:Schiffner2018, Sect. IV.B. Types of Incident Waves (subsec:syn_p_in_types)
% - Modern \ac{UI} systems typically WEIGHT AND DELAY A COMMON REFERENCE VOLTAGE SIGNAL TO PRODUCE THE INDIVIDUAL EXCITATION VOLTAGES.
% - This paper thus specifies excitation voltages of the general form [...] for all transducer elements, where
%   $u^{(\text{tx})} \in \C$ is the amplitude of the COMMON REFERENCE VOLTAGE SIGNAL at
%   the angular frequency $\omega$ IDENTICALLY EXCITING THESE ELEMENTS,
%   $a_{m} \in \R$ are frequency-independent apodization weights, and
%   $\mathcal{Q}$ denotes an operator implementing the quantization of the nonnegative time delays $\Delta t_{m} \in \Rnonneg$.
% article:Schiffner2018, Sect. IV: Syntheses of the Incident Waves (sec:syn_p_in)
% - The \ac{UI} system excites
%   the individual physical elements of the planar transducer array by
%   specified voltage signals to synthesize
%   various types of incident waves.
% - Modern \ac{UI} systems typically generate the individual excitation voltages by applying
%   specified apodization weights, time delays, or combinations thereof to
%   a common reference voltage signal.
The proposed syntheses of
random waves applied
% 1.) random apodization weights
random apodization weights,
% 2.) random time delays
time delays, or
% 3.) combinations thereof
combinations thereof to
% 4.) reference excitation voltages
reference excitation voltages and aided in meeting
% 5.) central requirement of the CS framework in fast compressed pulse-echo UI
a central requirement of
the \ac{CS} framework in
% 6.) fast compressed pulse-echo UI
fast compressed pulse-echo \ac{UI}.
% b) proposed syntheses decorrelated the pulse echoes of the admissible structural building blocks composing the lossy heterogeneous object to be imaged relative to the steered QPWs
% article:Schiffner2018, Sect. II: Compressed Sensing
% - This latter constraint effectively reduces the total number of unknown components to
%   a relatively small number of unknown coefficients associated with
%   the INDIVIDUAL BASIS FUNCTIONS REPRESENTING THE RELEVANT STRUCTURAL BUILDING BLOCKS.
% - Let the unitary matrix $\mat{\Psi} \in \C^{ N \times N }$ represent a suitable orthonormal basis of $\C^{ N }$, i.e.
%   $\mat{\Psi} \herm{ \mat{\Psi} } = \herm{ \mat{\Psi} } \mat{\Psi} = \mat{I}$, e.g. the \name{Fourier}, a wavelet, or the canonical basis.
% - Its column vectors $\vectsym{\psi}_{n} \in \C^{ N }$, $n \in \setcons{ N }$, DEFINE THE ADMISSIBLE STRUCTURAL BUILDING BLOCKS.
They decorrelated
the pulse echoes of
% 1.) admissible structural building blocks
the admissible structural building blocks composing
% 2.) lossy heterogeneous object to be imaged
the lossy heterogeneous object to be imaged relative to
% 3.) steered QPWs
the prevalent steered \acp{QPW}.
% c) structural building blocks equaled the individual basis functions in a nearly-sparse representation of the spatial compressibility fluctuations
These blocks equaled
the individual basis functions in
a nearly-sparse representation of
the spatial compressibility fluctuations.
% d) proposed method aims at recovering the compressibility fluctuations inside the specified FOV from only a few sequential pulse-echo measurements of the received RF voltage signals
% 1.) sparsity-promoting lq-minimization method
A sparsity-promoting $\ell_{q}$-minimization method enabled both
% 2.) structural recovery
the structural and
% 3.) quantitative recovery
the quantitative recovery of
% 4.) two phantoms
two phantoms from
synthetic \ac{RF} voltage signals.
% e)
These were generated by
single realizations of
the random waves in
numerical simulations of
the pulse-echo measurement process in
the two-dimensional Euclidean space.
% e) spatial variations in the incident acoustic energies caused residual errors for the sparse wire phantom
Although
the spatial variations in
% 1.) incident acoustic energies (multiple pulse-echo measurements, multifrequent)
the incident acoustic energies caused
residual errors at
low \acp{SNR} for
the sparse wire phantom,
% f) discrete Fourier basis converts the erratic spatial variations in the incident acoustic energy into beneficial enlarged passbands
they improved
the identification of
the spatially extended structural building blocks defined by
the discrete \name{Fourier} basis and significantly enlarged
the passbands of
the sensing matrices for
the tissue-mimicking phantom.
% f)
%spatially extended structural building blocks of
%the tissue-mimicking phantom thus increase
%the robustness against
%the additive errors.
% g) focus on the efficient implementation using the FMM
The \ac{FMM} enabled
an efficient \ac{GPU}-based implementation of
both types of
matrix-vector products required by
the iterative algorithms.

%---------------------------------------------------------------------------------------------------------------
% 2.) outlook and current research
%---------------------------------------------------------------------------------------------------------------
% article:KruizingaSciAdv2017: Compressive 3D ultrasound imaging using a single sensor
% DISCUSSION
% - Furthermore, we think that a coded aperture mask could potentially be used in conjunction with conventional 1D ultrasound arrays to better estimate
%   the out-of-plane signals and possibly extend the normal 2D imaging capabilities to 3D. (p. 9)
% a) proposed physical models provide the flexibility to investigate alternative types of incident waves
The proposed physical models for
% 1.) linear physical model for the pulse-echo measurement process
the pulse-echo measurement process and
% 2.) syntheses of the incident waves
the syntheses of
the incident waves provide
the flexibility to
investigate alternative types of
incident waves, e.g.
superpositions of
multiple individually coded quasi-$(d-1)$-spherical waves
\cite{article:MisaridisITUFFC2005},
% 2.) random focused beams
random focused beams,
% 3.) structured sequences
% article:IlovitshNatComBio2018: Acoustical structured illumination for super-resolution ultrasound imaging
structured waves
\cite{article:IlovitshNatComBio2018}, and even
% 4.) arbitrary experimentally measured waves
arbitrary experimentally measured waves.
% sparse arrays
% article:RouxSciRep2018: Experimental 3-D Ultrasound Imaging with 2-D Sparse Arrays using Focused and Diverging Waves
% Abstract
% - However, the experimental test of new 3-D US approaches is contrasted by
%   the NEED OF CONTROLLING VERY LARGE NUMBERS OF PROBE ELEMENTS.
% - Although this problem [need of controlling very large numbers of probe elements] may be overcome by
%   the use of 2-D SPARSE ARRAYS, just a few experimental results have so far corroborated the validity of this approach. (p. 1)
% - In this paper, we experimentally compare the performance of
%   [1.)] a FULLY WIRED 1024-ELEMENT (32 × 32) ARRAY, assumed as reference, to that of
%   [2.)] a 256-ELEMENT RANDOM and of
%   [3.)] an “OPTIMIZED” 2-D SPARSE ARRAY, in both
%   FOCUSED AND COMPOUNDED DIVERGING WAVE (DW) TRANSMISSION MODES. (p. 1)
% - Furthermore,
%   the experimental results in 3-D DW mode and 3-D focused mode are also compared for the first time and they show that both
%   the contrast and the resolution performance are higher when using the 3-D DW at volume rates up to 90/second which represent
%   a 36x speed up factor compared to the focused mode. (p. 1)
\TODO{mixing schemes}
% b) author is currently investigating the superpositions of multiple individually coded quasi-(d-1)-spherical waves to further decorrelate the pulse echoes
The author is currently investigating
% 1.) superpositions of multiple individually coded quasi-(d-1)-spherical waves
the superpositions of
multiple individually coded quasi-$(d-1)$-spherical waves to further decorrelate
the pulse echoes.
% c) author is simultaneously extending the implementation to the three-dimensional Euclidean space
% book:Devaney2012, Chapter 6: Scattering theory / Sect. 6.7.1: The Born approximation
% - The ADVANTAGE OF THE LINEARIZED BORN MODEL is that
%   we can employ a SYSTEMATIC PROCEDURE TO DEVELOP ANALYTIC INVERSION SCHEMES that
%   CAN LATER BE GENERALIZED TO INCLUDE NON-LINEAR EFFECTS in the actual scattering experiments. (p. 256)
% - For example, we can EXTEND ALL OF THE ANALYSIS contained in this and the following chapter TO
%   NON-CONSTANT BACKGROUNDS CHARACTERIZED BY A (KNOWN) WAVENUMBER k0 = k0(r) THAT VARIES WITH POSITION. (p. 256)
% - Such a generalization is based on the so-called distorted-wave Born approximation (DWBA), which
%   will be developed in Chapter 9. (p. 256)
%Its solution can potentially be generalized to include
%the nonlinear effects
%\cite[256]{book:Devaney2012}.
He is simultaneously extending
% 1.) three-dimensional Euclidean space
the implementation to
the three-dimensional Euclidean space and exploring
% 2.) nonlinear CS to regularize the nonlinear ISP
nonlinear \ac{CS} to regularize
the nonlinear \ac{ISP} generalizing
the \name{Born} approximation.
% d) alternative orthonormal bases and customized redundant dictionaries deserve additional studies
Alternative orthonormal bases, e.g.
wavelet bases, and
% 2.) customized redundant dictionaries
customized redundant dictionaries
\cite{article:CandesACHA2011}, which can be learned from
the ultrasound images of
interest
\cite{article:LorintiuITMI2015}, potentially further improve
% TODO: improves the quasi-continuous tradeoff between
fast compressed pulse-echo \ac{UI} and deserve
additional studies.

%%%%%%%%%%%%%%%%%%%%%%%%%%%%%%%%%%%%%%%%%%%%%%%%%%%%%%%%%%%%%%%%%%%%%%%%%%%%%%%%%%%%%%%%%%%%%%%%%%%%%%%%%%%%%%%%
% acknowledgment
%%%%%%%%%%%%%%%%%%%%%%%%%%%%%%%%%%%%%%%%%%%%%%%%%%%%%%%%%%%%%%%%%%%%%%%%%%%%%%%%%%%%%%%%%%%%%%%%%%%%%%%%%%%%%%%%
\section*{Acknowledgment}
%---------------------------------------------------------------------------------------------------------------
% acknowledgment
%---------------------------------------------------------------------------------------------------------------
% a) author thanks Georg Schmitz
The author thanks
Prof. Dr.-Ing. Georg Schmitz, who kindly provided
access to
the laboratory and
the computer equipment.
% b) author gratefully acknowledges
He gratefully acknowledges
the financial support provided by
German Academic Exchange Service (DAAD) for
his invited talk at
Acoustics '17
\cite{article:SchiffnerJASA2017}.

%%%%%%%%%%%%%%%%%%%%%%%%%%%%%%%%%%%%%%%%%%%%%%%%%%%%%%%%%%%%%%%%%%%%%%%%%%%%%%%%%%%%%%%%%%%%%%%%%%%%%%%%%%%%%%%%
% appendices
%%%%%%%%%%%%%%%%%%%%%%%%%%%%%%%%%%%%%%%%%%%%%%%%%%%%%%%%%%%%%%%%%%%%%%%%%%%%%%%%%%%%%%%%%%%%%%%%%%%%%%%%%%%%%%%%
\appendices
%\appendix

%---------------------------------------------------------------------------------------------------------------
% Appendix A: The EBVP for the Helmholtz Equation, Green's Functions, and Their Angular Spectra
%---------------------------------------------------------------------------------------------------------------
\section{}
\label{app:helmholtz_green}
%---------------------------------------------------------------------------------------------------------------
% change format for equation numbers
%---------------------------------------------------------------------------------------------------------------
\renewcommand{\theequation}{A.\arabic{equation}}

The outgoing free-space \name{Green}'s functions
%(unit: $\si{\meter}^{2-d}$)
(cf. e.g.
\cite[(2.14) and (2.19)]{book:Devaney2012},
\cite[(3.14) and (3.15)]{book:Natterer2001}%
)
\begin{equation}
 %--------------------------------------------------------------------------------------------------------------
 % outgoing free-space Green's functions (two- and three-dimensional Euclidean spaces)
 %--------------------------------------------------------------------------------------------------------------
  g_{l}( \vect{r} )
  =
  \begin{cases}
   %------------------------------------------------------------------------------------------------------------
   % a) two-dimensional Euclidean space
   %------------------------------------------------------------------------------------------------------------
    j \hankel{0}{2}{ \munderbar{k}_{l} \norm{ \vect{r} }{2} } / 4
    & \text{for } d = 2,\\
   %------------------------------------------------------------------------------------------------------------
   % b) three-dimensional Euclidean space
   %------------------------------------------------------------------------------------------------------------
    -
    e^{ - j \munderbar{k}_{l} \norm{ \vect{r} }{2} } / ( 4 \pi \norm{ \vect{r} }{2} )
    & \text{for } d = 3,
  \end{cases}
 \label{eqn:app_helmholtz_green_free_space_2_3_dim}
\end{equation}
uniquely solve
% 1.) fundamental inhomogeneous Helmholtz equations (d-dimensional Euclidean space)
the fundamental inhomogeneous \name{Helmholtz} equations
\begin{equation*}
 %--------------------------------------------------------------------------------------------------------------
 % fundamental inhomogeneous Helmholtz equations (d-dimensional Euclidean space)
 %--------------------------------------------------------------------------------------------------------------
  \left( \Delta + {\munderbar{k}_{l}}^{2} \right)
  g_{l}( \vect{r} )
  =
  \delta( \vect{r} )
 \label{eqn:app_helmholtz_fund_n_dim}
\end{equation*}
subject to
% 2.) Sommerfeld radiation conditions (d-dimensional Euclidean space)
% book:Devaney2012, Chapter 2: Radiation and boundary-value problems in the frequency domain / Sect. 2.1: Frequency-domain formulation of the radiation problem / Sect. 2.1.4: The Sommerfeld radiation condition in dispersive media
% - The SRC can be stated in either of the two forms (cf. Eqs. (1.45a) and (1.48))
%   [ \lim_{ r \rightarrow \infty } r \left[ \partial{ U_{+}( \vect{r}, \omega ) }{ r } - j k U_{+}( \vect{r}, \omega ) \right] \rightarrow 0, ] (2.8a)
%   [ U_{+}( \vect{r}, \omega ) \sim f( \vect{s}, \omega ) \frac{ e^{ j k r } }{ r }, ] (2.8b)
%   where \vect{s} = \vect{r} / r is the unit vector along the \vect{r} direction and, as usual,
%   we have used the subscript + to denote the field that satisfies the SRC. (p. 48)
% book:Devaney2012, Chapter 1: Radiation and initial-value problems for the wave equation / Sect. 1.5: Frequency-domain solution of the radiation problem / Sect. 1.5.1: The radiation pattern and the Sommerfeld radiation condition
% - The asymptotic expression Eq. (1.45a) is one form of the famed SOMMERFELD RADIATION CONDITION (SRC). (p. 24)
% - An alternative, and the most often quoted, form of the SRC is given by
%   [ \lim_{ r \rightarrow \infty } r \left[ \partial{ U_{+}( \vect{r}, \omega ) }{ r } - j k U_{+}( \vect{r}, \omega ) \right] \rightarrow 0, ] (1.48)
%   with a similar expression holding for G_{+}. (p. 24)
% - The equivalence of the two forms of the SRC is easily established. (p. 24)
% book:Natterer2001
the \acp{SRC}
(cf. e.g.
\cite[(1.48) or (2.8)]{book:Devaney2012},		% complex-valued k, 3-dimensional, neg. sign convention
\cite[(7.61)]{book:Natterer2001}%			% real-valued k, n-dimensional, neg. sign convention, existence of unique solution is mentioned explicitly
)
\begin{equation}
 %--------------------------------------------------------------------------------------------------------------
 % Sommerfeld radiation conditions (d-dimensional Euclidean space)
 %--------------------------------------------------------------------------------------------------------------
  \underset{ r \rightarrow \infty }{ \lim }
  \underset{ \norm{ \vect{r} }{2} = r }{ \max }
    \norm{ \vect{r} }{2}^{ \frac{ d - 1 }{ 2 } }
    \bigl[
      \inprod{ \nabla g_{l}( \vect{r} ) }{ \uvect{r}( \vect{r} ) }
      +
      j \munderbar{k}_{l} g_{l}( \vect{r} )
    \bigr]
  = 0
 \label{eqn:app_helmholtz_src}
\end{equation}
for
% 3.) all relevant discrete frequencies
all $l \in \setsymbol{L}_{ \text{BP} }^{(n)}$, where
% 4.) zero-order Hankel function of the second kind
% book:OlverNHMF2010, §10.4 Connection Formulas
% - [ \hankel{\nu}{2}{ z } = \bessel{\nu}{ z } - j \neumann{\nu}{ z } ] (10.4.3)
% book:OlverNHMF2010, §10.2(ii) Standard Solutions / Bessel Functions of the Third Kind (Hankel Functions)
% - These solutions of (10.2.1) [Bessel’s Equation] are denoted by Hν(1)⁡(z) and Hν(2)⁡(z), and their defining properties are given by 10.2.5 [...] and 10.2.6 [...].
$\hankelsymbol{0}{2}$ denotes
the zero-order \name{Hankel} function of
the second kind
\cite[§10.2(ii) and 10.4.3]{book:OlverNHMF2010},
% 5.) Dirac delta distribution
$\delta$ indicates
the \name{Dirac} delta distribution, and
% 6.) radial unit vector
$\uvect{r}( \vect{r} ) = \vect{r} / \tnorm{ \vect{r} }{2}$ for
all $\vect{r} \in \R^{d} \setminus \{ \vect{0} \}$ is
the radial unit vector.
The \acp{SRC}
\eqref{eqn:app_helmholtz_src} account for
% 1.) lossy homogeneous fluid of infinite extent
a lossy homogeneous fluid of
infinite extent and ensure
% 2.) causality in the time domain
the causality in
the time domain
\cite[Sect. 2.1.4]{book:Devaney2012}.

%textwidth: \the\textwidth\par
%columnwidth: \the\columnwidth

%---------------------------------------------------------------------------------------------------------------
% Appendix B: adjoint sensing matrix
%---------------------------------------------------------------------------------------------------------------
\section{}
\label{app:adjoint}
%---------------------------------------------------------------------------------------------------------------
% 1.) left multiplication by the adjoint normalized sensing matrix
%---------------------------------------------------------------------------------------------------------------
% a) left multiplication of the normalized recorded RF voltage signals by the adjoint normalized sensing matrix
The left multiplication of
% 1.) normalized linear algebraic system (all pulse-echo measurements, multifrequent, all array elements, additive errors)
the normalized recorded \ac{RF} voltage signals
\eqref{eqn:recovery_reg_norm_obs_trans_coef_error} by
% 2.) adjoint normalized sensing matrix (all pulse-echo measurements, multifrequent, all array elements)
the adjoint normalized sensing matrix
\eqref{eqn:recon_reg_norm_sensing_matrix} yields
\begin{equation*}
 %--------------------------------------------------------------------------------------------------------------
 % left multiplication by the adjoint normalized sensing matrix
 %--------------------------------------------------------------------------------------------------------------
  \check{\bar{\vectsym{\theta}}}_{\xi}^{(\kappa)}
  =
  \herm{ \bar{\mat{A}} }_{\xi}\bigl[ p^{(\text{in})} \bigr]
  \bar{\vect{u}}^{(\text{rx})}
  =
  \frac{ 1 }{ \dnorm{ \vect{u}^{(\text{rx})} }{2}{1} }
  \mat{W}_{\xi}^{-1}
  \underbrace{
    \herm{ \mat{A} }\bigl[ p^{(\text{in})} \bigr]
    \vect{u}^{(\text{rx})}
  }_{ = \check{\vectsym{\theta}}^{(\kappa)} }.
\end{equation*}
% b) components further combine the TPSFs
Its components linearly combine
% 1.) transform point spread function (TPSF)
the \acp{TPSF}
\eqref{eqn:cs_math_tpsf} according to
\begin{equation}
\begin{split}
 %--------------------------------------------------------------------------------------------------------------
 % components of the left multiplication by the adjoint normalized sensing matrix [using TPSFs]
 %--------------------------------------------------------------------------------------------------------------
  \check{\bar{\theta}}_{\xi, i}^{(\kappa)}
  &=
  \dinprod{
    \underbrace{
      \bar{\mat{A}}_{\xi}\bigl[ p^{(\text{in})} \bigr]
      \bar{\vectsym{\theta}}_{\xi}^{(\kappa)}
      +
      \bar{\vectsym{\eta}}
    }_{ = \bar{\vect{u}}^{(\text{rx})} }
  }{
    \frac{
      \vect{a}_{i}\left[ p^{(\text{in})} \right]
    }{
      a_{ \xi, i }\left[ p^{(\text{in})} \right]
    }
  }{2}\\
  &=
  \dinprod{
    \underbrace{
      \sum_{ q \in \setcons{ N_{\text{lat}} } }
        \frac{
          \vect{a}_{q}\left[ p^{(\text{in})} \right]
        }{
          a_{ \xi, q }\left[ p^{(\text{in})} \right]
        }
        \bar{\theta}_{\xi, q}^{(\kappa)}
    }_{ = \bar{\mat{A}}_{\xi}[ p^{(\text{in})} ] \bar{\vectsym{\theta}}_{\xi}^{(\kappa)} }    
  }{
    \frac{
      \vect{a}_{i}\left[ p^{(\text{in})} \right]
    }{
      a_{ \xi, i }\left[ p^{(\text{in})} \right]
    }
  }{2}
  +
  \frac{
    \inprod{ \bar{\vectsym{\eta}} }{ \vect{a}_{i}\left[ p^{(\text{in})} \right] }
  }{
    a_{ \xi, i }\left[ p^{(\text{in})} \right]
  }\\
  &=
  \sum_{ q \in \setcons{ N_{\text{lat}} } }
    \frac{
      \norm{ \vect{a}_{q} }{2}
      \norm{ \vect{a}_{i} }{2}
      \bar{\theta}_{\xi, q}^{(\kappa)}
      \tpsf{ \mat{A} }{ q }{ i }
    }{
      a_{ \xi, q } a_{ \xi, i }
    }
  +
  \frac{
    \inprod{ \bar{\vectsym{\eta}} }{ \vect{a}_{i} }
  }{
    a_{ \xi, i }
  }
\end{split}
\label{eqn:app_adjoint_tpsf}
\end{equation}
for
% 2.) all structural building blocks
all $i \in \setconsnonneg{ N_{\text{lat}} - 1 }$, where
% 3.) dependence on p^{(\text{in})}
the dependence on
$p^{(\text{in})}$ was omitted in
% 4.) last line
the last line.
% c) components of the latter product contain the zero-lag cross-correlations
Inserting
% 1.) Fourier coefficients of the recorded RF voltage signals
the \name{Fourier} coefficients
\eqref{eqn:recovery_disc_freq_v_rx_Fourier_series_coef},
% 2.) components
the components contain
% 3.) zero-lag time-domain cross-correlations
the zero-lag cross-correlations
\begin{equation}
\begin{split}
 %--------------------------------------------------------------------------------------------------------------
 % zero-lag time-domain cross-correlations
 %--------------------------------------------------------------------------------------------------------------
  \check{\theta}_{i}^{(\kappa)}
  &=
  \sum_{ n = 0 }^{ N_{\text{in}} - 1 }
    \sum_{ l \in \setsymbol{L}_{ \text{BP} }^{(n)} }
    \sum_{ m = 0 }^{ N_{\text{el}} - 1 }
      \conj{ a_{ m, l, i }^{(n)}{} }
      \underbrace{
        \frac{ 1 }{ T_{ \text{rec} }^{(n)} }
        \int_{ \setsymbol{T}_{ \text{rec} }^{(n)} }
          \tilde{u}_{m}^{(\text{rx}, n)}( t )
          e^{ -j \omega_{l} t }
        \text{d} t
      }_{ = u_{ m, l }^{(\text{rx}, n)} }\\
  &=
  \sum_{ n = 0 }^{ N_{\text{in}} - 1 }
    \sum_{ m = 0 }^{ N_{\text{el}} - 1 }
      \frac{ 1 }{ T_{ \text{rec} }^{(n)} }
      \int_{ \setsymbol{T}_{ \text{rec} }^{(n)} }
        \tilde{u}_{m}^{(\text{rx}, n)}( t )
        \conj{
          \Bigl[
            \underbrace{
              \sum_{ l \in \setsymbol{L}_{ \text{BP} }^{(n)} }
                a_{ m, l, i }^{(n)}
                e^{ j \omega_{l} t }
            }_{ = \tilde{a}_{ m, i }^{(n)}( t ) }
          \Bigr]
        }
      \text{d} t\\
  &=
  \sum_{ n = 0 }^{ N_{\text{in}} - 1 }
    \sum_{ m = 0 }^{ N_{\text{el}} - 1 }
      \frac{ 1 }{ T_{ \text{rec} }^{(n)} }
      \int_{ \setsymbol{T}_{ \text{rec} }^{(n)} }
        \tilde{u}_{m}^{(\text{rx}, n)}( t )
        \conj{ \tilde{a}_{ m, i }^{(n)}{} }( t )
      \text{d} t\\
\end{split}
\label{eqn:app_adjoint_xcorr_td}
\end{equation}
for
% 5.) all structural building blocks
all $i \in \setconsnonneg{ N_{\text{lat}} - 1 }$.

% include bibliography
\bibliographystyle{ieeetr}

\end{document}